# OPERATIONAL DECISION SUPPORT IN THE PRESENCE OF UNCERTAINTIES

## Water Distribution Systems

Corneliu T. C. Arsene

Operations Research, Information Science and Statistics

Corneliu T. C. Arsene

# Operational Decision Support in the Presence of Uncertainties

## Water Distribution Systems

Corneliu T. C. Arsene, PhD
CorneliuArsene@gmail.com

# Preface

This book addresses the scientific domains of operations research, information science and statistics with a focus on engineering applications. The purpose of this book is to report on the implications of the loop equations formulation of the state estimation procedure of the network systems, for the purpose of the implementation of Decision Support (DS) systems for the operational control of the network systems. In general an operational DS comprises a series of standalone applications from which the mathematical modeling and simulation of the distribution systems and the managing of the uncertainty in the decision-making process are essential in order to obtain efficient control and monitoring of the distribution systems. The mathematical modeling and simulation forms the basis for detailed optimization of the network operations and the second one uses uncertainty based reasoning in order to reduce the complexity of the network system and to increase the credibility of its model. This book reports on the integration of the two aspects of operational DS into a single computational framework of loop network equations.

The proposed DS system will be validated using case studies taken from the water industry. The optimal control of water distribution systems is an important problem because the models are non-linear and large-scale and measurements are prone to errors and very often they are incomplete.

The problem of steady state analysis of water distribution systems is studied in the context of a co-tree flows simulator algorithm that is derived from the basic loop corrective flows algorithm. It is shown that the co-tree formulation has several inherent advantages over the original formulation due to the use of the spanning trees. This allows a rapid determination of the necessary input data for the simulator (the loop and the topological incidence matrices and the initial flows) as well as the fast calculus of the nodal heads at the end of the simulation.

A novel Least Square (LS) state estimator that is suitable for on-line monitoring of the water distribution systems is presented. The state variables are both the loop corrective flows and the variation of nodal demands. It is shown that the input data necessary to build the network equations can be derived from the spanning tree obtained for the co-tree flows simulator and so there is a natural connection between the novel state estimator and the simulator algorithm. In spite of the increased size of the state vector, a satisfactory convergence is obtained through an enhancement in the Jacobian matrix. Furthermore a fine-tuning of the inverse of the tree incidence matrix is carried out in order to avoid the lack of numerical stability characteristic to the nodal heads state estimators. A very efficient and effective loop flows LS state estimator is developed that is tested successfully on realistic water networks.

Based on the novel state estimation technique, new algorithms for quantifying the measurement uncertainty impact on the state estimates are developed. The Confidence Limit Analysis (CLA) algorithms include a formulation of an Experimental Sensitivity Matrix (ESM) method, a sensitivity matrix method within the loop equations framework and an Error Maximization technique (EM). The performances of these algorithms are assessed in terms of their computational complexity and the accuracy of the results that they produce. It is shown that the computational efficiency and the accuracy of results of the EM method renders it suitable for online DS applications.

Finally, it is shown that the novel state estimation technique and the chosen CLA algorithm (EM method) are connected to an existent pattern classification module. The overall system is used for fault detection and identification for a realistic 34-node water network. Both, the "loop-equations based" state estimates and the variations of the nodal demands, together with their confidence limits are used as input data to the classification module in order to decide on the operational status of the 34-node water network. The extensive performance analyses for 24 hour of water network operations with particular stress on the detection and the correct location of leakages are carried out.

## Acknowledgments

The author would like to thank to Prof. Bogdan Gabrys from Bournemouth University who initially developed the neural network presented in Chapter 6 and also for his previous PhD work which allowed the new developments presented in this book. The author would like to thank to several other people who have helped with proofreading draft material and providing comments and suggestions, including Prof. Andrzej Bargiela from the University of Nottingham in United Kingdom (UK), Prof. David Al-Dabass and Dr. Johanne Hartley from Nottingham Trent University (UK) and Prof. Bogumil Ulanicki from De Montfort University also in UK.
This book is based on the PhD thesis of the author which was defended successfully in April 2004 at the Nottingham Trent University, United Kingdom.

Corneliu Arsene
Nottingham
United Kingdom
April 2011

# Contents

# Chapter 1

# Introduction

## 1.1. Book outline

The overall aim of this book is to investigate the implications of the loop equations formulation of the state estimation procedure for the implementation of Decision Support (DS) systems in operational control of distribution networks with a clear application to the water networks.

*Operational decision support* includes a couple of standalone applications from which the mathematical modelling and simulation of the water systems and the managing of the uncertainty in the decision-making process are essential in order to supply safely, treated water to consumers. The mathematical modelling and simulation forms the basis for detailed optimization of the water network operations and the second one uses uncertainty based reasoning in order to reduce the complexity of the water distribution system and to increase the credibility of its model. This book attempts to integrate the two aspects of operational decision support into a single computational framework of loop equations. The book will build on previous research results in the area of operational decision support of water systems in the context of uncertainties, and extends other researchers' work that confirmed the feasibility of using the loop equations in water networks modelling and simulation.

The great novelty of this book lays in developing more efficient tools for decision support of water networks, as well as enhancing the existing ones, and then combining them into a coherent environment for optimal control of complex water networks.

The prototype DS system will be validated using case studies taken from the water industry. The optimal control of water systems is a challenging problem because the models are non-linear and large-scale and measurements are noisy and frequently incomplete. Loop equations are used also in the modelling and simulation of the gas networks (Osiadacz, 1987; Osiadacz & Salimi, 1988) and power systems (Exposito et al., 1995; Shirmohammadi et al., 1988; Goswami & Basu, 1994) and within these respects the results of this book can have similar correspondents in other utility systems.

All parts of the book and their interconnections can be presented in a form of the block diagram presented at Figure 1-1.

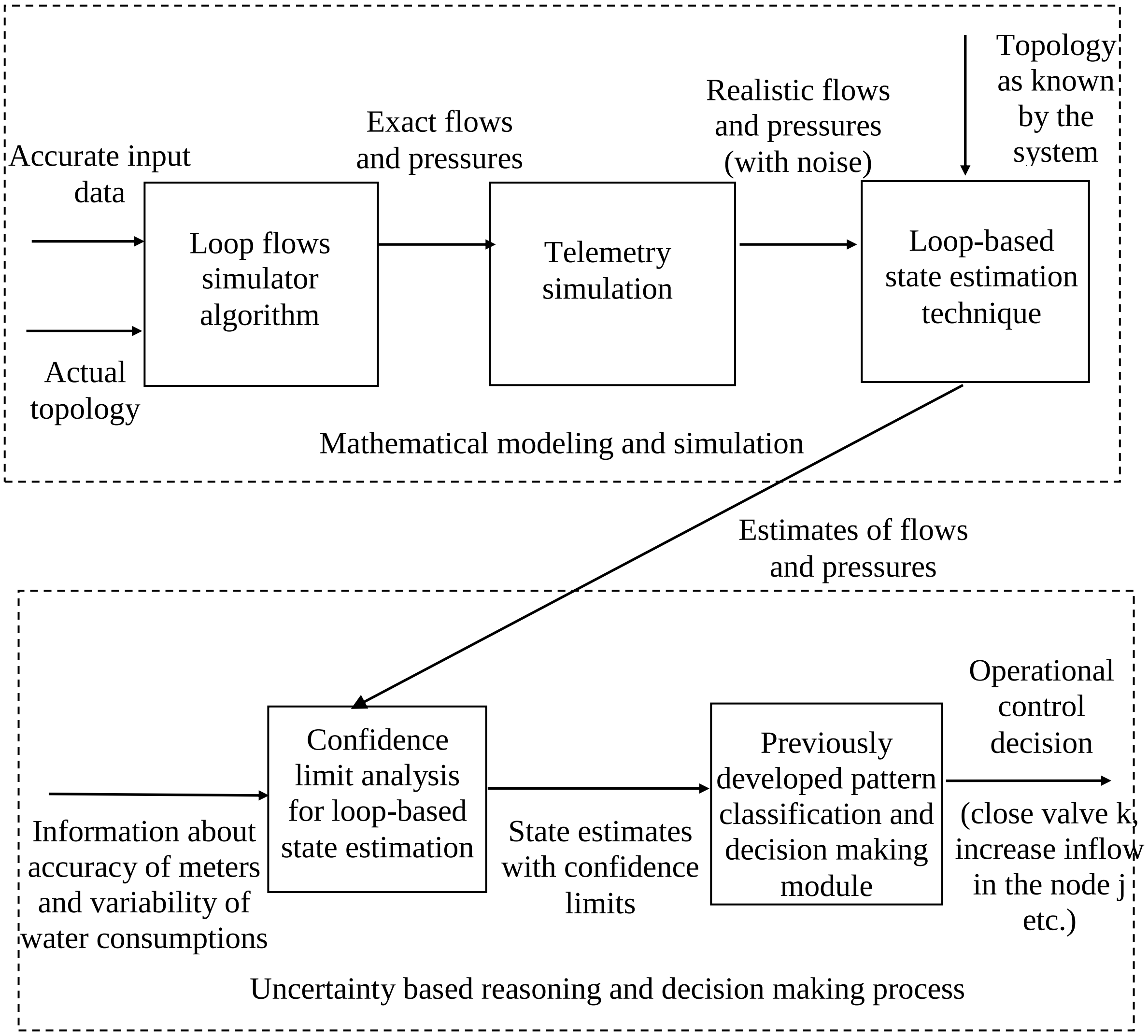

Figure 1-1: Operational decision support for water distribution systems using the loop equations in the numerical algorithms.

## 1.2. Water distribution systems

Modern *water distribution systems*, as the other two major utility systems – electricity and gas, are characterized by its complexity and large scale.

Additionally, it is quite difficult to identify a typical water distribution system. Each one has some unique characteristics due to the water source, service area topography, history of the system, etc. In general, all that can be said is that there are water sources and water users and that they are connected by pipes. The pipes can be made of many different materials (cast iron, steel, concrete etc.) and may be connected in many configurations.

There may be a single source such as a central pump station, or water may be supplied by a large number of wells. While pumps are a common component of many systems, there are networks which do not have any pumping and the water is supplied from some sources at high elevation.

Most systems contain some storage capacity in the form of tanks which are connected directly to the system, from which water must be pumped or which hold water under pressure.

Valves are required to shut off lines, suppress surges, release air, drain pipes, or control pressure.

Booster pumping may be required to provide adequate pressure in certain portions of a system where there is significant variation in elevation or use rate. On the other hand, pressure reducing valves serving just the opposite purpose may be needed.

Thus, water distribution systems consisting of large number of pipes, pumps, valves etc. are indeed complex hydraulic systems.

As indicated by Figure 1-1 our interest in water distribution systems as a case study is directed towards operational control rather than design or management of such systems. In fact in water distribution systems one can identify three levels of *decision making* that differ by time horizon, planning and decision variables (Bargiela, 1993; Urbaniak, 1998):

- *system development decisions*: decisions about the planning and sequencing of the investments in building new elements of a distribution system.
- *strategic planning decisions*: decisions concerning the utilization of water resources and their preservation , as well as the legislative measures.
- *operational control decisions*: decisions concerning water pumping schedules, pressure control measures, leakage monitoring an the co-ordination of the leakage remedial actions.

This book deals with the area of operational control which comprises of those decisions and actions that need to be varied in time, in response to current operating conditions and the actual state of the water distribution system. This is the area where the system dynamics play a dominant role, where the randomness of inputs cannot be neglected and where the physical and system management constraints make the problem of control both difficult and challenging.

Therefore *efficient control* of a water distribution system requires accurate information about its operating state. At present in the water industry, modern telemetry hardware and software systems are being installed to meet these needs. Unfortunately, due to financial constraints, it is not practical to measure all variables of interest. By variables of interest, we mean here heads at all network nodes and inflows at fixed-head nodes which are the components of the state vector of the system because given this information and the static parameters of the network, all other variable such as pipe flows or consumptions, may be calculated immediately. As this information is not entirely achievable, consequently any advanced operational control of the water distribution network needs to rely on the mathematical modeling and subsequent simulation of the system.

The system state is obtained by solving a set of equations which are defined using the network topology data, the measured or estimated water consumptions and the inflows into the system. Different sets of independent variables can be employed in order to build the water network equations. The most frequently used variables are the *nodal heads*, the *pipe flows* and the *loop corrective flows*.

Regardless the independent variables used to build the network equations, from on-line control point of view this method has two major drawbacks. Firstly, if one measurement is incorrect or lost, this approach gives incorrect results or no results at all. And secondly, the method uses only the system inflows and consumptions which, as in the case of predicted values, may carry considerable errors, while other more accurate and readily available measurements are not used.

A method that overcomes these drawback is known as *state estimation* procedure and over the last two decades has been the key point for the implementation of monitoring and control of large scale public utility systems. Its strength lays in processing all available measurements and formulating the problem in terms of redundant equations. This redundancy is essential for the successful performance of state estimation procedure since it enables the erroneous information to be filtered out. In water systems the degree of redundancy is achieved by combining the measurement information with the *pseudomeasurements* (i.e. predictions about water consumptions).

Thus, by increasing the number of measurements it is possible to improve both the reliability and accuracy of state estimation.

The simulation of any complex engineering system will always include a degree of *uncertainty*. No meter can be fully accurate, no mathematical model can fully reflect the intricacies of a real system's behaviour and no engineer's knowledge is complete. Water distribution systems are no exception to this rule.

This measurement uncertainty has clearly an impact on the accuracy to which state estimates can be calculated. It is, therefore, very important that the level of uncertainty present in state estimates can be quantified in some way if these estimates are to be used as the basis for making control decisions. The process of quantifying the uncertainty in the state estimates is known in water distribution systems as *confidence limit analysis* (CLA). In the effect of applying this procedure, the lower and upper limits for each state estimate value are produced and the state vector is rather presented with corresponding confidence limits than in deterministic form.

Although the knowledge of the current operating state, and how accurate the estimates are, is absolutely essential, it is the task of classifying the current state of the network (i.e. normal operating state, leakage in area *i*, etc.) and subsequently, on the basis of this classification, making an *operational decision* (i.e. close valve *k*, do nothing, etc.) that is paramount in an operational decision system. In the process of simulation of faults in the distribution system the leakages will not be pressure dependent. Finally, *fault detection* and *diagnosis* in water systems, based on *patterns recognition* that involves *neural networks* is one of the latest attempts to build effective and efficient operational decision support systems for water networks.

The material presented so far, concerning *water networks simulation* and state estimation, CLA and *fault diagnosis* is well researched and documented. However most of these algorithms are based on *network equations* that employ the nodal heads as state variables. This raises the question of potential benefits of using other set of variables, such as the loop corrective flows, in numerical simulations. This implies not only reformulation of the mathematical foundations of the existent algorithms but also development of new methods that match the new sets of variables.

## 1.3. Why loop equations?

The practice of transporting water for human consumption has been around for several millennia. From the first pipes in Crete about 3500 years ago, to today's sophisticated and complex hydraulic models, the history of water distribution technology is an impressive story. Over time people have understood that supplying water requires an understanding of the basic hydraulics and water network flow problems. While *hydraulics* problems include fluid properties and fluid flow characteristics in pipes, the water network problems consist mainly of water network definition and optimization, and network flow analysis.

*Network flow analysis* for real water distribution systems, that do not consist of a single pipe and cannot be described by a single equation, consists of solving a system of equations. The first systematic approach for solving these equations was developed by *Hardy Cross* (Cross, 1936). The invention of digital computers, however, allowed more powerful numerical technique to be developed. These techniques set up and solve the system of equations describing the hydraulics of the network in matrix form. These *numerical methods* can be classified in the following way: the *numerical minimization* methods (Collins et al., 1978; Contro & Franzetti, 1982), the *Hardy-Cross* method (Chenoweth & Crawford, 1974; Eggener & Polkowski, 1976), the *Newton-Raphson* method (Martin & Peters, 1963; Lemieux, 1973; Donachie, 1974) and the *Linear Theory* method (Collins & Johnson, 1975; Fietz, 1973). The last three classes include methods used for the solution of systems of non-linear equations, while the first deals with the search of minimum of a non-linear *convex function* under linear equality and inequality constraints.

Irrespective of the numerical methods used, the solution of network flow analysis has led to the development of many methods of analysis using various types of decompositions. Each decomposition expresses the resulting system of equations in terms of a specific type of independent variables. The most common methods are those in which the independent variables are expressed in terms of the *link flow* $Q$ (Wood & Charles, 1972), the *loop corrective flows* $\Delta Q_l$ (Epp & Fowler, 1970; Gofman & Rodeh, 1982), and the *nodal heads* $H$ (Shamir & Howard, 1968; Jeppson, 1975).

The overall methods for network flow analysis are summarized at Figure 1-2.

The methods for network flow analysis can be classified in terms of

The **independent variables** used to build the system of equations that describes the distribution network.

1. Nodal heads H.

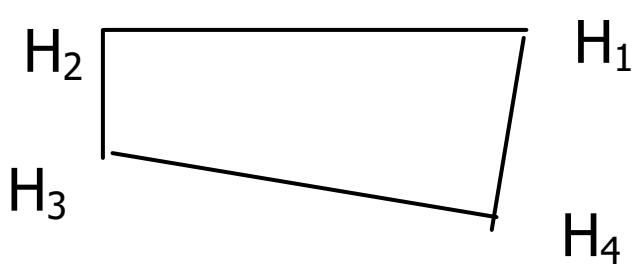

2. Loop corrective flows $\Delta Q_l$.

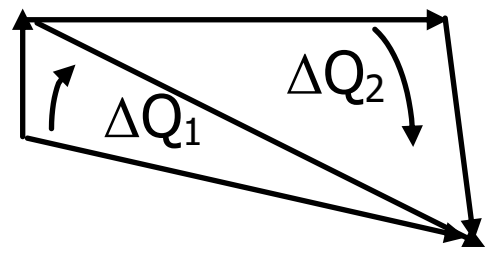

3. Pipe flows *Q*.

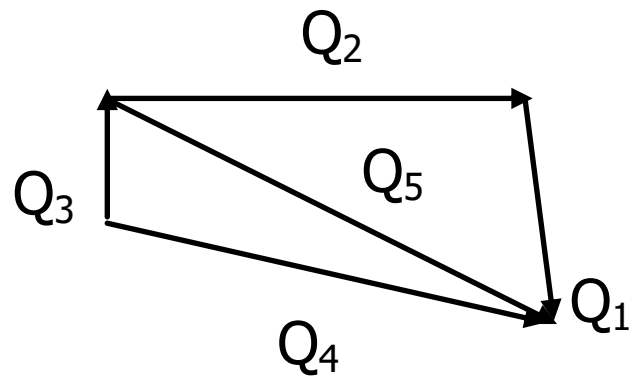

4. Hybrid node-loop approach used in the software developed by the Environmental Protection Agency of USA – *EPANET* (Rossman, 1994).

The **numerical methods** used to find the solution of the system of equations.

1. The Hardy-Cross method.
   The drawback of this method is the slow rate of convergence.

2. The Newton-Raphson method.
   A powerful numerical method for solving systems of non-linear equations; most frequently used.

3. The Linear Theory method.
   It requires the solution of a larger system of equations (number of loops + number of nodes).

4. Numerical minimization
   Not practical when dealing with large complex networks.

Figure 1-2: Network flow analysis.

It has been shown (Todini & Pilati, 1988; Todini, 2000; Ulanicka et al., 1998) that the Newton-Raphson method applied in the node or loop sub-domain and the Linear Theory method applied in the loops domain have equivalent convergence properties. However, they involve solving systems of linear equations of different size. In order to asses the relative merits of the different formulations for solving large pipe network problems, the comparison can be made in terms of simplicity of input, initial solution, size of the system of linear equations and efficiency of solution of the system of equations.

The balance of these merits made the combination of the nodal heads and the Newton-Raphson algorithm to be the most frequently used procedure for solving water networks (Usman et al., 1987; Rahal, 1980; Powell, 1992). However the use of nodal equations in network flow analysis has disclosed a couple of weaknesses. In (Nielsen, 1989) it has been reported that nodal heads based algorithms have *weak convergence* for the parts of water network containing low pipe flows, while in (Gabrys & Bargiela, 1995; Sterling & Bargiela, 1984; Powell et al., 1988; Hartley, 1996) the sensitivity of the nodal heads based estimation procedure to the measurement errors has been recognized and reported. Moreover practical problems in modelling and simulations of particular networks have been envisaged (Hartley & Bargiela, 1993).

The combination of the Newton-Raphson method and the loop corrective flows is called the *loop system of equations*. In the last years the numerical simulations based on loop equations have received an increased attention. It has been shown that using the loop equations for the simulator algorithm is a suitable framework for the inclusion of *pressure-controlling elements* without specifying the operational state (Andersen & Powell, 1999b). Moreover a *rapid convergence* has been reported (Andersen & Powell, 1999a; Rahal, 1995) for the *steady state analysis* of water networks based on the co-tree formulation that is derived from the loop equations method. Although the results were encouraging, no further efforts have been made for developing an integrated operational decision support system for *on-line monitoring* of water networks based on loop equations. This book can be regarded as a major contribution towards this goal.

## 1.4. Organization of the book

Chapter 2 is wholly devoted to mathematical modeling in water distribution systems. First the basic principles of hydraulics that are frequently employed in water distribution system modeling process are described. Next the mathematical models of the simple elements that are parts of the water systems and the methods of combining those models of elements with the basic hydraulics laws to produce concise model of the water network are presented. Since consumption of water, also known as water demand, is the driving force behind the hydraulic dynamics occurring in water distribution systems, some questions will be answered such as how much water is being used or how does the usage change as a function of time. Following this, since the uncertainty is an inherent part of water systems the ways of introducing it into the network model are also discussed. Finally having developed the network model and addressed some of the connected problems, the task of operational control of large scale systems is tackled.

Chapter 3 introduces the loop corrective flows simulator algorithm. It defines what is a steady-state simulation and it presents methods for finding the loops and calculation of initial flows with case studies taken from water industry. The performance of the simulator algorithm will be enhanced through the modification of the Jacobian matrix.

Chapter 4 can be regarded as one of the major contributions of this book. The chapter discusses a state estimation technique suitable for on-line monitoring of water networks. The mathematical formulation of the state estimator is based on the loop equations and the optimality of the state estimation vector is addressed. The accuracy of results is improved through the modelling of the topological incidence matrix $A_{np}$ as an upper form tree incidence matrix $T$ and a co-tree incidence matrix $C$. The advantages of this state estimator over the nodal heads based state estimator are presented.

Chapter 5 is another major contribution. This chapter presents confidence limit analysis algorithms based on the novel loop flows state estimator. These include a formulation of an experimental sensitivity matrix method, a sensitivity matrix method within the loop equations framework and an error maximization technique. The performances of these algorithms are assessed in terms of their computational complexity and the accuracy of the results that they produce.

Chapter 6 reviews the fuzzy state classification and clustering techniques used in pattern recognition problems. Having found the state estimates with their corresponding confidence limits, the next task, usually carried out by a human operator, is to classify the current operating state (e.g. normal operating status, leakage in pipe *i*) before any control action can be taken. This chapter describes a previously developed fuzzy min-max clustering and classification neural network that is capable of clustering as well as classifying the state estimation vector with its confidence limits.

Chapter 7 integrates all the modules developed in the previous chapters into an effective and efficient operational decision support for water distribution systems. It is shown that the loop flows state estimation technique and the confidence limits analysis algorithms are connected to a pattern classification module presented in the previous chapter. The overall system is used for fault detection and identification for a realistic 34-node water network. Both, the nodal heads and the variation of nodal demands, together with their confidence limits are used as input data to the classification module in order to decide on the operational status of the 34-node water network.

Chapter 8 presents the main conclusions of this book and some suggestions for further research.

# Chapter 2

# Mathematical modeling of water distribution networks

## 2.1. Introduction

*Water network simulation* is the process of using the mathematical representation of the real system, or the *water network model*, in order to replicate the dynamics of the existing water system when it is not practical for the real water system to be directly subjected to experimentation, or for the purpose of evaluating a water system before it is actually built. In addition, for situations in which water quality is an issue, directly testing a system may be costly and a potentially hazardous risk to public health. Simulations can be used to predict system responses to events under a wide range of conditions without disrupting the actual system. Using simulations, problems can be anticipated in proposed or existing systems, and solutions can be evaluated before time, money, and materials are invested in a real-world project.

Mathematical modeling and simulation of many water distribution systems requires an understanding of the hydraulics and pipe network flow problems. *Hydraulics problems* include fluid properties, fluid flow characteristics in pipes, and pipe network problems which consists mainly of two different stages: network definition and optimization, and network flow analysis.

Properties which influence the flow behavior of fluids include density, viscosity and surface tension. These properties affect the fluid flow in a pipe, fluid flow which can be described by the following equations: continuity, momentum and energy equation, and for solving practical problems the energy equation must be coupled with one of the equations that predict head loss.

Since water distribution systems consist of combination of pipes, pumps and other hydraulic control systems, specific techniques should be used for finding the solution of network problems. These techniques can be classified in methods for *network definition and optimization* (e.g. rules for using equivalent pipe and pump to simplify the problems, modeling distributed consumer demands) and methods for *network flow analysis* (e.g. setting up flow equations, loop equations and head equations, methods used to solve the set of equations that describes the water distribution system).

Although the mathematical model may be accurate, the simulation process is based on input data that contain a significant amount of uncertainty. Therefore the uncertainty in the network model and solution of flows and pressures will be also discussed. Finally, the topic of operational decision support for large scale and complex water distribution systems based on uncertain measurement data will be addressed.

## 2.2. Review of Closed Conduit Hydraulics

Usually the solution process for modeling and simulation of a water distribution system involves simultaneous consideration of a couple of equations describing the fluid flow in pipes and some independent relationships describing what is so called head loss. These equations depend on the properties and flow characteristics of the fluids met in water supply systems.

### 2.2.1. Types of flow and head loss formulas

Pipes that are most frequently used for the conveyance of fluids are produced from a variety of materials, including steel, cast iron, concrete, plastic and glass. In such pipes the flow can be described as *steady* at a particular location if the velocity vector $V$ [m/s] at the location does not change with time; it is described as unsteady if the velocity vector changes with time.

In mathematical terms these definitions are written as follows:

(1) *steady flow*

$$\left(\frac{\partial V}{\partial t}\right)_{x_0 y_0 z_0} = 0 \qquad \text{(Eq. 2.1)}$$

(2) *unsteady flow*

$$\left(\frac{\partial V}{\partial t}\right)_{x_0 y_0 z_0} \neq 0 \qquad \text{(Eq. 2.2)}$$

Flow is said to be uniform if the velocity vector is constant along the flow path or streamline. Conversely, flow is described as non-uniform if the velocity vector varies along the flow path. These definitions are expressed in mathematical terms as follows:

(1) *uniform flow*

$$\left(\frac{\partial V}{\partial S}\right)_{t_0} = 0 \qquad \text{(Eq. 2.3)}$$

(2) *non-uniform flow*

$$\left(\frac{\partial V}{\partial S}\right)_{t_0} \neq 0 \qquad \text{(Eq. 2.4)}$$

where $S$ [m$^2$] is the cross-sectional area through which the flow occur.

The most regular of the foregoing flow types is steady uniform flow such as that which occurs in pipes of fixed diameter having a constant discharge rate.

In their new condition, the internal wall surfaces of the pipes vary considerably in roughness depending on the material from which they are fabricated: from the very smooth glass or plastic surface to the relatively rough concrete surface. Also depending on the fluid transported and the pipe material, the condition of the pipe wall may vary with time, either due to corrosion, as in steel pipes, or deposition, as in hard water areas. When fluid flow is confined by solid boundaries, such that random lateral mixing in a direction perpendicular to that flow is suppressed, flow is described as *laminar*, that is, flowing in separate layers with minimal lateral momentum transfer between layers. Where there is significant lateral mixing and momentum transfer in a direction normal to the flow direction, flow is classified as *turbulent*. Reynolds (1885) carried out extensive pipe flow tests from which he was able to define the flow regime as either laminar, transitional, or turbulent. Reynolds found that transition from one type flow to another occurs at a critical velocity for a given pipe and fluid. He expressed his results in terms of the dimensionless parameter, $R_e$, called *Reynolds number*:

$$R_e = \frac{VD}{v} \qquad \text{(Eq. 2.5)}$$

where: $V$ - the average velocity of flow [m/s].

$D$ - the pipe diameter [m].

$v$ - the kinematic viscosity [m$^2$/s].

He found that for $R_e$ less than about 2000 the flow was laminar, and that for $R_e$ greater than about 4000 the flow was always turbulent. For $R_e$ between 2000 and 4000, he found that the flow could be either laminar or turbulent, and termed this the *transition region*.

In a further set of experiments, he found that for laminar flow the frictional head loss in a pipe was proportional to the velocity, and that for turbulent flow the head loss was proportional to the square of the velocity.

These two results have been previously determined by *Hagen and Poiseille* (h~V) and *Reynolds* (h~$V^2$) who put these equations in the context of laminar and turbulent flow.

Since most of flows which are encountered in water distribution systems are turbulent flow we restrict our considerations to turbulent flows.

The Darcy-Weisbach head loss equation for turbulent flows can be written as follows:

$$h = \lambda \frac{LV^2}{2gD} \qquad \text{(Eq. 2.6)}$$

where: $\lambda$ - pipe friction factor [dimensionless factor].

$L$ - pipe length [m].

$g$ - gravitational acceleration [m/s$^2$].

The original investigators presumed that the friction factor was constant. Nikuradse (Nikuradse, 1932), however, found that the turbulent flow could be divided into three regions and that the value of friction factor depends on relative roughness (*k/D*) of the pipe and $R_e$. These three kinds of turbulent flow can be described as follows:

- *Smooth turbulence* – the limiting line of turbulent flow that is approached by all values of relative roughness (*k/D*) as $R_e$ decreases.
- *Transitional turbulence* – the region in which $\lambda$ remains constant for a given *k/D*. In practice, most of pipe flow lies within this region.
- *Rough turbulence* – the region in which $\lambda$ remains constant for a given *k/D*, and is independent of $R_e$.

The following equation:

$$\frac{1}{\sqrt{\lambda}} = -2\log\left(\frac{k}{3.7D} + \frac{2.51}{Re\sqrt{\lambda}}\right) \qquad \text{(Eq. 2.7)}$$

that relates the friction factor to $k/D$ and $R_e$ is known as *Colebrook-White transition formula*. It is applicable to the whole of the turbulent region for commercial pipes, using effective roughness values determined experimentally for each type of pipe.

Although the Darcy-Weisbach equation using Colebrook-White formula is the most accurate for the head loss assessment it had not been easy to use for engineers hand calculations. There was a need for simpler empirical formulae. For water distribution system analysis one the most commonly used of empirical formulas is the *Hazen-Williams equation* (Williams & Hazen, 1920; ASCE, 1992).

$$q_{ij} = 0.27746 C_{ij} D_{ij}^{2.63} \left( \frac{|H_j - H_i|}{L_{ij}} \right)^{0.54} \qquad \text{(Eq. 2.8)}$$

where: $q_{ij}$ - flow from node $j$ to node $i$ [m$^3$/s].

$C_{ij}$ - Hazen-Williams coefficient for pipe [dimensionless factor].

$D_{ij}$ - diameter of pipe [m].

$L_{ij}$ - length of pipe [m].

$H_j$ - head at node $j$ [m].

$H_i$ - head at node $I$ [m].

$H_j > H_i$.

or for computer program implementation:

$$q_{ij} = R_{ij}^{-0.54} \left( H_j - H_i \right) \left| H_j - H_i \right|^{-0.46} \qquad \text{(Eq. 2.9)}$$

where $R_{ij}$ is the resistance between nodes $i$ and $j$ given by:

$$R_{ij} = 10.742 C_{ij}^{-1.85} L_{ij} D_{ij}^{-4.87} \qquad \text{(Eq. 2.10)}$$

The use of the Hazen-Williams versus Darcy Weisbach equations is one of the most frequently discusses in practical hydraulics. In general Hazen-Williams equation gives the same results as the Darcy-Weisbach for smooth flow and in the transition range. It is not until the flow becomes rough turbulent when the Hazen-Williams runs into somewhat significant errors (Liou, 1998; Walski, 1984; Karney, 2000). However in typical water networks the flows are in the transitional range, and therefore the changes in fluid density and viscosity does not matter too much (Walski, 2002).

Nevertheless for the purposes of this book the head loss has been calculated using the Hazen-Williams equation.

## 2.2.2. Continuity equation

During any time interval $\Delta t$, the principle of conservation of mass implies that for any control volume the mass flow entering minus the mass flow leaving equals the change of mass within the control volume. It can be written as follows:

$$m_{in} - m_{out} = \Delta m_{store} \qquad \text{(Eq. 2.11)}$$

where: $m_{in}$ - mass flow entering [kg].

$m_{out}$- mass flow leaving [kg].

$\Delta m_{store}$- change of mass stored [kg].

If the flow is steady, then the mass must be entering (or leaving) the volume at a constant rate. If we further restrict our attention to incompressible flow, then the mass of fluid within the control volume must remain fixed. In other words, the change of mass within the control volume is zero.

Taking these assumptions into consideration and knowing that:

$$q = \frac{m}{\rho t} \qquad \text{(Eq. 2.12)}$$

where: $\rho$ - density of fluid (water) [kg/m$^3$].

$t$ - unit of time [s].

$m$ - mass flow [kg] .

$q$ - volumetric flow [m$^3$/s].

The equation (Eq. 2.11) can be written as:

$$Q_{in} - Q_{out} = \frac{\Delta S}{\Delta t} \qquad \text{(Eq. 2.13)}$$

where: $Q_{in}$ - flow in.

$Q_{out}$ - flow out.

$\Delta S$ - change of volume stored.

$\Delta t$ - time period.

For $p$ pipes meeting at a point, if we will consider the inflow positive, outflows to be negative and let $t$ to be small, equation (Eq. 2.13) reduces to

$$Q_1 + Q_2 + Q_3 + \ldots + Q_p = dS/dt \qquad \text{(Eq. 2.14)}$$

where: $Q_1$ - flow in through pipe 1.

$Q_2$ - flow in through pipe 2.

$Q_p$ - flow in through pipe $p$.

$dS/dT$ - time rate of change storage.

The equation (Eq.2.14) is the *continuity equation* that will be used for water distribution problems.

### 2.2.3. Energy equation

In general terms for the fluid (water) flow in a pipe the *energy equation* states that, given the energy at point 1, the energy at point 2 equals the energy at point 1, plus the net work done on the fluid (work done on water minus work done by water), minus any energy losses due to friction. Mathematically this can be written as follows:

$$E_2 = E_1 + W - H \qquad \text{(Eq. 2.15)}$$

where: $E_2$ - energy at point 2 [N·m].

$E_1$ - energy at the entry point [N·m].

$W$ - net work done on the fluid [N·m].

$H$ - friction energy loss [N·m].

The energy in the fluid exists in three forms: *kinetic energy*, *potential energy* due to elevation, and *internal energy* (pressure). The *total energy* may be expressed as:

$$E = (mV^2/2) + mgz + Pm/\rho \qquad \text{(Eq. 2.16)}$$

where: $m$ - mass of the fluid [kg].

$V$ - velocity [m/s].

$g$ - acceleration due to gravity [m/s$^2$].

$z$ - elevation [m].

$P$ - pressure [N/m$^2$].

$\rho$ - density [kg/m$^3$].

The term $mV^2/2$ refers to the kinetic energy and in water distribution system problems is usually small in comparison to the other terms. The $mgz$ term refers to the energy the fluid has because of its position in the gravitational field. The $Pm/\rho$ term refers to the amount of energy stored in the form of pressure.

Dividing equation (Eq. 2.16) by *gm* and noting $\gamma$ for $\rho g$ (specific weight of water), will allow all the energy terms to be expressed in units of length:

$$\frac{V_1^2 - V_2^2}{2g} + z_1 - z_2 + \frac{P_1 - P_2}{\gamma} = -w + h \qquad \text{(Eq. 2.17)}$$

where: $w$ - net work done on fluid [m].

$h$ - friction energy loss [m].

$\gamma$ - specific weight [N/m$^3$].

The terms from the equation (Eq. 2.17) have the following names:

$\frac{V^2}{2g}$ - velocity head [m].

$\frac{P}{\gamma}$ - pressure head [m].

$z$ - elevation head [m].

$w$ - lift (when referring to pumps) [m].

$h$ - friction head [m].

The *momentum equation*, that may be used directly to evaluate the force causing a change of momentum in a fluid, has been omitted. The reason for this omission is the fact that the applications where the momentum equation is used include:

- determining forces on pipes bends and junctions, nozzles and hydraulic machines - useful for designing the water network; or
- solving problems when the flow is unsteady.

The problems described in this book are not concerned with either of these applications. More information on dynamic behaviour of the fluids can be found in (Walski et al., 2000; Casey, 1992; Chadwick & Morfett, 1986; Meritt, 1967).

## 2.3. Water Distribution Systems Modeling and Simulation

As discussed at the beginning of this chapter, a water distribution model is a mathematical description of a real-world water system. The network model contains all of the various components of the system, and defines how those elements are interconnected. Network models are comprised of *nodes*, which represent features at specific locations within the system, and *links*, which define relationships between nodes. *Water distribution models* have many different types of nodal elements, including junctions nodes where pipes connect, storage tanks and reservoirs nodes, while examples of link elements are pipes, pumps and valves. Sometimes valves and pumps are classified as nodes depending on the modelling method.

A *boundary node* is a network element used to represent locations with known hydraulic grade elevations. A boundary condition imposes a requirement within the network that simulates flows entering or exiting the system in agreement with that hydraulic grade. A *reservoir* represents a boundary node in a water distribution model, that can supply or accept water with such a large capacity that theoretically can handle any inflow or outflow rate, for any length of time, without running dry or overflowing. A *tank* is also a boundary node, but unlike a reservoir, the hydraulic grade of a tank fluctuates according to the inflow and outflow of water.

The *junction node* role, as the term implies, is to provide a location for two or more pipes to meet. The other use is to settle the place to withdraw water demanded from the system or inject inflows into the system. A *pipe* conveys flow as it moves from one junction to another in a network. A *pump* is an element that adds energy to the system in the form of an increased hydraulic grade while a *valve* is an element that can be opened and closed to different extents in order to control the movement of water through a pipeline.

During the constructing of the water network model, a *Skeletonization* process (Ulanicki et al., 1996; Shamir & Hamberg, 1988a; Shamir & Hamberg, 1988b; Eggener & Polkowski, 1976) is performed that is the selection for inclusion in the water network model only of the parts of the hydraulic network that have a significant impact on the behaviour of the overall system. Attempting to include each individual component of a large system in a model could be a huge undertaking without a significant impact on the model results. Moreover capturing every feature of a system would also result in tremendous amounts of data, enough to make managing, using, and troubleshooting the water network model an overwhelming and error-prone task. An example of a water network model is depicted at Figure 2-1.

Once the basic elements have been defined and the water network model has been assembled, further refinement of the model can be done depending on its intended purpose. There are various types of simulations that a model may perform, depending on what the modeler is trying to observe or predict. The two most basic types are:

- *steady-state simulation* (Nogueira, 1993; Collins & Johnson, 1975; Rahal, 1994) – computes the state of the system (flows, pressures, valve position, etc.) assuming that the water demands and boundary conditions do not change with respect to time.
- *extended period simulation* (Rao et al., 1977a; Rao et al., 1977b) – determines the dynamic behaviour of a system over a period of time, computing the state of the system as a series of steady-state simulations in which water demands and boundary conditions do change with respect to time.

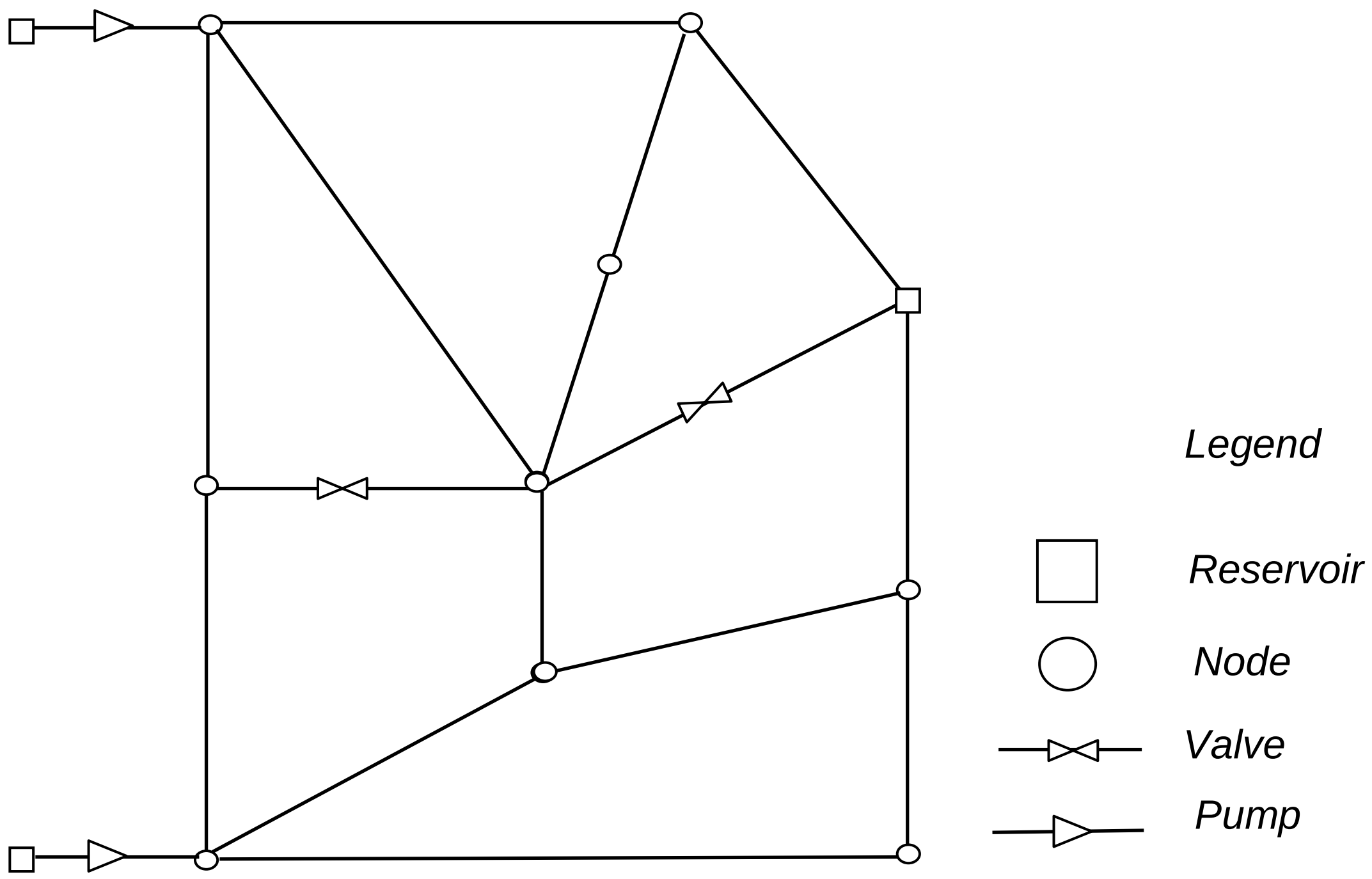

Figure 2-1: Water network model.

Other types of simulations are *water quality simulations* (Rossman et al., 1994; Rossman, 2000; Clark & Grayman, 1998; Harding & Walski, 2000; Grayman et al., 1996) used to ascertain chemical or biological constituent levels within a system or to determine the age or source of water, automated fire flows analyses (Boulos, 1996; Moore & Boulos, 1998) that establishes the suitability of a system for fire protection needs or cost analyses that are used for looking at the monetary impact of operations and improvements.

The simulation of a water distribution system that does not consist of a single pipe, but combinations of pipes, pumps and other elements, depends to a great extent on whether the network is branched or looped. *Looped network problems* are the problems in which there are one or more loops in the network or if the head is specified at more than one location, in which case the energy and continuity equations cannot be solved independently. Even a single pipe with head known at both ends fall into this category. *Branch network problems* will refer to networks with no loops and one or no fixed head point and for this kind of network the continuity equation can be solved to yield flows in individual pipes, and then the energy equation can be used to calculate heads. In the following section, the continuity and momentum equations will be written in a more detailed form for a water network model consisting of $p$ pipes, $n$ nodes and $l$ loops.

### 2.3.1. Continuity Equation for *n* nodes

Applying the continuity equation (Eq. 2.14) for $n$ nodes we obtain a set of equations called the nodal equations:

$$\sum_{j\in\Omega_i} q_{ij} + u_i = d_i \qquad \text{(Eq. 2.18)}$$

where: $n$ - number of nodes in the network.

$\Omega_i$ - set of nodes connected at node $i$.

$u_i$ - inflow at node $i$.

$d_i$ - demand at node $i$.

$q_{ij}$ - flow from node $j$ to node $i$.

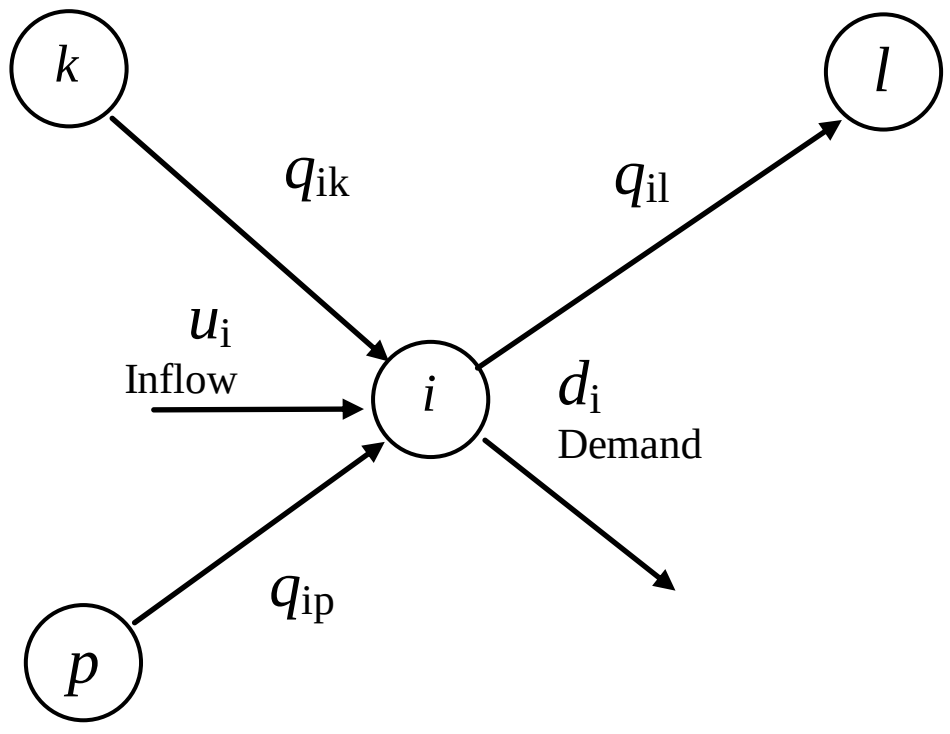

Figure 2-2: Continuity equation for *i*-th node.

The continuity equation for the *i*-th node is as follows:

$$q_{ik} - q_{il} + q_{ip} + u_i = d_i \qquad \text{(Eq. 2.19)}$$

## 2.3.2. Energy Equation for a loop

Applying the energy equation (Eq. 2.15) written for a single pipe to a pipe loop, since the beginning and ending points are the same ($E_2=E_1$), this means that the net work done on the fluid in a loop, must equal the head loss in the loop:

$$\sum h_{pj} - \sum h_i = 0 \qquad \text{(Eq. 2.20)}$$

where: $h_{pj}$ - head provided by $j$-th pump.

$h_i$ - head loss in $i$-th pipe.

In solving a loop problem, it is essential to define a direction in which flow is considered as positive (e.g. clockwise or in a pseudo-loop from higher to lower tank). The continuity equation (Eq. 2.18) and energy equation (Eq. 2.20) can now be coupled with a suitable head-flow formula (e.g. Hazen-Williams or Darcy-Weisbach/Colebrook-White) to construct a set of network equations.

Sometimes when a network is fairly complicated, it becomes difficult to identify loops without double counting and thus, over specifying the problem. For preventing this problem, it is necessary to apply the energy equation only for independent loop (i.e. loop for which the energy equation can not be derived from the energy equations written for the other loops).

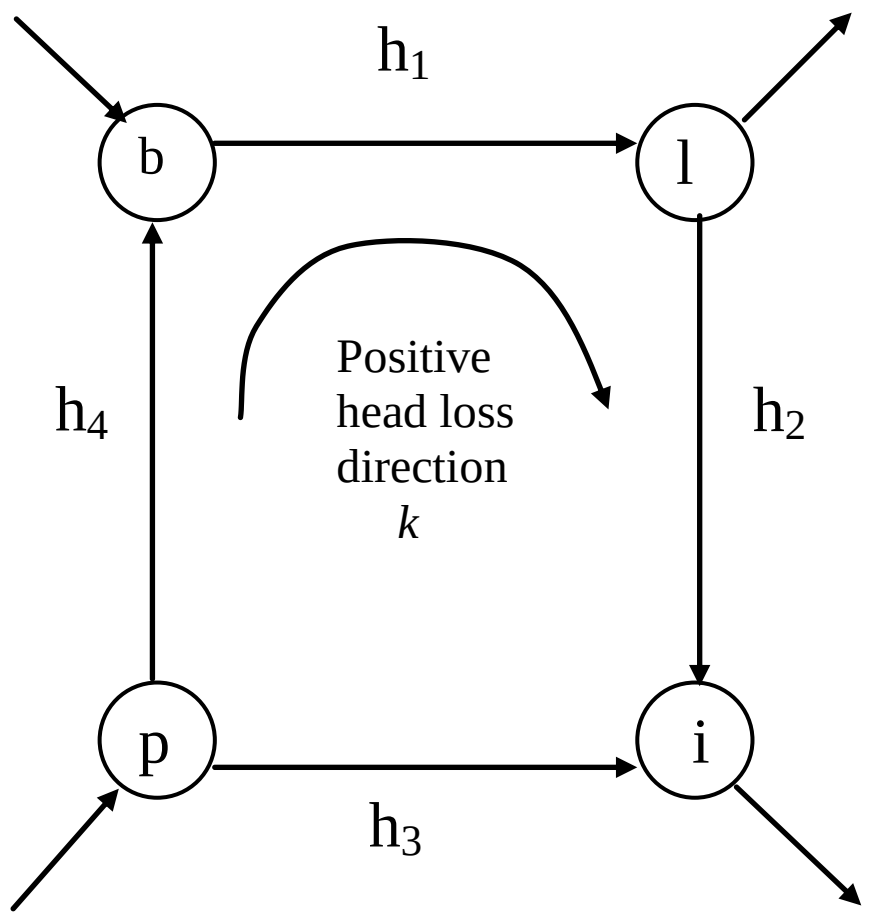

Figure 2-3: Energy equation for *k*-th loop.

$$- h_1 - h_2 + h_3 - h_4 = 0 \qquad \text{(Eq. 2.21)}$$

The number of independent loops $l$ for a water network with $n$ nodes and $p$ pipes can be determined from the following rule which is called also *Euler relationship*:

$$l = p - n + 1 \qquad \text{(Eq. 2.22)}$$

where: $l$ - number of basic loops.

$p$ - number of pipes.

$n$ - number of nodes.

The number of pseudo-loops in a network is given by $t$-1 where $t$ is the number of constant head nodes (i.e. boundary nodes) in the network (e.g. tanks, reservoirs).

## 2.3.3. Network flow analysis

*Network flow analysis* of real water distribution systems implies writing one continuity equation for each node in the system, and one energy equation for each pipe (or loop). For real systems, these equations can number in the thousands. Moreover energy equations are non-linear in terms of flow and head, they cannot be solved directly. Therefore some powerful numerical techniques for solving non-linear equations must be employed. Definitely more powerful and faster method is the Newton-Raphson method. It obtains the solution to a system of non-linear equations by linearization and iteratively solving a system of linear equations. This method has been adopted in this work for solving the network equation.

The network equations can be written in the following way:

$$g(x) = z \qquad \text{(Eq. 2.23)}$$

where $x$ represents the state vector that needs to be determined and $z$ is the measurement vector. An important observation is that vectors $x$ and $z$ can comprise different combinations of variables and measurements, depending of the method used to construct the network equations.

There are three main ways of constructing the network equations that have been mentioned earlier: the *flows,* the *nodal heads,* and the *loop corrective flows*.

In the first of these methods the network equations are set up by using the flow rates as unknowns (state variables) and writing one energy equation for each independent loop and one continuity equation for each node. It results in deriving *p* equations (where *p* is the number of pipes) called the flow equations.

In the second method the network equations are derived by combining the continuity equation for each node with head loss equations. The state vector in this method consists of nodal pressures and inflows into fixed head nodes if any of such nodes are in the system. This method is called the nodal equations and it has been used largely in water networks flow analysis. The size of the system of equations is equal with *n* that is the number of nodes of the water network.

Third approach to setting up the network equation is to write energy equation in such a way that, for an initial solution, the continuity equation is not violated. This can be done by adding a correction to the flow to every pipe in the loop. These corrections are the unknown in a set of *l* equations – one for each loop. The simulation problem based on the loop equations is discussed in mode detail in Chapter 3.

Some more details about each of these methods can be found elsewhere (Bhave, 1991; Lansey & Mays, 2000; Larock et al., 1999).

Once the network equation has been established, it is solved iteratively with the Newton-Raphson numerical method. Hence, more detailed explanation of *the Newton-Raphson* method in a context of the network equation is given below.

Expanding *g(x)* by an initial guess of the state vector $x^{(0)}$, using a first-order *Taylor series* and defining $z^{(0)} = g(x^{(0)})$, we obtain:

$$z = z^{(0)} + \Delta z \qquad \text{(Eq. 2.24)}$$

$$g(x) = g(x^{(0)}) + J^{(0)}\Delta x \qquad \text{(Eq. 2.25)}$$

Using the linearized models (Eq. 2.24), (Eq. 2.25) and the network equation (Eq. 2.23) we obtain the following set of linear equations:

$$J^{(k)}\,\Delta x^{(k)} = z - g(x^{(k)}) \qquad \text{(Eq. 2.26)}$$

where: $J^{(k)} = \left.\frac{\partial g}{\partial x}\right|_{x=x^{(k)}}$ - Jacobian matrix evaluated at $x^{(k)}$.

$\Delta x^{(k)}$ - the correction vector.

$x^{(k)}$ - the current estimate of the state vector.

*z* - the measurement vector.

$g(x^{(k)})$ - the network function evaluated at $x^{(k)}$.

Since the network equation is nonlinear, the solution finding is an iterative process with the consecutive state estimates calculated by under-relaxation of the linear solution:

$$x^{(k+1)} = x^{(k)} + \Delta x^{(k)} \qquad \text{(Eq. 2.27)}$$

If all elements of $\Delta x^{(k)}$ in *k*-th iteration are lower or equal to a predefined convergence accuracy, the iteration procedure stops. Otherwise, a new correction vector is calculated using equation (Eq. 2.26) with $x^{(k+1)}$ instead of $x^{(k)}$.

### 2.3.4. Uncertainty of the network model and solution

Even though all the required data have been collected and entered into a hydraulic simulation software package, the human operator cannot assume that the model is an accurate mathematical representation of the system. This is because the simulation software solves the equations of continuity and energy using the supplied data. In consequence, the quality of the results will depend on the quality of the data that is used in the simulation software.

Usually, the accuracy of data is the weakest link in the modeling process. Therefore the model results need to be compared with the field observations, and, if necessary, adjusting the input data describing the system so that the model results to agree with the measured system performance over a wide range of operating conditions. This process is called *Calibration* (Walski, 1990; Walski, 1983; Ormsbee & Lingireddy, 1997; Ormsbee & Woods, 1986; Ormsbee, 1989) and any and all input data that have uncertainty associated with them are candidates for adjustment during the calibration process, in order to obtain reasonable agreement between model-predicted behavior and actual field behavior.

The scope of carrying out such a calibration procedure is to increase the confidence of the engineer in the model's ability to predict the real water system behavior. Results provided by such a computer model are frequently used to aid in the operational decisions taken for the hydraulic system. Other reasons are a better understanding of the behavior and performance of the system for better capital improvements or operational changes or identifying errors caused by mistakes during the model-building process.

There are two main sources of uncertainty and errors in water systems modeling and simulations, for which we need to employ the calibration procedures. First, topological errors, are associated with the modeling of the physical elements and represent static inaccuracy of network model. The second source of uncertainty has dynamic nature and is associated with inaccurate predictions of consumptions and inaccuracies of measured pressure and flow values. While, for instance, the pipe's C factor or friction factor usually varies gradually over years or decades, consumptions and flows in the network change from minute to minute and are of unpredictable nature.

There has been a lot of research work done in order to develop methods for improving the network model accuracy. One of the sources of inaccuracies of network model is a simplified representation of the physical system and it can appear during the scheletonization process (Eggener & Polkowski, 1976). Then the correct estimation of the *C*-values and friction factors has been considered by many authors (Cesario & Davis, 1984; Coulbeck, 1984; Lansey, 1988; Ormsbee & Chase, 1988; Walski, 1984). The calibration of roughness coefficients can be based on manual calibration approaches (Herrin, 1997; Walski, 1983) or computing based calibration methods. The last category implies solving an optimization problem that is trying to minimize the discrepancy between the model heads and flows and the observed heads and flows. The optimization approaches include gradient-based methods (Ormsbee & Lingireddy, 2000; Lansey & Banset, 1991) and stochastic search method, more commonly referred to as genetic algorithms (Savic & Walters, 1995, 1997; Walters et al, 1998).

In all the publications dealing with methods of assessing and improving the accuracy of network models it has been emphasized that models must be calibrated and recalibrated regularly. On the basis of the above briefly discussed research work it can be said that there will be only a small amount of residual inaccuracy if the network model is constructed and calibrated properly.

The problem of measurement and pseudo-measurements uncertainty and their influence on accuracy of the network equation solution has received, also, a significant attention in the literature (Gabrys & Bargiela, 1995; Powell et al., 1988; Sterling & Bargiela, 1984; Bargiela, 1984; Bargiela & Hainsworth, 1989; Carpentier & Cohen, 1993; Powell, 1992).

One major source of inaccuracy in water network modeling and simulation is how to allocate the water demand through the system. Determining nodal demands is not a straightforward process like collecting data on the physical characteristics of a system. Some data, such as billing and production records, can be collected directly from the utility but are usually not in a form that can be directly entered into the model. There have been many suggestions for how nodal consumptions should be modeled and predicted (Suter & Newsome, 1988; Wright & Cleverly, 1988) that involves modeling different types of consumption (e.g. domestic use, industrial use etc.) separately and combining them to represent the overall nodal consumption. Short-time water demand prediction methods based on time series analysis can be found (Chen, 1988; Quevedo et al., 1988). However there are water systems where instead of metering individual customers, the distribution systems are divided up into smaller zones, called *District Metered Areas* (DMAs). These zones are isolated by valving and are fed through a smaller number of inlet and outlet meters (WRc, 1985). The number of properties in a DMA is known fairly precisely and the flows are recorded using data logging technology or telemetered to a central location. Within those DMA by studying the demand of a large number of consumers then is possible to determinate reasonably the water demands for other groups of consumers that have the same characteristics (i.e. type of consumer, socioeconomic background, etc.) as the first group. A similar study is carried out in the first part of Chapter 3 for the nodal consumptions of a general water network.

Although it is possible to study the patterns of water consumption of a few customers in detail and extend the conclusions of that study to the rest of the system, this type of data extrapolation can carry a significant amount of inaccuracies. Therefore another way of diminishing the effect of inaccuracies in measurements and pseudo-measurements on the solution is to redefine the method of solving the network equations. Since the number of unknowns is equal to the number of equations in the linearized model (Eq. 2.25) of the network equations (Eq. 2.23) used in the Newton-Raphson method, each inaccurate data has a huge influence on the solution. It could even lead to the case when there would be no solution to the set of equations (Eq. 2.23) at all. The more robust method, known as the state estimation procedure (Gabrys & Bargiela, 1995; Powell et al., 1988; Sterling & Bargiela, 1984; Bargiela, 1984; Arsene & Bargiela, 2001; Arsene & Bargiela, 2002a), utilizing all available measurements and pseudo-measurements is, therefore, used. Using all available information results in constructing overdetermined set of equation (number of equations is grater than the number of unknowns) and the solution can no longer be found by simple solving a square set of equations. The solution finding problem has to be defined as the optimization of a suitable chosen cost function. On the other hand, these additional measurements, also known as redundant measurements since they are not absolutely necessary to arrive at some solution, allow the corrupted data to be rejected and to obtain a more reliable solution. The estimation procedure will be based on the loop corrective flows variables and is discussed in more detail in Chapter 4.

The methodology and algorithms for quantifying the impact of measurements and pseudo-measurements inaccuracies on the state estimate vector in water distribution systems were first introduced in (Bargiela & Hainsworth, 1989; Hainsworth, 1988) under the name of *confidence limit analysis*. The concept has been further investigated in (Gabrys & Bargiela, 1996; Gabrys, 1997) yielding a number of confidence algorithms that were based on the nodal heads variables for building the network equations. In Chapter 5 new confidence limit procedures are developed (Arsene & Bargiela, 2002b) using the loop corrective flows and the accuracy and efficiency of the results will be compared with the solution from the nodal heads based algorithms.

## 2.4. Operational decision support of water distribution systems

The design and development of techniques for operational control and analysis of large-scale water distribution systems has captured the attention of researchers for the last twenty years. In the 1980s, general-purpose hydraulic simulators became commercially available, such as GINAS (Coulbeck et al., 1989), WATNET (Wright & Cleverly, 1988) and more recently EPANET (Rossman, 1994). Initially, the simulators were improved primarily in terms of their numerical computation. The simulators have been used for over twenty years for planning work, but have also been incorporated into on-line operating schemes (Orr et al., 1990; Rance et al., 1993). As the simulator became computationally mature and robust, its software organization, data model and user interface became more important. This was made particularly acute by the rapid advances in computer hardware and information technology. However the theoretical models for these simulators often assumed idealized pipe networks, thus hiding the inherent discontinuous operation of pressure sustained valves and pressure reduction valves. Realistic operational support system must allow the inclusion in the network flows analysis of such elements.

At the beginning of 90's, water companies have come under increasing pressure to improve the performance of their physical assets and their management organization. Many companies started to invest in information systems for automation of decision activities and data processing such as *Geographical Information Systems* (GIS), *Supervisory Control*, *Automation and Data Acquisition Systems* (SCADA), databases and customer billing systems (Johnson, 1993; Johnson et al., 1993).

Modern water production and distribution system have required systematic handling of systems complexities so that more efficient guidelines to be provided for overall operations. Consequently, the requirements on decision support software have greatly increased and were found to be deficient in a number of ways. Most of the water software packages did not communicate with any other package, especially one developed by another vendor. Consequently, they did not exchange or share data and so the user could not readily share the services of other packages. Secondly, the software packages were not designed to interface to any new application or the decision support software usually considered a single network.

*Integrated Operational Systems* have emerged as the key to successful decision support systems for operational control of water distribution systems (Morris, 1998; Ulanicki & Rance, 1998). Tenant, Rance, Ulanicki and Bounds (1998) report on an architecture for integrating water network applications. They developed an on-line operational water network environment for both simulation and optimal scheduling that include a module for communicating with a SCADA system via a third party supervisory system. More recently additional applications were integrated into a new version of this environment (Rance et al., 2001). Other very good examples of such integrated operational systems in water industry can be found elsewhere (Cameron et al., 1998; Vigus, 2001; Alzamora et al., 2001).

A neurocomputing system for operational decision support in water distribution networks has been outlined (Gabrys & Bargiela, 1998; Gabrys, 1997; Gabrys & Bargiela, 1999; Bargiela et al., 2002). An analog neural network was used to calculate the water network state estimates. This was followed by the confidence limit analysis of the calculated state estimates and a General Fuzzy Min-max neural pattern classification/clustering system (Gabrys & Bargiela, 2000). The overall system was used for fault detection and identification in water networks based on the examination of patterns of state estimates. In Chapter 7 the neuro-fuzzy system is adapted and extended so that to include the new procedures developed for state estimation and confidence limit analysis based on loop equations. The integration of the basic water network simulation program with the confidence limit analysis and neural classification/clustering modules will deliver operational decision support functionality to water systems operators.

# Chapter 3

# Loop-based simulation of water networks using nodal modeled water consumptions

## 3.1. Introduction

Water network simulation provides a fast and efficient way of predicting the network behavior, calculating flows, velocities, head-losses, pressures and heads, reservoir levels, reservoir inflows and outflows and operating costs.

In the last twenty years the capabilities of the hydraulic simulation module have been continuously extended based on improved mathematical and computer-programming techniques as well as advances in computer hardware. The first computer programs for water network simulation were developed and implemented in the 1960s. Two programs that were widely used are Shamir-Howard's program (1968) and the Epp-Fowler's program (1969). Both programs made use of the Newton-Raphson technique for calculations but while the first was based on a node-continuity formula, the Epp-Fowler's program was loop oriented. The historical development of the simulator algorithms continued with the programs developed by Wood and Charles (1972), Jepson and Davis (1976). Significant improvements in the capabilities of network simulation packages began to appear in 1970s. The software could simulate all of the components of a water system, including pump stations, pressure-regulating valves, check valves and reservoirs. Extended time simulations became available as well (Rao et al., 1977a; Rao et al., 1977b).

Microcomputers of the mid-1980s were more powerful than large mainframe computers of the 1950s and 1960s and, therefore, were capable of solving complex mathematical problems. Popular network simulation programs of the 1960s and 1970s were converted to run on microcomputers, and many new programs were introduced by software vendors and consultants (Orr & Coulbeck, 1988; Coulbeck & Orr, 1984; Coulbeck et al., 1985; Wright & Cleverly, 1988). Attention began to shift to expanded capabilities, such as database management, network graphics, pumping-energy cost calculations, sizing optimization routines (Orr et al., 1988; Coulbeck & Orr, 1982).

In the last two decades, hydraulic simulators became powerful decision-making tools enabling the engineers and the scientists to analyze and manage distribution networks with unprecedented accuracy and efficiency. The hydraulic simulators come now integrated with a complete geographic information system, they provide data management tools necessary for handling the enormous amount of data required for hydraulic modeling, they hold as well very good graphical presentations and diagnostics capabilities, analyze and perform steady-state and extended time simulations for very large water distribution systems. Example of such complete commercial simulators are EPANET (Rossman, 1994), *FINESSE* (Rance et al., 2001), *WATERCAD* (Haestad, 2002) or *InfoWorks* (HR Wallingford, 2002).

The numerical method that is used to solve the flow continuity and headloss equations in some of these simulators is based on a nodal heads formulation (Todini & Pilati, 1988; Salgado et al., 1988; Rossman, 2000) that for small and medium size water networks proved to be numerically stable. Moreover the simulator algorithms necessitate some input data that include network topology and components parameters, and water demands. One would like to minimize the amount of information to be provided to the simulator algorithm and this made the nodal heads formulation to be preferred to other formulations when developing network flows analysis packages (Todini & Pilati, 1988).

However, with the increased computational power of the computers in the last decade, the requirements of the simulator algorithms in terms of input data, has stopped to represent a major drawback in the modeling and simulation of water networks. This produced a revival of the simulators based on the loop corrective flows variables and combined with the Newton-Raphson numerical method. This combination is called the *loop simulator* (Epp & Fowler, 1970; Gofman & Rodeh, 1982). The input information, that is generally required by this class of simulators, is obtained now with computer based graph search operations and stack-oriented data structures (Rahal, 1995; Arsene & Bargiela, 2001; Andersen & Powell, 1999b). The aim of using the loop simulators in network flows analysis is to benefit from the *smaller solution matrices* which relate to the loop structure rather than the nodal structure and to avoid the poor convergence reported in some of the case studies where the nodal heads simulators were employed (Nielsen, 1989; Donachie, 1974; Hartley & Bargiela, 1993; Hartley, 1996).

While the full potential of hydraulic simulators for water networks can only be realized with appropriate computer hardware and software solutions, their numerical results can be fully exploited by the human operator only if great level of accuracy exists in the data provided to the algorithm with special regard to the predictions of water consumptions.

The consumption of water is the driving force behind the hydraulic dynamics occurring in water distribution systems. Determining the water consumptions is not a straightforward process like collecting data on the physical characteristics of a system. For example billing and production records can be collected directly from the utility but are usually not in a form that can be directly entered into the simulator algorithm. Moreover establishing *water consumptions* is a process requiring study of the past and present usage trends, and, sometimes, the projection of future ones. While the water consumptions are determined, the water use has to be spatially distributed as demands assigned to the network nodes. It can be concluded that in order to obtain an accurate representation of the system performances, water consumptions must be allocated geographically and throughout the day that is diurnal-demand allocation.

In the next section the modeling of the water consumptions is performed for a general distribution system that will enhance the ulterior results obtained with the loop simulator algorithm.

## 3.2. Water consumption

In network simulation software, nodal demand values are assumed known and are used as the given variables from which the other hydraulic variables are derived.

Usually the process of modeling the nodal water consumption starts with determining the *baseline demand* that is commonly represented by the average day demand in the current year. To this baseline demand a dimensionless demand factor is applied at each time increment that represents the amount of time between measurements and has a direct correlation to the resolution and construction of the diurnal-demand curve: as for example if measurements are available hourly, then hourly averages can be used to define the pattern over the entire day.

The *demand factors* are derived from information such as property counts and consumer type, and they can be reused at nodes with similar characteristics. The baseline demand is found from several sources such as the water utility's existing data (previous studies or already existing models), system operational records (flow and pressure measurements) and customer meters and billing records.

There are water systems where *District Metered Areas* (DMAs) are dividing the networks up into separately metered areas that are giving information about demands based on the total inflow pumped into these smaller units of the distribution networks. Within those DMAs, the distribution factors can be used to allocate the overall water consumption through the network. Although the distribution factors are only proportional guesses, this information can be sometimes more reliable than the baseline demand from the non-DMA networks. This is because the baseline demands typically include both customer demands and flows that are lost in the system such as leakage or flow measurement errors. These are called unaccounted-for water and it means that the user does not know where to place it in the network.

Methods for determining the baseline demand have been developed (Cesario, 1995; AWWA, 1989). They are based on information coming from the full billing records and the overall water production. However in spite of this data, sometimes it is impossible to know with absolute certainty how much water is being used in a short period of time and how much water has been lost. Therefore studies have been carried out in order to develop stochastic models for the various types of water consumptions such as residential water consumptions (running the washing machines, using the shower or the sink). The methods have been verified with data collected from individual customers (Buchberger & Wu, 1995; Buchberger & Wells, 1996; Bowen et. al, 1993). Other investigations into the patterns of the water use have been realized (Flack, 1982; Linaweaver et al., 1966; Butler, 1991). Such an approach is presented in the next section where the patterns of water usage of a group of customers are investigated and then applied to other customers with similar social and economical characteristics. For a residential house the socio-economic characteristics can be the number of occupants, household income or water price.

### 3.2.1. Modeling the water network nodal demands

The total water consumption allocated to a node can be determined by flow measurements, customer meters, billing records or even previous studies of the respective water network. If such data is available on a node-by-node basis, then the modeler will be faced with a simple task of using this data without any other preparations. However due to the cost of metering, obtaining such information is justifiable only for large customers and a sampling of smaller ones. Based on this sampling, the results can be extended further to other similar water consumers, increasing the accuracy of the overall nodal consumptions.

In Figure 3-1 is displayed a typical *diurnal-demand curve* for a residential area of 1000 thousand people that is plotted as flow rate versus time and is modeled as an average nodal consumption of 2.4 l/s.

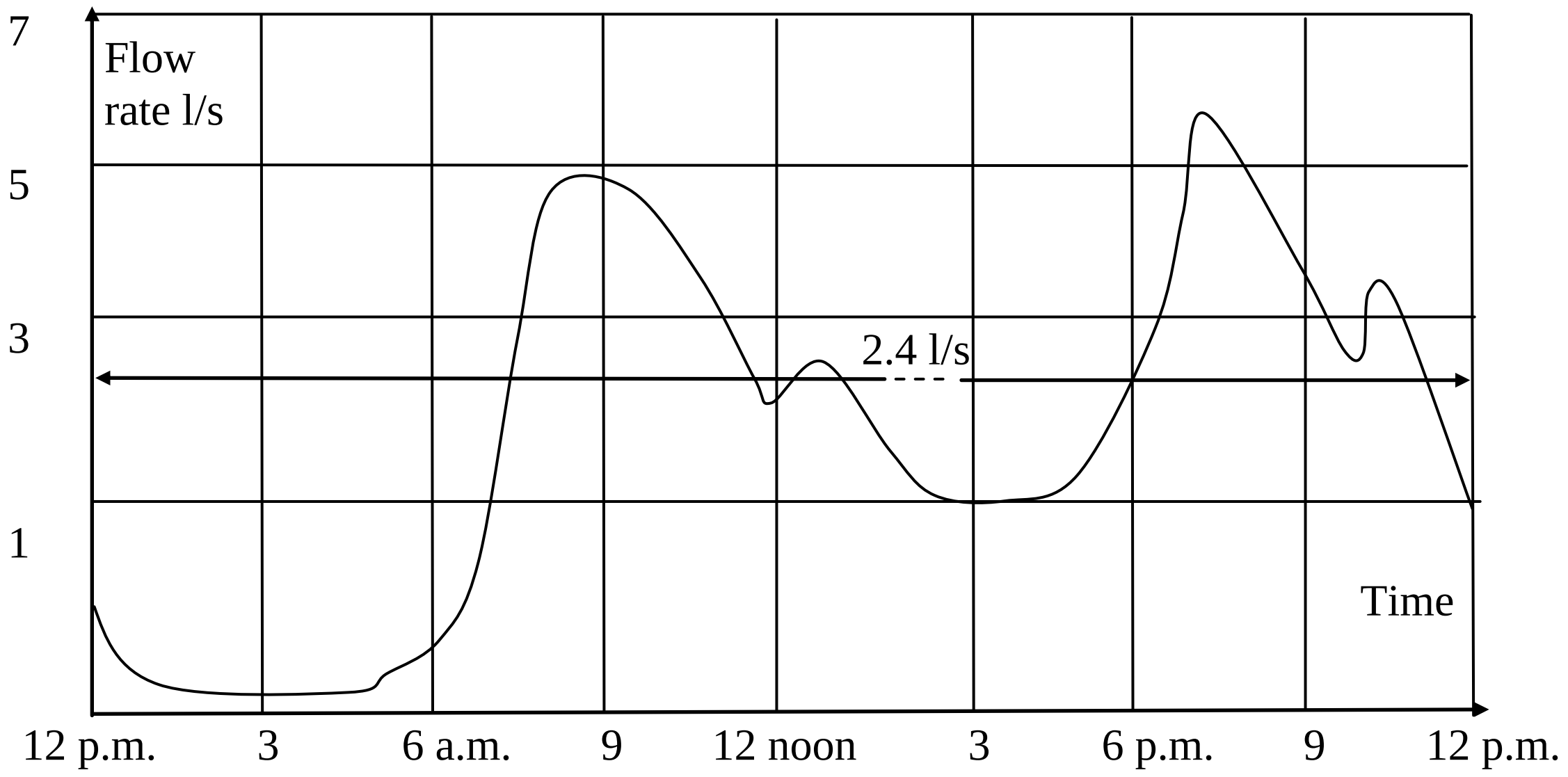

Figure 3-1: Typical diurnal demand for a residential area of 1000 thousand people.

By carrying out a comprehensive survey as well as using customer meters is possible to characterize the intensity, the duration and the frequency of the *water use* for the various fixtures and appliances that exist at a residential house. In this work there have been chosen three types of distributions in order to represent the water use: rectangular, triangular and trapezoidal distribution. Each of the distributions is described by the following parameters: the duration and the intensity of flow, and the frequency of a given distribution within a time interval. In Figure 3-2 are displayed the rectangular, triangular and trapezoidal distributions together with their parameters.

The modeling of the water consuming process by the three distributions is based on the characteristics of the water consuming process itself. As for example, the water use while flushing a toilet can be described as a short time process that involves a significant amount of water. Therefore a triangular distribution with a peak flow of 0.3 l/s and a total duration of 8 seconds is suitable to represent the water use in this case.

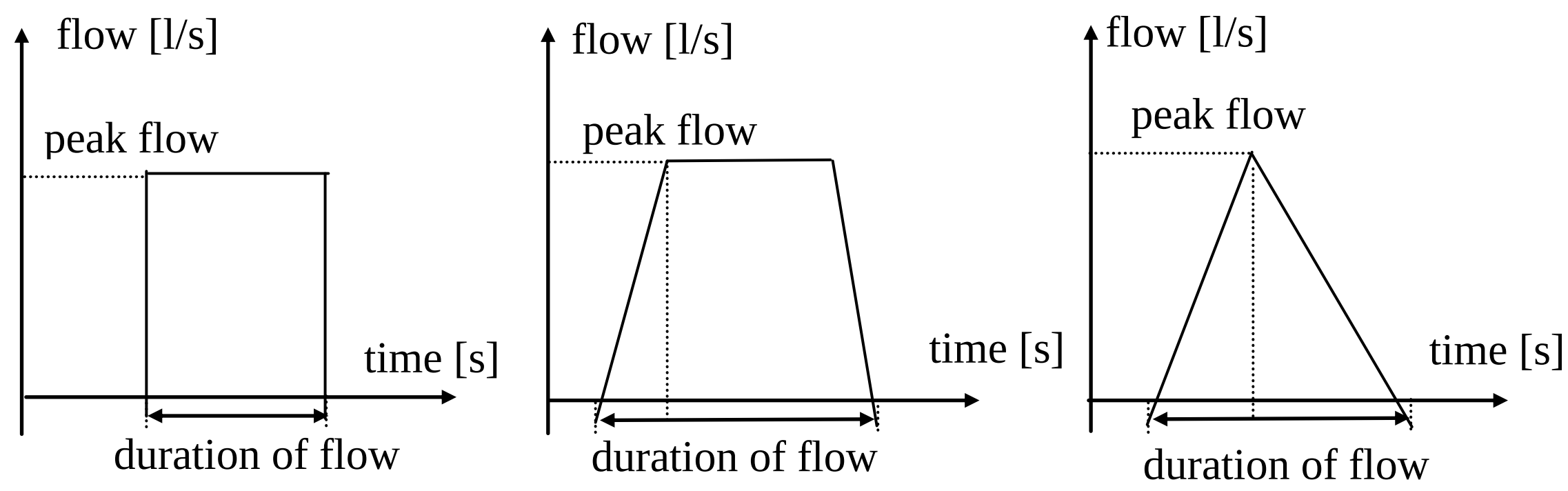

a) Rectangular distribution b) Trapezoidal distribution c) Triangular distribution

Figure 3-2: Types of distributions.

In order to investigate how the overall water consumption can be described by these distributions, a group of 6 people living together in a house had their water use habits scrutinized. In Table 3-1 are displayed the water consumptions for the group of 6 people, together with the number and the frequency of the distributions characterizing their water use habits, as well as the time intervals when they occurred.

| Use Category | Rectangular distributions | | | | Trapezoidal distributions | | | | Triangular distributions | | | |
|---|---|---|---|---|---|---|---|---|---|---|---|---|
| | night | morning | afternoon | evening | night | morning | afternoon | evening | night | morning | afternoon | evening |
| Sink | 2 | 5 | 3 | 5 | 2 | 3 | 2 | 4 | - | - | - | - |
| Shower | 1 | 5 | - | 3 | - | - | - | - | - | - | - | - |
| Toilet | - | - | - | - | - | - | - | - | 4 | 14 | 2 | 8 |
| Leakage | 2 | 2 | 2 | 2 | 3 | 4 | - | 4 | - | - | - | - |

Table 3-1: Modeling the residential water use with rectangular, trapezoidal and triangular random distributions.

A rectangular distribution with the intensity of flow of 0.15 l/s and the duration of 2 minutes can represent the use of a sink during the morning. It can model also a continuous leakage for 24 hours at the flow rate of 0.0005 l/s. The same logic is followed for the other types of distributions. Therefore not only the water use habits can be described by different distributions, but the distributions of the same type can have different values for their parameters (i.e. duration and intensity of flows). This makes the modeling and simulation of the water use versatile and corresponds to the real life situations where for example a leakage can be spread throughout the day at a constant flow rate, and therefore modeled as a rectangular distribution, but can also be associated with the temporarily used of a sink and then represented as a trapezoidal distribution. The entire data obtained from the survey, have been translated into groups of distributions that were assembled together and randomly spread within the given time intervals. By virtue of a random process, it is unlikely that more than one distribution will start at the same time. Moreover, owing to the finite duration of each distribution, it is possible that two or more distributions with different starting and ending times, and different intensity of flow, to overlap for a limited period. When this happens, the total water use at the residence is the sum of the individual intensities from the coincident distributions. Finally, this produces a stochastic model for estimating the overall house water consumption and the resulted diurnal demand curve is shown at Figure 3-3.

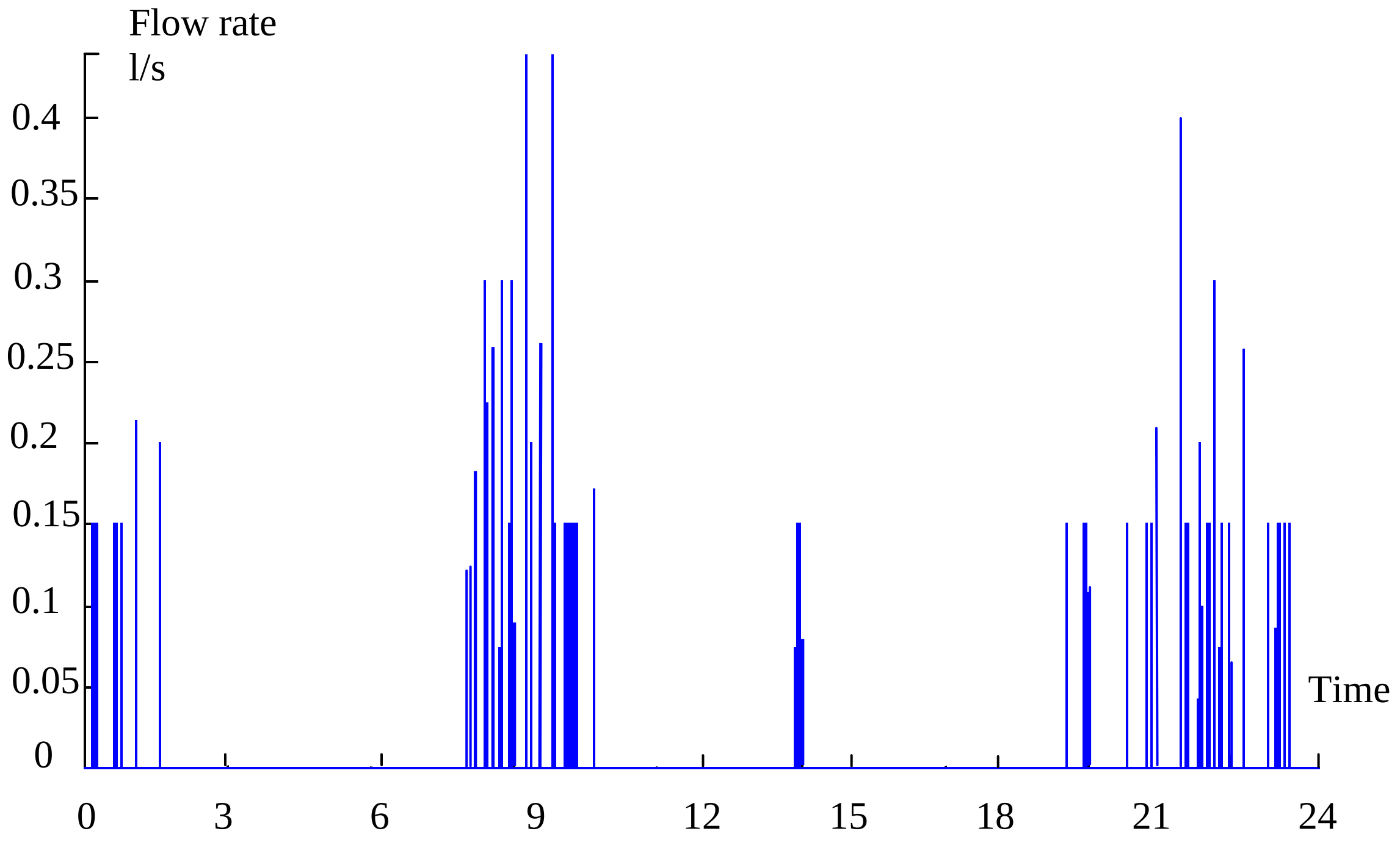

Figure 3-3: Daily pattern of water use at a 6 people residence (Vertical lines are hypothetical water consumptions).

The plot shows about 55 demands with the maximum rate approaching 18 l/min. Periods of peak activity are clustered around 9 A.M. and again around 22 P.M. Most of the time, however, this residence uses no water.

The 6 people residence is selected as the elemental unit for analysis and the results are extended to multiple residences that in total count 1000 people. The predicted demand profile for this large community is shown with continuous line at Figure 3-4.

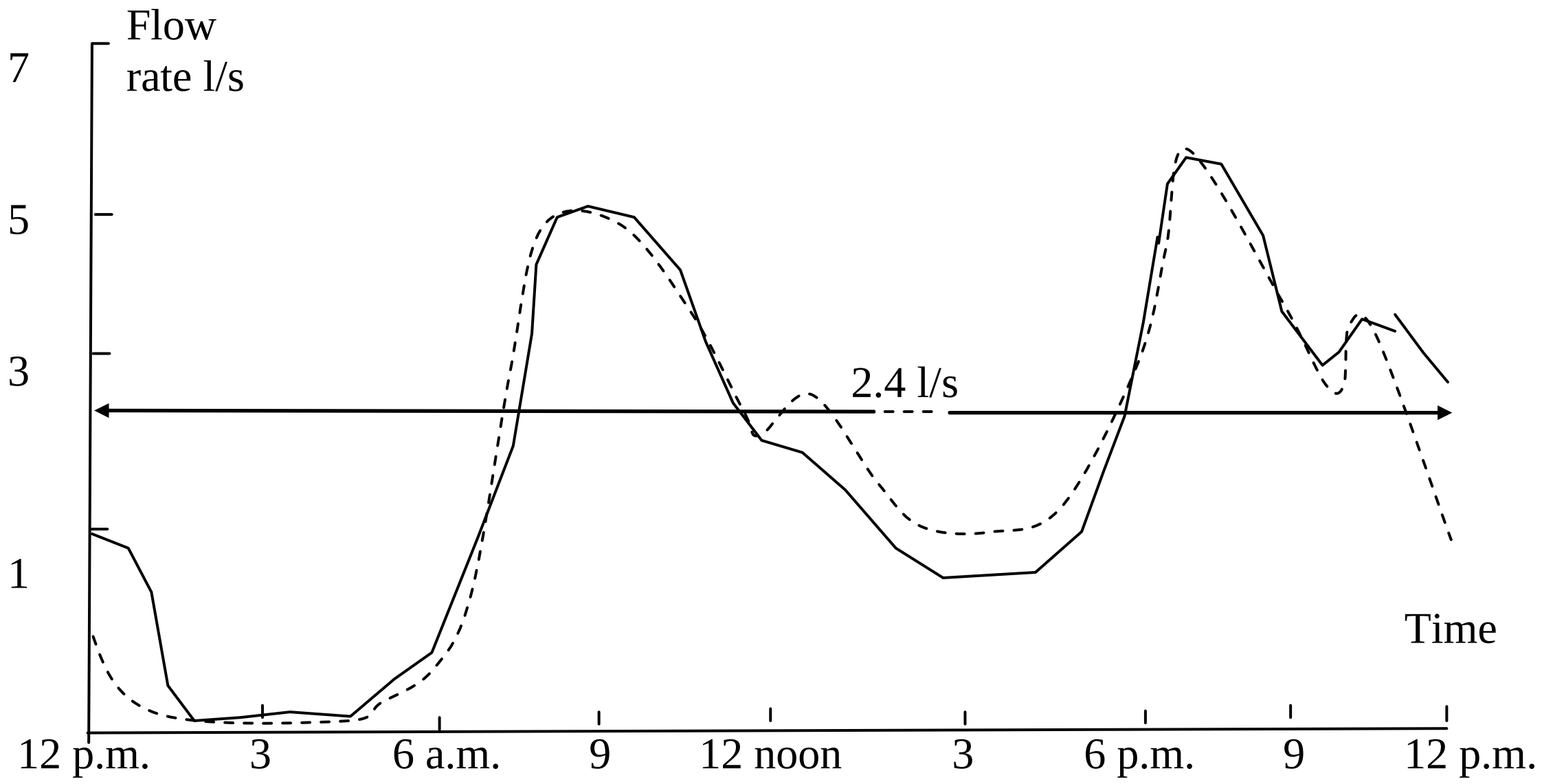

Figure 3-4: Estimated diurnal demand for 1000 people.

The derived domestic demand profile for this large area of 1000 people will require further modification to reconcile measured demands shown with dashed line at Figure 3-4 and predicted consumptions obtained through our study. This will require a reduction of the number of distributions in the interval 12 pm to 2 am and a larger number of the distributions for the afternoon. Finally the modified adopted domestic profile will contain reliable information about the characteristics of the daily water use of the group of 1000 people.

The final scope of building a stochastic model is to apply it for other customers from the distribution system that exhibits similar patterns of water consumption. Through the comparison of the simulated water demands by the stochastic model, versus the information coming from available meters or customer bills, it will be able to increase the accuracy of the final values of the nodal consumptions and to localize the problems to the parts of the water network where they are coming from.

For example, if the metering information shows an increasing of water consumption in a network node during the night (i.e. continuous line 12 p.m. to 6 a.m. at Figure 3-5) when compared to the values expected from the previous studies and the stochastic models (i.e. dashed line at Figure 3-5), then this could be a good indication that there exist a leakage in the area modeled by the network node.

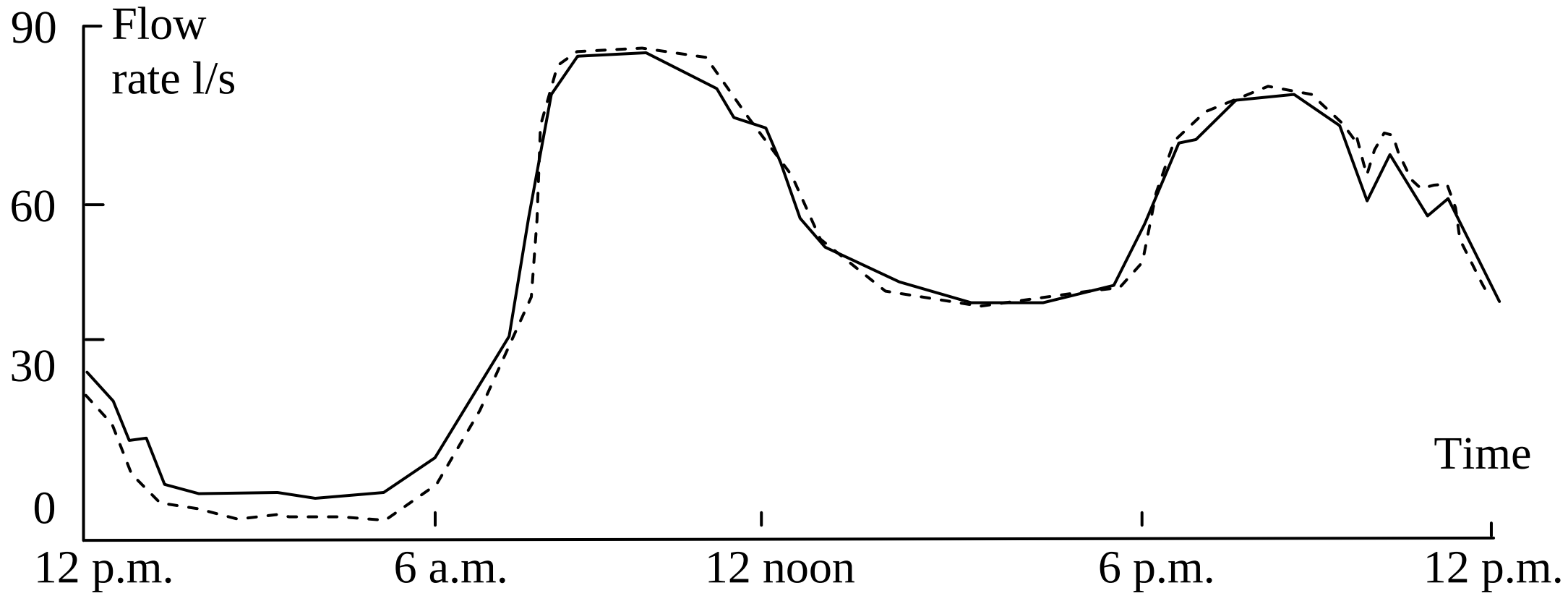

Figure 3-5: Differences between the predicted water consumption and metered demand.

The work presented here is being motivated by the need to provide a more realistic picture of the nodal consumptions of a distribution system. Such information can improve the reliability of the water network model that through simulation intends to predict the behavior of the real network.

The present formulation is limited to the use of simple distributions randomly spread within some time intervals as a way of describing the water consumptions for a 6 people residence house. The results are then extended for multiple residences with the same social and economic characteristics. The discrepancy between the expected values for the water use and the available metered information can represent an indication of some anomalies (e.g. leaks) that might exist in some parts of the network.

Although it is possible to study a few customers in detail and extend the conclusions of that study to the rest of the system, this generalization has some inherent dangers. The probability of selecting the “best” representative customer for all the system is small, and any deviation from the norm of error in nodal consumptions will sum up when it is applied to an entire water system. Moreover there are cases in which the use of a representative customer is inappropriate under any circumstances: demands for large consumers such as hospitals, hotels should be individually determined and the volume and the patterns of their usage should not be applied to other consumers just because they are neighbors within a geographical zone. Therefore some other specific methods, state estimation and confidence limit analysis, will be used later in order to improve the accuracy of the overall operational status of the water system. As for now, this study showed how the uncertainty in the water nodal consumptions can be reduced by comparison of groups of consumers with the same socio-economic background. In the next section the water consumptions together with the other physical characteristics of the system will be used as input data in the simulation process of a water network.

## 3.3. The simulator algorithm

A *water system simulation* represents a snapshot in time and is used to determine the operating behaviour of the water network for a set of nodal demands. The simulation algorithm shown here is based on a *co-tree flows* formulation, which is derived from the loop corrective flows algorithm, defined for a water distribution system with *n*-nodes, *l*-loops, and *p*-pipes.

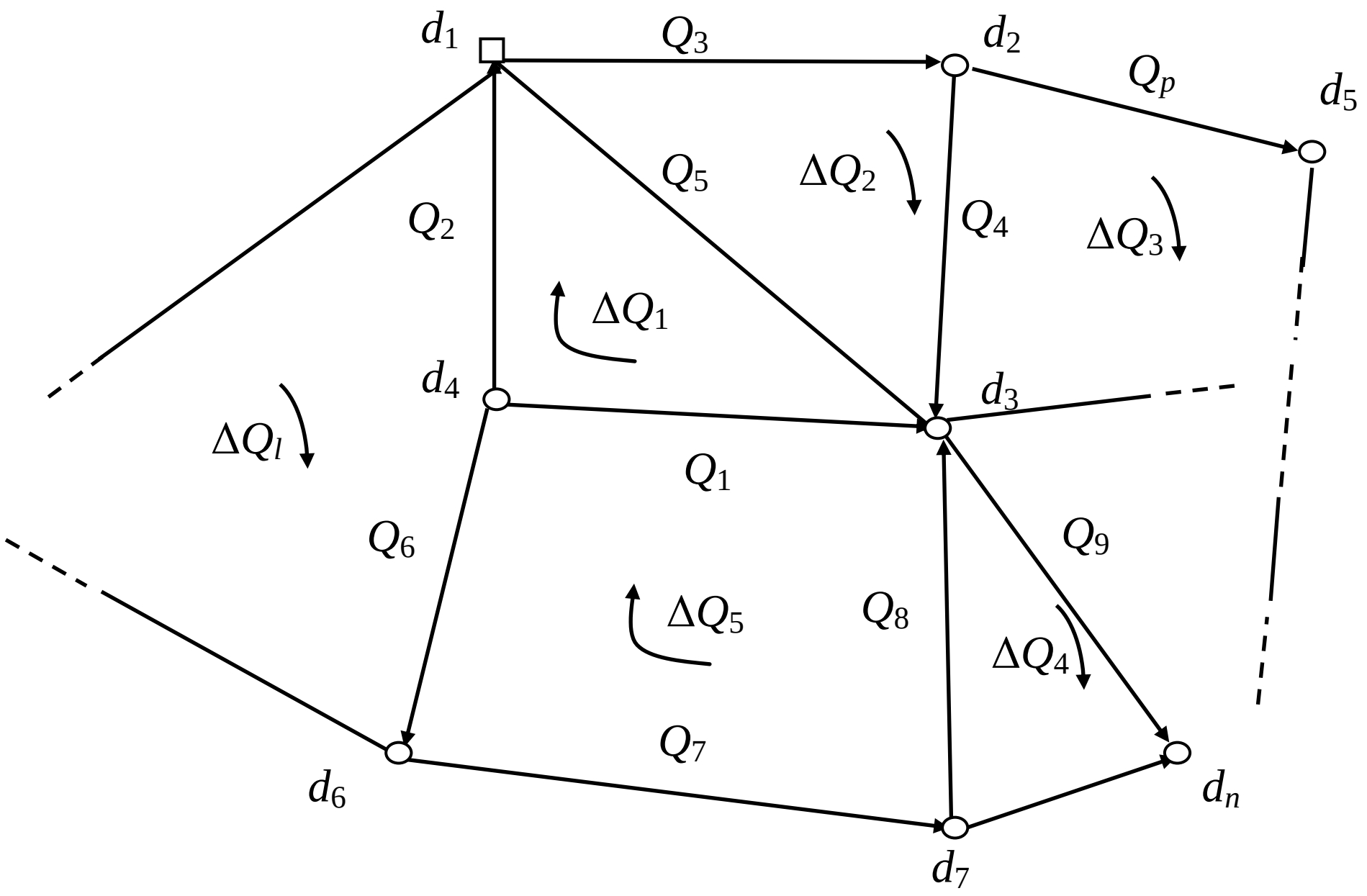

Figure 3-6: General water network with *n* - nodes, *p* - pipes and *l* - loops.

An *initial solution* $Q_i$ that satisfies the continuity equation is calculated as:

$$A_{np}Q_i=d \qquad \text{(Eq. 3.1)}$$

where $d$ ($n$ x 1) is the *vector of nodal demands* and $A_{np}$ ($n$ x $p$) is the *topological incidence matrix* that has a row for every node and a column for every branch (component) of the network. The non-zero entries for each row +1 and –1 indicate that the flow in pipe $j$ enters or leaves node $i$.

$$A_{np}(i,j)=\begin{cases} 1 & \text{if flow of pipe } j \text{ enters node } i \\ 0 & \text{if pipe } j \text{ is not connected with node } i \\ -1 & \text{if flow of pipe } j \text{ leaves node } i \end{cases}$$

The energy equation has to be satisfied, that is the *vector of loop head losses* residuals $\Delta H$ ($l$ x 1) must be equal to zero:

$$\Delta H = 0 \qquad \text{(Eq. 3.2)}$$

The vector of loop head losses residuals $\Delta H$ is calculated as:

$$\Delta H = M_{lp}\, h \qquad \text{(Eq. 3.3)}$$

where $h$ ($p$ x 1) is the vector of *pipe head losses* described by the Hazen-Williams equation:

$$h=k\tilde{Q}^n \qquad \text{(Eq. 3.4)}$$

Here $k$ ($p$ x 1) is the vector of *pipe resistance coefficients* and $\tilde{Q}$ ($p$ x 1) is the *vector of flows* that has to be determined. The *loop incidence matrix* $M_{lp}$ is the ($l$ x $p$) matrix with the following properties:

$$M_{lp}(j,k)=\begin{cases} 1 & \text{if flow of pipe } k \text{ flows clockwise in loop } j \\ 0 & \text{if pipe } k \text{ does not pertain to loop } j \\ -1 & \text{if flow of pipe } k \text{ flows anti-clockwise in loop } j \end{cases}$$

Solving equation (Eq. 3.2) with the Newton-Raphson iteration method, the loop corrective flows at the step $t$+1 of the iteration method are:

$$\Delta Q_{l_{t+1}} = \Delta Q_{l_t} - \left[\frac{\partial \Delta H}{\partial \Delta Q_{l_t}}\right]^{-1} \Delta H \quad \text{(Eq. 3.5)}$$

where $\frac{\partial \Delta H}{\partial \Delta Q_{l_t}}$ ($l$ x $l$) is the Jacobian matrix, i.e. the derivatives of the loop head losses residuals $\Delta H$ with respect to the loop corrective flows $\Delta Q_{l_t}$ at the $t$-th step of the Newton-Raphson process. The final solution is obtained for the vector of flows $\tilde{Q}$ :

$$\tilde{Q} = Q_i + M_{pl} \Delta Q_l \quad \text{(Eq. 3.6)}$$

where $M_{pl}$ ($p$ x $l$) is the transpose of the loop incidence matrix and $\Delta Q_l$ ($l$ x 1) is the vector of loop corrective flows.

The *Jacobian matrix* can be expressed also as:

$$J = M_{lp} A\, M_{pl} \quad \text{(Eq. 3.7)}$$

The matrix $A$ ($p$ x $p$) is the diagonal matrix with the property:

$$A = \begin{pmatrix} n\,k_1 |Q_1|^{n-1} & 0... & 0 \\ 0 & n\,k_2 |Q_2|^{n-1} & 0 \\ 0 & 0... & n\,k_p |Q_p|^{n-1} \end{pmatrix} \quad \text{(Eq. 3.8)}$$

where $k_{1,2,..p}$ are the pipe head losses coefficients, and $n$ is the exponent in the Hazen-Williams equation.

### 3.3.1. Loop defining algorithm and calculation of the initial flows

The loop method requires the computation of the loop incidence matrix $M_{lp}$ and the initial flows $Q_i$. There are many ways to choose the loops and to define the loop incidence matrix. These loops can be considered either *fundamental* (Gofman & Rodeh, 1981) or *natural* (Epp & Fowler, 1970). The natural loops are, most likely, what a designer could pick visually, because they are formed by the most neighbouring links to a loop. The fundamental set of loops are the independent loops found in a *spanning tree* and defined by the non-tree pipes.

A spanning tree for a network (with $n$ nodes and $p$ pipes) is a subnetwork (with $n^1$ nodes and $p^1$ pipes) such that $n = n^1$ and the subnetwork contains at least one pipe and no loops, and is said to be connected. A network is connected if for every pair of different nodes $n_1$ and $n_2$, there is a *path* between them. A path represents a finite sequence of nodes and pipes between the initial node $n_1$ and the terminal node $n_2$ and no node or pipe is repeated in the path.

Computer experiments for generating the loops showed that the fundamental loop generator is faster than the natural loop generator (Rahal, 1995). Furthermore the loop flows correction algorithm requires the determination of the nodal heads when the convergence has been reached. The fastest way to determine the nodal heads is to follow the tree structure of the spanning tree by starting with the main source.

The loop method that is using a fundamental loop generator is called the *co-tree method* from the pipes that do not belong to the spanning tree and are called *co-tree pipes* or *chords* (e.g. dashed arrows in Figure 3-7), and are providing the loop information. The co-tree method is employed in this presentation. For the general water network from Figure 3-6, a spanning tree is shown at Figure 3-7.

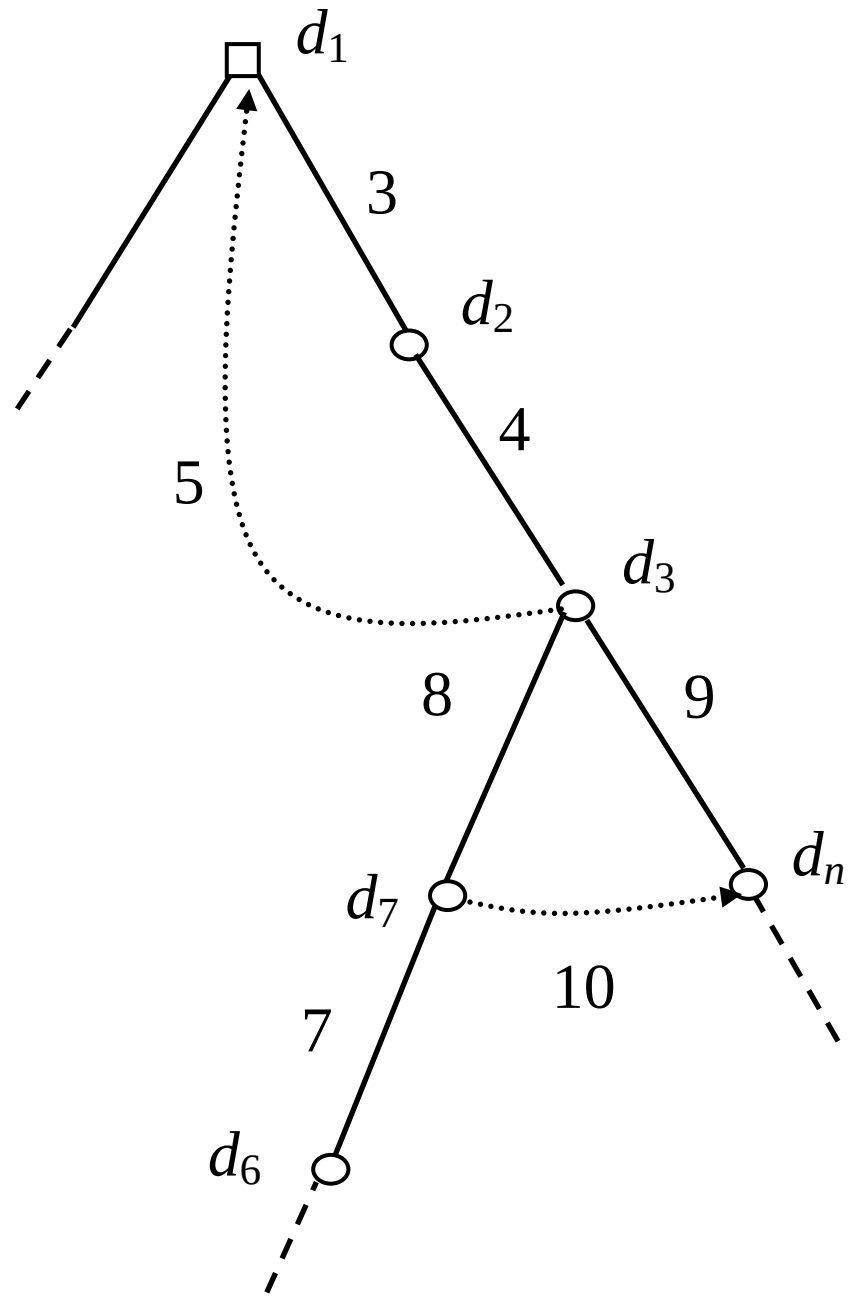

Figure 3-7: Spanning tree for the water network shown at Figure 3-6.

The *fundamental loop generator* works in the following way: all the pipes are assigned an initial ordering according with a hydraulic criterion such as the head loss coefficient. This order is followed while a searching algorithm is used to visit the entire network and construct the spanning tree. During the visiting of the network, the information about nodes and pipes is recorded. The spanning tree is built from a *root node* that has to be a fixed-head node (e.g. node $d_1$ in Figure 3-7). The root node has zero depth. As the search moves to a new node, the respective node is marked as visited. The search is continued and if a previously visited node is encountered it means that a loop has been detected (e.g. pipes 3, 4 and 5 in Figure 3-7).

Different search strategies can be employed in order to produce the spanning tree: *Depth First* (DF) search or *Breadth First* (BF) search. Choosing the right search depends on the particulars of the tree. However, for water networks it has been shown that Depth First (DF) search is more suitable for finding the fundamental set of loops (Gofman & Rodeh, 1982; Andersen & Powell, 1999b; Bounds, 2001); this is based on the property of the DF search that always the pipe that does not belong to the tree (i.e. co-tree pipe), connects a node with one of its predecessor in the tree. Moreover in the DF search algorithm, the nodes adjacent to an already visited node are recursively traversed and therefore the DF search can be implemented as a recursive algorithm.

On the water network used in (Gofman & Rodeh, 1982) new labels are assigned to pipes and nodes during the search of the water network.

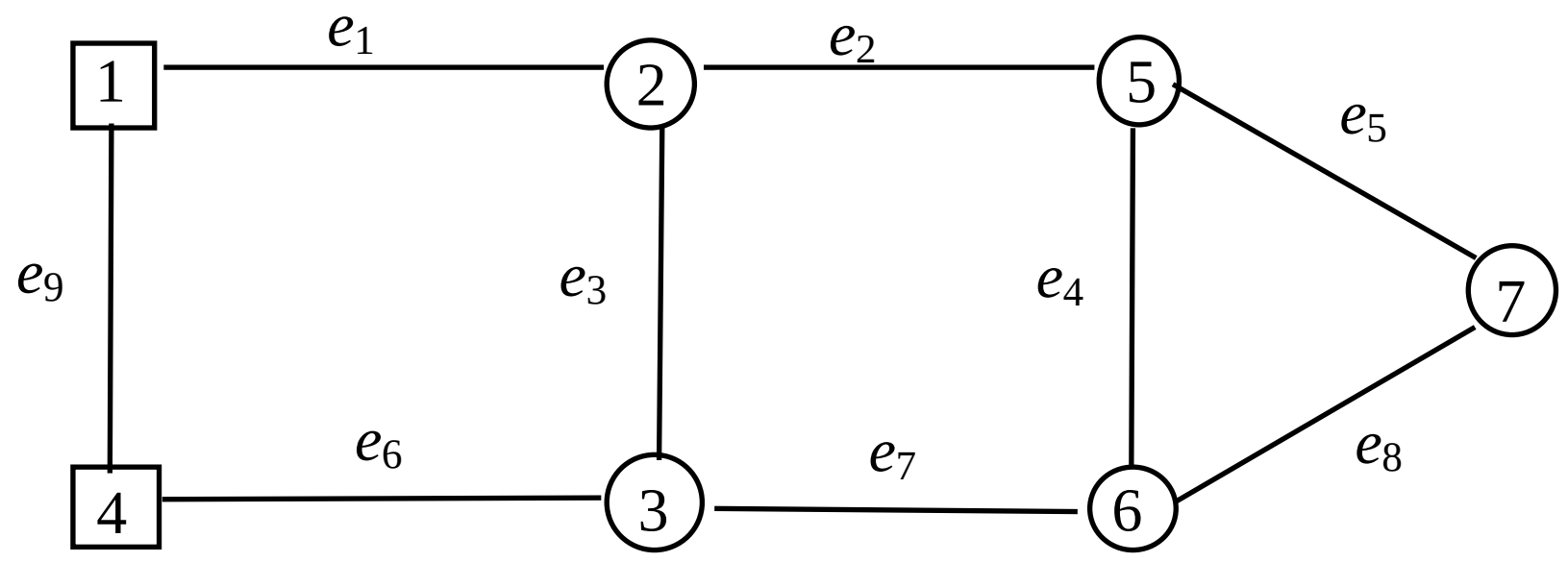

Figure 3-8: Water Network from Gofman and Rodeh (1982).

The spanning tree for the water network from Figure 3-8 is at Figure 3-9. Starting with the fixed-head node 1 the DF search traverses tree pipe $e_1$, and arrives at node 2, further visiting nodes 3 and 4. Edge $e_9$ goes back to the already visited node 1 and it represents a *chord pipe* (or a co-tree pipe). Assuming the direction of pipes the one given in the figure, tracing the loop becomes an easy task: starting with the down node of the chord pipe, the loop tracing algorithm is traversing all the tree edges up, that is pipes $e_6$, $e_3$ and $e_1$, until the upper node 1 of the chord pipe is encountered (Figure 3-9a).

While building the spanning tree and tracing the loops, new labels are assigned to nodes and pipes (Figure 3-9b). Based on the new labels, the DF search produces a one-column matrix: the index of the matrix points to the current node, while the element of the matrix points to the previous visited node from the tree.

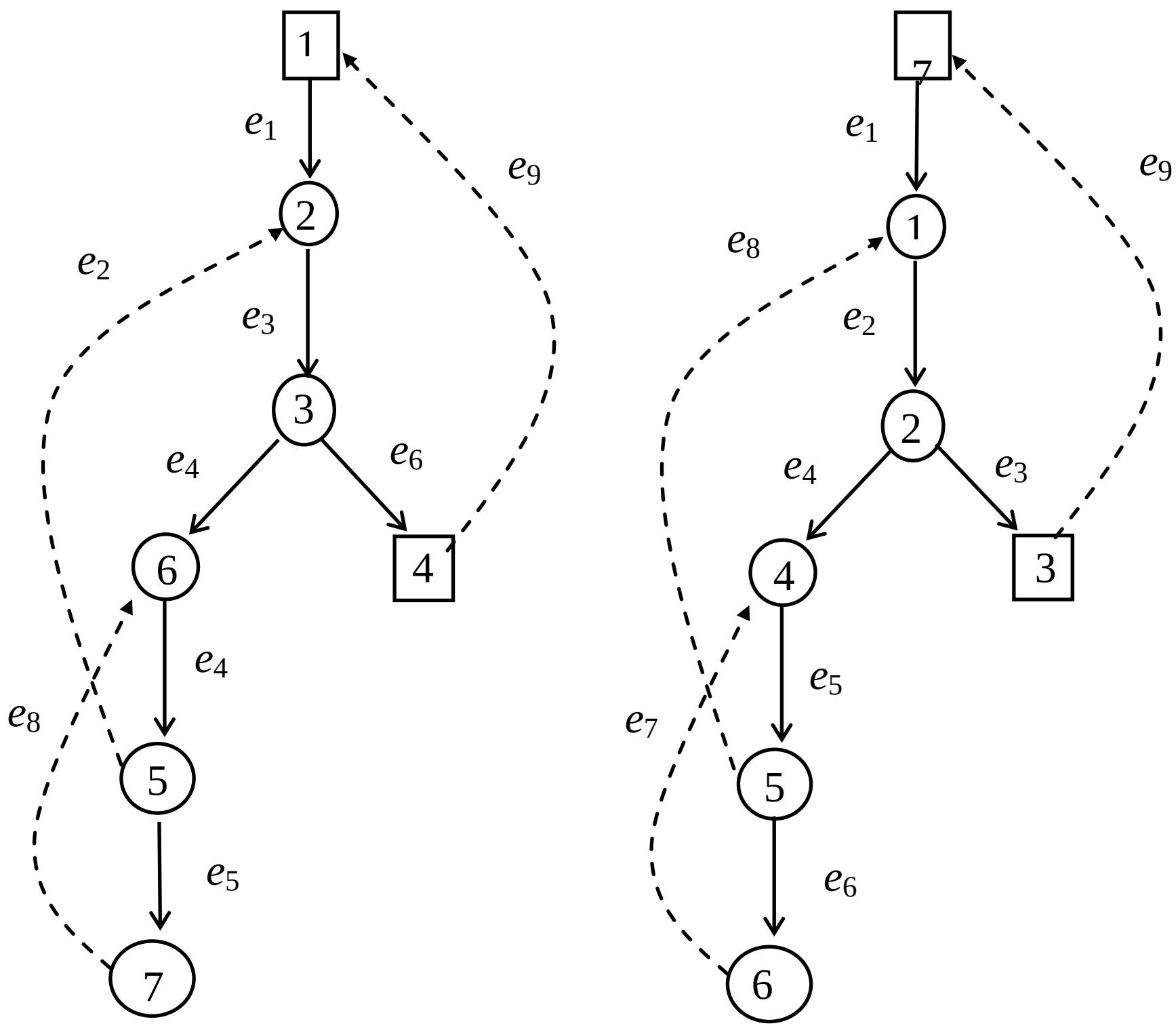

a) b)

Figure 3-9: a) Tree Network, b) Tree Network with new labels for pipes and nodes.

For the spanning tree from Figure 3-9b, the following one-column matrix stands:

$$L = [7\ 1\ 2\ 2\ 4\ 5]^T \qquad \text{(Eq. 3.9)}$$

If the search detects a chord pipe (e.g. pipe $e_9$ connects node 3 to node 7 at Figure 3-9b), we start in matrix $L$ at index 3 (i.e. the down node of the chord pipe $e_9$) which points to node 2. We move then to index 2 in matrix $L$ which points to node 1 and further index 1 points to node 7. Node 7 represents the upper node of the chord pipe $e_9$, and the loop has been obtained, that is the tree pipes $e_3$, $e_2$ and $e_1$, and chord pipe $e_9$. As the tree is built, the loop incidence matrix $M_{lp}$ is obtained together with the topological incidence matrix $A_{np}$: the topological incidence matrix contains an upper-form tree incidence matrix $T$ and a co-tree incidence matrix $C$. Because the head of the root node (i.e. node *7* in Figure 3-9b) is known, it is not included in the tree incidence matrix $T$.

$$A_{np} = [T \quad C] \qquad \text{(Eq. 3.10)}$$

where $T$ ($n$ x $n$) is the *upper tree incidence matrix* and $C$ ($n$ x $l$) is the *co-tree incidence matrix*. The upper form of the tree incidence matrix will have a threefold importance for the numerical algorithms: it will make easier the calculation of the initial flows, it will alleviate the modelling of a leak in a pipe and will improve the accuracy of the state estimator results.

Once the topological and the loop incidence matrixes have been obtained, the initial flows $Q_i$ must be determined. The initial solution in the co-tree method must respect the mass continuity, because the governing system of equations does not account either implicitly or explicitly for continuity. While the initial co-tree flows are zero, the initial flows $Q_i$ in tree pipes can be calculated as follows:

$$Q_i = T^{-1} d \qquad \text{(Eq. 3.11)}$$

A number of authors have investigated the performances of the loop method with respect to the initial flows: closer are the initial flows to the final solution, less iterations are necessary in order that Newton-Raphson method converges.

In (Nielsen, 1989) there were reported difficulties in starting the Newton-Raphson iterations due to the existence of a *singular Jacobian matrix*. If the initial pipe flows in a loop are zero then the sum of the head losses around that loop is equal to zero which gives a singular Jacobian matrix. An algorithm had been developed which was not using an initial flow solution that had to respect the continuity equation. Instead the numerical algorithm was starting with a single Linear Theory Method (LTM) iteration followed by the Newton-Raphson iterations. Rahal (1994) was using a linear head-loss formula during the first iteration of the Newton-Raphson algorithm and the Colebrook-White head loss formula for subsequent iterations.

In order to avoid a singular Jacobian matrix, initial small flows of less than 0.1 l/s are considered in the co-tree pipes that belong to the loops with the head losses equal to zero. Furthermore an enhancement (Arsene & Bargiela, 2002a) is introduced in the Jacobian matrix that speeds-up the convergence of the Newton-Raphson algorithm regardless the initial solution. The enhancement is introduced in the chord flows $\tilde{Q}_C$ that are on the diagonal of matrix $A$ and are calculated with the following formula:

$$\tilde{Q}'_{C_r} = \left[ \frac{\tilde{Q}_{C_r} + sign(H_i - H_j)\left(\frac{\left|H_i - H_j\right|}{k}\right)^{1/n}}{2} \right] \qquad \text{(Eq. 3.12)}$$

where $H_{i,j}$ are the nodal heads at the two end nodes *i* and *j* of the chord pipe, $\tilde{Q}_{C_r}$ is the loop corrective flow in chord pipe *r* calculated from equation (Eq. 3-5) and $sign(H_i - H_j)\left(\frac{|H_i - H_j|}{k}\right)^{1/n}$ represents the Hazen-Williams equation written as flow function of pipe head loss.

Instead of using the Newton-Raphson formula for solving iteratively the loop head losses equations, we calculate a partial solution by solving one equation at a time. Each equation represents the head losses in a single loop and it has to be equal to zero. In the Figure 3-10 is shown how at iteration *t*+1 the corrective flows $\Delta Q^1_{l_{t+1}}$ are calculated for each loop with the following equation:

$$\Delta Q^1_{l_{t+1}} = sign(H_i - H_j)\left(\frac{|H_i - H_j|}{k}\right)^{1/n} \qquad \text{(Eq. 3.13)}$$

*ΔH* has been written as the difference between the entry and the exit nodes in the loops (Figure 3-10). It is obvious that the influence that the loops with common pipes can have one onto the other is overlooked in equation (Eq. 3.13).

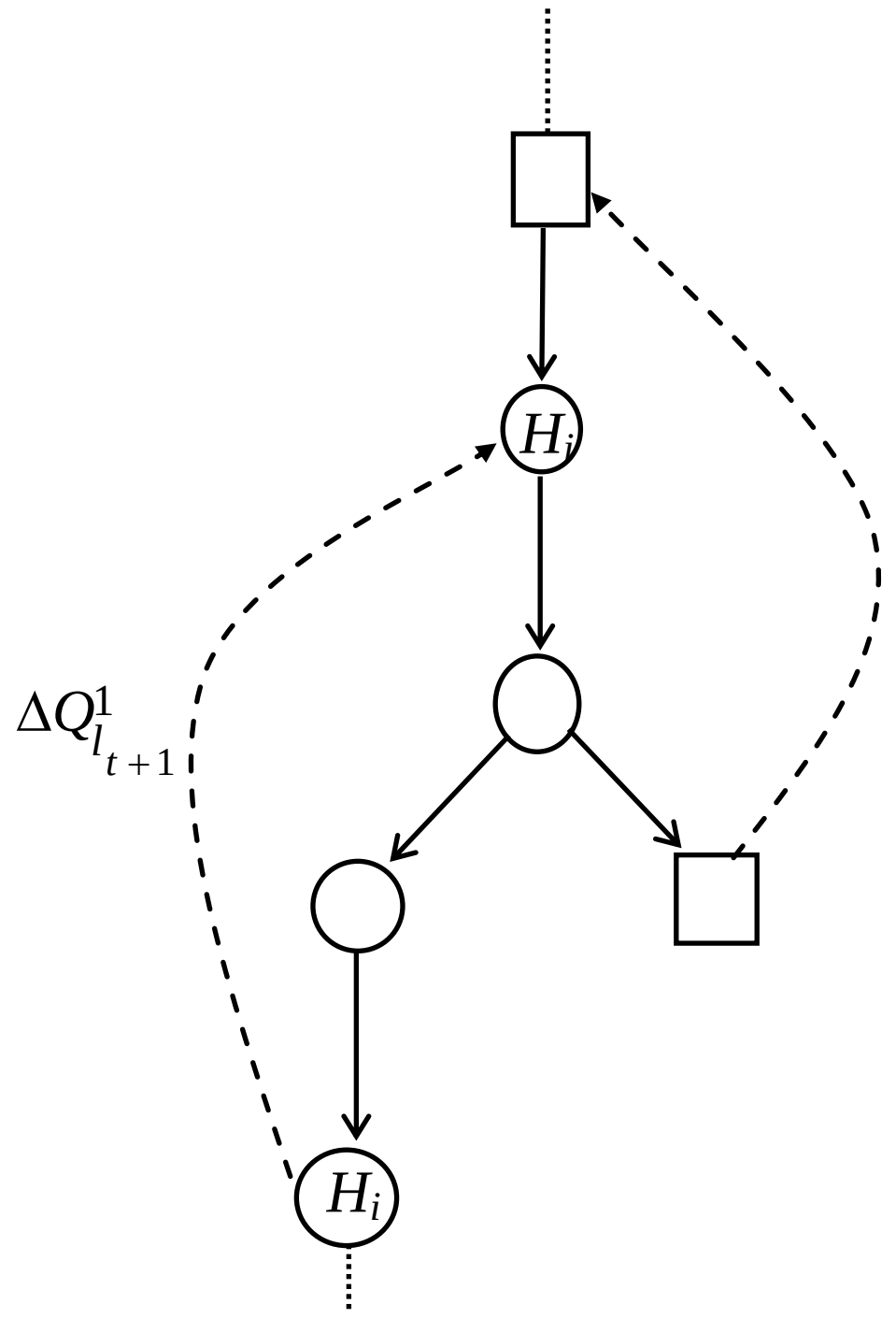

Figure 3-10: The solution $\Delta Q^1_{l_{t+1}}$ of the system of equations $\Delta H = 0$ is found individually for each loop.

We observed experimentally that if we introduced the average sum of $\Delta Q_{l_{t+1}}$ (calculated from Newton-Raphson method) and $\Delta Q^{1}_{l_{t+1}}$ in the Jacobian matrix we could then improve the convergence of the co-tree flow simulator algorithm:

$$\Delta Q^{2}_{l_{t+1}} = \frac{\Delta Q_{l_{t+1}} + \Delta Q^{1}_{l_{t+1}}}{2} \qquad \text{(Eq. 3.14)}$$

A graphical interpretation of the previous equation is shown at Figure 3-11.

Perhaps we should mention here that other authors have used such enhancements in their algorithms that solve water networks. Wood and Charles (1972) in developing their linear theory method used for each pipe flow the average flow rate for the pipe from the past two solutions (i.e. $Q_t = \frac{Q_{t-1} + Q_{t-2}}{2}$). This resulted in a stable algorithm when the successive iterative solutions approached the final solution.

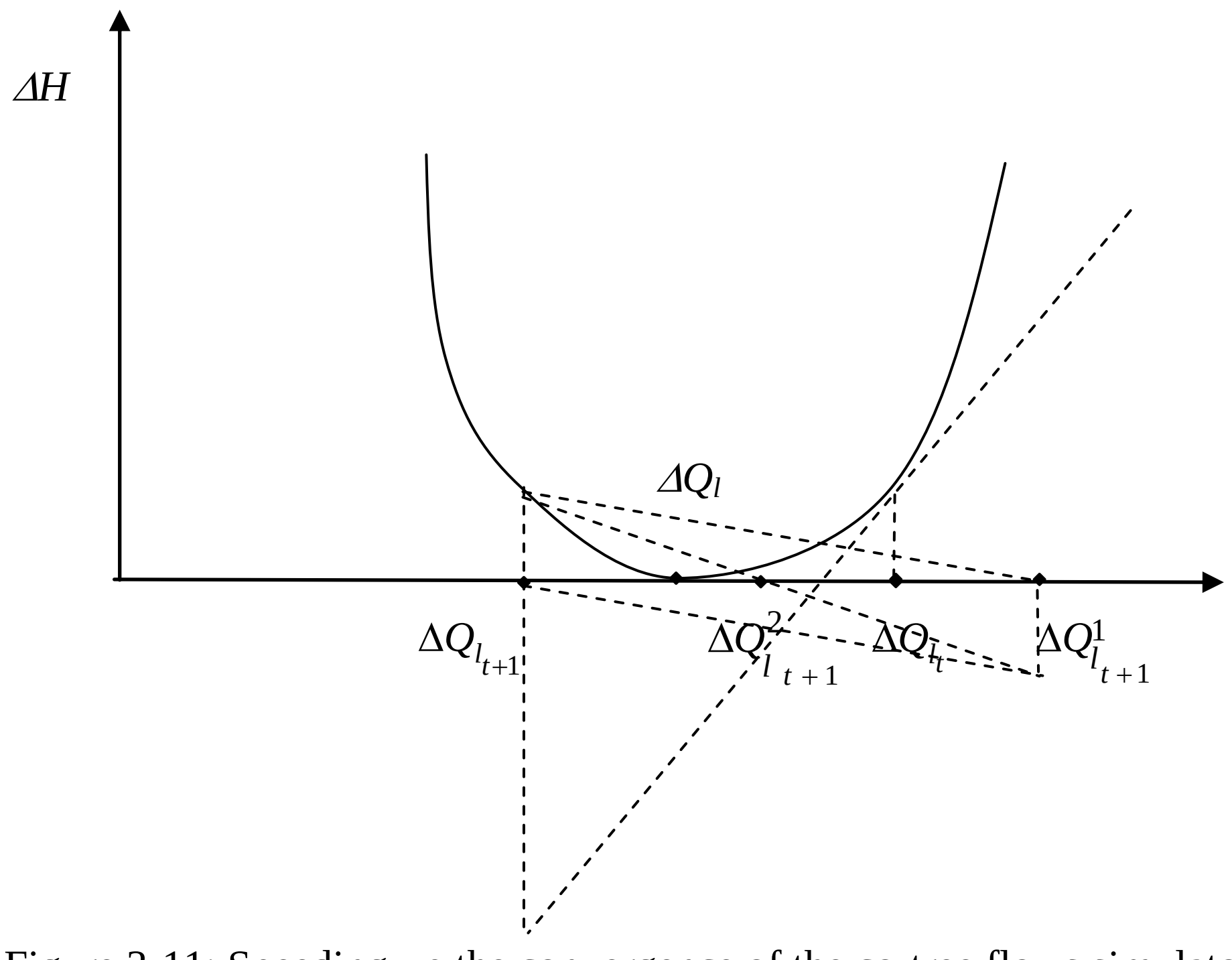

Figure 3-11: Speeding-up the convergence of the co-tree flows simulator.

The software developed by Rossman (1994) (i.e. EPANET) for simulating water networks is based on the algorithm presented by Todini and Pilati (1988) and later Slagado et al. (1988) and called the "Gradient Method". This algorithm updates the pipe flow by using the average flow rate for the pipe from the present and the last solution. To formulate the network equations, a hybrid node-pipe approach is used.

A similar enhancement to the (Eq. 3.12) has been previously presented by Andersen and Powell (1999b) in their co-tree flows simulator algorithm which was expressed as a function of the head losses in the co-tree pipes.

Having defined the topological and the loop incidence matrices and calculated the initial flows, the modified Newton-Raphson algorithm is used to determine the pipe flows. Following this, the nodal heads are obtained with the following equation:

$$H = \tilde{H}_0 - (T^T)^{-1} h_T \qquad \text{(Eq. 3.15)}$$

where $\tilde{H}_0$ is the vector of ($n$ x 1) size with the fixed-head root node $H_0$ on each entry in the vector. Matrix $(T^T)^{-1}$ (i.e. the inverse of the tree incidence transpose matrix) describes the tree structure between the root node $H_0$ and the nodal heads $H$ (e.g. nodal heads 1 to 6 at Figure 3-12). Vector $h_T$ represents the head losses in tree pipes.

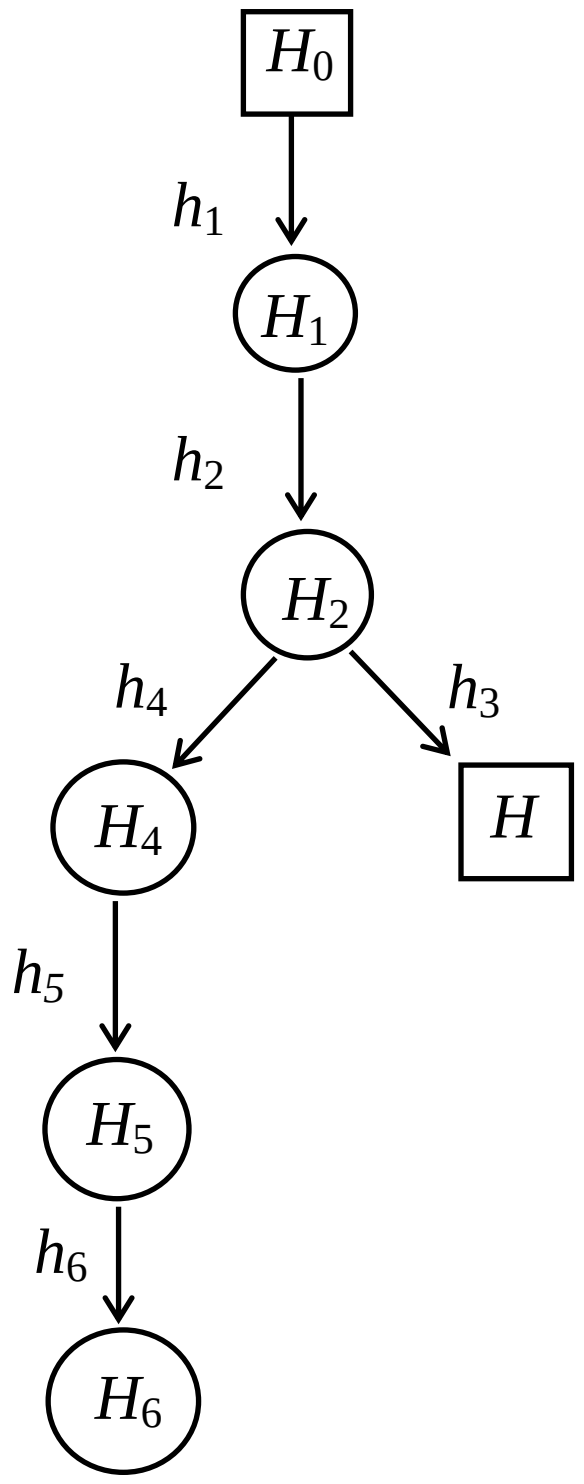

Figure 3-12: Calculation of the nodal heads based on the tree structure.

Before presenting some numerical results, let us point out the reasons for which the co-tree flows algorithm is superior to the nodal heads simulator.

During the Newton-Raphson iterative process, it may happen that one or more pipe flows to be close to zero $\tilde{Q} = 0$ or, equivalently the pipe head loss $h=0$. In the nodal heads formulation, the Jacobian matrix expresses the continuity equations at the nodes of a network and zero pipe flows may result in a singular Jacobian matrix. By contrast, in the co-tree flows simulator the effect of one or more vanishing pipe flows is simply that the contribution from the pipe with the zero flow is ignored in the system of equations (Eq. 3.5). As long as not all the pipe flows are zero in a loop, the Jacobian matrix $J=J(\Delta Q_l)$ is non-singular and we can find a well-defined solution. However, even when the head losses around a loop are equal to zero, the respective loop can be excluded from the system of equations and a solution can be found without the loop with the zero head loss.

A more serious situation can appear in the nodal heads simulator due to the limits on the numerical precision (e.g. the condition number of the Jacobian matrix is big). In the nodal heads formulation the Newton-Raphson method iterates in the space of nodes and therefore the variation of pipe flow at the step $t$ of the iteration method can follow the direction from node $n_1$ to node $n_2$ while at the next iteration the direction of flow reverses from node $n_2$ to $n_1$. This oscillating behaviour may appear when the iteration gets close to the solution. It results in a numerically oscillating algorithm with poor convergence properties which represents a disadvantage for the simulators based on the nodal heads equations.

In the loop equations framework the Newton-Raphson algorithm iterates in the space of loops where the number of unknowns is equal to the number of independent loops that is typically half the number of nodes. Thus the dimension of the matrices involved in the numerical computations is much smaller than the one used in the nodal equations framework. Therefore not only the Jacobian matrix has a reduced size but also it is denser than its counterpart from the nodal heads equations. In fact it has been observed that more looped is the Jacobian matrix, more stable is the co-tree flows simulator algorithm.

Worth mentioning here that the DF search procedure finds the most “depth” loops compared to the other search methods. This means that the loops obtained by the DF search tend to interact with each other more than in other search procedures, and this assures again a denser Jacobian matrix.

We conclude that the loop simulator is numerically stable when the iteration approaches the solution (Nielsen, 1989) and the error at the steps *t* and *t*+1 satisfies the relationship:

$$|\varepsilon_{t+1}| < |\varepsilon_t| \qquad \text{(Eq. 3.16)}$$

The errors $\varepsilon_{t+1}$, $\varepsilon_t$ can be written as:

$$\varepsilon_t = \Delta Q_{l_t} - \Delta Q_{l_{t-1}} \qquad \text{(Eq. 3.17)}$$

$$\varepsilon_{t+1} = \Delta Q_{l_{t+1}} - \Delta Q_{l_t} \qquad \text{(Eq. 3.18)}$$

### 3.3.2. Numerical results

To illustrate the preceding analytical expressions, the water network from Figure 3-13 is considered. The initial data that is the topological and the loop incidence matrixes and the initial flows are obtained with the DF search.

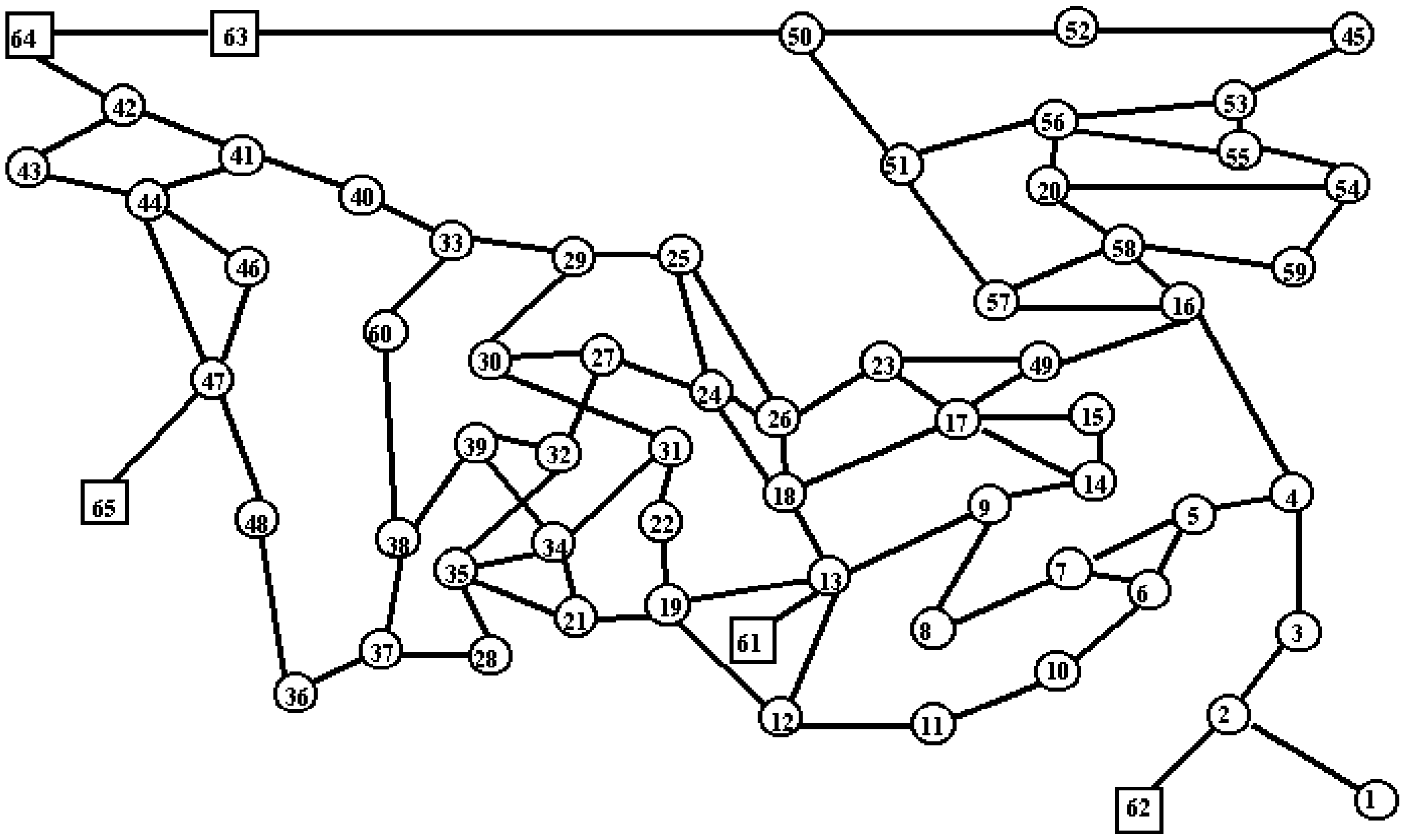

Figure 3-13. Water network.

A new labelling for pipes and nodes is produced so that to obtain an upper form for the tree incidence matrix and to be able to calculate the initial flows and the nodal heads (Figure 3-14).

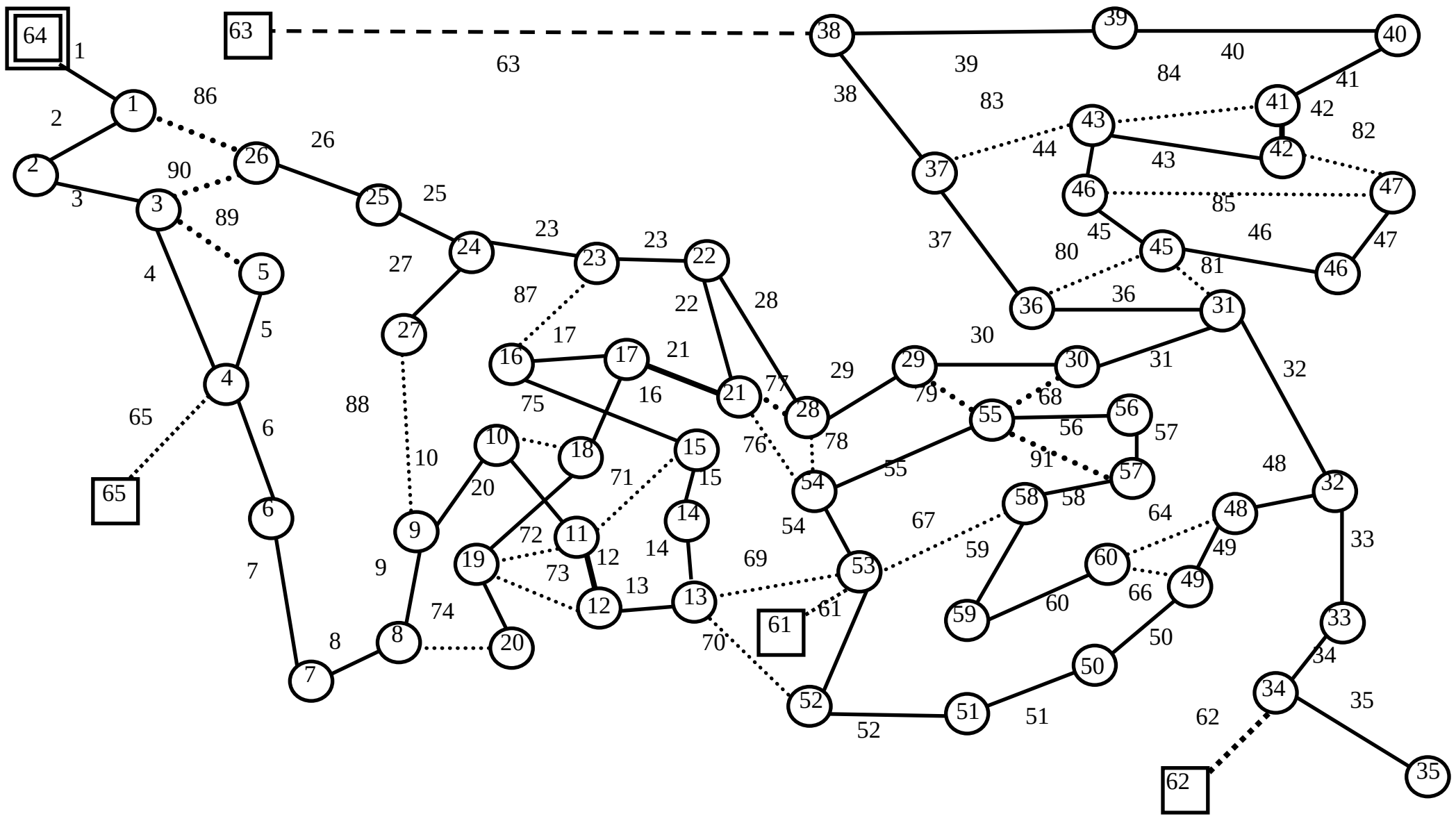

Figure 3-14: DF search with new labels assigned to nodes and pipes.

As expected the DF first search together with the enhancement in the Jacobian matrix exhibits the best convergence (Figure 3-15). It is generally accepted that the convergence is achieved when the loop head losses residues are smaller than $10^{-3}$. This corresponds to 7 iterations for the DF search with enhancement. The minimum value of $10^{-14}$ for the loop head loss residues is obtained in 9 iterations.

The convergence of the co-tree flows formulation for the steady state simulation compares very well with what other authors (Bargiela, 1984; Powell et. al, 1988) have reported for the same testing conditions (i.e. loading condition, network pipes data) using nodal heads network equations.

This formulation uses a smaller set of equations since the number of loops in a water network is likely to be far smaller than the number of nodal heads and/or pipe flows. As it will be shown in the next chapter, the reduced set of equations will improve the numerical stability of the loop-based state estimator.

Finally, the smaller size of the set of equations together with the enhancement in the Jacobian matrix does not require any new conditions on the initial solution estimate, unlike the original formulation of loop corrective flows simulator.

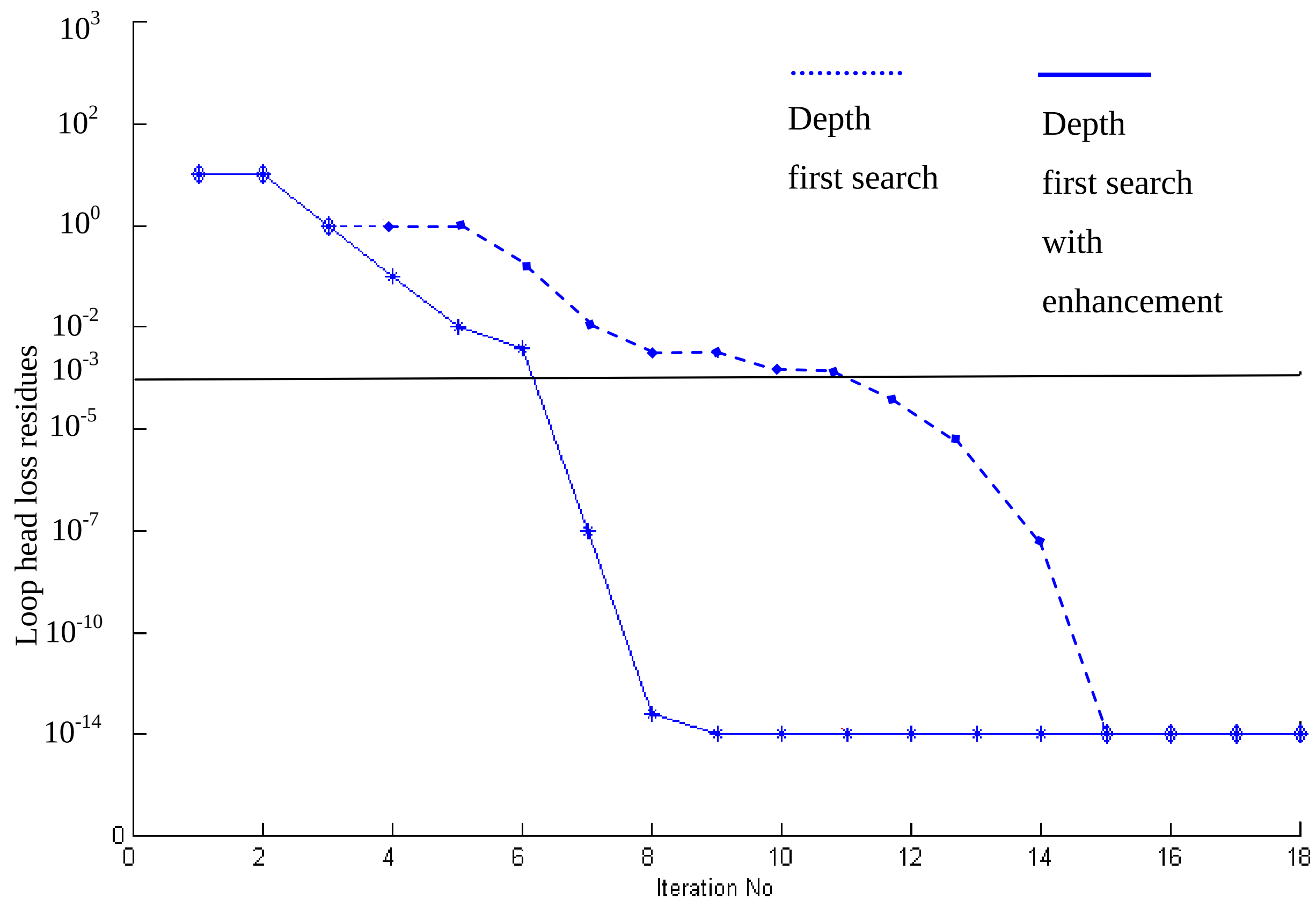

Figure 3-15: Convergence of the co-tree flows simulator for the water network from Figure 3-13.

## 3.4. Concluding remarks

In this Chapter, the problem of steady state analysis of water distribution systems is studied in the context of a co-tree flows simulator algorithm that is derived from the basic loop corrective flows algorithm. It is shown that the co-tree formulation has some inherent advantages over the original formulation due to the use of the spanning trees. This allows a rapid determination of the necessary input data for the simulator, the loop incidence matrix and the initial flows, as well as the fast calculus of the nodal heads at the end of the simulation. Furthermore an enhancement introduced in the Jacobian matrix speeds-up the overall convergence of the algorithm regardless of the initial testing conditions.

The co-tree flows simulator algorithm uses nodal water consumptions that are obtained from the information available from different sources such as customer's meters or billing records. A simple stochastic model for prediction of nodal consumptions is developed and the results are extended to other consumers that exhibit similar patterns of water use. It can represent a way to reconcile the discrepancy between the simulated water consumptions obtained by the stochastic model and the metered information in order to increase the accuracy of the final values of the nodal consumptions. However, it has been stressed that such an approach can produce in certain circumstances completely wrong results and therefore more effective operational decision tools will be developed in the next chapters.

The final result is a robust co-tree flows simulator algorithm that is using more consistent values for the nodal water consumptions. To the author's knowledge this unique approach to the modeling and simulation of the water systems based on a co-tree flows formulation has not been treated in the literature before and in this sense it is an original contribution to this work as well as a first step towards building an effective operational decision system for water networks.

# Chapter 4.

# Loop flows state estimation technique

## 4.1. Introduction

In the operational decision support of water networks, *state estimation* is an important element that enables processing of inconsistent measurements and facilitates calculation of the best approximation of the operational state of the system (Sterling & Bargiela, 1984; Powell et al., 1988; Gabrys & Bargiela, 1995). The measurements can be real flow or pressure measurements obtained through telemetry systems from the real water network, as well as the less accurate predictions of the water consumptions at the network nodes. These predictions are frequently referred to as *pseudomeasurements*.

As the complexity of modern water networks increases, the state estimators became an accepted tool assisting operators in their operational decision-making (Gabrys & Bargiela, 1996; Gabrys, 1997). However, the adoption of state estimators as operational control tools puts new requirements with respect of their efficiency and effectiveness. The state estimators are required to be both capable of real-time data processing and be relatively immune to numerical convergence problems that might be caused by incomplete or inaccurate data.

State estimation can be viewed as a process of optimisation of a suitably chosen cost function (also called energy function). The choice of the optimisation criterion characterizes different state estimators. According to the criterion used the state estimation procedure can be divided into the following three major groups: *least square* (LS) *criterion* where the sum of the squared differences between the measured and estimated values is minimised, *least absolute value* (LAV) *criterion* where the sum of the absolute differences between the measured and estimated values is minimized and *minimax criterion* where the maximum difference between the measured and estimated values is minimised.

The state estimators became these days the key utility for the implementation of monitoring and control of large public utility systems, not only in water distribution systems but also for gas or electric distribution systems. Within these applications, the proper choice of the estimation criterion depends greatly on the type of errors that are likely to occur in the system. Due to this fact, that will be discussed in more detail in the following sections of the chapter, only the first two criterions (LS and LAV) and their variations have been practically used in water systems' state estimation problem. The review of the state estimation methods will include also a perspective over the independent sets of variables that can be employed in the mathematical formulation of the water network equations while using an adequate state estimation criterion.

## 4.2. State estimation in water distribution networks

### 4.2.1. Review of state estimation methods

Before we begin the presentation of our loop flows state estimation method for water systems let us take a closer look at the optimization criterions mentioned above and find out why the LS and LAV criterions have been so popular amongst water systems researchers.

From robust statistics (Hampel et al., 1987; Huber, 1981) it is known that the LS, LAV and minimax criterions are optimal for certain error distributions. The standard LS criterion, that has been the most popular one and in use for a long time, is optimal for the Gaussian (normal) distribution only. However, in many applications the assumption that the distribution of measurement errors is Gaussian is unrealistic. For a non-Gaussian error distribution a standard LS estimation may be very poor, especially where measurements contain large errors called "measurement outliers". In order to reduce the influence of the outliers the more robust iteratively re-weighted LS or LAV estimator can be used. The LAV criterion may be also preferable when very little is known about the distribution of errors. The LAV criterion produces optimal results for an error distribution having long tails, i.e. the Cauchy distribution. If the error distribution has sharply defined transitions, such as the uniform distribution, the *Chebyshev criterion* can be the most suitable choice. Because the maximum deviation is minimized in the minimax criterion it is an appropriate one to be used when the data are relatively free from outliers.

When we now look at the most likely errors to occur in water networks we find that there are two main errors: these associated with transducer noise, A/D conversions etc. that can be classified as having Gaussian distribution and these associated with topological anomalies, caused by the physical system deviating from the original system as modeled (e.g. due to a new pipe burst) and meter malfunction, that can be classified as gross errors or outliers. In view of these facts the choice of LS and LAV estimators for water systems state estimation seems to be justified.

Although the choice of appropriate optimality criterion is absolutely crucial the algorithms used to solve these optimization problems are also very important. As a matter of fact the problems with using the LAV and minimax criterions, mainly due to the non-differentiability of the objective function which may and have caused some analytical and numerical problems, have been another reason why the LS criterion is so popular. In water systems different algorithms such as linear programming, non-linear programming, unconstrained optimization have been used to solve the state estimation problem.

The comparison of the *Weighted Least Squares* (WLS) problem solved using the augmented matrix approach with the LAV problem solved using linear programming technique can be found in (Bargiela, 1984). Via simulation results Bargiela (1984) found that the LS estimator in its augmented matrix formulation is computationally efficient and exhibits very good numerical stability characteristics, especially in the case of structurally ill-conditioned systems. However, the LS approach was found to be intrinsically sensitive to measurement outliers thus requiring further bad data processing followed by re-estimation of the state variables. In contrast, the LAV produced unbiased estimates automatically rejecting the bad data but the solution time of the linear programming technique dramatically increased with the size of the network preventing its on-line application to large-scale problems. To enhance the efficiency of implementations both methods utilized the sparsity of matrices involved in problem formulations.

The sensitivity of the LS estimators to the outliers has been recognised and reported in many other publications dealing with the utility systems state estimation problem (Dopazo et al., 1970; Falcao et al., 1981; Gabrys & Bargiela, 1995; Handschin et al., 1974; Hartley, 1996; Merill & Schweppe, 1971; Powell et al., 1988; Powell, 1992; Schweppe et al., 1970; Sterling & Bargiela, 1984). In order to diminish the influence of the bad data on the final solution several techniques used with the LS estimators have been developed.

In (Schweppe et al., 1970) two tests were used: observing the weighted sum of squared residuals for detection of bad data and using the list of largest normalised residuals as a guide for the identification of bad data points.

Another approach to improvement of the LS estimates in presence of the bad data is the use of methods penalizing the largest residuals so that the potential bad data have a reduced influence on the final estimates. In these methods of state estimation the non-quadratic cost functions, which approximate to a standard LS criteria when all the data are good, are often used (Falcao et al., 1981; Handschin et al., 1974; Merill & Schweppe, 1971). Another set of examples of iteratively reweighted LS estimators, based on detecting the largest residuals, can be found in (Hartley, 1996; Powell et al., 1988; Powell, 1992).

An alternative formulation to the LS criterion, that can be classified as another case of a non-quadratic criterion and is known as the weighted least absolute method (WLAV), has been proposed by a number of authors for utility systems applications (Bargiela, 1984; Falcao et al., 1981; Gabrys & Bargiela, 1995; Hartley, 1996; Kotiuga & Vidyasagar, 1982; Sterling & Bargiela, 1984).

To conclude, the method of LS state estimation is sensitive to the presence of large errors. An alternative to this problem can be the weighted variant of the LS method or the LAV method. However, the LAV method requires more time than the usual LS method for solving large water networks. A solution would be to use the LS criterion that will assure computational efficiency, while constraining the state estimation to the regions of the water network for which there exists reliable measurement information so that to avoid the influence that bad data can pose on the final state estimates.

The application of the LS/WLS state estimators for on-line monitoring of water networks has been studied at an extent level in the past. The nodal heads were in most of these studies the independent variables (Gabrys, 1997). Although the mathematical model is accurate, it leads sometime to difficulties in modeling and simulation of realistic water networks (Nielsen, 1989; Gabrys & Bargiela, 1995; Sterling & Bargiela, 1984; Powell et al., 1988; Hartley, 1996; Hartley and Bargiela, 1993).

An alternative to the nodal heads as state variables is the loop corrective flows. This can be an advantage because of the smaller size of the matrixes involved in the numerical computations. A WLS state estimator based on the loop equations and the state variables were the unknown nodal demands was presented in Andersen and Powell (1999a). The minimization problem was solved using a Lagrangian approach. A LS state estimator was shown in Arsene and Bargiela (2001) that employs both the variation of nodal demands and the loop corrective flows as independent variables. A leakage detection scheme was envisaged based on this state estimation technique.

The novel state estimator is depicted in the next section. Following this, the global optimality of the LS solution is demonstrated.

### 4.2.2. Formulation of the loop flows state estimation problem

In the previous chapter, it has been shown how the co-tree flows simulator algorithm can benefit from using the spanning trees: the loop incidence matrix and the initial flows were easily obtained from a spanning tree built for a water network. Moreover, at the end of the Newton-Raphson method the nodal heads could be calculated based on the same spanning tree.

The *loop flows state estimator* presented here is using once again the spanning trees and their properties. The state estimator has been received the loop flows denomination because of the loop corrective flows variable that is one of the sets of variables used to construct the network equations:

$$z = g(x) + r \qquad \text{(Eq. 4.1)}$$

where $z$ is the vector of measurements contaminated by errors and disturbances; $r$ is the unknown vector, called the vector of residuals, that accounts for measurement noise, model errors and disturbances; $g()$ is the nonlinear function (also) called network function describing the system; $x$ represents the state variables used to build the network function $g$. At this stage in our presentation no decision has been taken with regard to the state variables.

If the loop corrective flows were the state variables, then the state estimation problem would not be completely defined.

The *continuity equation* for the initial flows can be written as follows:

$$A_{np}\, Q_i = d_i \qquad \text{(Eq. 4.2)}$$

where $A_{np}$ is the topological incidence matrix, $d_i$ are the nodal demands at the beginning of the Newton-Raphson method, and $Q_i$ are the initial pipe flows. At this stage in our presentation the topological incidence matrix has not taken any special structure (e.g. upper form).

The continuity equation at the end of the Newton-Raphson method can be written as follows:

$$A_{np}\, \tilde{Q} = d_f \qquad \text{(Eq. 4.3)}$$

where $\tilde{Q}$ are the pipe flows that are updated at each iteration step and are expressed in the following form:

$$\tilde{Q} = Q_i + M_{pl} \Delta Q_l \qquad \text{(Eq. 4.4)}$$

From the basic hydraulics theory, between the topological and the loop incidence matrixes stands the following equation:

$$A_{np}\, M_{pl} = 0 \qquad \text{(Eq. 4.5)}$$

Therefore in the continuity equation (Eq. 4.4) the loop corrective flows would not carry anymore because the product between the topological incidence matrix and the loop incidence matrix is equal to zero:

$$A_{np}\, \tilde{Q} = A_{np}\, Q_i + A_{np}\, M_{pl} \Delta Q_l \qquad \text{(Eq. 4.6)}$$

Equations (Eq. 4.3) and (Eq. 4.6) gives:

$$A_{np}\, Q_i = d_f \qquad \text{(Eq. 4.7)}$$

The topological incidence matrix multiplied by the initial pipe flows equals the nodal demands $d_f$ obtained at the end of the Newton-Raphson iteration method. Therefore from equations (Eq. 4.2) and (Eq. 4.7) an identity has been obtained (i.e. $d_f = d_i$ ) which means that the continuity equation as written at equation (Eq. 4.2) can have an infinite of flow solutions $\tilde{Q}$. In order to avoid these difficulties an additional set of variables has been considered, the variation of nodal demands $\Delta d$. Hence the hydraulic model *g*() for head, flow and demand measurements becomes a function of both the loop corrective flows (*y*) and the variation of nodal demands (*x*).

The advantage of using the variation of nodal demands is that we are able to write the network equations based on the topological information obtained from the spanning tree. By making use of the spanning tree, the topological incidence matrix $A_{np}$ has been modeled in the previous chapter as an upper form *tree incidence matrix* $T$ ($n$ x $n$) and a *co-tree incidence matrix* $C$ ($n$ x $l$):

$$A_{np} = [T \quad C] \qquad \text{(Eq. 4.8)}$$

In order to build the *hydraulic model g*() the spanning tree from Figure 4-1 is used.

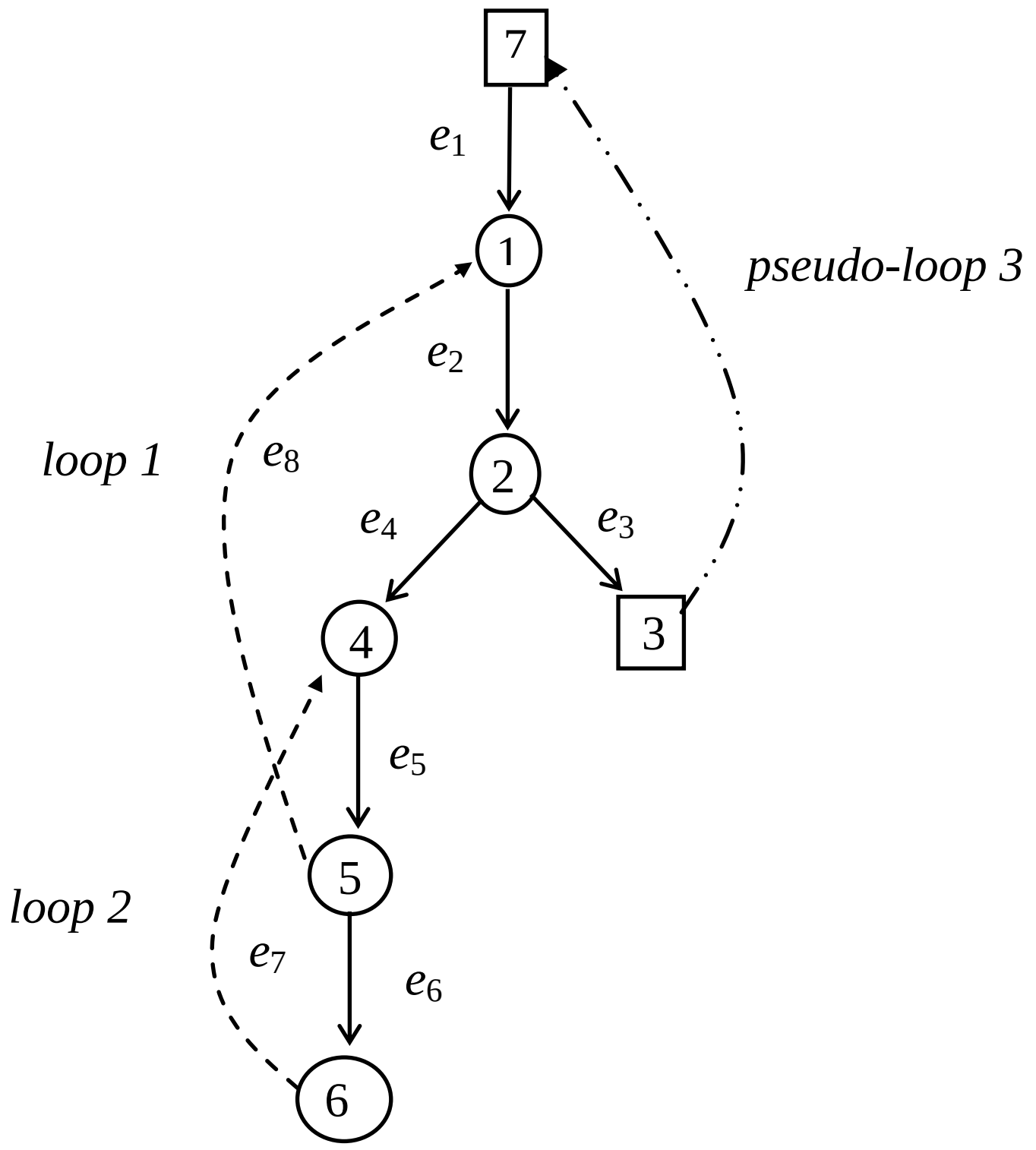

Figure 4-1: Example of a spanning tree.

The transpose of the loop incidence matrix can be written as follows:

$$M_{pl} = \begin{bmatrix} M_{nl} \\ M_{ll} \end{bmatrix} \qquad \text{(Eq. 4.9)}$$

The matrix $M_{ll}$ ($l$ x $l$) is a diagonal matrix (each loop corrective flow corresponds to a single loop) that together with the matrix $M_{nl}$ ($n$ x $l$) resembles the transpose of the loop incidence matrix $M_{pl}$ ($p=n+l$).

Since an additional set of independent variables has been introduced, the pipe flows will be written function of the loop corrective flows $\Delta Q_l$ and the variation of nodal demands $\Delta d$:

$$\tilde{Q} = Q_i - A^* \Delta d + M_{pl} \Delta Q_l \qquad \text{(Eq. 4.10)}$$

where $\tilde{Q}$ are the pipe flows in tree and co-tree pipes, $Q_i = \begin{bmatrix} Q_{T_i} \\ 0_l \end{bmatrix}$ are the initial flows in tree pipes and the initial flows in the co-tree pipes are zero (i.e. zero vector (1 $x$ $l$) $0_l$ ), and matrix $A^*$ is the matrix with the property $A^* = \begin{bmatrix} T^{-1} \\ 0_{\ln} \end{bmatrix}$.

There are two sets of equations which will be used to describe the hydraulics of the water network.

The first set of equations states that the loop head losses around the loops are equal to zero:

$$\Delta H(\Delta Q_l, \Delta d) = 0 \qquad \text{(Eq. 4.11)}$$

where the loop head losses residuals $\Delta H$ are a function of the loop corrective flows $y$ ($\Delta Q_l$) and the variation of nodal demands $x$ ($\Delta d$) and are calculated from equation (Eq. 3.3).

The second set of equations states that the total amount of inflow/outflow from the water network carried out through the fixed-head nodes should equal the variation of nodal demands. This equation can be written as follows:

$$\Delta d = B_{nl} \Delta Q_l \qquad \text{(Eq. 4.12)}$$

The matrix $B_{nl}$ ($n$ x $l$) from equation (Eq. 4.12) has a non-zero element equal to 1 which corresponds to the main root node and -1 for each of the fixed-head nodes. For the spanning tree from Figure 4-1 there is a pseudo-loop between the fixed-head node 3 and the main root node which corresponds to the 3rd column in the matrix $B_{nl}$. The matrix is shown in full form below:

$$
\begin{array}{rc}
 & \textit{Pseudo-loops} \\
\begin{array}{r} 1 \\ 2 \\ \textit{fixed-head node} \;\; 3 \\ 4 \\ 5 \\ 6 \\ \textit{main root node} \;\; 7 \end{array} &
\begin{pmatrix} 0 & 0 & 0 \\ 0 & 0 & 0 \\ 0 & 0 & -1 \\ 0 & 0 & 0 \\ 0 & 0 & 0 \\ 0 & 0 & 0 \\ 0 & 0 & 1 \end{pmatrix}
\end{array}
$$

The equations (Eq. 4.11) and (Eq. 4.12) represents the hydraulic function that describes the water network. It can be written as a system of equations:

$$
\begin{cases} \Delta H(\Delta Q_l, \Delta d) = 0 \\ \\ B_{nl} \Delta Q_l - \Delta d = 0 \end{cases} \qquad \text{(Eq. 4.13)}
$$

If $\hat{x}$ ($n$ x 1) and $\hat{y}$ ($l$ x 1) are the estimates of the vectors $x$ (*variation of nodal demands*) and $y$ (*the loop corrective flows*), and $g(\hat{x}, \hat{y})$ is the *non-linear function* to be minimized (Eq. 4.13), by using the first-order Taylor series $g(\hat{x}, \hat{y})$ becomes:

$$
g(\hat{x}, \hat{y}) = g(\hat{x}, \hat{y})^{(0)} + \left[ \left. \frac{\partial g(\hat{x}, \hat{y})}{\partial \hat{x}} \right|_{\hat{x} = \hat{x}^{(0)}} \quad \left. \frac{\partial g(\hat{x}, \hat{y})}{\partial \hat{y}} \right|_{\hat{y} = \hat{y}^{(0)}} \right] \begin{bmatrix} (\hat{x} - \hat{x}^{(0)}) \\ (\hat{y} - \hat{y}^{(0)}) \end{bmatrix}
$$

(Eq. 4.14)

where [ $\hat{x}^{(0)}$ $\hat{y}^{(0)}$ ] is the initial guess of the state vector. The Jacobian matrix is:

$$
J = \begin{bmatrix} \dfrac{\partial g(\hat{x}, \hat{y})}{\partial \hat{x}} & \dfrac{\partial g(\hat{x}, \hat{y})}{\partial \hat{y}} \end{bmatrix} \qquad \text{(Eq. 4.15)}
$$

where $\dfrac{\partial g(\hat{x}, \hat{y})}{\partial \hat{x}}$ and $\dfrac{\partial g(\hat{x}, \hat{y})}{\partial \hat{y}}$ are the partial derivatives of the function $g(\hat{x}, \hat{y})$ with respect to the vectors $\hat{x}$ and $\hat{y}$.

The Jacobian becomes:

$$
J = \begin{bmatrix} \dfrac{\partial \Delta H}{\partial \hat{x}} & \dfrac{\partial \Delta H}{\partial \hat{y}} \\ -\dfrac{\partial (\hat{x})}{\partial \hat{x}} & B_{nl} \dfrac{\partial (\hat{y})}{\partial \hat{y}} \end{bmatrix} \qquad \text{(Eq. 4.16)}
$$

The Jacobian matrix can be further written as:

$$J = \begin{bmatrix} \frac{\partial \Delta H}{\partial \hat{x}} & \frac{\partial \Delta H}{\partial \hat{y}} \\ -I_{nn} & B_{nl} \end{bmatrix} \qquad \text{(Eq. 4.17)}$$

where $I_{nn}$ is the ($n$ x $n$) identity matrix.

By suppressing the line and the row corresponding to the main root node in matrixes $I_{nn}$ and $B_{nl}$ and calculating the inflows/outflows in the water network at the end of the Newton-Raphson method (i.e. by subtracting the variation of the nodal demands at the fixed-head nodes from the loop corrective flows corresponding to the pseudo-loops) then matrix $B_{nl}$ can be modified to become the zero matrix of size ($n$ x $l$). Following this, the Jacobian matrix becomes:

$$J = \begin{bmatrix} \frac{\partial \Delta H}{\partial \hat{x}} & \frac{\partial \Delta H}{\partial \hat{y}} \\ -I_{nn} & 0_{nl} \end{bmatrix} \qquad \text{(Eq. 4.18)}$$

The Jacobian matrix $J$ resembles the one presented in Arsene and Barigela (2001).

If we denote with $g(\hat{x}, \hat{y})$ the non-linear equation describing the residuals in the loop head losses and the variation of nodal demands then the variation of the state variables $\begin{bmatrix} \Delta \hat{x} \\ \Delta \hat{y} \end{bmatrix}$ during the Newton-Raphson method is calculated as:

$$\begin{bmatrix} \Delta \hat{x} \\ \Delta \hat{y} \end{bmatrix} = (J^T J)^{-1} J^T \; g(\hat{x}, \hat{y}) \qquad \text{(Eq. 4.19)}$$

The LS estimate of $\begin{bmatrix} \hat{x} \\ \hat{y} \end{bmatrix}$ is found by an iterative process with the consecutive state estimates calculated with the following equation:

$$\begin{bmatrix} \hat{x}^{(k+1)} \\ \hat{y}^{(k+1)} \end{bmatrix} = \begin{bmatrix} \hat{x}^{(k)} \\ \hat{y}^{(k)} \end{bmatrix} + \begin{bmatrix} \Delta \hat{x}^{(k)} \\ \Delta \hat{y}^{(k)} \end{bmatrix} \qquad \text{(Eq. 4.20)}$$

If all elements of $\begin{bmatrix} \Delta\hat{x}^{(k)} \\ \Delta\hat{y}^{(k)} \end{bmatrix}$ at *k*-th step of the estimation process are lower or equal to a predefined convergence accuracy, the iteration procedure stops. Otherwise, a new correction vector is calculated using (Eq. 4.20) with $\begin{bmatrix} \hat{x}^{(k+1)} \\ \hat{y}^{(k+1)} \end{bmatrix}$ instead of $\begin{bmatrix} \hat{x}^{(k)} \\ \hat{y}^{(k)} \end{bmatrix}$.

Solving $g(\hat{x}, \hat{y})$ with the LS method may result sometimes in incorrect water network physical solutions such as nodal demands smaller than 0 (i.e. $d_f$ < 0) or loop head losses $\Delta H$ which are not zero. In this case the positive values of the resulted (i.e. state estimated) nodal demands $d_f$ are used in a final simulation (i.e not state estimation) of the water network in order to calculate the final pipe flows and nodal heads solution. Therefore this is an additional computational step which could be required after running the LS loop flows state estimator, that is to check that a correct water network physical solution is obtained and if not to run an additional simulation based on the positive values of the state estimated nodal demands $d_f$ . This solution is the correct one as opposed to the one presented in Arsene and Gabrys (2014), that is the Newton-Raphson algorithm to be constrained iteratively (i.e. do not use Algorithm 1st from Arsene and Gabrys, 2014), which does not converge.

## 4.2.3. Optimality of the LS estimate

The non-linear equation $g(\hat{x}, \hat{y})$ is a function of the head losses residuals in the co-tree pipes and the variation of nodal demands. This *non-linear function* can be split in two different equations which can be solved by a *Lagrangian approach* where the head losses residuals in the co-tree pipes represent the constrains (i.e. the head losses residuals are equal to zero) and the variation of nodal demands represents the function to be minimized (Andersen & Powell, 1999a). However, if the *Hessian matrix* of the non-linear function $g(\hat{x}, \hat{y})$ is (semi) positive definite then any local minimum is a global minimizer as well. Moreover, when the LS method is used, it is guaranteed that one single and global minimum point is found and hence the LS method is used herein.

The *first derivative* of the loop head losses residuals with respect to the loop corrective flows is:

$$\frac{\partial \Delta H}{\partial \hat{y}} = M_{lp} A M_{pl} \qquad \text{(Eq. 4.21)}$$

where matrix $A$ is given by the equation (Eq. 3.8) from the co-tree flows simulator. For the exponent $n$ equals 2, matrix $A$ is:

$$A = \begin{pmatrix} 2k_1\, abs(Q_1) & 0 & \ldots & 0 \\ 0 & 2k_2\, abs(Q_2) & \ldots & 0 \\ 0 & 0 & \ldots & 2k_p\, abs(Q_p) \end{pmatrix} \quad \text{(Eq. 4.22)}$$

where the pipe flows $Q_{j=\overline{1,p}}$ are written as:

$$Q_j = Q_{i_j} + M_{pl}(j,1:l)\Delta Q_l - A^*(j,1:n)\Delta d \quad \text{(Eq. 4.23)}$$

where $Q_{i_j}$ is the initial flow in pipe $j$ and $M_{pl}(j,1:l)$ is the $j$-th row of the transpose of the loop incidence matrix.

The *second derivative* becomes:

$$\frac{\partial^2 \Delta H}{\partial \hat{y}^2} = M_{lp}\, K\, M_{pl} \quad \text{(Eq. 4.24)}$$

where $K$ is the $p$-diagonal matrix with the positive coefficients $2k_{1,\ldots p}$ on the diagonal. The second derivative of the loop head losses residuals with respect to $\hat{y}$ gives a positive definite matrix.

In the Jacobian matrix $J$ the identity matrix $I_{nn}$ is a constant and the derivative with respect to the variation of nodal demands $\hat{x}$ is zero. It remains to calculate the first and the second derivative of the loop head losses residuals with respect to the variation of nodal demands. The *first derivative* is:

$$\frac{\partial \Delta H}{\partial \hat{x}} = M_{ln} A(1:n,1:n)(-T^{-1}) \quad \text{(Eq. 4.25)}$$

The *second derivative* is:

$$\frac{\partial^2 \Delta H}{\partial \hat{x}^2} = M_{ln} K(1:n,1:n)(-T^{-1})^2 \quad \text{(Eq. 4.26)}$$

The square of the inverse of the tree incidence matrix $T$ is positive-definite. This is also true for the first $n$-rows and $n$-columns in matrix $K$. In the spanning tree from Figure 4-1 by choosing the clockwise direction in the loops then matrix $M_{nl}$ will contain greater than 0 elements. The *Hessian matrix* of the function $g(\hat{x},\hat{y})$ becomes:

$$\tilde{H} = \begin{bmatrix} \frac{\partial^2 \Delta H}{\partial \hat{x}^2} & \frac{\partial^2 \Delta H}{\partial \hat{y}^2} \\ 0_{nn} & 0_{nl} \end{bmatrix} \qquad \text{(Eq. 4.27)}$$

Matrix $\tilde{H}$ is square but there are block matrixes which are zero (i.e. $0_{nn}$ and $0_{nl}$), however it is a semi-positive matrix. Moreover, since the LS method is used herein, the LS estimate $\begin{bmatrix} \hat{x} \\ \hat{y} \end{bmatrix}$ of the function $g(\hat{x},\hat{y})$ when there are no pressure and flow measurements is a global and single minimum solution point for this function.

The function $g(\hat{x},\hat{y})$ can be augmented with pressure and flow measurements. In the residual form a *flow measurement* $Q_r$ is written as:

$$Q_r = Q_j - Q_{i_j} - M_{pl}(j,1:l)\Delta Q_l + A^*(j,1:n)\Delta d \qquad \text{(Eq. 4.28)}$$

where $Q_j$ is the flow measurement at pipe *j*. In the equation (Eq. 4.28) the second derivatives with respect to the loop corrective flows and the variation of nodal demands are zero. Since LS method is used, a single and global minimum solution point for the non-linear function $g(\hat{x},\hat{y})$ can be found again when a flow measurement is added to *g* system of non-linear equations.

A *pressure measurement* in the residual form $H_r$ can be written as:

$$H_r = H_j - H_{i_j} - \begin{bmatrix} T_1(j,1:n) & 0_l \end{bmatrix} h \qquad \text{(Eq. 4.29)}$$

where $H_j$ is the pressure measurement at node *j*, $H_{i_j}$ is the initial pressure value at node *j* calculated at the beginning of the Newton-Raphson method, matrix $T_1$ is the inverse of the transpose of the tree incidence matrix (i.e. $(T^T)^{-1}$), *h* is the vector of pipe head losses and $0_l$ is the zero vector of size *l*. The *first derivative* of the equation (Eq. 4.29) with respect to the loop corrective flows is:

$$\frac{\partial H_r}{\partial \hat{y}} = \begin{bmatrix} T_1(j,1:n) & 0_l \end{bmatrix} A\,M_{pl} \qquad \text{(Eq. 4.30)}$$

The *second derivative* of the equation (Eq. 4.29) with respect to the loop corrective flows is:

$$\frac{\partial H_r}{\partial \hat{y}} = \left[ T_1(j, 1:n) \quad 0_l \right] K\, M_{pl} \qquad \text{(Eq. 4.31)}$$

where matrix *K* is from equation (Eq. 4.26). The elements of the transpose of the tree incidence matrix are equal to -1 and the elements of the transpose of the loop incidence matrix are equal to 1 if the direction of the loops and the tree flows is taken as in Figure (4-1). Again since LS method is used, a single and global minimum solution point for the non-linear function $g(\hat{x}, \hat{y})$ can be found when a pressure measurement is added to the *g* system of non-linear equations. The benefit of using the loop corrective flows and the variation of nodal demands together with the LS method stays in the simplicity of formulating the equations for the non-linear function $g(\hat{x}, \hat{y})$. This allows for different combination of pressure and flow measurements to be used. However, the superiority of this method over the nodal heads LS state estimator in terms of convergence and numerical stability has still to be shown.

### 4.2.4. Advantages of the LS loop flows state estimator over the LS nodal heads formulation

The state estimators based on the nodal heads equations have brought to attention their poor convergence for the regions of the water networks characterized by *low flow condition* (i.e. pipe flows tend to be zero) as early as Donachie (1974). Therefore the scope of using the loop corrective flows in the state estimator is to obtain an improvement of the performances of the state estimators based on the nodal heads.

In the previous chapter it has been observed experimentally that the convergence of the co-tree flows simulator has been improved if we introduced an enhancement in the Jacobian matrix. The same enhancement is used in the Jacobian matrix of the state estimator described by the equation (Eq. 4.18). The loop flows state estimator has been tested on the medium-sized water network from Figure 3-12 that is shown again below. Combinations of pressure and flow measurements have been used in conjunction with the pseudo-measurements. The convergence has been in most of the simulations between six to eight iterations. This compares very well with the convergence of the state estimators based on the nodal heads equations.

A water consumption of 10 l/s has been considered at node 6 in addition to the initial consumption of 1.6 l/s. This represents a major leakage at node 6. The estimated pressure nodes and pipe flows are shown at Figure 4-3. It can be observed that the pressure at node 6 and the area around the tested node is decreasing which is due to the additional water consumption. The twin effect to the decreasing of pressure is the increasing of the flows in the pipes adjacent to the tested node (i.e. pipes 6 and 65).

In the first experiment no pressure or flow measurements have been used. From the convergence's point of view 7 iterations were necessary in order to obtain a relative variation of the loop corrective flows and variation of nodal demands smaller than an error of $10^{-4}$.

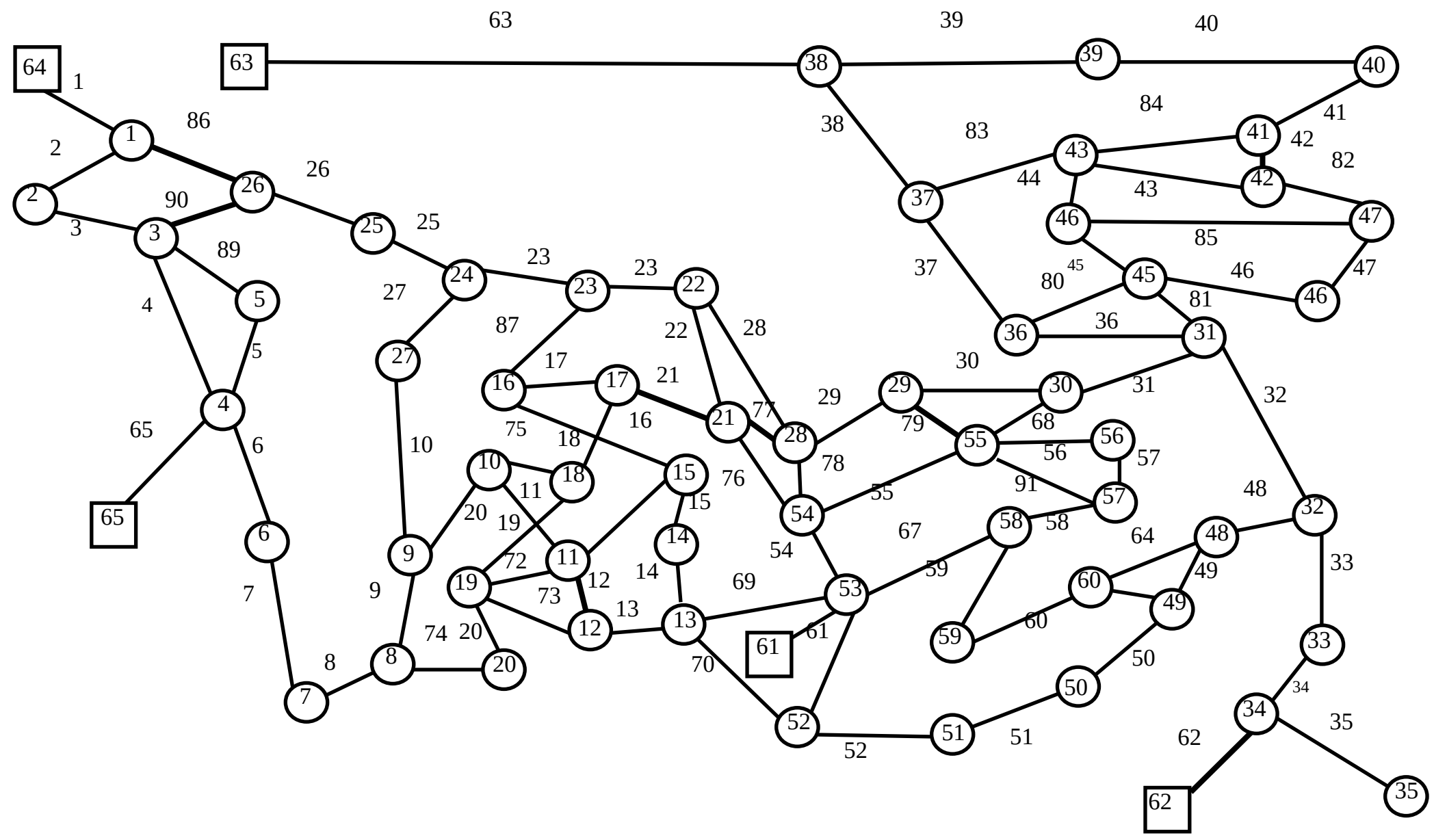

Figure 4-2: Water network.

Let see how the state estimator behaves in the following experiment where the set of measurements has been augmented with two pressure measurements at nodes 5 and 8 and a flow measurement at pipe 8. By increasing the number of real measurements it is expected that the loop flows estimator to be able to reject the additional water consumption from node 6. This can be actually noticed at Figure 4-4 where the pressure and flow values tend to approach the horizontal axes that are representing the *normal operating status* of the water network obtained with the co-tree flows simulator algorithm.

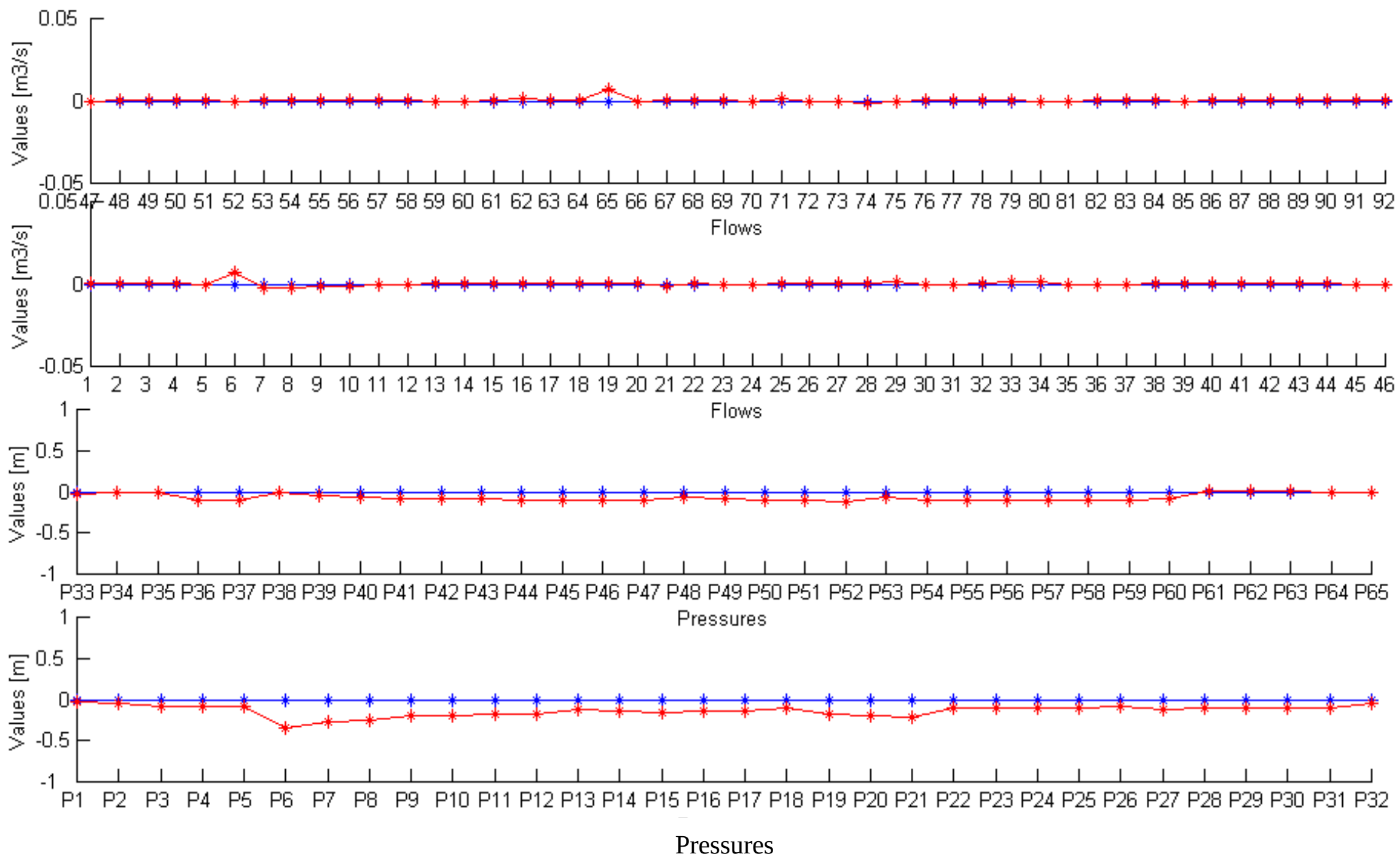

Figure 4-3: Additional water consumption at node 6 affecting pressure and flow values.

$P_1 - P_{65}$ : Pressure values at nodes 1 to 65.

1- 92: Pipe indexes.

In the last simulation we will look to the ability of the state estimator to make the best use of the available measurements. This corresponds in our series of experiments to reject entirely the supplementary water consumption of 10 l/s. Therefore in addition to the existent flow and pressure measurements a pressure measurement is placed at the tested node 6.

The result displayed at Figure 4-5 shows that the pressure and flow values coincide with the horizontal axes that is what the state estimator should deliver. This also compares very well with the numerical results of the nodal heads formulation of the state estimator both in terms of the convergence (i.e. 6 to 8 iterations) and the values of the state estimates.

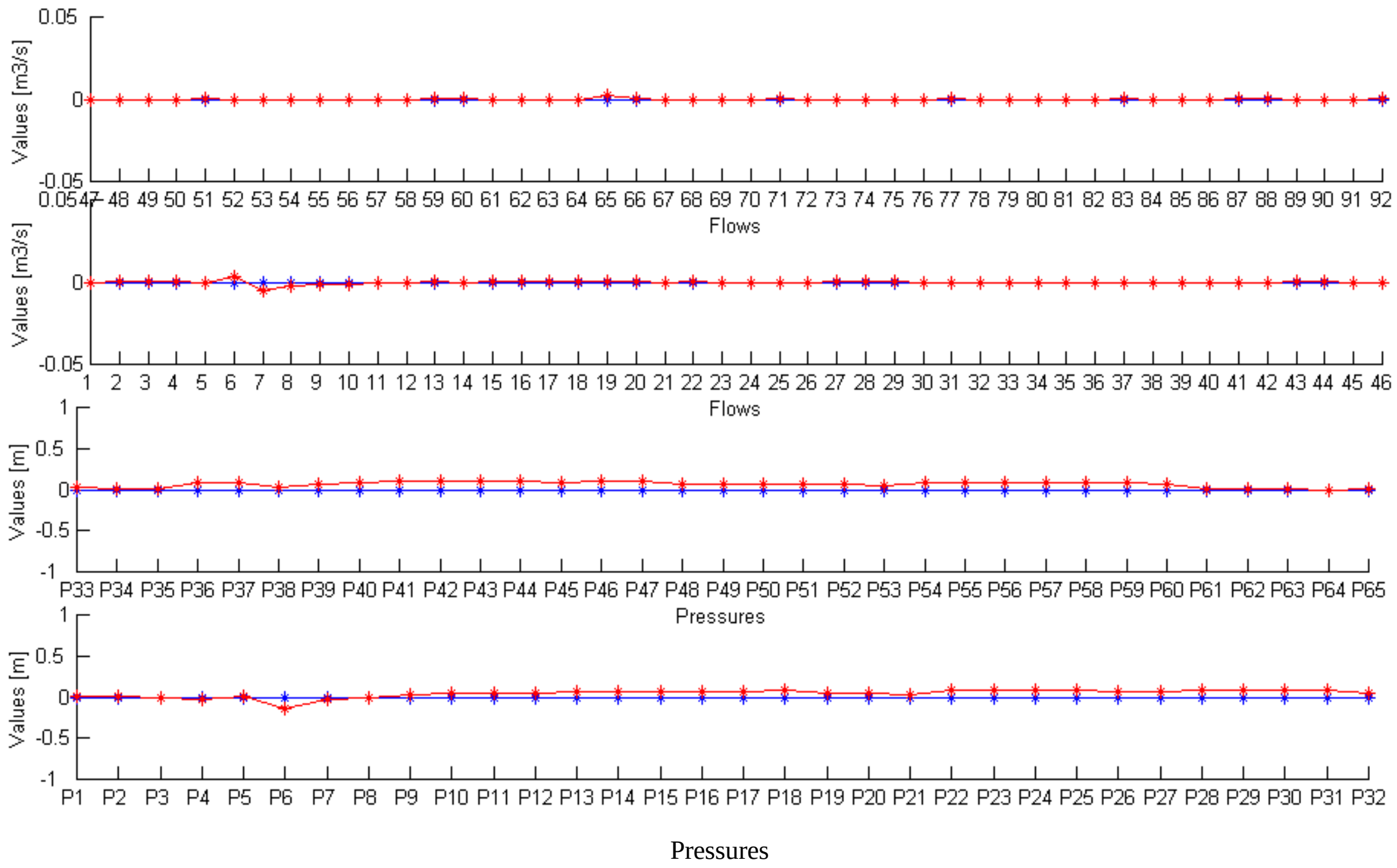

Figure 4-4: Pressure and flow measurements improving the state estimates.

$P_1 - P_{65}$ : Pressure values at nodes 1 to 65.

1- 92: Pipe indexes.

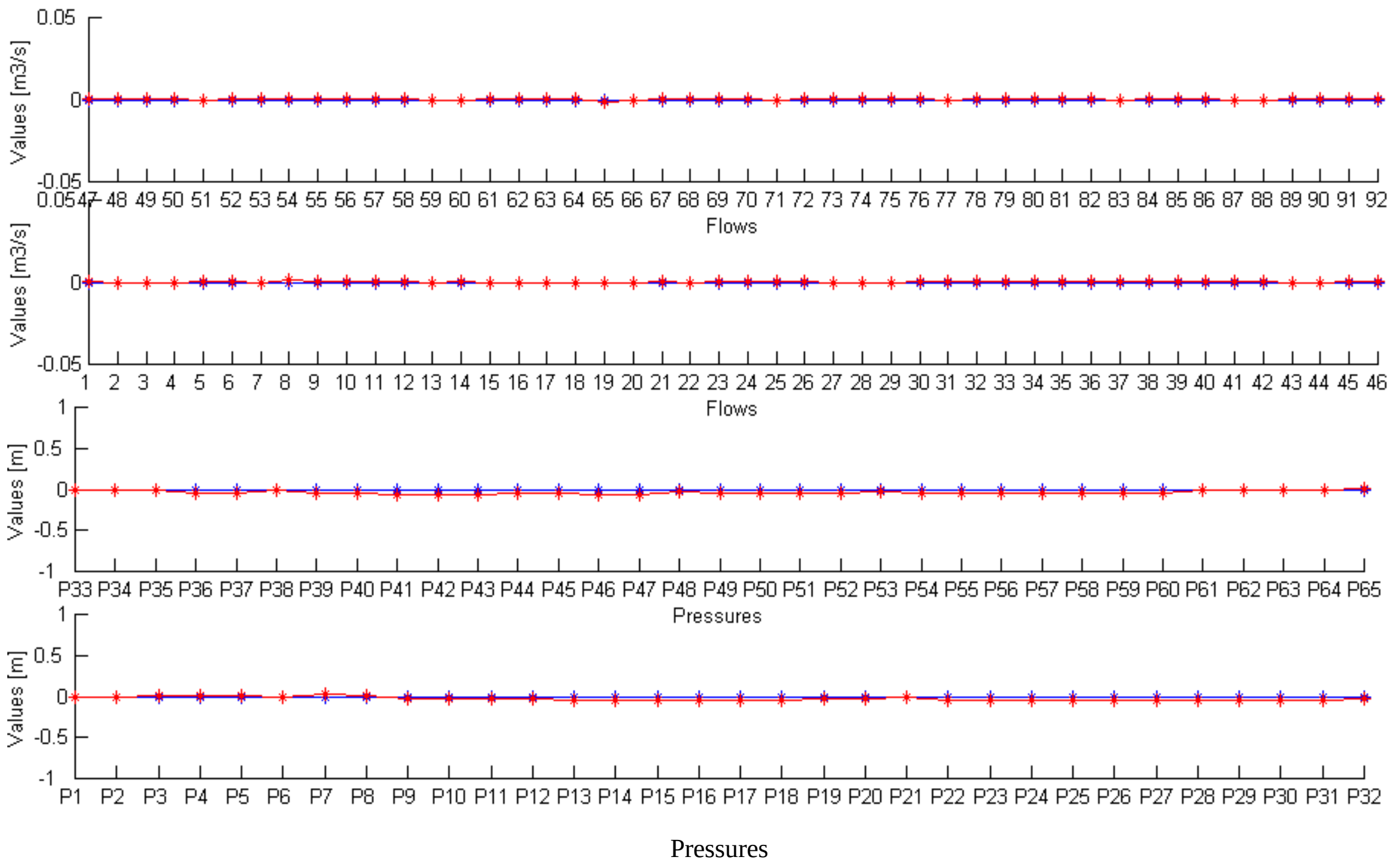

Figure 4-5: Pressure and flow measurements improving the state estimates.

$P_1 - P_{65}$ : Pressure values at nodes 1 to 65.

1- 92: Pipe indexes.

Until this stage in our presentation no significant benefits in terms of convergence or numerical stability have been pointed out for the loop flows state estimator. In order to search for such possible improvements the loop flows state estimator has been used extensively for the water network from Figure 4-2. A *low flow condition* (i.e. pipe flows tend to be zero) was identified in the area delimited by nodes 22, 28, 29, 30 and 31 (Figure 4-6).

It is well known that for the regions of water network with low pipe flows the numerical stability of the LS nodal heads state estimator may suffer. This is due to the oscillating behaviour of the low pipe flow that at the *k*-th iteration during the Newton-Raphson method follows the direction from node $n_1$ to node $n_2$ and at next iteration $k$+1 the direction of flow changes from node $n_2$ to node $n_1$. The outcome of the oscillation is a poor performance of the state estimator in terms of convergence.

Having been tested for the region of the water network shown at Figure 4-6, the loop flows state estimator exhibited the same lack of numerical stability as the nodal heads counter part.

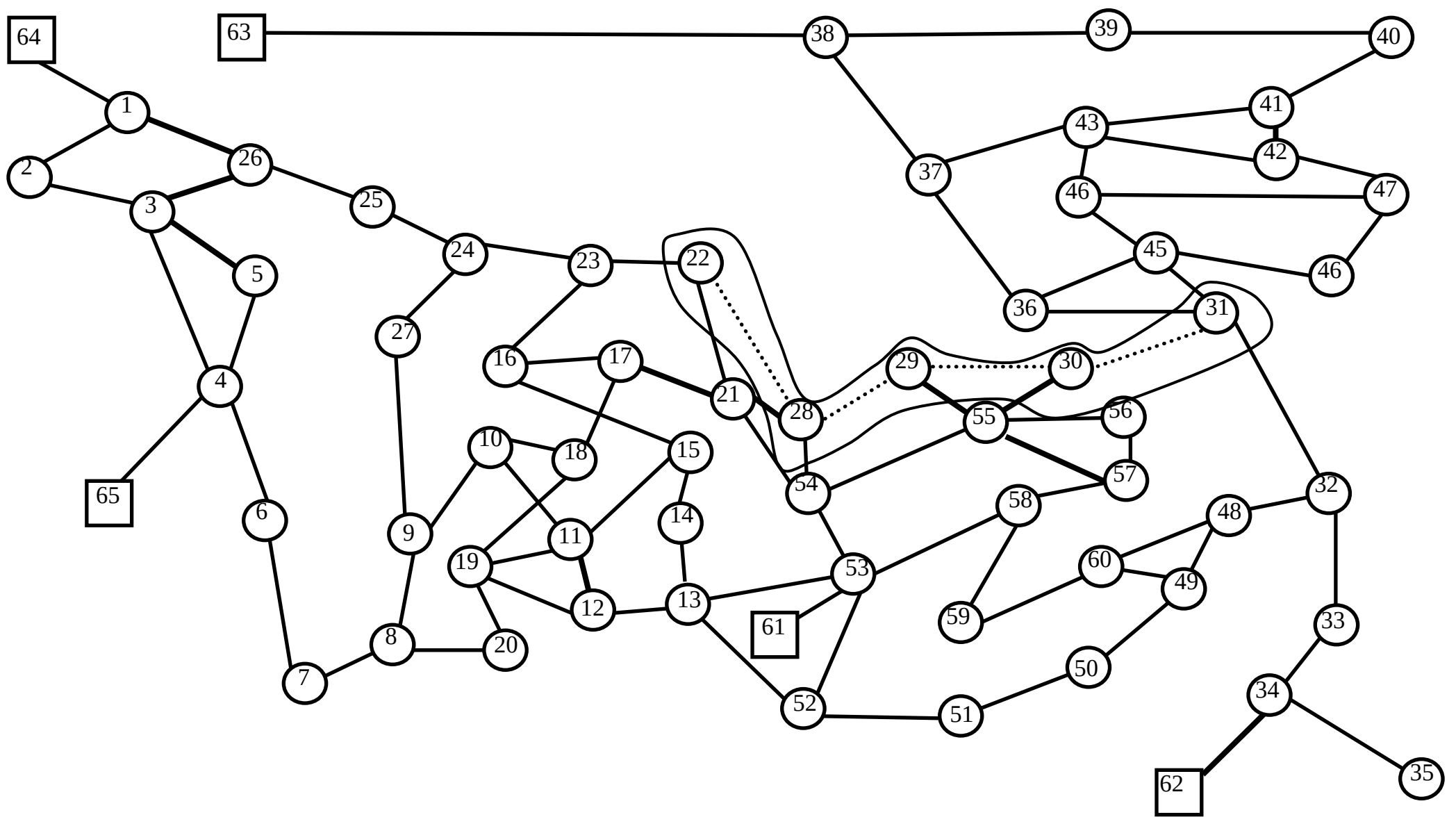

Figure 4-6: Region of the water network characterized by low pipe flows.

The Jacobian matrix from equation (Eq. 4.18) can be rewritten as:

$$J = \begin{bmatrix} -M_{lp} A A^{*} & M_{lp} A M_{pl} \\ -I_{nn} & 0_{nl} \end{bmatrix} \qquad \text{(Eq. 4.32)}$$

The matrix $A^*$ relates the variation of nodal demands to the tree flows. If this matrix is zero then the derivatives of the loop head losses with respect to the variation of nodal demands become zero:

$$M_{lp} A A^* = 0 \tag{Eq. 4.33}$$

Furthermore the Jacobian matrix becomes:

$$J = \begin{bmatrix} 0_{ln} & M_{lp} A M_{pl} \\ -I_{nn} & 0_{nl} \end{bmatrix} \tag{Eq. 4.34}$$

Since $I_{nn}$ is the $n$-identity matrix and the initial value of the variation of nodal demands is zero then in the Newton-Raphson method the loop corrective flows would be updated while the variation of the nodal demands would maintain the same values equal to zero. It is equivalent to saying that the state estimation problem has been transformed into a simulation one by zeroing in the Jacobian matrix (Eq. 4.32) the derivatives of the loop head losses with respect to the variation of nodal demands (Eq. 4.33).

The inverse of the tree incidence matrix $T^{-1}$ has the ability to constrain the state estimation procedure to some regions in the water network while for the rest of the water network the simulation problem is carried out.

In order to be more explicit in our presentation, let us consider some numerical examples. In the first part of the following experiment, the water consumption at node 6 is increased and a pressure measurement is used at node 8. The purpose of the simulation is to investigate the influence of the pressure measurement onto the nodal demands situated in the vicinity of the tested node and throughout the network.

In the figure below it is shown that the pressure measurement at node 8 has a limited influence on the pressure value at node 6 (Figure 4-7): that is the pressure at node 6 does not coincide with the point on the horizontal axis. Moreover the rest of the pressure nodes are somehow increasing. It suggests that the additional water consumption at node 6 has not been rejected by the pressure measurement and is satisfied by the reservoirs or the other nodal demands which exist in the system. The numerical results for the fist part of the simulation are displayed at Table 4-1 (rows 2, 5, 8, 11 and 14). It can be observed that the variation of nodal demands has spread to the most part of the water network (i.e. the majority of nodes have changed their demands).

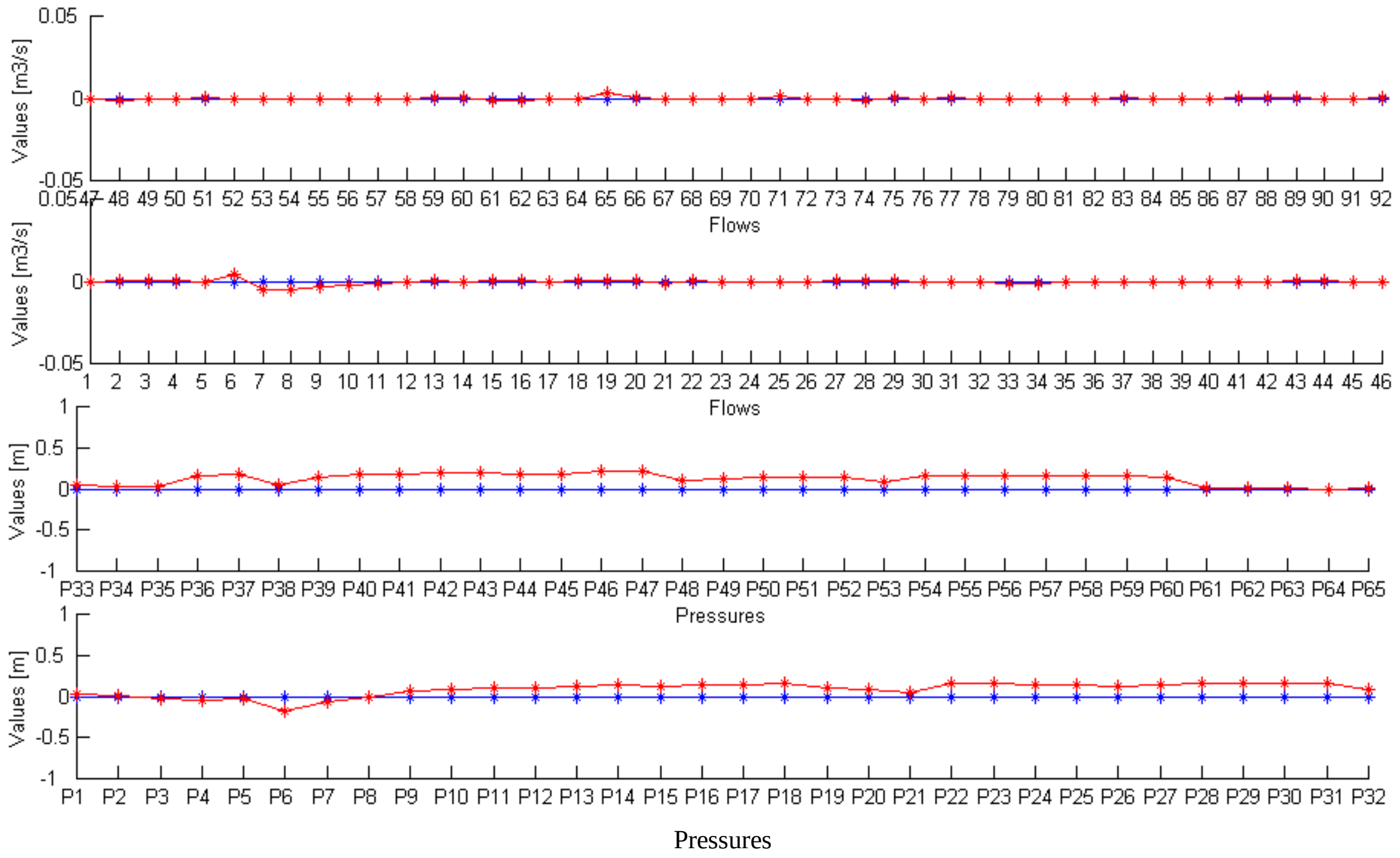

Figure 4-7: Increased water consumption at node 6 and pressure measurement at node 8.

$P_1 – P_{65}$ : Pressure values at nodes 1 to 65.

1- 92: Pipe indexes.

The variation of nodal demands due to the presence of a pressure measurement at node 8 can be avoided by zeroing some of the columns of matrix $A^*$. This is because matrix $A^*$ makes the connection between the variation of nodal demands and the tree pipe flows (Eq. 4.32 – Eq. 4.34).

Therefore, in the second part of the experiment, the columns 1 to 3 and 10 to 65 have been zeroed. This results in *constraining the variation of the nodal demands* to the area around the node with the increased water consumption (i.e. node 6) as well as the node with the pressure measurement (i.e. node 8) (see Table 4-1 rows 3, 6, 9, 12 and 15). The state estimation constrained to a region of the water network is illustrated at Figure 4-8:

| Node | 1 | 2 | 3 | 4 | 5 | 6 | 7 | 8 | 9 | 10 | 11 | 12 | 13 |
|---|---|---|---|---|---|---|---|---|---|---|---|---|---|
| **$\Delta d$(I)** | 0 | 0 | 0.1 | 0.1 | 0.1 | 0.2 | 0.4 | 0.4 | 0.4 | 0.3 | 0.3 | 0.3 | 0.2 |
| **$\Delta d$(II)** | 0 | 0 | 0 | 0.3 | 0.3 | 1.2 | 1.7 | 2 | 1.7 | 0 | 0 | 0 | 0 |
| **Node** | **14** | **15** | **16** | **17** | **18** | **19** | **20** | **21** | **22** | **23** | **24** | **25** | **26** |
| **$\Delta d$(I)** | 0.2 | 0.3 | 0.2 | 0.2 | 0.2 | 0.3 | 0.3 | 0.4 | 0.2 | 0.2 | 0.2 | 0.2 | 0.1 |
| **$\Delta d$(II)** | 0 | 0 | 0 | 0 | 0 | 0 | 0 | 0 | 0 | 0 | 0 | 0 | 0 |
| **Node** | **27** | **28** | **29** | **30** | **31** | **32** | **33** | **34** | **35** | **36** | **37** | **38** | **39** |
| **$\Delta d$(I)** | 0.2 | 0.2 | 0.2 | 0.2 | 0.2 | 0.1 | 0 | 0 | 0 | 0.2 | 0.2 | 0 | 0.1 |
| **$\Delta d$(II)** | 0 | 0 | 0 | 0 | 0 | 0 | 0 | 0 | 0 | 0 | 0 | 0 | 0 |
| **Node** | **40** | **41** | **42** | **43** | **44** | **45** | **46** | **47** | **48** | **49** | **50** | **51** | **52** |
| **$\Delta d$(I)** | 0.1 | 0.1 | 0.1 | 0.1 | 0.2 | 0.2 | 0.2 | 0.2 | 0.1 | 0.1 | 0.2 | 0.2 | 0.2 |
| **$\Delta d$(II)** | 0 | 0 | 0 | 0 | 0 | 0 | 0 | 0 | 0 | 0 | 0 | 0 | 0 |
| **Node** | **53** | **54** | **55** | **56** | **57** | **58** | **59** | **60** | **61** | **62** | **63** | **64** | **65** |
| **$\Delta d$(I)** | 0.1 | 0.2 | 0.2 | 0 | 0.2 | 0.2 | 0.2 | 0.1 | 0 | 0 | 0 | 0 | 0 |
| **$\Delta d$(II)** | 0 | 0 | 0 | 0 | 0 | 0 | 0 | 0 | 0 | 0 | 0 | 0 | 0 |

Table 4-1: Spreading of the variation of nodal demands through the water network (I) followed by a limitation of this spreading around node 6 (II).

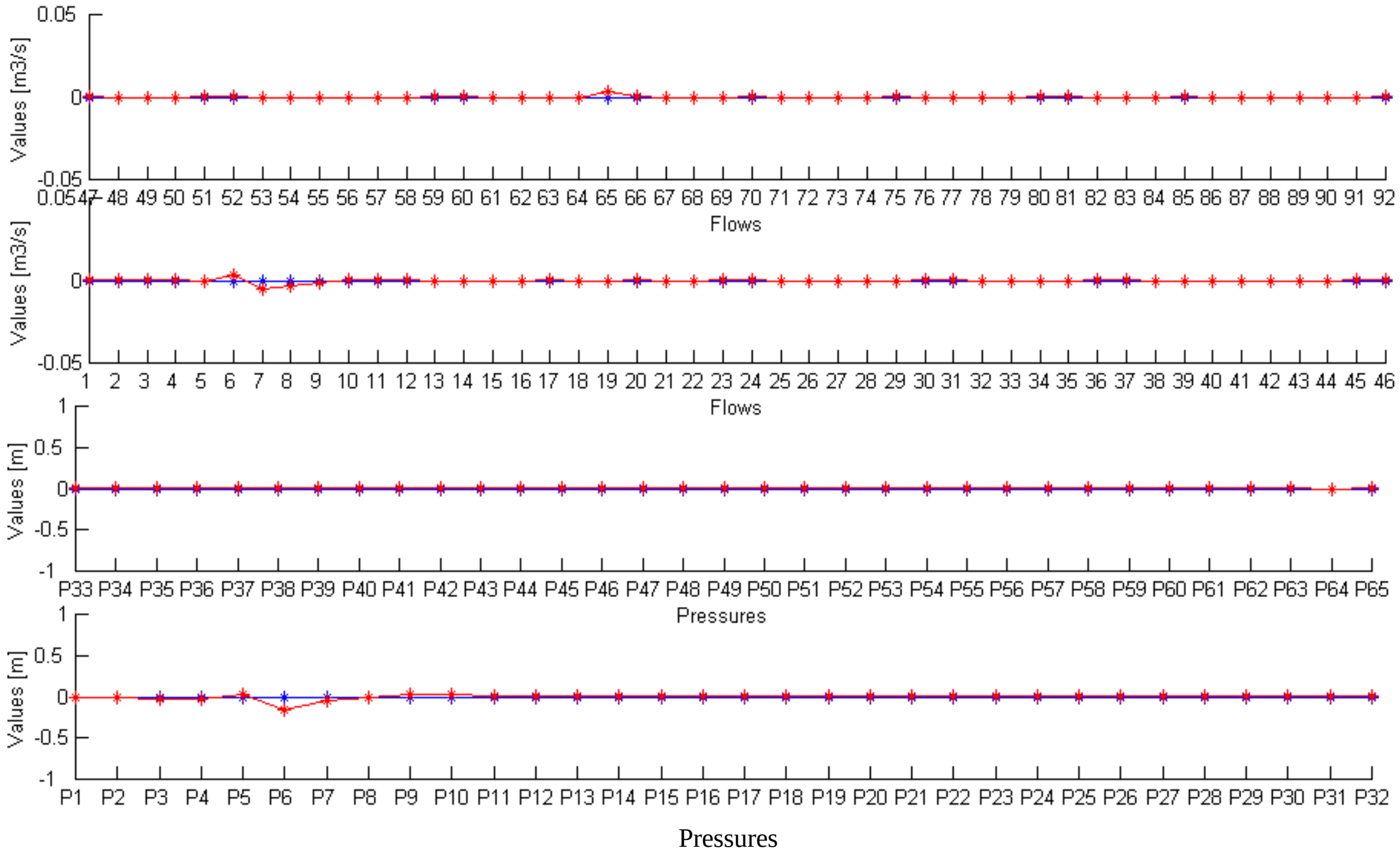

Figure 4-8: Constraining the state estimation to the area delimited by nodes 4 to 9. $P_1 - P_{65}$ : Pressure values at nodes 1 to 65, 1- 92: Pipe indexes.

The matrices shown at Figure 4-9 have been obtained with the function *spy* from the MATLAB simulation software.

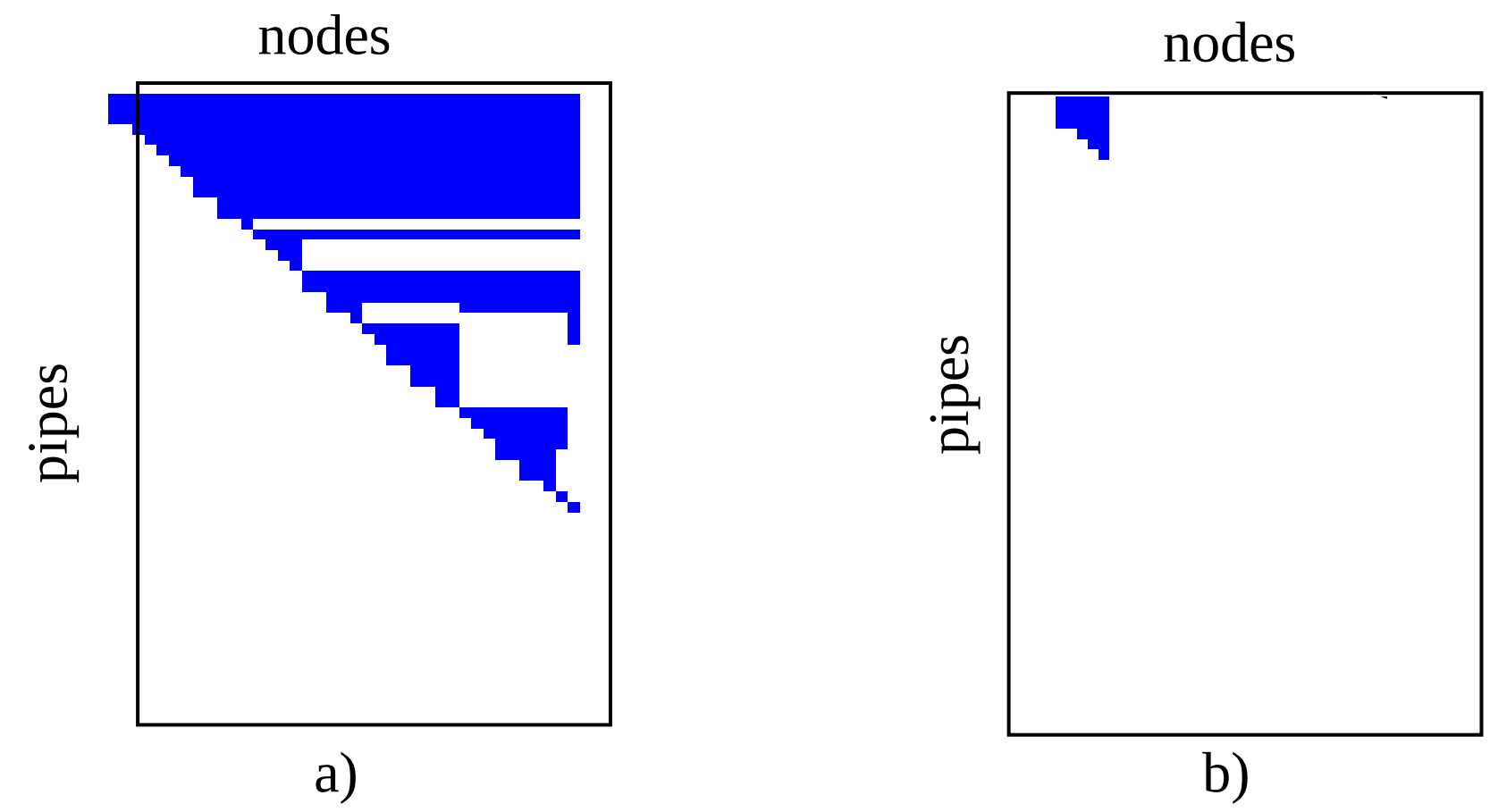

Figure 4-9: Visualization of the matrix $A^*$ used in the simulations shown at Table 4-1.

The dark areas within each rectangle represent the non-zero elements 1 or –1. The state estimation procedure is carried out for the regions of the water network that corresponds to the non-zero columns shown at Figure 4-9b (i.e. columns 4 to 9). For the rest of the network we say that a simulation has been performed. This can be observed at Figure 4-8 where the estimated pressures (nodes 10 to 65) and flows are identical with the simulated values represented by the horizontal axes. Hence, the state estimation problem can be constrained to the regions of the water network that are of interest (i.e. where the real meters lays). By zeroing the *j*-columns of matrix $A^*$, which corresponds to the *j*-entries in the vector of variation of nodal consumptions, then the influence of the variables $\Delta d_j$ on the tree flows is zeroed (Figure 4-10). The *j*-tree pipe flows are calculated from the loop corrective flows $\Delta Q_l$ while a simulation is carried out for the nodal consumptions $d_j$.

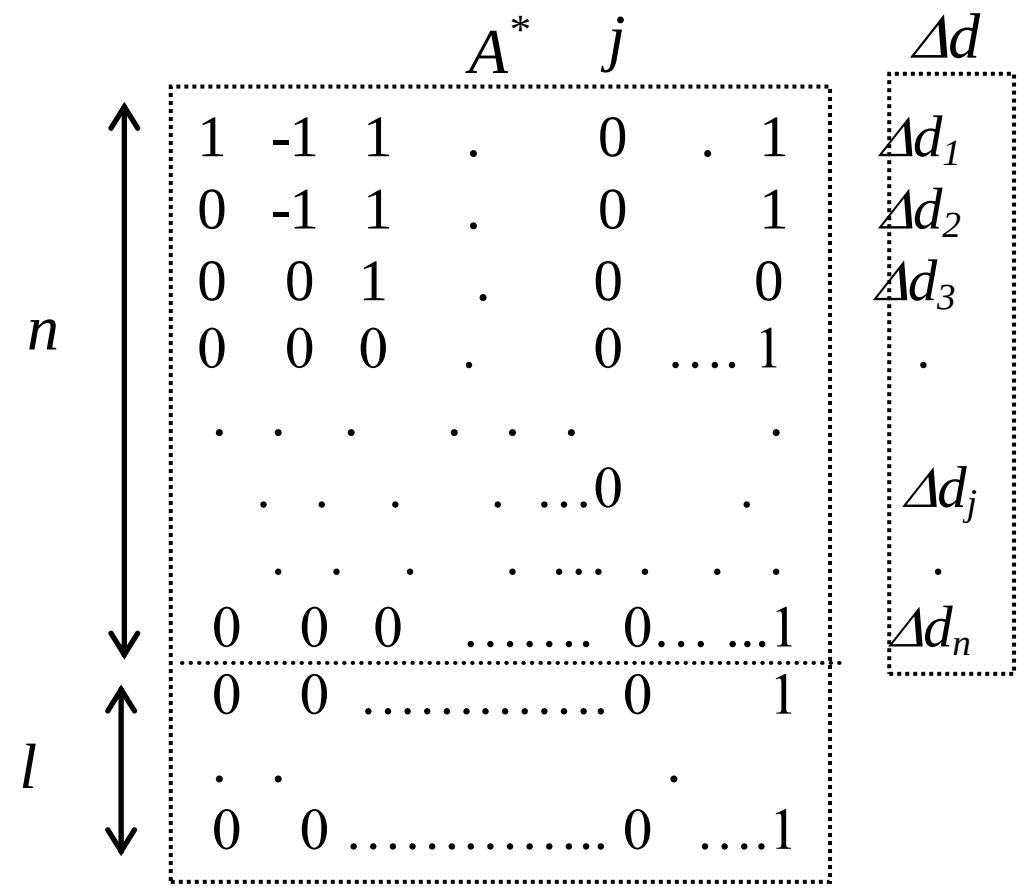

Figure 4-10: Zeroing the *j*-th column of matrix $A^*$ (matrix $A^*$ contains the inverse of the tree incidence matrix $T$ and a zero block matrix of size ($l$ x $n$)).

By constraining the state estimation procedure to the regions of the water network where there exist real measurements a limitation of the changes in the nodal demands is obtained. Actually this can be regarded as a special case of the *observability problem* in water systems (Bargiela, 1984). Therefore the solution is to apply the LS procedure to the network areas where the real meters (pressure and flow measurements) are located.

It has to be remembered that for the water network a spanning tree has been built. The tree incidence matrix describes the incidence of nodes and pipes in the spanning tree. Therefore we can relate to the spanning tree in order to decide the extent of the network area around a real measurement that will be scrutinized (i.e. the state estimation problem will be solved for this part of the water network).

In the spanning tree shown at Figure 4-11 there is a pressure measurement at node 32. It is of common sense to suppose that the pressure measurement will improve the measurements accuracy in all the nodes except node 37 which is the remotest node from the pressure measurement. Therefore the 37-th column of matrix $A^*$ associated to the spanning tree will be zeroed. In the Newton-Raphson method the pressure measurement at node 32 will not affect the pseudo-measurement value at node 37.

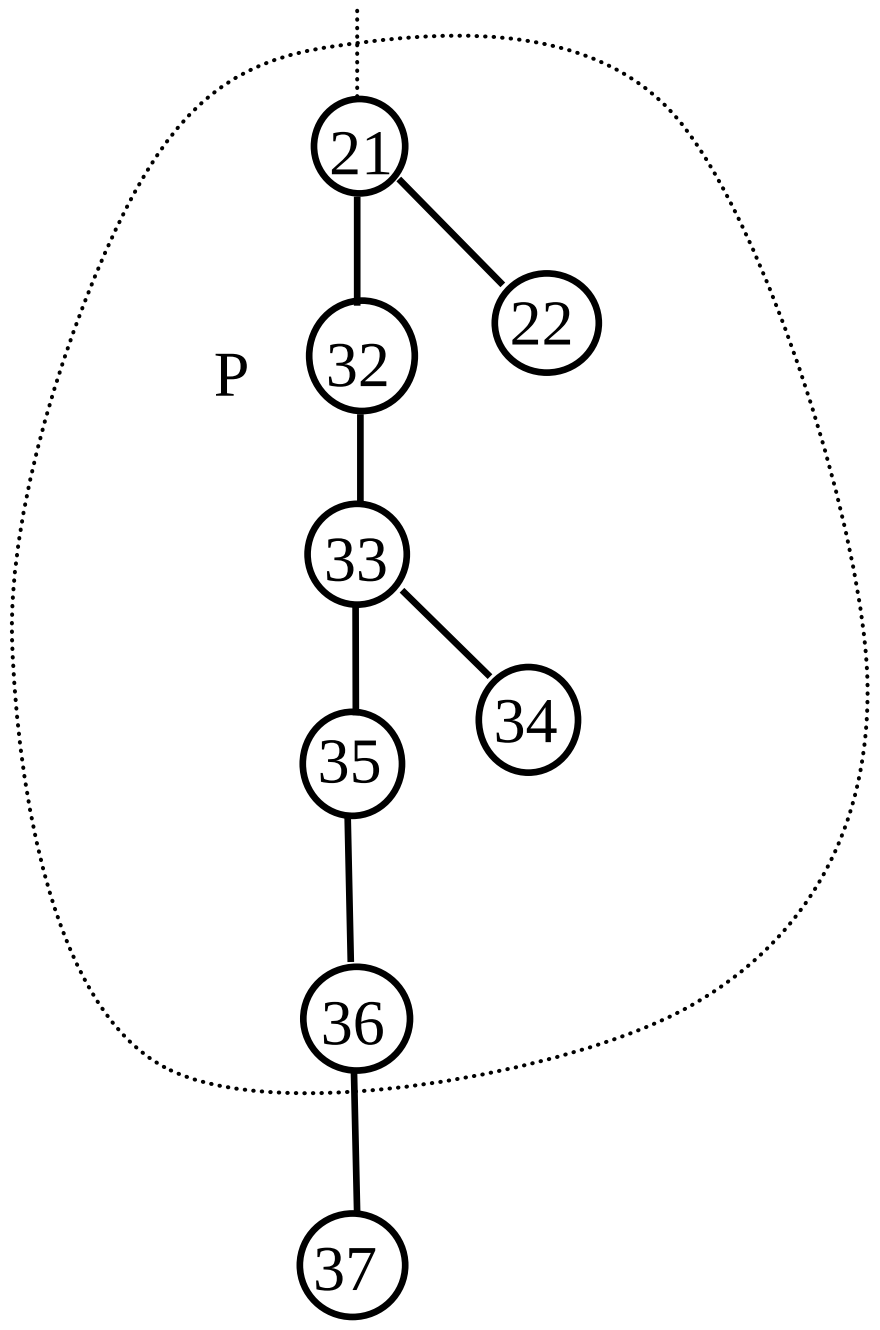

Figure 4-11: Pressure meter in the spanning tree.

We can base on the topological distance in the spanning tree in order to decide the extent around the real meter that is state estimated. Other factors that can be taken into account are the size of the water network or the depth of the spanning tree and the position of the real meter in the spanning tree.

If we go back to the water network from Figure 4-2, flow and/or pressure measurements are introduced in the same time with constraining the state estimation procedure to those regions of the water network where the low pipe flows have been detected (Figure 4.12). This scheme results in a highly stable state estimator even for the zones of the water network with low flow condition such as the one discussed here. No ‘a-priori’ information is needed about the low flow condition zones and only the topological distance in the spanning tree is used to delimit the regions in the water network which are state estimated.

It is interesting to notice that in our experiment, flow measurements introduced in the co-tree pipes did not affect the convergence of the state estimator even for the low flow regions of the water network. It suggests that the error calculated as the difference between the loop corrective flows at successive iterations is decreasing monotonically towards zero.

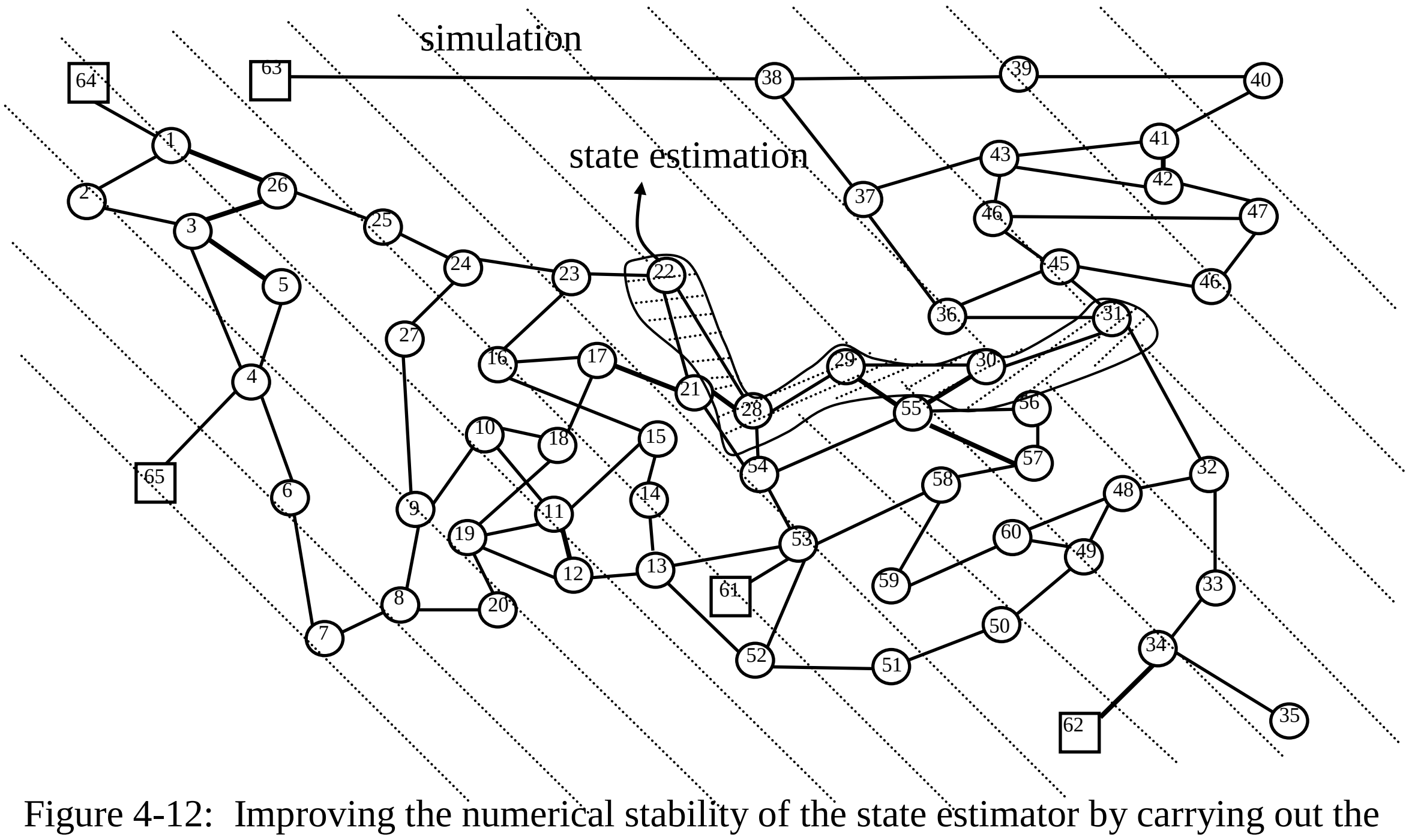

Figure 4-12: Improving the numerical stability of the state estimator by carrying out the state estimation procedure to a region of the water network, that is shown with horizontal dashed line above.

Finally, the state estimator showed a very good convergence and numerical stability for all the simulations carried out on the water network shown at Figure 4-12.

### 4.2.5. The nature of the numerical results of the LS loop flows state estimator

The loop flows state estimator will be the heart of a decision system for the operational control of water networks. Therefore it is important not only to acknowledge the good convergence and numerical stability of the state estimator but also to have an insight into the numerical results that it produces it.

As for example the proposed decision system will contain a module for quantification of the uncertainty in the state estimates due to the presence of inaccuracies in the input data. This module is called *Confidence Limit Analysis* (CLA) and will make use of the state estimator in order to calculate the confidence intervals on the state estimates. The sizes of these intervals depend on the properties of the state estimator and the confidence algorithm employed in calculations. It might be the case that state estimators using different sets of variables in order to construct the network equations will deliver different state vectors and subsequently different confidence intervals.

In general a state estimate achieved with the LS criterion represents a point in the space of feasible solutions minimizing the sum of squares of distances between itself and the measurement hyperplanes. Therefore the position of the solution point in the space depends on the existing *measurement hyperplanes*. Those measurements hyperplanes are subsequently depending on the set of independent variables (loop corrective flows or nodal heads) used to build the network equations.

***Inflows and fixed-head nodes***

One of the main characteristics of the *loop flows state estimator* is the main root node from where the spanning tree is constructed and the loops are obtained. The pseudo-loops are the loops that connect the fixed-head nodes to the main root node (Figure 4-13a).

It will be one equation in the estimator for each loop corrective flow, which will hold for the pressure difference between the main root node and each of the fixed-head nodes from the spanning tree. Therefore in the loop flows state estimator we can introduce as measurement data the value of the *fixed-head node* (pressure value at node 5 in Figure 4-13a) but we can not use in the same time the *inflow* at the same node since the loop corrective flow between the respective node and the main source will modify the inflow value according to the water consumptions and the other boundary nodes (i.e. fixed-head nodes) existing in the network.

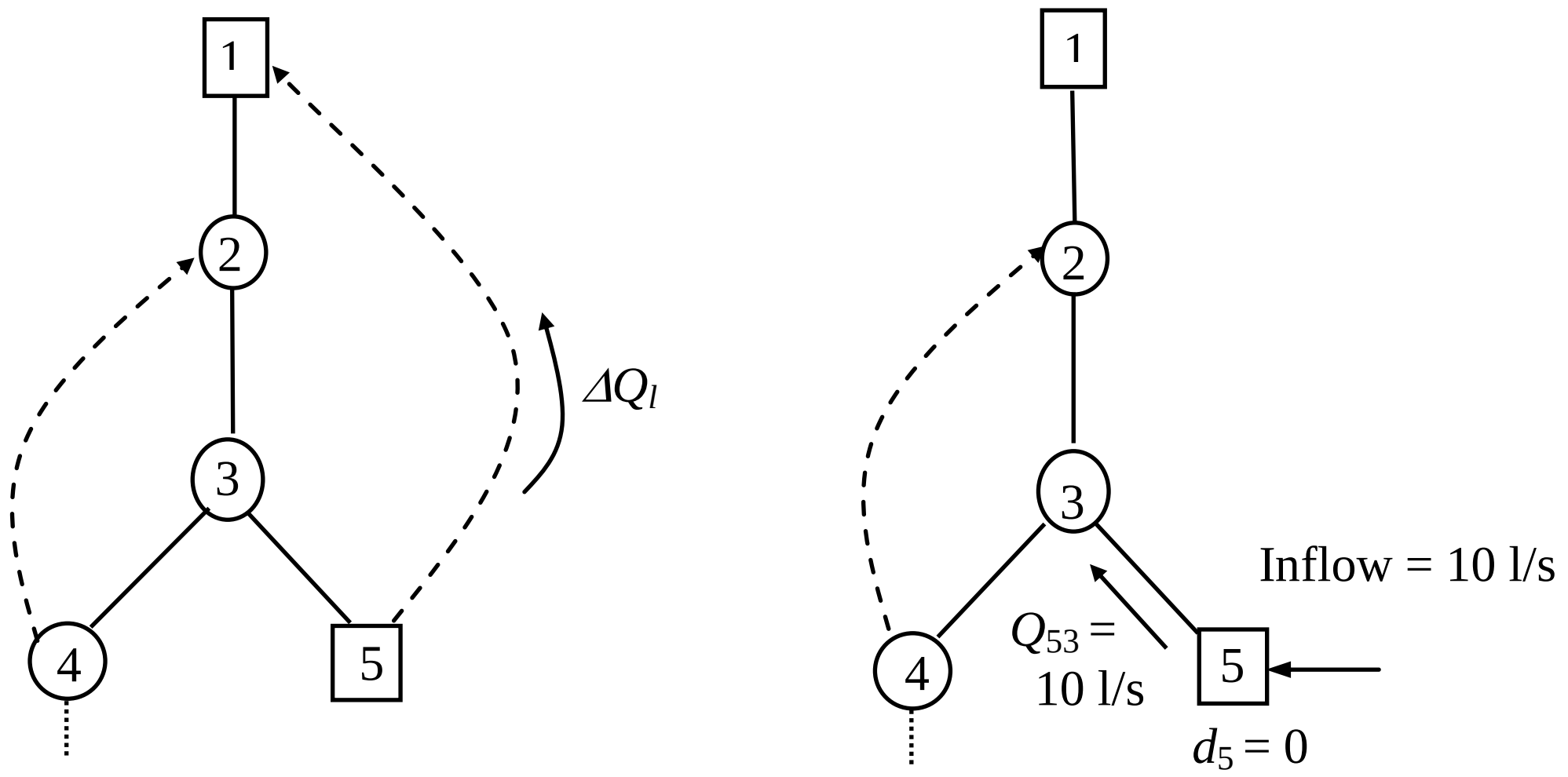

Figure 4-13: a) Pseudo-loop between node 5 and node 1, b) Inflow measurement.

On the other hand if we know the inflow but not the head at node 5 then we can calculate the initial flow solution considering the *inflow* as measurement data and discarding the pseudo-loop (Figure 4-13b).

In Figure 4-13, $\Delta Q_l$ is the loop corrective flow determined from the pseudo-loop that connects the fixed-head node 5 to the main source node 1. In this case the head at node 5 is considered known and the inflow is calculated as follows:

$$Inflow_{\text{new}} = Inflow_{\text{initial}} - \Delta Q_l \qquad \text{(Eq. 4.35)}$$

where $\text{Inflow}_{\text{initial}}$ is the inflow used at the beginning of the Newton-Raphson method to calculate the initial pipe flows and $\text{Inflow}_{\text{new}}$ is the inflow calculated at the end of the Newton-Raphson method when $\Delta Q_l$ becomes available.

If the inflow at node 5 is the measurement data then the pseudo-loop is discarded. The initial flow in the pipe 5-3 is equal to the value of the inflow since the demand $d_5$ is zero. As there are no other loops that include pipe 5-3, the flow through the pipe will remain constant when solving the system of non-linear equations with the Newton-Raphson method. The head at node 5 will be calculated as the difference between the head value at the main source and the head losses in the pipes that are on the path to node 5.

It can be observed that in the examples shown at Figure 4-13 we used as measurements the head of the *main source* and the head or the inflow at node 5. However, it is not possible to consider as measurement data, in whichever combination of pressure and flow measurements, the inflow at the main source node 1. The condition to have a main source node where the inflow can not be considered known *a priori* may sound as a disadvantage. As we will see in the following sections, this will actually help to detect the topological error (e.g. leakages) and to determine efficiently the confidence intervals.

***Pressure and flow measurements***

For the water network shown at Figure 4-14 we use the nodal heads and the loop flows state estimators and then we compare the results. The measurement data used in the simulations is displayed at Table 4-2.

| | |
|---|---|
| Head measurements | 1, 2, 4, 8, 11, 15, 17, 19, 22 |
| Fixed-head inflow measurements | 27, 28, 29, 30, 31, 32, 33, 34 |
| Water consumptions | All nodes |
| Fixed-head measurements | 27, 28, 29, 30, 31, 32, 33, 34 |

Table 4-2: Measurement data for the 34-node water network.

We build the spanning tree from the node 30 which is the main source node. A pseudo-loop is added between the fixed-head node 31 and the *main source* node 30. The inflow at the fixed-head nodes 27, 28, 29, 32, 33 and 34 is considered constant and used as measurement data in the loop flows state estimator. Consequently, no pseudo-loops are added for these nodes. In Table 4-3 on the second column are displayed the simulated nodal heads when a leakage is modeled as an additional water consumption of 15 l/s in the pipe 14-15. On the third column are shown the state estimates obtained with the loop flows state estimator and on the fourth column are shown the state estimates obtained with the *nodal heads state estimator*. We compare the numerical results of the two state estimators by looking to the pressure measurement at node 15, which is the closest pressure measurement to the location of the leakage. It is obvious that the loop flows state estimator gives ‘better’ results compared to the nodal heads state estimator. ‘Better’ results means that the loop flows state estimator is able to deliver the set of state estimates that entirely satisfy the pressure measurement at node 15. This is not the case with the nodal heads state estimator.

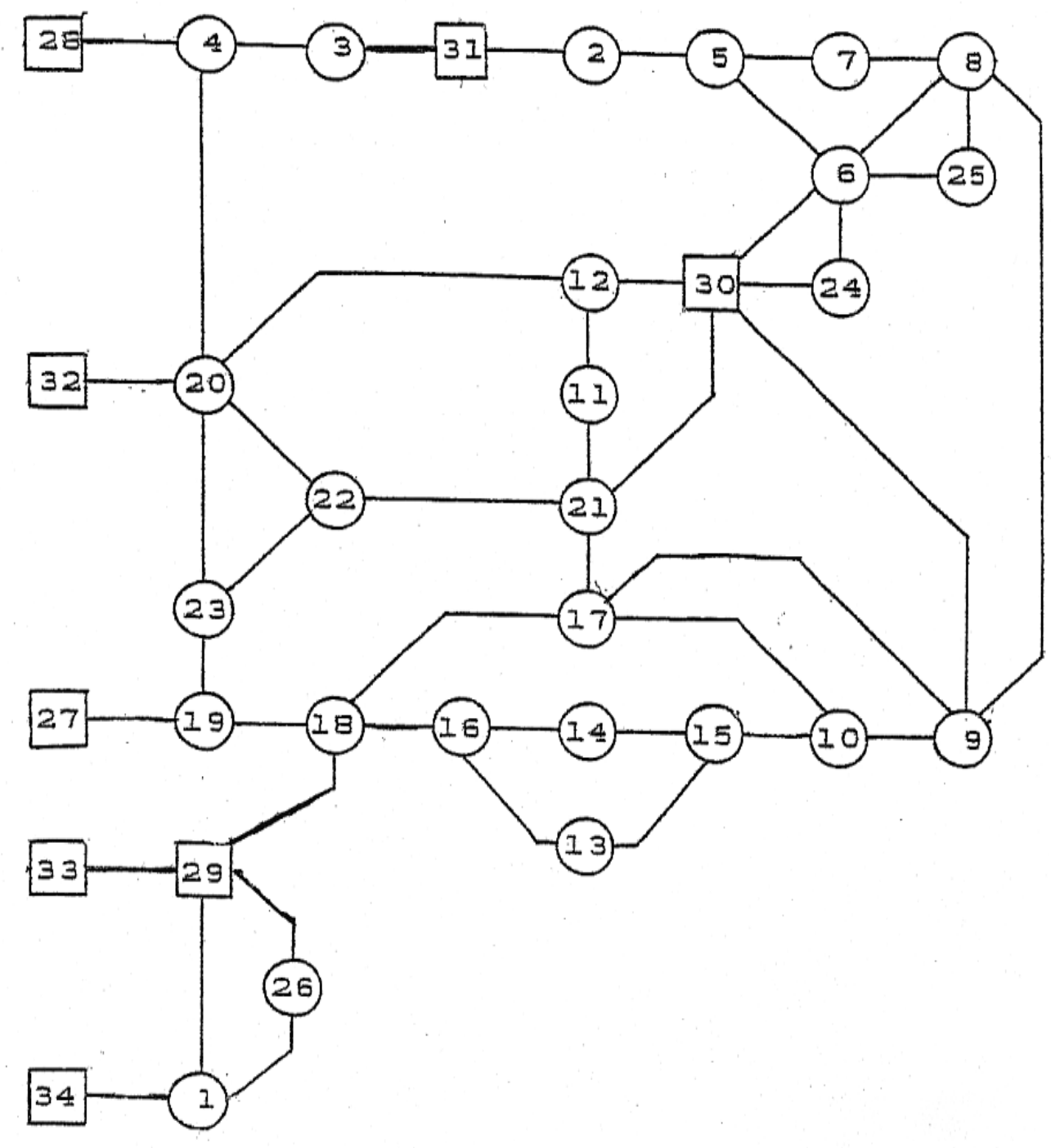

Figure 4-14: 34-node water network.

The explanation for the above situation is discussed next. The Jacobian and Hessian matrixes can be partitioned according to the state estimator (loop flows or nodal heads) used in calculations and the type of measurements.

$$J_l = \begin{bmatrix} \frac{\partial \Delta H}{\partial \Delta d} & \frac{\partial \Delta H}{\partial \Delta Q_l} \\ -\frac{\partial \Delta d}{\partial \Delta d} & 0 \\ 0 & \frac{\partial H}{\partial \Delta Q_l} \\ \frac{\partial F}{\partial \Delta d} & \frac{\partial F}{\partial \Delta Q_l} \end{bmatrix}$$

a)

$$J_n = \begin{bmatrix} \frac{\partial L}{\partial H} & \frac{\partial L}{\partial Inflow} \\ \frac{\partial H}{\partial H} & 0 \\ \frac{\partial F}{\partial H} & 0 \\ 0 & \frac{\partial Inflow}{\partial Inflow} \end{bmatrix}$$

b)

$$H_l = \begin{bmatrix} \frac{\partial^2 \Delta H}{\partial \Delta d^2} & \frac{\partial^2 \Delta H}{\partial \Delta Q_l^2} \\ 0 & 0 \\ \frac{\partial^2 H}{\partial \Delta d^2} & \frac{\partial^2 H}{\partial \Delta Q_l^2} \\ 0 & 0 \end{bmatrix}$$

c)

$$H_n = \begin{bmatrix} \frac{\partial^2 L}{\partial H^2} & 0 \\ 0 & 0 \\ \frac{\partial^2 F}{\partial H^2} & 0 \\ 0 & 0 \end{bmatrix}$$

d)

Figure 4-15: The Jacobian and the Hessian matrixes of the loop flows state estimator ($l$) and the nodal heads state estimator ($n$).

| State variable | Simulated values | S.E. obtained with the loop state estimator. | S.E. obtained with the nodal heads state estimator. |
|---|---|---|---|
| 1 | 31.5653 | 31.5560 | 31.5100 |
| 2 | 43.7285 | 43.7271 | 43.7270 |
| 3 | 45.8230 | 45.8218 | 45.7986 |
| 4 | 46.3367 | 46.3352 | 46.3079 |
| 5 | 43.2182 | 43.2169 | 43.2188 |
| 6 | 42.9314 | 42.9304 | 42.9182 |
| 7 | 42.3066 | 42.3052 | 42.3044 |
| 8 | 42.0317 | 42.0303 | 42.0243 |
| 9 | 43.6634 | 43.6457 | 43.6370 |
| 10 | 45.7147 | 45.3114 | 45.6821 |
| 11 | 44.5385 | 44.5370 | 44.5224 |
| 12 | 43.9563 | 43.9546 | 43.9362 |
| 13 | 47.9004 | 47.8206 | 47.8014 |
| 14 | 45.7848 | 45.7330 | 46.1468 |
| **15** | **45.6675** | **45.6675** | **46.0849** |
| 16 | 48.0820 | 48.0468 | 48.0099 |
| 17 | 46.8565 | 46.8550 | 46.7829 |
| 18 | 48.2814 | 48.2721 | 48.2137 |
| 19 | 48.1606 | 48.1606 | 48.0983 |
| 20 | 46.3327 | 46.3312 | 46.3034 |
| 21 | 45.3763 | 45.3753 | 45.3468 |
| 22 | 46.2572 | 46.2572 | 46.2250 |
| 23 | 47.6503 | 47.6530 | 47.5946 |
| 24 | 43.1938 | 43.1929 | 43.1785 |
| 25 | 42.4350 | 42.4337 | 42.4254 |
| 26 | 31.0197 | 31.0104 | 30.9631 |
| 27 | -16.1085 | -16.1086 | -16.0806 |
| 28 | -33.8088 | -33.8103 | -33.8088 |
| 29 | 30.6295 | 30.6203 | 30.5700 |
| 30 | 43.5820 | 43.5820 | 43.5609 |
| 31 | 44.1879 | 44.1865 | 44.1746 |
| 32 | -46.6972 | -46.6987 | -46.6972 |
| 33 | -37.6454 | -37.6546 | -36.6454 |
| 34 | -13.2992 | -13.3003 | -13.2991 |
| 35 | 0.0723 | 0.0723 | 0.0723 |
| 36 | 0.0927 | 0.0927 | 0.0927 |
| 37 | -0.0229 | -0.0229 | -0.0229 |
| 38 | -0.0384 | -0.0379 | -0.0384 |
| 39 | -0.0325 | -0.0325 | -0.0323 |
| 40 | 0.0254 | 0.0254 | 0.0254 |
| 41 | 0.0614 | 0.0614 | 0.0614 |
| 42 | 0.1063 | 0.1063 | 0.1063 |

Table 4-3: State estimates obtained by the loop flows and the nodal heads LS state estimators for an additional water consumption of 15 l/s in pipe 14-15.

The *state variables* 1 to 34 are the nodal heads [m] at nodes 1-34 and the state variables 35-42 are the fixed-head nodes in/out flows [$m^3$/s] at nodes 27-34.

We denoted with $J_l$ and $H_l$ the Jacobian and the Hessian matrixes of the loop flows state estimator. $J_n$ and $H_n$ are the Jacobian and the Hessian matrixes of the nodal heads state estimator. $\Delta H$ are the residual loop head losses.

*H* and *F* are the pressure and flow measurements in the loop flow state estimator. Their formulas are given by the equations (Eq. 4.29 - Eq. 4.30) for the pressure measurement and equation (Eq. 4.28) for the flow measurement. As it has been previously shown, the second derivatives of a flow measurement are zero and this has been shown in the Hessian matrix (i.e. the fourth row in the matrix at Figure 4.15c).

In the nodal heads state estimator *L* are the *n*-continuity equations and the *Inflow* variables represent the fixed-head node inflows. The formulas of the first and second derivatives of the real measurements used in the nodal heads state estimator have been presented in the literature (Bargiela, 1984).

The Hessian matrix gives the direction of minimizing the *objective function* that in our case comprises the loop head losses and the additional pressure and flow measurements. If we compare the Hessian matrix of the loop flows state estimator (Figure 4-15c) with the Hessian matrix of the nodal heads state estimator (Figure 4-15d), then we can observe that the second derivatives of the pressure measurements have vanished in the Hessian matrix of the nodal heads state estimator while they are present in the Hessian matrix of the loop flows state estimator. This explains why the loop flows state estimator is able to calculate the nodal demands for which the pressure measurements are satisfied and the loop head losses are zero.

It is important to observe that introducing a pressure measurement in the loop flows state estimator does not come in contradiction to finding the loop corrective flows for which the loop head losses are zero. This is because a pressure measurement implies the alteration of the nodal demands while the loop flows state estimator modifies the inflows into the fixed-head nodes so that the sum of the new nodal demands to match the amount of inflow into the network. We have previously mentioned the existence of the main source node where the inflow can not be considered known beforehand and can not be maintained constant during the iteration method. This is now explained by the recalculation of the inflows during the Newton-Raphson method and the modification of the nodal demands due to the presence of pressure and flow measurements.

With regard to the flow measurements in the loop flows state estimator: they can not be entirely accounted in the network model because their second derivatives in the state estimator Hessian matrix are equal to zero. However, we observed that if accurate measurements are available then the differences between the pipe flows calculated from the mathematical model and the flow measurements are constantly decreasing. This in itself is a very useful result because it proves that the loop flows state estimator has the tendency to fully solve the flow measurements when enough real measurements are available. The case in which there is a pressure measurement at one of the two end nodes of the pipe where a flow measurement is located, is a good example when not only a pressure measurement may be fully accounted for in the network model, but also a flow measurement (i.e. similar situation for the case with 2 pressure measurements).

Let us now turn the attention onto the nodal heads state estimator. As it has been reported in the literature, the nodal heads state estimator does not fully solve any of the n-continuity equations but reaches a point in the space of *feasible solutions* that minimizes the sum of squares of distances between the derived solution and all the measurement hyperplanes. This is because when a pressure measurement is affected by non-Gaussian noise (e.g. as would be the case with a leakage in the vicinity of the measurement) then minimizing the discrepancies in the first *n*-continuity equations may come in contradiction with the respective pressure measurement affected by the non-Gaussian noise.

The first four columns of Table 4-4 are the same as in Table 4-3 and on the fifth column are shown the state estimates obtained with the nodal heads state estimator for a flow measurement at pipe 15-10. The value of the flow measurement is 11.9 [l/s] while the estimated pipe flow obtained with the nodal heads state estimator is 5.4 [l/s]. Therefore we can observe that although the second derivatives of a flow measurement are still available in the Hessian matrix (Figure 4-15), it may not be fully solved by using the LS nodal heads state estimator.

| State variable | Simulated values | S.E. obtained with the loop state estimator. | S.E. obtained with the nodal heads state estimator. | S.E. obtained with the nodal heads state estimator (flow measurement at pipe 15-10) |
|---|---|---|---|---|
| 1 | 31.5653 | 31.5560 | 31.5100 | 31.5135 |
| 2 | 43.7285 | 43.7271 | 43.7270 | 43.7271 |
| 3 | 45.8230 | 45.8218 | 45.7986 | 45.7994 |
| 4 | 46.3367 | 46.3352 | 46.3079 | 46.3086 |
| 5 | 43.2182 | 43.2169 | 43.2188 | 43.2184 |
| 6 | 42.9314 | 42.9304 | 42.9182 | 42.9165 |
| 7 | 42.3066 | 42.3052 | 42.3044 | 42.3034 |
| 8 | 42.0317 | 42.0303 | 42.0243 | 42.0229 |
| 9 | 43.6634 | 43.6457 | 43.6370 | 43.6323 |
| 10 | 45.7147 | 45.3114 | 45.6821 | 45.6113 |
| 11 | 44.5385 | 44.5370 | 44.5224 | 44.5216 |
| 12 | 43.9563 | 43.9546 | 43.9362 | 43.9346 |
| 13 | 47.9004 | 47.8206 | 47.8014 | 47.8053 |
| 14 | 45.7848 | 45.7330 | 46.1468 | 46.1841 |
| **15** | **45.6675** | **45.6675** | **46.0849** | **46.0794** |
| 16 | 48.0820 | 48.0468 | 48.0099 | 48.0162 |
| 17 | 46.8565 | 46.8550 | 46.7829 | 46.7744 |
| 18 | 48.2814 | 48.2721 | 48.2137 | 48.2176 |
| 19 | 48.1606 | 48.1606 | 48.0983 | 48.1018 |
| 20 | 46.3327 | 46.3312 | 46.3034 | 46.3042 |
| 21 | 45.3763 | 45.3753 | 45.3468 | 45.3464 |
| 22 | 46.2572 | 46.2572 | 46.2250 | 46.2257 |
| 23 | 47.6503 | 47.6530 | 47.5946 | 47.5975 |
| 24 | 43.1938 | 43.1929 | 43.1785 | 43.1767 |
| 25 | 42.4350 | 42.4337 | 42.4254 | 42.4239 |
| 26 | 31.0197 | 31.0104 | 30.9631 | 30.9667 |
| 27 | -16.1085 | -16.1086 | -16.0806 | -16.0804 |
| 28 | -33.8088 | -33.8103 | -33.8088 | -33.8088 |
| 29 | 30.6295 | 30.6203 | 30.5700 | 30.5738 |
| 30 | 43.5820 | 43.5820 | 43.5609 | 43.5590 |
| 31 | 44.1879 | 44.1865 | 44.1746 | 44.1752 |
| 32 | -46.6972 | -46.6987 | -46.6972 | -46.6972 |
| 33 | -37.6454 | -37.6546 | -36.6454 | -37.6454 |
| 34 | -13.2992 | -13.3003 | -13.2991 | -13.2991 |

Table 4-4: State estimates obtained with the loop flows and the nodal heads LS state estimators for an additional water consumption of 15 l/s in pipe 14-15 (on the last column are shown the state estimates with a flow measurement at pipe 15-10).

The following conclusion can be reached with regard to the characteristics of the loop flows state estimator: if accurate pressure measurements exist, then the loop flows state estimator will give accurate results. Otherwise the corrupted pressure measurements will affect the final state estimates and eventually mislead the human operator.

We can also use the inflows at the fixed-head nodes as measurement data. Alternatively, the head values of the fixed-nodes can be used to form the pseudo-loops. In this last case the inflows are used only to calculate the initial pipe flows.

Flow measurements may improve the accuracy of the state estimates but not at the extent of a pressure measurement.

## 4.3. Concluding remarks

In this chapter, a novel LS state estimator that is suitable for on-line monitoring of the water distribution systems is presented. Present day deterministic state estimation techniques are very efficient, having small computational requirements and producing results of an acceptable level of accuracy. However for particular water networks, like ones displaying low pipe flows, the numerical stability of the algorithm may suffer. A solution to this problem is to employ the more stable loop flows state estimation techniques. Therefore a new formulation of the standard least squares (LS) criterion for water networks is developed. It is shown that the loop corrective flows do not provide enough basis to build the network equations of the state estimator. Therefore the state variables are both the loop corrective flows and the variation of nodal demands. Using the variation of nodal demands in addition to the loop corrective flows do not pose any problems on the input information that is needed in order to develop the mathematical model of the water network. It has been shown that this information can be derived from the spanning tree obtained for the co-tree flows simulator and so there is a natural connection between the simulator algorithm and the state estimator.In spite of the increased size of the state vector, a satisfactory convergence is obtained through an enhancement in the Jacobian matrix for the loop corrective flows. Furthermore a fine-tuning of the inverse of the tree incidence matrix is used in order to avoid the lack of numerical stability characteristic to the nodal heads state estimators. A very efficient and effective LS state estimator has been developed that has been tested successfully on realistic water networks. In the final part of the chapter a comparison has been made with the nodal heads state estimator and some of the intrinsic properties of the loop flow state estimator have been shown. The state estimator developed here represents one of the major contributions in this book and will serve as a central part for a decision support system for fault detection and preventive maintenance of water distribution systems.

# Chapter 5.

# Confidence limit analysis – a loop flows approach

## 5.1. Introduction

In order to supply water to consumers without any disruption in service, the state of the water distribution system has to be monitored. In the previous chapter it has been shown that this can be achieved by using state estimators that provide a means of combining diverse measurements by relating them to the mathematical model of the system (Bargiela, 1984; Sterling & Bargiela, 1984; Powell et al., 1988). Although the mathematical model may be accurate, the state estimates are based on *input data* that contain a significant amount of *uncertainty*. The uncertainty in input data associated with the real measurements, flows and pressures, and the pseudo-measurements, estimation of the water consumptions, is discussed here in the context of the loop flows state estimation technique.

The measurement uncertainty has an impact on the accuracy with which the state estimates are calculated. It is important, therefore, that the system operators are given not only the values of flows and pressures in the network at any instant of time but also that they have some indications of how reliable these values are. The procedure for the quantification of the inaccuracy of the state estimates caused by the input data uncertainty was developed in the late 1980s and termed *Confidence Limit Analysis* (CLA) (Bargiela & Hainsworth, 1989). Rather than a single deterministic state estimate, the CLA enables the calculation of a set of all feasible states corresponding to a given level of measurement uncertainty. The set is presented in the form of upper and lower *bounds* for individual variables and hence provide limits on the potential error of each variable. A decision system build on the concept of confidence limit analysis has been further developed (Gabrys & Bargiela, 1999). It performed like a fault detection and identification system being able to distinguish between different types of errors that are occurring in water networks. Although a great amount of work has been done, and significant results have been delivered in the area of uncertainty analysis of water networks, they were obtained with the nodal network equations. This raises the question of the potential benefits of using the loop equations for CLA.

Employing the loop corrective flows variables for the numerical simulations, has received an increased attention in the last years. As it has been shown in the first chapters as well as reported in the literature of speciality, satisfactory convergence and good numerical stability have been achieved for the loop flows based simulations (Arsene & Bargiela, 2001; Arsene & Bargiela, 2002a; Andersen & Powell, 1999a). In spite of this work, the results in the area of CLA when using the loop flows algorithms are scarce (Nagar et al., 2002).

This chapter addresses the problem of CLA based on the loop flows state estimator and the co-tree flows simulator algorithm shown in the previous chapters. It investigates both the relationship between the quality of measurement data and the quality of the confidence limits for the individual state estimates, as well as the nature of the relationship itself. It shows several CLA algorithms and compares them with the CLA algorithms that are using a nodal heads formulation.

The chapter is organized as follows: next section presents the review of previous research in uncertainty analysis of water networks followed by the description of the Experimental Sensitivity Matrix method for CLA. This method will make use of the loop flows state estimator. Although computationally inefficient, the Experimental Sensitivity Matrix method will give a useful reference point for interpreting further results. A Sensitivity Matrix method within the loop equations framework and an Error Maximization technique will be then developed. The performances of these algorithms will be assessed in terms of their computational complexity and the accuracy of the results that they produce.

## 5.2. Uncertainty analysis in water networks

In the previous chapter it has been shown that for a given set of input data and estimation criterion there is one optimal solution. However due to the inaccuracies in the input data, there are many possible, different combinations of such input data and therefore there are many feasible, different state estimate vectors. As a result the uncertainty analysis becomes an inevitable part of the water distribution systems since it is very important, from the safety of the system operational control point of view, to know how the inaccuracies can affect the estimated solution.

Extensive work on the quantification of the influence of measurement and pseudomeasurement uncertainties in water distribution system has been done in Bargiela and Hainsworth (1989) and carried out further in Gabrys and Bargiela (1996), Brdys and Chen (1993) and Gabrys (1997). It is called *Confidence Limit Analysis* and is based on the principle of *unknown-but-bounded errors* for the set of measurements:

$$z=g(x)+r\ ,\ |r_i| \leq |e_i|\ ,\ i=1,\dots,m \qquad \text{(Eq. 5.1)}$$

where $e$ is the vector representing the maximum expected measurement errors, $z$ is the measurement vector, $g$ is the network function and $x$ are the state variables. The knowledge of statistical properties of errors is not required and the only restriction imposed was the one of errors falling within a range bounded by $e$. A several CLA algorithms were proposed but the most successful ones in terms of computational complexity were based on the linearized model of the water network. From those, the Sensitivity Matrix method proved to be efficient enough to be used in real time decision support.

The linearized model of the water network was used to obtain the *sensitivity matrix* $S$. The sensitivity matrix was the pseudo-inverse of the Jacobian matrix calculated for the state estimates by using a deterministic state estimator. A state estimate was produced on the assumption that the measurement vector $z^t$ is correct and the possible error of the measurement set $\Delta z$ was considered and used together with the sensitivity matrix $S$ in order to predict the resulting error in the state vector.

This approach was facilitated by the use of the nodal heads equations in the state estimator. Because of this, the $(i,j)$-th element $s_{ij}$ of the pseudo-inverse of the Jacobian matrix relates the sensitivity of the $i$-th element, $x_i$, of the state vector, $x^t$, to the $j$-th element, $z_j$, of the measurement vector. Calculating the confidence limits for the state variables $x_i$ was produced as:

$$x_{cl_i} = max\ s_i \Delta z \qquad \text{(Eq. 5.2)}$$

where $s_i$ is the $i$-th row of the sensitivity matrix $S$ and $\Delta z$ represents the perturbations in the vector of measurements. The underlying principle of the CLA is to consider the worst possible case for the perturbations in the vector of measurements (i.e. the maximum variability of consumptions and inaccuracies for real meters).

The method was very efficient and flexible since in the state estimator any combination of real measurements could be used in conjunction with the water consumptions in order to obtain the confidence limits on the state variables.

Unfortunately the Sensitivity Matrix method as described above is difficult to follow within the loop equations framework.

Let us rewrite the equation (Eq. 3.5) that gives the loop corrective flows in the simulator algorithm:

$$\Delta Q_l = -\left[\frac{\partial \Delta H}{\partial \Delta Q_l}\right]^{-1} \Delta H \qquad \text{(Eq. 5.3)}$$

The element $s_{ij}$ of the inverse of the Jacobian matrix $\left[\frac{\partial \Delta H}{\partial \Delta Q_l}\right]^{-1}$ relates the *sensitivity* of the *i*-th element, $\Delta Q_{l_i}$, of the vector of loop corrective flows, $\Delta Q_l$, to the *j*-th element, $\Delta H_j$, of the vector of loop head losses residuals.

The relationship carried out by the inverse of the Jacobian matrix between the measurement vector (only water consumptions in the previous equation) and the loop corrective flows is not straight forward as in nodal heads based algorithms but is realized through the mean of the loop head losses residuals (i.e. vector $\Delta H$).

Introducing an error, $\Delta d$, in the pseudo-measurement vector, $d$, then a perturbation, $\delta(\Delta H)$, in the vector of loop head losses is obtained, and further an alteration of the loop corrective flows from equation (Eq. 5.3). Therefore the value $\delta(\Delta H)$ is important to determine for a given level of errors in the vector of measurements: providing that $\delta(\Delta H)$ is determinable, the confidence limits on the loop corrective flows would be obtained and the other variables of interest (i.e. tree flows and nodal heads).

However, we will see in this chapter that the correct value of $\delta(\Delta H)$ is difficult to calculate for the variability of *water consumptions*. The analysis can become even more tedious if real measurements are included. In the next section an Experimental Sensitivity Matrix is constructed that has the properties of the pseudo-inverse of the Jacobian matrix from the nodal heads based simulations.

## 5.3. Experimental Sensitivity Matrix

In normal use, deterministic state estimators produce one set of state variables for one measurement vector. Used in this way, they give no indication of how the state variables may be affected by the fuzziness of input data. Alternatively, if a deterministic state estimator is used repeatedly for each measurement modified with its defined *maximum variability*, then a matrix $S^e$ can be determined as:

$$S^e = \frac{\Delta x_i}{\Delta z_j} \quad i=1,\ldots n;\, j=1,\ldots m \qquad \text{(Eq. 5.4)}$$

where $i=1,\ldots n$ is the index for the state vector $x^t$, that is nodal heads and in/out flows, and $j=1,\ldots m$ is the index for the measurement vector $z^t$.

The measurement vector $z^t$ comprises the estimates for the water consumptions and the fixed-head nodes. It can be augmented with real pressure and flow meters. The loop flows state estimator will be the deterministic state estimator used to obtain the *experimental sensitivity matrix*.

Matrix $S^e$ is called the Experimental Sensitivity Matrix (ESM) since resembles the characteristics of the pseudo-inverse of the Jacobian matrix from the nodal heads state estimator and is obtained through a number of successive simulations. It expresses the variation, $\Delta x$, of the $i$-th element, $x_i$, of the true state vector, $x^t$, because of a perturbation, $\Delta z$, in the $j$-th element, $z_j$, of the true measurement vector $z^t$.

The true state of the system is not known but instead the best state vector available $\hat{x}$ is used in the process of determining the sensitivity matrix and the confidence limits.

The method is applied for the water network shown at Figure 5-1. For real measurements and pseudo-measurements an interval is defined $[z_l, z_u]$ according to the relative variability of $z^t$. The variability of the pseudo-measurements is ±20% and the accuracy of the fixed-head nodes is ±0.01 [m].

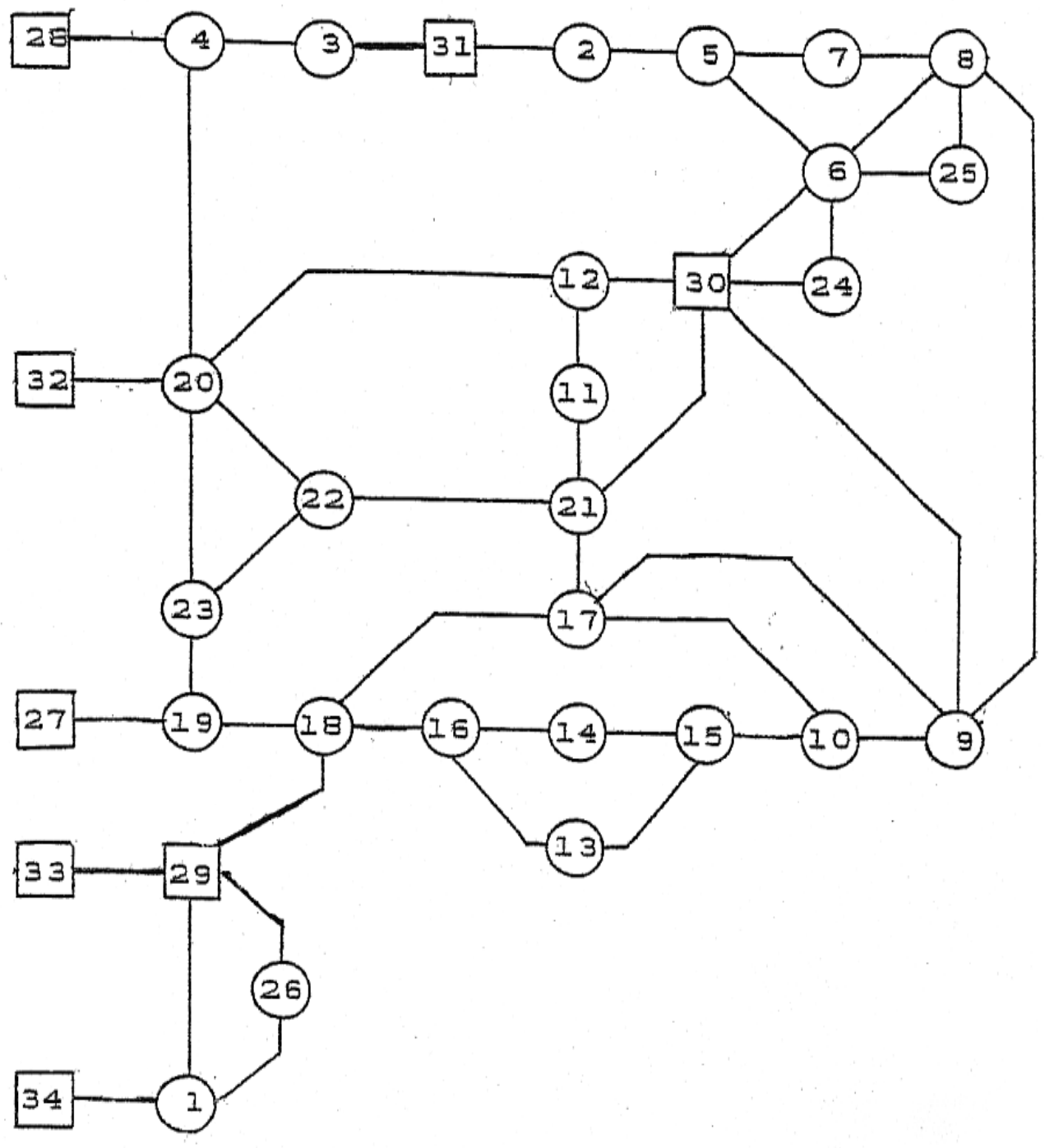

Figure 5-1: Realistic water network.

In real water networks, the *true measurement vector* $z^t$ rarely coincides with the *observed measurement vector* $z^o$. This discrepancy is caused by meter noise or meter error in the case of real measurements, and because of the difficulty in predicting demand in the case of nodal consumptions. Because of this, the measurement values used (i.e. the observed measurement values in Table 5-1 and Table 5-2) are not the same as the true measurement values that would be expected for the true operating state and are listed in the 2-nd column of Table 5-1 and the 2-nd and 5-th column of Table 5-2.

| Fixed-head nodes [m] | | |
|---|---|---|
| Node | True | Obs. |
| 27 | -15.1991 | -15.1991 |
| 28 | -33.4879 | -33.4978 |
| 29 | 31.7221 | 31.7221 |
| 30 | 43.5619 | 43.5819 |
| 31 | 44.1710 | 44.1703 |
| 32 | -46.3814 | -46.3814 |
| 33 | -36.5470 | -36.5478 |
| 34 | -12.1990 | -12.1963 |

Table 5-1: True and observed fixed-head nodes.

| Nodal consumptions [l/s] | | | | | |
|---|---|---|---|---|---|
| Node | True | Obs. | Node | True | Obs. |
| 1 | 52.6 | 57.5 | 18 | 12.1 | 13.2 |
| 2 | 2.7 | 3.0 | 19 | 4.5 | 4.9 |
| 3 | 19.2 | 21 | 20 | 12.1 | 13.2 |
| 4 | 5.9 | 6.5 | 21 | 22.3 | 24.4 |
| 5 | 1.1 | 1.23 | 22 | 32.4 | 35.4 |
| 6 | 2.1 | 2.3 | 23 | 38.2 | 41.7 |
| 7 | 3.0 | 3.3 | 24 | 5.0 | 5.5 |
| 8 | 69.4 | 75.8 | 25 | 9.0 | 9.8 |
| 9 | 8.1 | 8.9 | 26 | 11.1 | 12.1 |
| 10 | 3.8 | 4.2 | 27 | 6.2 | 6.8 |
| 11 | 1.9 | 2.1 | 28 | 0 | 0 |
| 12 | 10.2 | 11.1 | 29 | 22.9 | 25 |
| 13 | 21.2 | 23.2 | 30 | 39.5 | 43.1 |
| 14 | 10.3 | 11.21 | 31 | 39.3 | 42.9 |
| 15 | 22.2 | 24.3 | 32 | 0 | 0 |
| 16 | 4.7 | 5.12 | 33 | 0 | 0 |
| 17 | 2.4 | 2.6 | 34 | 0 | 0 |

Table 5-2: True and observed nodal consumptions.

The observed measurement values, $z^o$, were selected randomly from within the range specified by the following upper and lower limits:

$$z_l = z^t\text{-}\Delta z_l \qquad \text{(Eq. 5.5)}$$

$$z_u = z^t\text{-}\Delta z_u. \qquad \text{(Eq. 5.6)}$$

An interval has been determined for the true measurement vector, $z^t\text{-}\Delta z_l \leq z^t \leq z^t\text{-}\Delta z_u$, that corresponds to the real-life situation where measurement values are not exact but are contained in a range specified by the accuracy of the real meters and the accuracy of the pseudo-measurements values.

The *state vector* $\hat{x}$ shown in columns 3 and 6 of Table 5-3 are the *state variables* (nodal heads and in/out flows) calculated for the observed measurement vector using the loop flows state estimator.

The difference between the observed state variable $\hat{x}$ and the true state $x^t$ should be noted. It is caused solely by the addition of the simulated measurement errors and shows how corrupted measurement data can affect deterministic state variables.

| True and observed state variables | | | | | |
|---|---|---|---|---|---|
| Nodal pressures [m] | | | Nodal pressures [m] | | |
| Node | True | Obs. | Node | True | Obs. |
| 1 | 31.1852 | 31.0577 | 23 | 44.0663 | 43.9127 |
| 2 | 43.3886 | 43.2835 | 24 | 42.9028 | 42.7773 |
| 3 | 44.2289 | 44.1968 | 25 | 42.0751 | 41.7974 |
| 4 | 44.3191 | 44.2706 | 26 | 31.3306 | 31.2399 |
| 5 | 42.8133 | 42.6358 | 27 | -15.1991 | -15.1991 |
| 6 | 42.6765 | 42.5082 | 28 | -33.4879 | -33.4966 |
| 7 | 41.8478 | 41.5228 | 29 | 31.7221 | 31.7242 |
| 8 | 41.7190 | 41.3762 | 30 | 43.5619 | 43.5819 |
| 9 | 43.0165 | 42.8746 | 31 | 44.1710 | 44.1715 |
| 10 | 41.6933 | 41.1195 | 32 | -46.3814 | -46.3798 |
| 11 | 43.5925 | 43.5813 | 33 | -36.5470 | -36.5457 |
| 12 | 43.5845 | 43.5817 | 34 | -12.1990 | -12.1942 |
| 13 | 45.3550 | 45.2569 | Inflows [l/s] | | |
| 14 | 40.1661 | 39.2083 | Node | True | Obs. |
| 15 | 43.0940 | 39.1235 | 27 | 34.0 | 35.2 |
| 16 | 43.4858 | 43.0441 | 28 | 96.5 | 96.6 |
| 17 | 43.9047 | 43.7263 | 29 | 64.3 | 73.4 |
| 18 | 44.7605 | 44.5342 | 30 | 106.3 | 130.2 |
| 19 | 44.3638 | 44.1934 | 31 | 38.9 | 48.7 |
| 20 | 44.1362 | 44.0702 | 32 | 6.0 | 6.0 |
| 21 | 43.6560 | 43.6053 | 33 | 121.7 | 121.7 |
| 22 | 43.8080 | 43.7161 | 34 | 21.6 | 22.8 |

Table 5-3: True and observed state variables.

If we consider vector $\hat{x}$ as being the *optimal state vector* (i.e. columns 3 and 6 of Table 5-3), then the observed measurement data $z^o$ is randomly modified $\Delta z$ according to the accuracy of the water consumptions and the fixed-head nodes. For the randomly generated measurement data the loop flows state estimator produces a variation of the state vector $\Delta\hat{x}_1$, which is shown at Figure 5-2:

$$\Delta\hat{x}_1 = \hat{x}_1 - \hat{x} \qquad \text{(Eq. 5.7)}$$

The observed measurement vector $z^o$ is used to obtain the experimental sensitivity matrix $S^e$. This matrix is multiplied with the variation of the measurement values $\Delta z$, which gives the variation of the state vector $\Delta\hat{x}_2$ shown at Figure 5-2:

$$\Delta\hat{x}_2 = S^e \Delta z \qquad \text{(Eq. 5.8)}$$

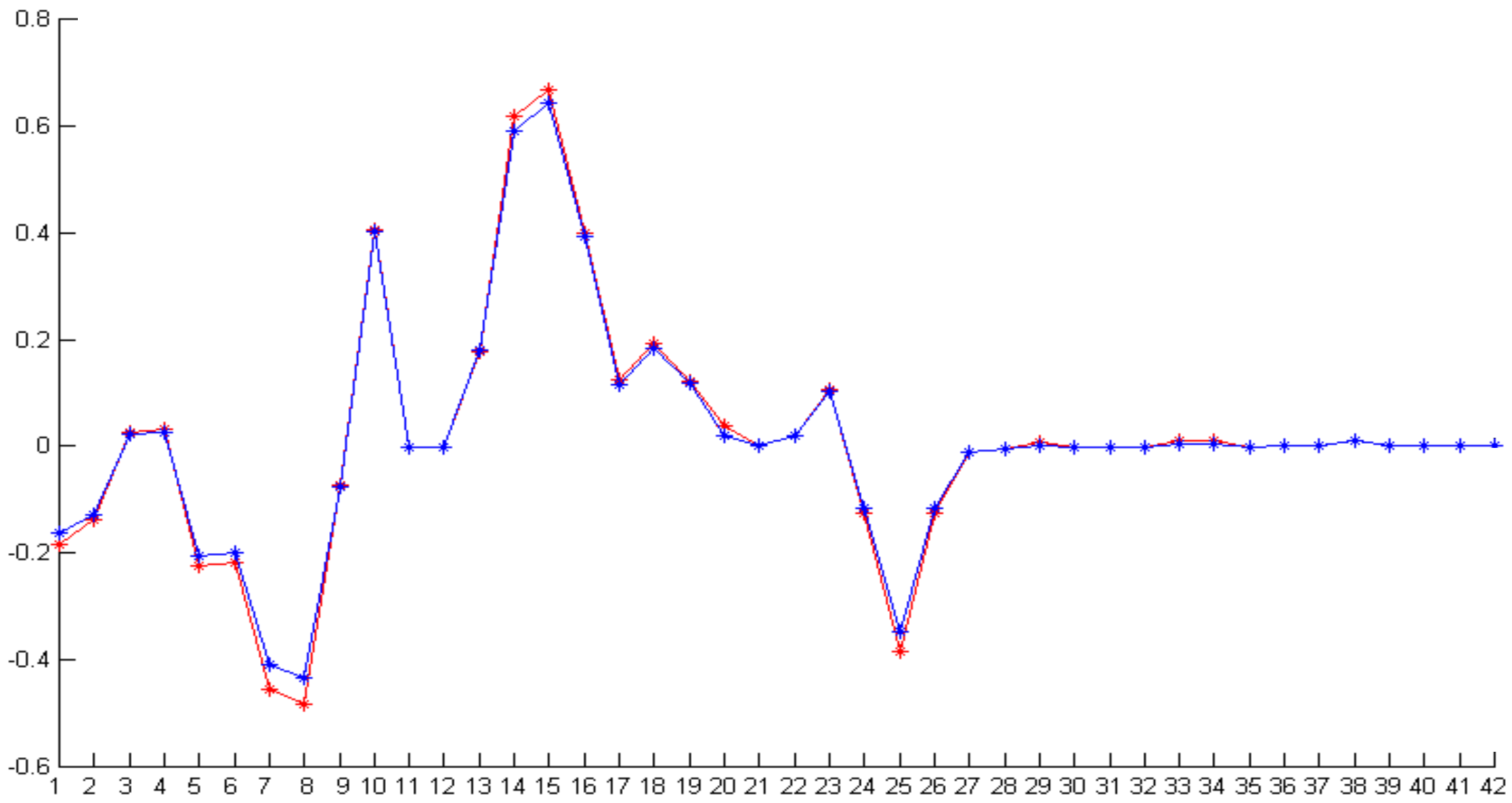

Figure 5-2: Variation of the state variables obtained with the ESM method and the loop flows state estimator.

1-34: variation of nodal heads [m] at nodes 1-34.

35-42: variation of fixed-head nodes in/out flows [$m^3$/s] at nodes 27-34.

The differences between the two sets of curves $\Delta\hat{x}_1$ and $\Delta\hat{x}_2$ are minimal and one can conclude that for a given set of measurements and level of errors associated with the set of measurements, the ESM method is an effective way of determining the state variables without running the loop flows state estimator.

### 5.3.1. Confidence limits based on Experimental Sensitivity Matrix method

Having found the matrix $S^e$, we can carry out the maximization process in order to obtain the confidence limits for the state variables (Eq. 5.2). For the *i*-th state variable, calculating its error bound is done by maximizing the product between the *i*-th row of the experimental sensitivity matrix $S^e$ and the vector $\Delta z$. The maximization process is performed separately for each row of the sensitivity matrix determined in the previous section. The confidence limits for the state variables (nodal heads and in/out flows) are shown on the 4-th column of Table 5-4.

The results have been obtained for the variability of consumptions ±20% and the accuracy of fixed-head nodes ±0.01 [m].

In the 5-th and 6-th column are shown the state variables and the confidence limits calculated with the Jacobian matrix from the state estimator based on the nodal heads equations. The confidence limits are comparable with the ones produced by the ESM method (i.e. the 4-th column of Table 5-4). It can be concluded that matrix $S^e$ resembles the properties of the pseudo-inverse of the Jacobian matrix from the nodal heads state estimator. It can be used as a substitute for the loop flows state estimator in order to determine the state variables and the confidence limits.

However, someone should perhaps observe that no pressure or flow measurements have been used in the example shown at Table 5-4. Nevertheless, in the absence of such real measurements the confidence limits obtained with the ESM method and the pseudo-inverse of the Jacobian matrix from the nodal heads state estimator show similarity.

Although the ESM method is effective in providing realistic state vectors and confidence limits, it requires as many simulations as the number of measurements and pseudo-measurements. Therefore the computational complexity tends to be a major drawback because even for a small-sized system, as discussed in this chapter, the number of feasible measurements is great, rendering this approach difficult to use in on-line decision support system. In view of these limitations, two alternative methods have been developed. In both methods an accurate linearization of the system model is used to reduce the mathematical complexity. The first uses the linearized network equations to construct a new sensitivity matrix that avoids the computational drawback from above. The second solves the linearized model of the water network for the maximum of errors in the estimated measurement vector.

| State variable | Exact State | State variables | Confidence limits | Nodal based s.e. | Nodal based c.l.a. |
|---|---|---|---|---|---|
| 1 | 31.1852 | 31.0577 | 0.3007 | 31.0566 | 0.3002 |
| 2 | 43.3886 | 43.2835 | 0.2557 | 43.2818 | 0.2604 |
| 3 | 44.2289 | 44.1968 | 0.0667 | 44.1960 | 0.0618 |
| 4 | 44.3191 | 44.2706 | 0.1013 | 44.2702 | 0.0965 |
| 5 | 42.8133 | 42.6358 | 0.4212 | 42.6336 | 0.4377 |
| 6 | 42.6765 | 42.5082 | 0.3971 | 42.5048 | 0.4166 |
| 7 | 41.8478 | 41.5228 | 0.7570 | 41.5214 | 0.8001 |
| 8 | 41.7190 | 41.3762 | 0.7925 | 41.3748 | 0.8441 |
| 9 | 43.0165 | 42.8746 | 0.3423 | 42.8282 | 0.3544 |
| 10 | 41.6933 | 41.0095 | 1.3561 | 40.9050 | 1.5754 |
| 11 | 43.5925 | 43.5813 | 0.0117 | 43.5815 | 0.0130 |
| 12 | 43.5845 | 43.5817 | 0.0033 | 43.5818 | 0.0035 |
| 13 | 45.3550 | 45.2569 | 1.3782 | 44.3555 | 1.9953 |
| 14 | 40.1661 | 39.3083 | 2.2626 | 39.5726 | 2.8137 |
| 15 | 43.0940 | 39.2235 | 2.2657 | 39.4799 | 2.8731 |
| 16 | 43.4858 | 43.0441 | 1.0413 | 43.1425 | 0.9928 |
| 17 | 43.9047 | 43.7263 | 0.4081 | 43.6956 | 0.3504 |
| 18 | 44.7605 | 44.5342 | 0.5268 | 44.5576 | 0.4893 |
| 19 | 44.3638 | 44.1934 | 0.3900 | 44.2076 | 0.3748 |
| 20 | 44.1362 | 44.0702 | 0.1410 | 44.0704 | 0.1360 |
| 21 | 43.6560 | 43.6053 | 0.1006 | 43.6056 | 0.0938 |
| 22 | 43.8080 | 43.7161 | 0.1993 | 43.7192 | 0.1899 |
| 23 | 44.0663 | 43.9127 | 0.3484 | 43.9223 | 0.3344 |
| 24 | 42.9028 | 42.7773 | 0.3027 | 42.7718 | 0.3119 |
| 25 | 42.0751 | 41.7974 | 0.6471 | 41.7951 | 0.6847 |
| 26 | 31.3306 | 31.2399 | 0.2259 | 31.2384 | 0.2096 |
| 27 | -15.1991 | -15.1991 | 0.0000 | -15.1991 | 0.0185 |
| 28 | -33.4879 | -33.4966 | 0.0151 | -33.4978 | 0.0191 |
| 29 | 31.7221 | 31.7242 | 0.0196 | 31.7221 | 0.0116 |
| 30 | 43.5619 | 43.5819 | 0.0004 | 43.5819 | 0.0102 |
| 31 | 44.1710 | 44.1715 | 0.0151 | 44.1703 | 0.0183 |
| 32 | -46.3814 | -46.3798 | 0.0151 | -46.3810 | 0.0197 |
| 33 | -36.5470 | -36.5457 | 0.0161 | -36.5478 | 0.0142 |
| 34 | -12.1990 | -12.1942 | 0.0135 | -12.1963 | 0.0159 |
| 35 | 34.0 | 35.2 | 3.1 | 35.6 | 3.7 |
| 36 | 96.5 | 96.6 | 0.1 | 97.0 | 0.1 |
| 37 | 64.3 | 73.4 | 21.4 | 73.4 | 21.5 |
| 38 | 106.3 | 130.2 | 56.9 | 130.1 | 56.2 |
| 39 | 38.9 | 48.7 | 22.8 | 48.3 | 22.0 |
| 40 | 6.0 | 6.0 | 0 | 6.0 | 0 |
| 41 | 121.7 | 121.7 | 0 | 121.7 | 0 |
| 42 | 21.6 | 22.8 | 2.6 | 22.8 | 2.6 |

Table 5-4: State variables and confidence limits for the 34-node water network;1-34: nodal heads at nodes 1-34; 35-42: fixed-head nodes in/out flows [$m^3/s$] at nodes 27-34.

## 5.4. Confidence limit analysis based on the linearized model of the water network

An alternative approach to the ESM method is the confidence limit algorithm based on the linearized model of the water network. If we use the loop flows state estimator, then the solution to the *linearized model* of the water network is:

$$\begin{bmatrix} \Delta d \\ \Delta Q_l \end{bmatrix} = (J_1^T J_1)^{-1} J_1^T \; g_1(\Delta d, \Delta H) \qquad \text{(Eq. 5.9)}$$

where $(J_1^T J_1)^{-1} J_1^T$ is the pseudo-inverse of the Jacobian matrix $J_1$ from the loop flows state estimator, and $g_1$ represent the loop head losses residuals and the variation of nodal demands.

In order to calculate the confidence limits for the nodal heads and pipe flows, we can use equation (Eq. 5.9) and derive the confidence limits on the variables $\Delta Q_l$ and $\Delta d$. Following this, by using equation (Eq. 4.10) from the loop flows state estimator, the confidence limits on pipe flows and subsequently on the nodal heads would be obtained:

$$\tilde{Q} = Q_i - A^* \Delta d + M_{pl} \Delta Q_l \qquad \text{(Eq. 5.10)}$$

Two algorithms are developed in the next sections based on the linearized model of the water network.

### 5.4.1. Sensitivity Matrix method within the loop framework

In the context of the nodal heads state estimator the elements of the pseudo-inverse of the Jacobian matrix were expressing a linear relationship between the elements of the state vector and the elements of the vector of measurements. Moreover the measurements together with their accuracy could be introduced explicitly in the linearized model of the water network. However this is not the case with the co-tree flows simulator algorithm or the loop flows state estimator.

Let us consider equation (Eq. 5.9) from the loop flows state estimator only for the loop corrective flows. We will then recall equation (3.5) from the simulator algorithm:

$$\Delta Q_l = -\left[\frac{\partial \Delta H}{\partial \Delta Q_l}\right]^{-1} \Delta H \qquad \text{(Eq. 5.11)}$$

where the matrix $\left[\frac{\partial \Delta H}{\partial \Delta Q_l}\right]^{-1}$ is the inverse of the Jacobian matrix from the simulator algorithm and $\Delta H$ is the vector of loop head losses residuals.

Equation (Eq. 5.11) can be used to determine the loop corrective flows for a given set of pseudo-measurements. No real meters are included at this stage. The vector of pseudo-measurements $d$ ($n$ x 1) is not introduced explicitly (as it is) in equation (Eq. 5.11) but through the mean of the vector of loop head losses residuals $\Delta H$ (Eq. 5.12 – Eq. 5.15):

$$Q_i = (T)^{-1} d \qquad \text{(Eq. 5.12)}$$

$$\tilde{Q} = Q_i + M_{lp}^T \Delta Q_l \qquad \text{(Eq. 5.13)}$$

$$h = k^n \tilde{Q} \qquad \text{(Eq. 5.14)}$$

$$\Delta H = M_{lp} h \qquad \text{(Eq. 5.15)}$$

where $h$ ($p$ x 1) represents the vector of pipe head losses and $k$ ($p$ x 1) is the pipe resistance coefficient.

A variation $\delta(\Delta H)$ of the vector of loop head losses residuals will be obtained if we modify with $\delta d$ the vector of pseudo-measurements and follow the set of equations (Eq. 5.12 – Eq. 5.15). This forms the basis of the first confidence limit algorithm based on the linearized model of the water network. The inverse of the Jacobian matrix of the co-tree flows simulator algorithm will act as a sensitivity matrix between the loop head losses residuals $\Delta H$ and the loop corrective flows $\Delta Q_l$.

The proposed algorithm comprises of a number of steps:

- By using the underlying principle of CLA of maximum errors in the pseudo-measurement vector $d$, that is [$d_l$  $d_u$], a confidence interval $\lfloor \Delta H_l \ \Delta H_u \rfloor$ is determined for the vector of loop head losses $\Delta H$.

- The confidence interval for the loop head losses together with the inverse of the Jacobian matrix calculated for the pseudo-measurement vector *d* will give the confidence limits on the loop corrective flows $\left[\Delta Q_l^l \ \Delta Q_l^u\right]$.

- The loop incidence matrix $M_{lp}$ and the limits $\left[\Delta Q_l^l \ \Delta Q_l^u\right]$ will produce the confidence limits on the tree flows and the nodal heads.

The scope of the algorithm is to calculate the confidence limits on the loop corrective flows without running the simulator algorithm but using the Jacobian matrix obtained for the pseudo-measurement vector *d* and the vector of loop head losses $\Delta H$.

Unfortunately the algorithm is inappropriate because of the impossibility of determining with accuracy of the confidence interval $\left[\Delta H_l \ \Delta H_u\right]$ for the vector $\Delta H$. This is because vectors $\Delta H_l$ and $\Delta H_u$ are calculated from the equations (Eq. 5.12 – Eq. 5.15) based on the non-realistic assumption that the initial co-tree pipes flows $Q_{C_i}$ are zero and the water consumptions are fed through the tree pipes.

At Figure 5-3 are displayed with dashed line the loop lead losses $\Delta H$ calculated from equation (Eq. 5.15) for the observed nodal consumptions. $\Delta H_t$ are the true loop head losses obtained from the equation:

$$\Delta H_t = -J \ \Delta Q_l \qquad \text{(Eq. 5.16)}$$

where *J* is the Jacobian matrix and $\Delta Q_l$ are the loop corrective flows obtained at the end of the Newton-Raphson iteration method for the observed pseudo-measurements *d*.

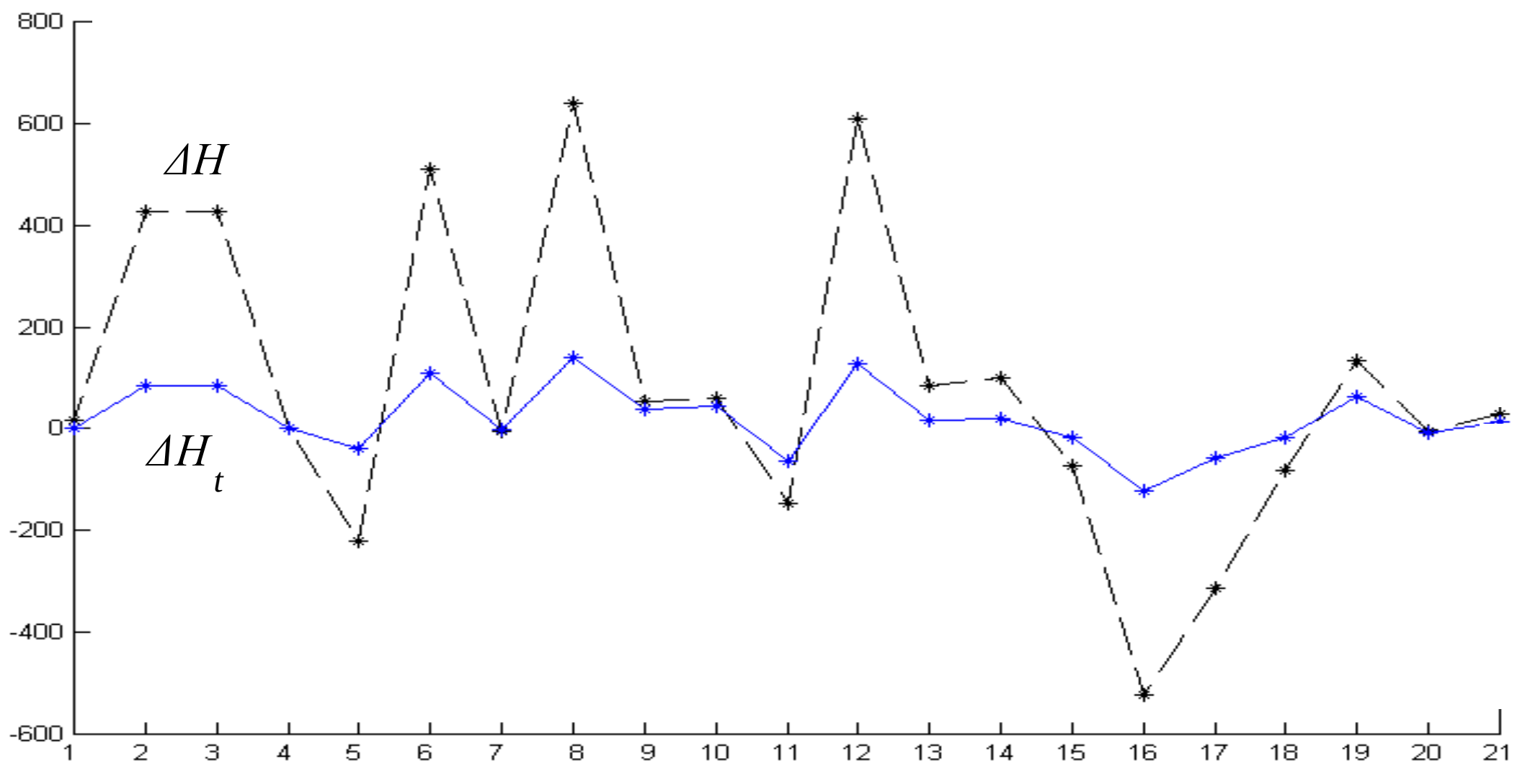

Figure 5-3: Comparison between vectors $\Delta H$ and $\Delta H_t$.

The correlation between $\Delta H$ and $\Delta H_t$ can be expressed by factor *r*:

$$r = \frac{\Delta H}{\Delta H_t} \qquad \text{(Eq. 5.17)}$$

The rationale behind determining factor *r* is to use it together with the confidence interval $\left[\Delta H_l \;\Delta H_u\right]$ and the Jacobian matrix in order to determine the confidence limits on the loop corrective flows.

An assumption is made that dividing the confidence interval $\left[\Delta H_l \;\Delta H_u\right]$ by factor *r* would give the same values as when simulating the water network for each pseudo-measurement vectors $d_l$ and $d_u$. At Figure 5-4 with dashed line are shown the loop head losses $\Delta H_l$ that are obtained from the equations (Eq. 5.12 – Eq. 5.15) for the measurement vector $d_l$ and subsequently divided by factor *r*.

The continuous line (Figure 5-4) is used to show the loop head losses $\Delta Ht_l$ calculated from the equation:

$$\Delta H_{t_l} = -J\,\Delta Q_l^{ll} \qquad \text{(Eq. 5.18)}$$

$$\Delta H_{t_u} = -J\,\Delta Q_l^{uu} \qquad \text{(Eq. 5.19)}$$

where $J$ and $\Delta Q_l^{ll}$ are the Jacobian matrix and the loop corrective flows obtained by simulating the water network for the pseudo-measurement vector $d_l$.

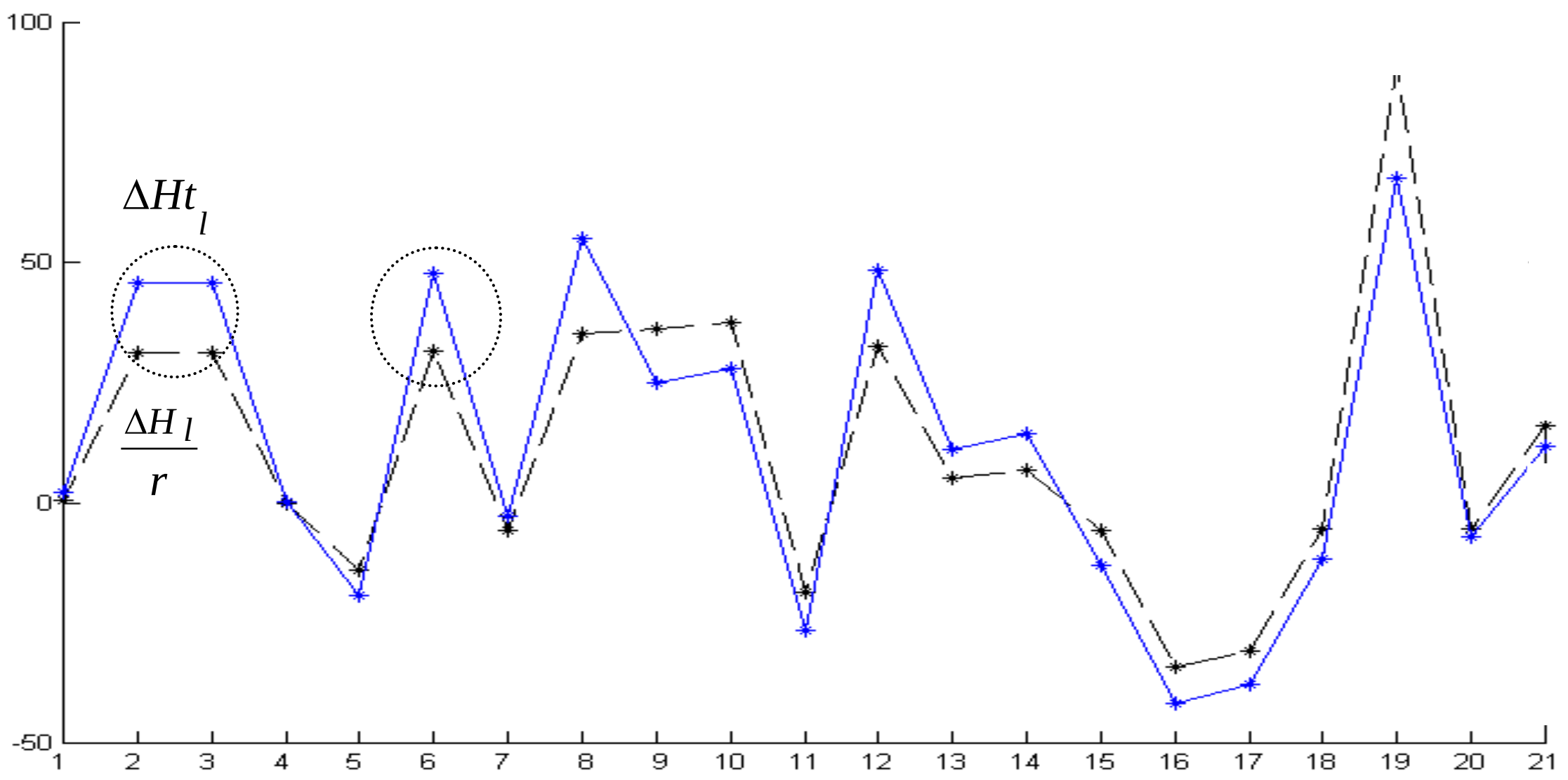

Figure 5-4: Comparison between the loop head losses calculated for the measurement vector $d_l$ and the loop head losses from equation (Eq. 5.18).

$\Delta H_{t_u}$ are the loop head losses residuals and $\Delta Q_l^{uu}$ are the loop corrective flows obtained by simulating the water network for the pseudo-measurement vector $d_u$

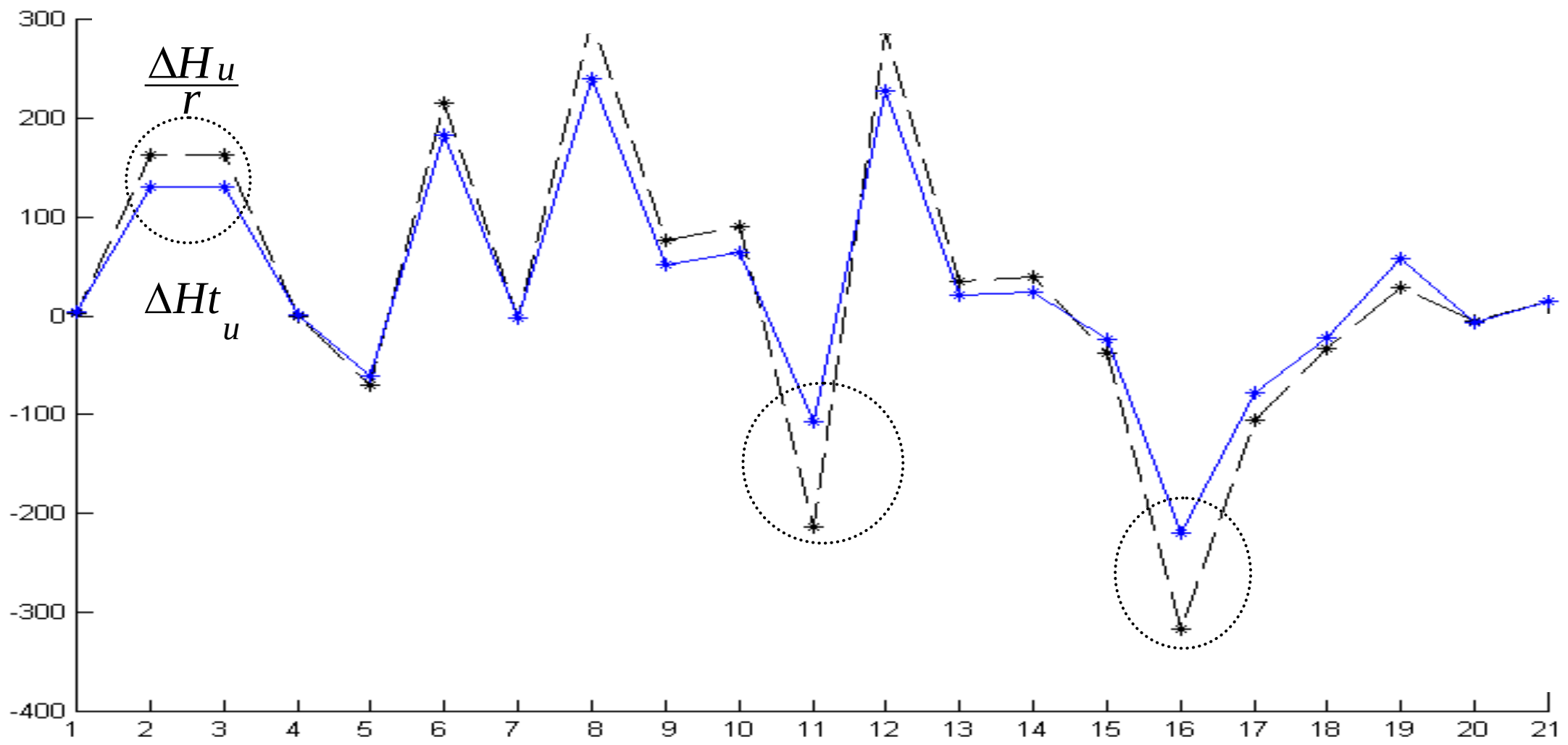

Figure 5-5: Loop head losses obtained for the pseudo-measurement vector $d_u$.

By applying the factor $r$ to the vectors of loop head losses $\Delta H_u$ and $\Delta H_l$, some inaccurate results are obtained. They are put in evidence by the dashed circles at Figures 5-4 and 5-5. These inaccuracies will propagate in the confidence limits $\Delta Q_l^l$ and $\Delta Q_l^u$ of the loop corrective flows $\Delta Q_l$ . The confidence limits are calculated with the Jacobian matrix $J$ and the ratios $\frac{\Delta H_u}{r}$ and $\frac{\Delta H_l}{r}$:

$$\Delta Q_l^l = -J^{-1} \frac{\Delta H_l}{r} \qquad \text{(Eq. 5.20)}$$

$$\Delta Q_l^u = -J^{-1} \frac{\Delta H_u}{r} \qquad \text{(Eq. 5.21)}$$

On the 3-rd and 4-th column of Table 5-5 are shown the confidence limits for the nodal heads. They have been obtained by using the upper and the lower limits of the loop corrective flows from equations (Eq. 5.20) and (Eq. 5.21). These values are compared with the confidence limits from the ESM method.

A CLA algorithm gives good results if the upper and lower confidence limits are identical. the size of the confidence intervals. More tight confidence intervals we have, more efficient the confidence algorithm is considered to be. The confidence limits shown on the 3-rd and 4-th column of Table 5-5 differ to a great extent from the confidence limits obtained by the ESM method. Moreover the upper and lower limits are not symmetrical.

| State variables | ESM method | Confidence limits ($\Delta Q_l^l$) | Confidence limits ($\Delta Q_l^u$) |
|---|---|---|---|
| 1 | 0.3007 | 0.9344 | 0.7840 |
| 2 | 0.2557 | 0.9645 | 0.8977 |
| 3 | 0.0667 | 1.2241 | 1.3777 |
| 4 | 0.1013 | 1.5303 | 1.4753 |
| 5 | 0.4212 | 1.4136 | 1.8550 |
| 6 | 0.3971 | 1.0208 | 2.0690 |
| 7 | 0.757 | 1.1580 | 2.4591 |
| 8 | 0.7925 | 1.1405 | 3.3654 |
| 9 | 0.3423 | 1.1436 | 3.3818 |
| 10 | 1.3561 | 1.5208 | 3.3300 |
| 11 | 0.0117 | 1.6088 | 3.1653 |
| 12 | 0.0033 | 2.0854 | 2.9140 |
| 13 | 1.0782 | 2.9216 | 2.6347 |
| 14 | 2.2626 | 3.1735 | 1.9935 |
| 15 | 2.2657 | 2.9216 | 1.8004 |
| 16 | 1.0413 | 2.6217 | 1.8174 |
| 17 | 0.4081 | 6.6122 | 3.8742 |
| 18 | 0.5268 | 3.5666 | 1.5654 |
| 19 | 0.3900 | 3.0151 | 1.7929 |
| 20 | 0.1410 | 2.4796 | 2.4794 |
| 21 | 0.1006 | 2.4796 | 2.4794 |
| 22 | 0.1993 | 2.4703 | 2.5374 |
| 23 | 0.3484 | 2.4575 | 2.6185 |
| 24 | 0.3027 | 2.2117 | 2.6931 |
| 25 | 0.6471 | 3.0133 | 1.7973 |
| 26 | 0.2259 | 3.0165 | 1.8051 |
| 27 | 0.0000 | 3.0381 | 1.8825 |
| 28 | 0.0151 | 3.0421 | 1.8727 |
| 29 | 0.0196 | 2.9339 | 2.1357 |
| 30 | 0.0004 | 1.2563 | 1.7632 |
| 31 | 0.0151 | 1.7321 | 1.0236 |
| 32 | 0.0151 | 1.8641 | 1.8973 |
| 33 | 0.0201 | 1.3045 | 1.2463 |
| 34 | 0.0199 | 1.8993 | 1.3172 |

Table 5-5: Inaccurate confidence limits on the nodal heads.
1-34: confidence limits for the nodal heads 1-34.

The inaccurate determination of the ratios $\frac{\Delta H_l}{r}$ and $\frac{\Delta H_u}{r}$, that have been used as confidence limits for the true loop head losses $\Delta H_t$, had caused the errors in the confidence limits of the nodal heads. This makes the algorithm impractical to use.

The previous discussion has taken into consideration only the water consumptions as the measurements. If real meters were to be considered, the Jacobian matrix from the state estimator should have been used. This can complicate even more the confidence limit analysis algorithm, since the structure of the loop flows state estimator Jacobian matrix is more complex than the Jacobian matrix from the co-tree flows simulator algorithm: it contains in addition the derivatives of the loop head losses with respect to the variation of nodal demands as well as the derivatives of the additional real pressure and flow measurements.

We can conclude that there are practical difficulties when using the inverse of the Jacobian matrix of the co-tree flows simulator algorithm as a sensitivity matrix between the loop head losses residuals and the pseudo-measurement vector *d*. New confidence limits algorithms are required that avoid the computational drawback of the ESM method and the inexact results of the last algorithm.

## 5.4.2. Error Maximization method

The ESM method uses a maximization step described as the multiplication of the absolute value of the experimental sensitivity matrix and the maximum level of errors in the observed measurement vector. This procedure is depicted again below and forms the basis of the new derived method called the *Error Maximization method* (EM).

$$z^t \longrightarrow z^o \xrightarrow{S^e} \hat{x} \qquad \hat{z}$$

| $z^t$ | $z^o$ | $\hat{x}$ | $\hat{z}$ |
| --- | --- | --- | --- |
| true measurement vector | observed measurement vector | state vector | estimated measurement vector |

$$ESM \;:\; S^e \quad \hat{x} \quad z^l < z^o < z^u$$

$$EM \;:\; x^1 \quad \hat{x} \quad \hat{z}^l < \hat{z} < \hat{z}^u$$

The experimental matrix $S^e$ calculated for the state vector $\hat{x}$ and the upper $z^u$ and the lower bounds $z^l$ of the observed measurement vector $z^o$ were the fundamental ingredients in the ESM method.

The EM method considers the maximum variability of consumptions and accuracy of meters for the estimated measurement vector $\hat{z}$ instead of the observed measurement vector $z^o$. Furthermore the upper or the lower measurement limits $[\hat{z}^l \quad \hat{z}^u]$ of the estimated measurement vector $\hat{z}$, is used in the loop flows state estimator. The resulted state vector $x^1$ is used for determining the confidence limits on the state variables (nodal heads, inflows) with the equation:

$$x_{cl_i} = abs(x^1 - \hat{x}) \qquad \text{(Eq. 5.22)}$$

where $x_{cl_i}$ is the confidence limit on the $i$-th state variable, $\hat{x}$ is the state vector obtained for the observed measurement vector $z^o$ and $x^1$ is the state vector obtained for the maximum level of errors in the estimated measurement vector $\hat{z}$.

The rationale beyond the equation (Eq. 5.22) lays in the properties of the loop flows state estimator, which were discussed at the end of the previous chapter. It is of paramount importance to emphasize that calculating the confidence limits with the EM method is characteristic to the loop flows state estimator and it will not work with other state estimators that are not based on the loop corrective flows such as the nodal heads state estimator.

It has been highlighted in the previous chapter the property of the loop flows state estimator of modifying the inflows into the fixed-head nodes so that to match the sum of the new nodal demands obtained with the loop flows state estimator. This means that if we modify the nodal consumptions to their lower or upper limit ($\hat{z}^l$, $\hat{z}^u$), then the mass balance of the network will still be satisfied. The inflows will be modified during the Newton-Raphson iteration method according to the nodal demands. In this case the fixed-head nodes are the measurement data and are used to form the pseudo-loops.

We can have a different case when the inflows into the fixed-head nodes represent the measurements. It means that they are known and kept constant during the iteration method, which if we consider the lower or the upper limits of the nodal consumption, may not satisfy the mass balance of the network. However, this will not happen because in the loop flows state estimator there is always the main source node (i.e. the root node from where the spanning tree is built) where the inflow can vary so that the sum of the nodal demands is equal to the amount of inflow.

The scheme can include pressure or flow measurements together with their measurement accuracy without loosing from generality.

The previous considerations are sketched in the following figure.

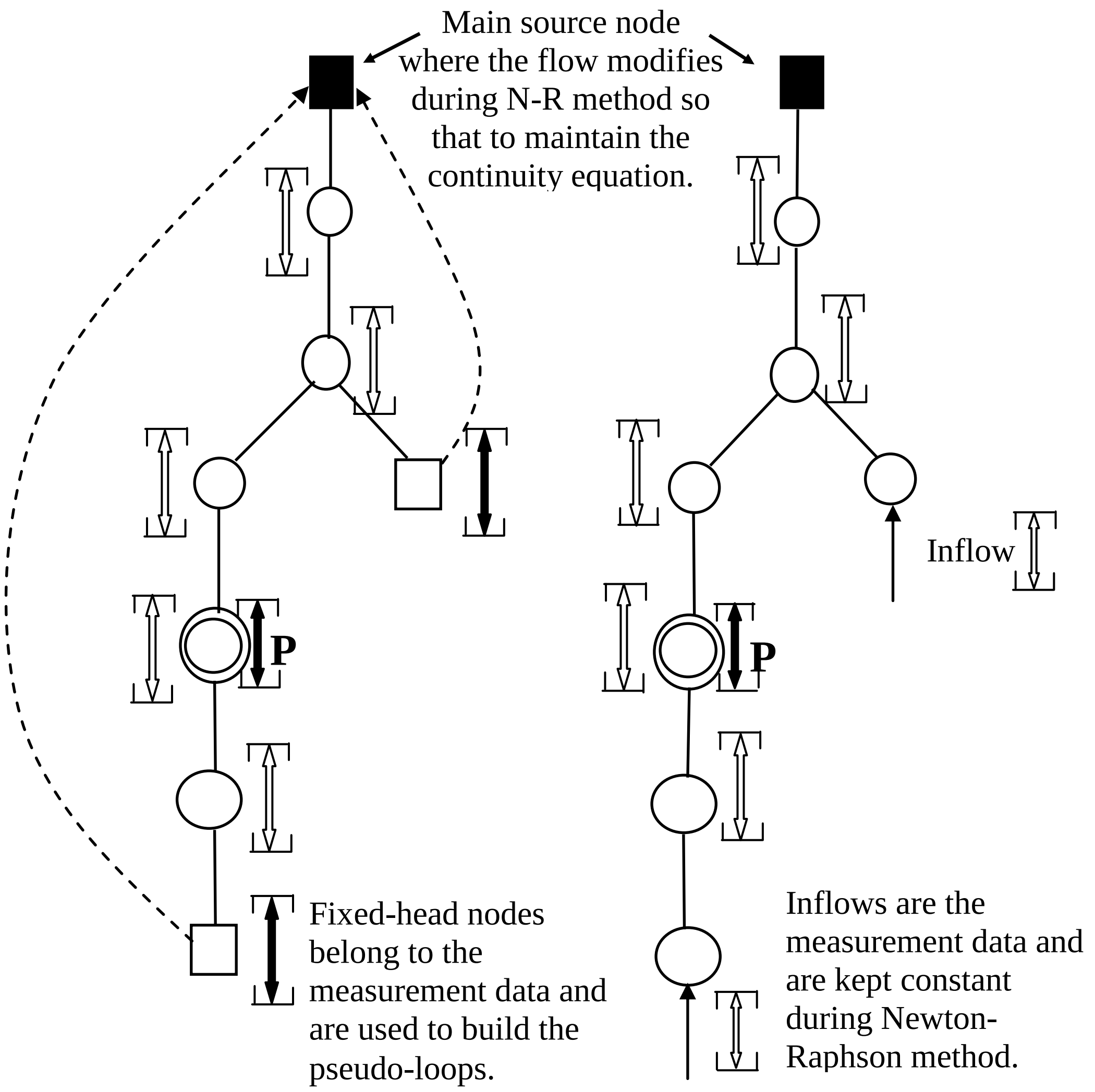

Figure 5-6: a) Fixed-head nodes used to form the pseudo-loops, b) Inflows are measurement data.

Figure 5-6 shows the spanning tree of a small water network. The black square represents the main source node where the inflow can not be controlled during a simulation. The circles are the network nodes while the empty squares are the fixed-head nodes. A pressure measurement is indicated with the letter *P*.

The measurement uncertainty has been represented in the figure by the means of an arrow with two wedge-shape ends. The white arrows define the accuracy of the nodal consumptions while the black arrows represent the maximum variability of the fixed-head nodes and the pressure measurements.

In Figure 5-6a is shown the first case where the fixed-head nodes are part of the measurement data and their head values are considered known and used to form the pseudo-loops (i.e. the dashed lines).

We take the lower or the upper limit on the nodal consumptions and then run the loop flows state estimator in order to obtain the confidence limits. The fixed-head nodes and the real meters can be also modified according to their maximum measurement accuracy.

In Figure 5-6b we show the second case where the values of the fixed-head nodes are not known and instead the inflows are used as measurements and kept constant during the Newton-Raphson method. If we bring now all the measurements at their lower or upper limit then the mass balance equation is satisfied by the inflow from the main source node.

It is obvious that the EM method works only with the loop flows state estimator. This is based on the existence of the main source node where the value of inflow can not be maintained at a fixed value and it varies according to the demands from the network. This is in contrast with the nodal heads state estimator where if we bring all the measurements to their lower or upper limit and then run the nodal heads state estimator, we do not obtain any valuable information for the confidence limits.

At Figure 5-7 is described the EM method in a form of a block diagram.

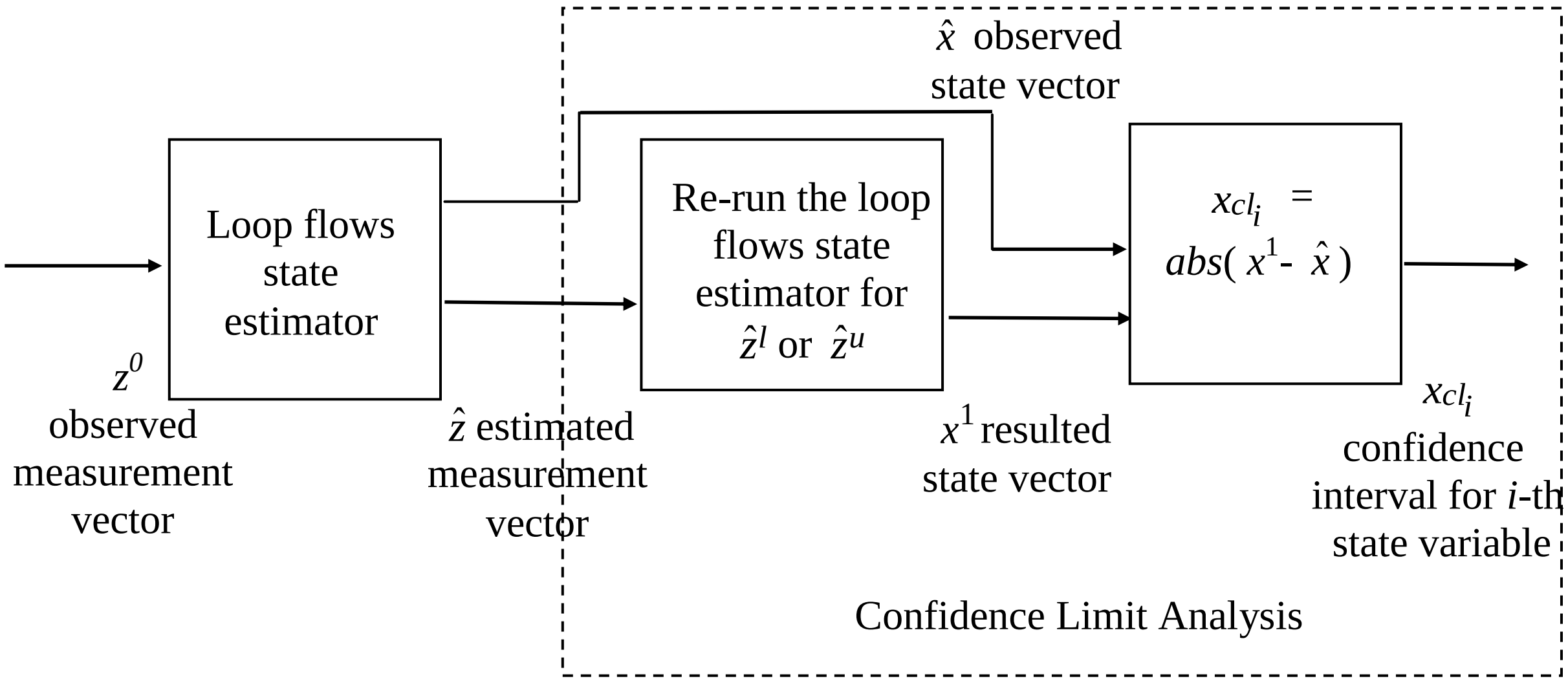

Figure 5-7: CLA based on EM method.

A comparison of the confidence limits produced by the Experimental Sensitivity Matrix method and the Error Maximization method is shown at Table 5-6. The confidence limits are calculated for the observed measurement vector shown on the columns 3 and 6 of Table 5-2 and column 3 of Table 5-1. The state variables 1 to 34 represent the nodal heads at nodes 1-34 and the state variables 35 to 42 represent the fixed-head nodes inflows at nodes 27-34.

| State variables | Exact State | State variables | C.L. with ESM method | C.L. with EM method |
|---|---|---|---|---|
| 1 | 31.1852 | 31.0577 | 0.3007 | 0.2893 |
| 2 | 43.3886 | 43.2835 | 0.2557 | 0.2381 |
| 3 | 44.2289 | 44.1968 | 0.0867 | 0.1056 |
| 4 | 44.3191 | 44.2706 | 0.1213 | 0.1505 |
| 5 | 42.8133 | 42.6358 | 0.4212 | 0.4010 |
| 6 | 42.6765 | 42.5082 | 0.3971 | 0.3785 |
| 7 | 41.8478 | 41.5228 | 0.7570 | 0.7302 |
| 8 | 41.7190 | 41.3762 | 0.7925 | 0.7700 |
| 9 | 43.0165 | 42.8746 | 0.3423 | 0.3233 |
| 10 | 41.6933 | 41.1195 | 1.3561 | 1.3034 |
| 11 | 43.5925 | 43.5813 | 0.0117 | 0.0511 |
| 12 | 43.5845 | 43.5817 | 0.0033 | 0.0152 |
| 13 | 45.3550 | 45.2569 | 1.0782 | 0.9456 |
| 14 | 40.1661 | 39.2083 | 2.2626 | 2.1782 |
| 15 | 43.0940 | 39.1235 | 2.2657 | 2.1947 |
| 16 | 43.4858 | 43.0441 | 1.0413 | 1.0164 |
| 17 | 43.9047 | 43.7263 | 0.4081 | 0.4438 |
| 18 | 44.7605 | 44.5342 | 0.5268 | 0.5347 |
| 19 | 44.3638 | 44.1934 | 0.3900 | 0.4137 |
| 20 | 44.1362 | 44.0702 | 0.1410 | 0.1951 |
| 21 | 43.6560 | 43.6053 | 0.1006 | 0.1476 |
| 22 | 43.8080 | 43.7161 | 0.1993 | 0.2445 |
| 23 | 44.0663 | 43.9127 | 0.3484 | 0.3746 |
| 24 | 42.9028 | 42.7773 | 0.3027 | 0.2826 |
| 25 | 42.0751 | 41.7974 | 0.6471 | 0.6240 |
| 26 | 31.3306 | 31.2399 | 0.2259 | 0.1880 |
| 27 | -15.1991 | -15.1991 | 0.0000 | 0.0000 |
| 28 | -33.4879 | -33.4966 | 0.0151 | 0.0112 |
| 29 | 31.7221 | 31.7242 | 0.0196 | 0.0112 |
| 30 | 43.5619 | 43.5819 | 0.0004 | 0.0000 |
| 31 | 44.1710 | 44.1715 | 0.0151 | 0.0141 |
| 32 | -46.3814 | -46.3798 | 0.0151 | 0.0139 |
| 33 | -36.5470 | -36.5457 | 0.0201 | 0.0121 |
| 34 | -12.1990 | -12.1942 | 0.0199 | 0.0141 |
| 35 | 34.0 | 35.2 | 3.1 | 2.9 |
| 36 | 96.5 | 96.6 | 0.1 | 0.2 |
| 37 | 64.3 | 73.4 | 21.4 | 21.9 |
| 38 | 106.3 | 130.2 | 56.9 | 55.5 |
| 39 | 38.9 | 48.7 | 22.8 | 23.9 |
| 40 | 6 | 6 | 0 | 0 |
| 41 | 121.7 | 121.7 | 0 | 0 |
| 42 | 21.6 | 22.8 | 2.6 | 2.6 |

Table 5-6: Confidence limits obtained with the ESM and EM methods.

The confidence limits obtained with the ESM and EM methods are similar. The computational load associated with the ESM method is over 15 seconds, which is far higher than the less of 0.5 second obtained with the EM method. This is due to the computational time required for calculating the experimental sensitivity matrix. In order to obtain the experimental sensitivity matrix, we had to run the state estimator for a number of times equal to the number of measurements. Therefore a higher number of measurements will require an equal amount of extra simulations which is posed to increase the computational time necessary to calculate the experimental sensitivity matrix in special for large water networks. However, what is important to notice here is the elegance in obtaining the confidence limits with the EM method that requires running the loop flows state estimator for the lower or the upper limits of the measurement data as opposed to the awkwardness of the ESM method that needs the calculation of the experimental sensitivity matrix.

The EM method can include pressure and flow measurements as well. A pressure measurement is introduced at node 14 of the water network from Figure 5-1. For simplicity we will consider the estimated measurement vector to be the same as the observed measurement data. The maximum variability of the pressure meters is ±30%.

On the 3-rd column of Table 5-7 are shown the confidence limits obtained for the variability of pseudo-measurements and accuracy of the fixed-head nodes but no other real meters are included.

The purpose of calculating the confidence limits is to obtain an information about how far from the real state the estimated values could be in the worst case. The requirement to have the state variables as close as possible to the real state is equivalent to the requirement of having the confidence limits as tight as possible. The means of achieving that is by the introduction of additional accurate real meters into the system. Therefore on the 4-th column of Table 5-7 are shown the confidence limits when a pressure measurement is located at node 14. There is an improvement not only in the node where the pressure measurement was introduced but also in the adjacent nodes.

The same logic is applied for the flow measurements. A flow measurement with the accuracy ±20% is introduced between nodes 22 and 23. On the 5-th column of Table 5-7 are shown the confidence limits. It can be observed an improvement in the region where the flow measurement was placed.

| State variables | State variables | C.L. (case 1) | C.L. (case 2) | C.L. (case 3) |
|---|---|---|---|---|
| 1 | 31.0577 | 0.2893 | 0.2801 | 0.2593 |
| 2 | 43.2835 | 0.2381 | 0.2448 | 0.2234 |
| 3 | 44.1968 | 0.1056 | 0.0842 | 0.0455 |
| 4 | 44.2706 | 0.1505 | 0.1123 | 0.0636 |
| 5 | 42.6358 | 0.4010 | 0.4042 | 0.3824 |
| 6 | 42.5082 | 0.3785 | 0.3832 | 0.3616 |
| 7 | 41.5228 | 0.7302 | 0.7254 | 0.7030 |
| 8 | 41.3762 | 0.7700 | 0.7641 | 0.7416 |
| 9 | 42.8746 | 0.3233 | 0.1553 | 0.1201 |
| 10 | 41.1195 | 1.3034 | 0.7980 | 0.8301 |
| 11 | 43.5813 | 0.0511 | 0.0401 | 0.0098 |
| 12 | 43.5817 | 0.0152 | 0.0199 | 0.0028 |
| 13 | 45.2569 | 0.9456 | 1.0041 | 1.4464 |
| 14 | 39.2083 | **2.1782** | **1.1753** | 1.1761 |
| 15 | 39.1235 | 2.1947 | 1.1631 | 1.1638 |
| 16 | 43.0441 | 1.0164 | 0.5342 | 0.4530 |
| 17 | 43.7263 | 0.4438 | 0.1043 | 0.0991 |
| 18 | 44.5342 | 0.5347 | 0.2145 | 0.0141 |
| 19 | 44.1934 | 0.4137 | 0.2307 | 0.0485 |
| 20 | 44.0702 | 0.1951 | 0.1397 | 0.0789 |
| 21 | 43.6053 | 0.1476 | 0.1026 | 0.0484 |
| 22 | 43.7161 | 0.2445 | **0.1737** | **0.0931** |
| 23 | 43.9127 | 0.3746 | **0.2432** | **0.0869** |
| 24 | 42.7773 | 0.2826 | 0.2916 | 0.2703 |
| 25 | 41.7974 | 0.6240 | 0.6224 | 0.6002 |
| 26 | 31.2399 | 0.1880 | 0.1988 | 0.1780 |
| 27 | -15.1991 | 0.0000 | 0.0100 | 0.0100 |
| 28 | -33.4966 | 0.0112 | 0.0095 | 0.0112 |
| 29 | 31.7242 | 0.0112 | 0.0087 | 0.0122 |
| 30 | 43.5819 | 0.0000 | 0.0100 | 0.0100 |
| 31 | 44.1715 | 0.0141 | 0.0095 | 0.0113 |
| 32 | -46.3798 | 0.0139 | 0.0095 | 0.0112 |
| 33 | -36.5457 | 0.0121 | 0.0087 | 0.0122 |
| 34 | -12.1942 | 0.0141 | 0.0087 | 0.0122 |
| 35 | 35.2 | 2.9 | 1.5 | 0.4 |
| 36 | 96.6 | 0.2 | 0.1 | 0.1 |
| 37 | 73.4 | 21.9 | 18.4 | 16.5 |
| 38 | 130.2 | 55.5 | 46.1 | 41.1 |
| 39 | 48.7 | 23.9 | 20.8 | 18.8 |
| 40 | 6 | 0 | 0 | 0 |
| 41 | 121.7 | 0 | 0 | 0 |
| 42 | 22.8 | 2.6 | 2.6 | 2.6 |

Table 5-7: Confidence limits obtained with EM method when real meters are present.

With bigger number of measurements the reliability of estimation increases. However, introducing a new measurement we introduce a new source of inconsistency which is given by the variability of the meter.

It can be concluded that the addition of a new measurement for the *i*-th state variable can have the tightening effect on the confidence limit of this variable only if the error resulted from the inaccuracy of the meter is smaller than the confidence limit calculated for the existing set of meters. If the previous condition is satisfied then the confidence limits for the inflows (i.e. state variable 35-42) become tighter as well.

The main contribution of this section is that effective CLA algorithms have been developed and applied for realistic water networks while using the loop equations in the numerical algorithms.

## 5.5. Concluding remarks

This chapter examines the problem of real measurements and pseudo-measurements uncertainty in water systems based on the loop equations framework for the numerical simulations. The loop flows approach for the CLA procedures has not been treated before in the literature and it represents a major contribution to the originality of this project.

Present day deterministic state estimation techniques are very efficient, having small computational requirements and producing results of an acceptable level of accuracy. However for particular water networks, like ones displaying low pipe flows, the convergence of the algorithm might suffer. A possible solution to this problem is to employ the more stable loop flows state estimation techniques.

On the other hand, in the process of state estimation, the inaccuracy of input data contributes greatly to the inaccuracy of system state estimates calculated from them. Due to the cost of metering, the water industry is constrained to make use of relatively inaccurate pseudo-measurements. For this reason, the computationally results of state estimators can be inaccurate when compared to the actual system state. Therefore the degree of confidence that can be put in these results must be calculated and presented with the state estimates themselves. Only then can the computational results be used in the operational control.

The calculation of these confidence limits for the state estimates has been largely investigated in the past within the nodal heads state estimators. In this chapter a similar investigation has been carried out but within the loop framework. It has been shown that the inverse of the Jacobian matrix from the co-tree flows simulator algorithm can not act as a sensitivity matrix between the loop corrective flows and pseudo-measurements because of the non-realistic way the initial loop head losses are calculated. This has a negative impact on the calculation of the confidence limits that are bigger than expect it. Instead, a sensible number of simulations can be used, one for each measurement modified with its defined maximum variability, in order to determine an experimental sensitivity matrix. This experimental sensitivity matrix can be later employed for determining the nodal heads and the inflows for a random error in the measurement data. Furthermore the confidence limits obtained with the experimental sensitivity matrix are comparable with the ones produced with the pseudo-inverse of the Jacobian matrix from the nodal heads state estimator. This mathematical result means that the Experimental Sensitivity Matrix (ESM) method provides a trusting reference point against which other confidence algorithms can be tested.

The ESM method requires a large number of simulations equal with the number of real measurements and pseudo-measurements. For this reason, the ESM method might be an unrealistic proposition for the real-time applications in special for large water networks. An alternative method is developed, the Error Maximization (EM) method, that requires only an extra simulation in order to derive the confidence limits. An additional simulation is carried out for the estimated measurement vector instead of the observed measurement vector, which is modified with the highest level of inaccuracies. Following this the confidence limits are calculated by subtracting the resulted state vector from the optimal state vector obtained for the observed measurement data. Finally the set of confidence limits are compared with the ESM method and it shows a very good similarity. The computational efficiency of the EM method renders it suitable for online decision support applications.

# Chapter 6.

# Pattern Recognition for Fault Detection in Water Networks

## 6.1. Introduction

The operational control of water systems is a challenging task because it requires that operators develop a *mental model* of operation of a large-scale non-linear system which is subject to random disturbances (fluctuation of consumption) and which is monitored using relatively few measurements.

Having found the state estimates with their corresponding confidence limits, the next task usually carried out by a human operator, is to classify current operating state before any control action can be taken. In the following chapter, the classification task is to be attempted by using an already developed neural algorithm (Gabrys & Bargiela, 2000) capable of clustering as well as classifying the state vector with its confidence limits. Therefore this chapter introduces the background information about what are the neural networks together with a special regard to the *General Fuzzy Min-Max* (*GFMM*) (Gabrys, 1997; Gabrys & Bargiela, 1999) neural network that will be used in the next chapter for fault detection and identification in operational control of water distribution systems.

The GFMM neural network has been originally developed by Gabrys (1997) and successfully applied to the *pattern recognition* in the water networks state identification task in order to distinguish between different patterns of 'nodal heads-equations based' state estimates. One of the main conclusions of his work was to study the recognition system performance in association with different state estimation procedures and this forms the premise for testing the GFMM neural network with the 'loop-based' state estimates and confidence intervals.

The chapter is organized in two main sections: first what are the neural networks and the fundamental features of a neural network are presented. Typical architectures of neural networks are shown together with methods for training and learning. The last section is dedicated to a special class of neural algorithms for classification and clustering. A review of the clustering and classification techniques used in pattern recognition, is given together with the description of the main characteristics of a general fuzzy min-max classification and clustering neural network (Gabrys & Bargiela, 1999).

## 6.2. Neural networks

There are various points of view as to the nature of an artificial neural net. For example, is it a specialized piece of hardware or a computer program? We shall take the view that *neural nets* are basically mathematical models of information processing. They provide a method of representing relationships that is quite different from Turing machines or computers with stored programs. As with other numerical methods, the availability of computer resources, either software or hardware, greatly enhances the usefulness of the approach, especially for large problems.

### 6.2.1. Biological neural systems

The human information processing system consists of the biological brain. The basic building block of the nervous system is the *neuron*, the cell that communicates information to and from the various parts of the body. Figure 6-1 shows a simplified representation of a biological neuron. The neuron consists of a cell body called *soma*, several spine-like extensions of the cell body called *dendrites*, and a single nerve fibre called the *axon* that branches out from the soma and connects to many other neurons.

The many dendrites receive signals from other neurons. The connections between neurons occur either on the cell body or on the dendrites at junctions called synapses. The signals are electric impulses that are transmitted across synaptic gap by means of chemical process. A helpful analogy is to view the axons and dendrites as insulated conductors of various impedances that transmit electrical signals to the neuron (Churchland, 1986; Kandel & Schwartz, 1985). The nervous system is constructed of billions of neurons with the axon from one neuron branching out and connecting to as many as 10,000 other neurons. All the neurons - interconnected by axons and dendrites that carry signals regulated by synapses - create a *neural network*.

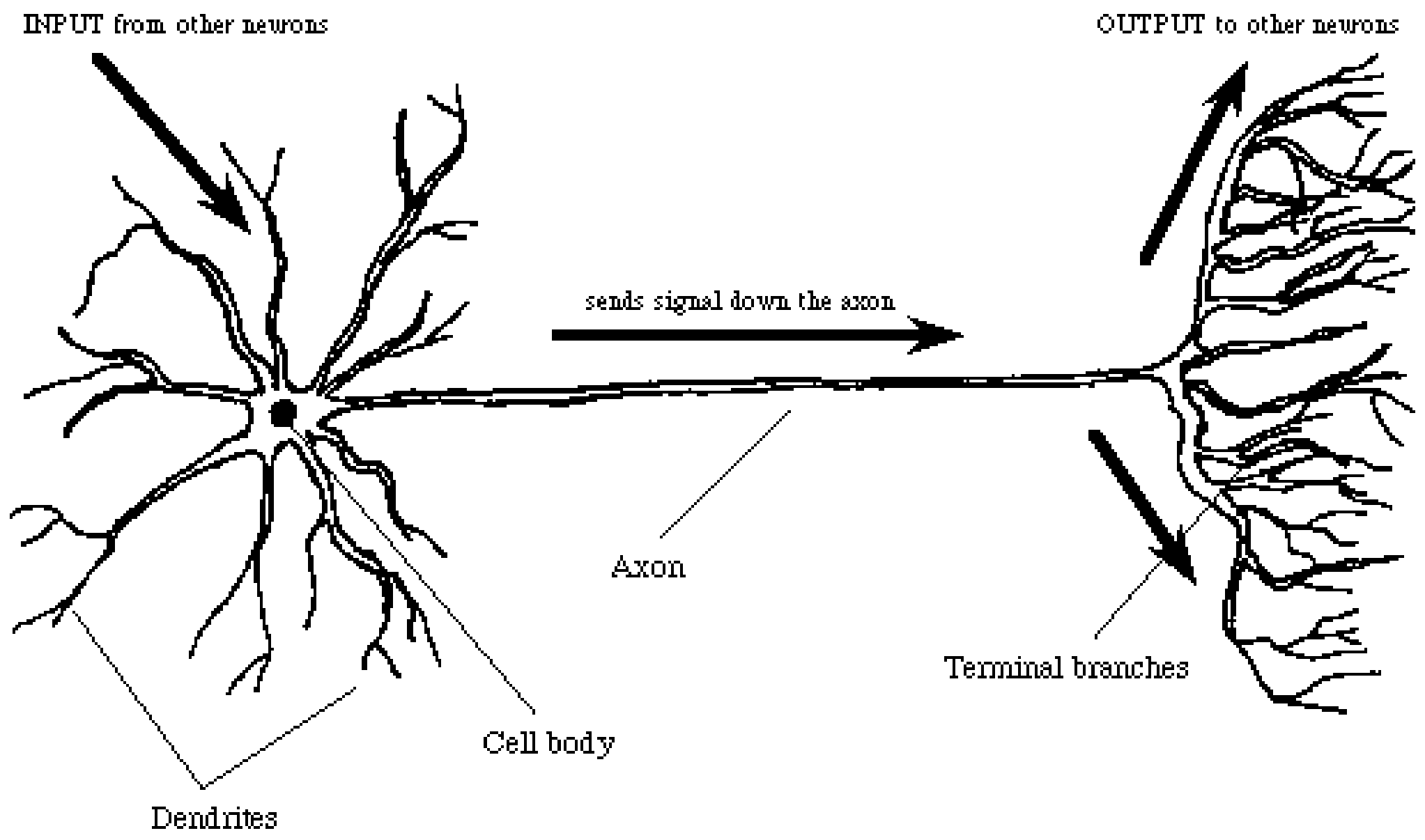

Figure 6-1: Biological neuron.

## 6.2.2. Artificial neural networks

An *artificial neural network* is an information-processing system that has certain performance characteristics in common with biological neural networks. The extent to which a neural network models a particular biological neural system varies. For some researchers, this is a primary concern, for others, the ability of the net to perform useful tasks (such as approximation of a function) is more important than the biological plausibility of the net. Although our interest lies almost exclusively in the computational capabilities of neural networks, we shall briefly present some features of biological neurons that may help to clarify the most important characteristics of artificial neural networks.

An artificial neural network is characterized by:

a) its topology of interconnected neurons with their non-linear activation functions (called its *architecture*),

b) its method of encoding information (called its *training* or *learning* algorithm).

Artificial neural networks are made up of large number of individual models of the biological neurons (artificial neurons). Each neuron is connected to other neurons by means of directional communication links, each with an associated weight. The neuron models that are used are typically much simplified versions of the actions of a real neuron. The weights represent the information used by the net in solving a particular problem.

Several key features of the processing elements of artificial neural networks are suggested by the properties of biological neurons:

- The processing element receives many signals.
- Signals may be modified by weight at the receiving synapse.
- The processing element sums the weighted inputs.
- Under appropriate circumstances (sufficient input), the neuron transmits a single output.
- The output from a particular neuron may go to many other neurons.
- Information processing is local.
- Memory is distributed: a) long memory resides in the neurons' synapses or weights, b) short-memory corresponds to the signals sent by the neurons.
- A synapse's strength may be modified by experience.
- Neurotransmitters for synapses may be excitatory or inhibitory.

Yet another important characteristic that artificial neural networks share with biological neural systems is *fault tolerance*. Biological neural systems are fault tolerant in two respects. First, they are able to recognize many input signals that are similar but not identical to any input that was seen before. Second, damage to individual neurons can occur in the brain without a severe degradation in its overall performance (Hopfield, 1982; Hopfield et al., 1983; Hopfield, 1984). If a portion of a brain is removed, the knowledge of the concept or idea is still retained through the redundant, distributed encoding of information. In a similar manner, artificial neural networks can be designed to be insensitive to small damage to the network, and the network can be retrained in cases of significant damage.

## 6.2.3. Fundamental features of ANNs

This section attempts to explain, in general terms, what an artificial neural network is and where the inspiration for neural computing came from, the subsequent sections will present typical neural network architectures and training algorithms.

***Artificial neurons***

Artificial neurons, also referred to as nodes or processing elements, are the ANN components where most, if not all, of the computing is done. The most commonly used neuron model is depicted in Figure 6-2 and is based on the model proposed by *McCulloch and Pitts* in 1943 (McCulloch & Pitts, 1943). Each neuron input, $x_1$ - $x_n$, is weighed by the adjustable values $w_1$ - $w_n$. A bias, or offset, in the node is characterized by an additional constant input of 1 weighted by the value of $w_0$. The output, $y$, is obtained by summing the weighted inputs to the neuron and passing the result through a non-linear activation function, f(). Mathematically this operation is defined as:

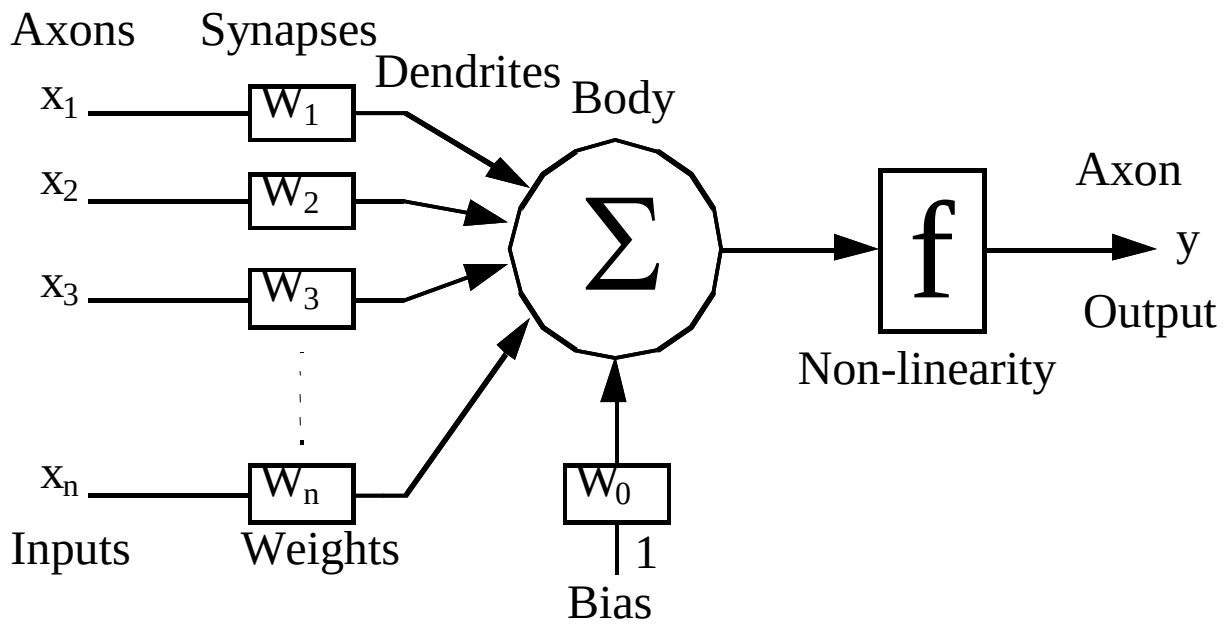

Figure 6-2: McCulloch-Pitts model of neuron.

$$y = f\left(\sum_{i=1}^{n} w_i x_i + w_0\right) \quad \text{(Eq. 6.1)}$$

Various types of non-linearity are possible and some of these are shown below

***Activation functions***

Activation functions, also called threshold functions or squashing functions, map the neuron's infinite domain (the input) to a prespecified range (the output). Four common activation functions are the linear, ramp, step, and sigmoid functions. Table 6-1 shows the mathematical equations describing these functions and their typical shapes:

| Name and mathematical description | Shape | Remarks |
|---|---|---|
| Linear function<br>$f(x) = \alpha x$ | 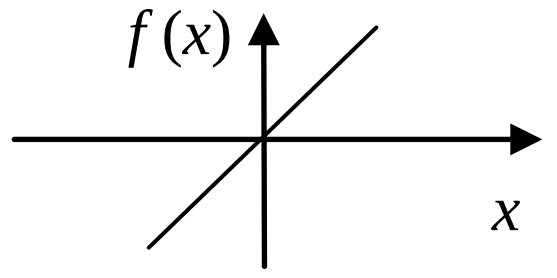  | $\alpha$ is a real-values constant that regulates the magnification of the neuron activity $x$. |
| Ramp function<br>$f(x)=\begin{cases}\gamma & if \;\; x \geq \gamma \\ x & if \;\; \lvert x \rvert < \gamma \\ -\gamma & if \;\; x \leq -\gamma\end{cases}$ | 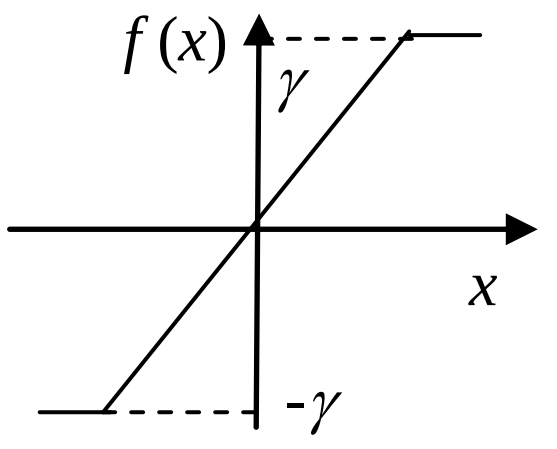  | The output is bounded to the range [-γ, +γ]. Values γ and -γ are commonly referred to as the saturation levels. |
| Step function<br>$f(x)=\begin{cases}\gamma & if \;\; x \geq 0 \\ -\delta & otherwise\end{cases}$ | 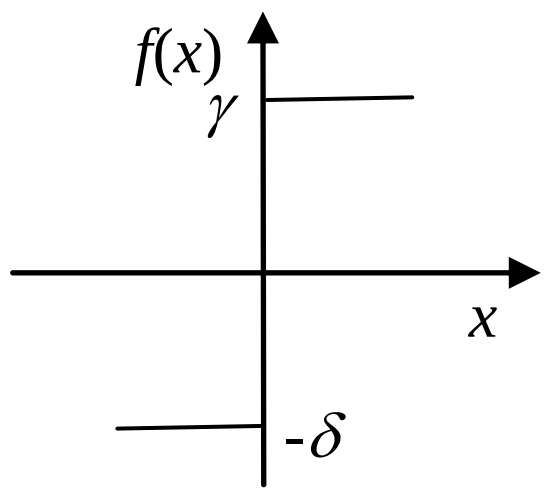  | Step function respond only to the sign of the input, emitting +γ if the input sum is positive and -δ if it is not. γ and δ are positive scalars. Often step function is binary in nature emitting a 1 if x > 0 and 0 otherwise. |
| Sigmoid function<br>$f(x)=(1+e^{-x})^{-1}$ | 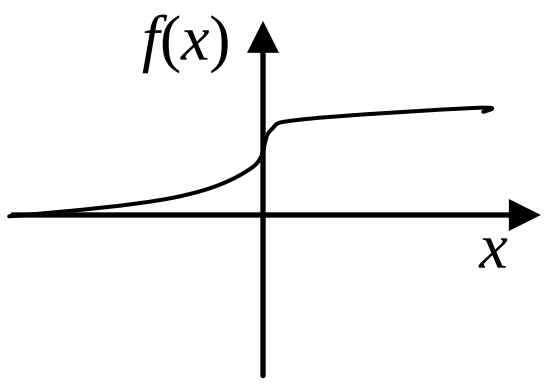  | Sigmoid function is bounded, monotonic, non-decreasing function that provides a graded, nonlinear response. The saturation levels are 0 and 1. |

Table 6-1: Four common activation functions.

## 6.2.4. Typical architectures

ANN architectures, or topologies, are formed by organizing neurons into layers (also called fields or slabs) and linking them with weighted interconnections.

There are three primary neuron interconnection schemes: lateral connections, inter-layer connections, and recurrent connections. *Lateral connections* are connections between neurons in the same layer of neurons. *Inter-layer connections* are connections between neurons in different layers. And finally, *recurrent connections* are connections that loop and connect back to the same neuron.

Interlayer connection signals propagate in one of two ways, either forward or feedback. *Feedforward* signals only allow information to flow amongst neurons in one direction. *Feedback* signals allow information to flow amongst neurons in either direction and/or recursively. On the basis of these two types of signal propagation the difference between two methods of information recall in ANNs can be defined as follow.

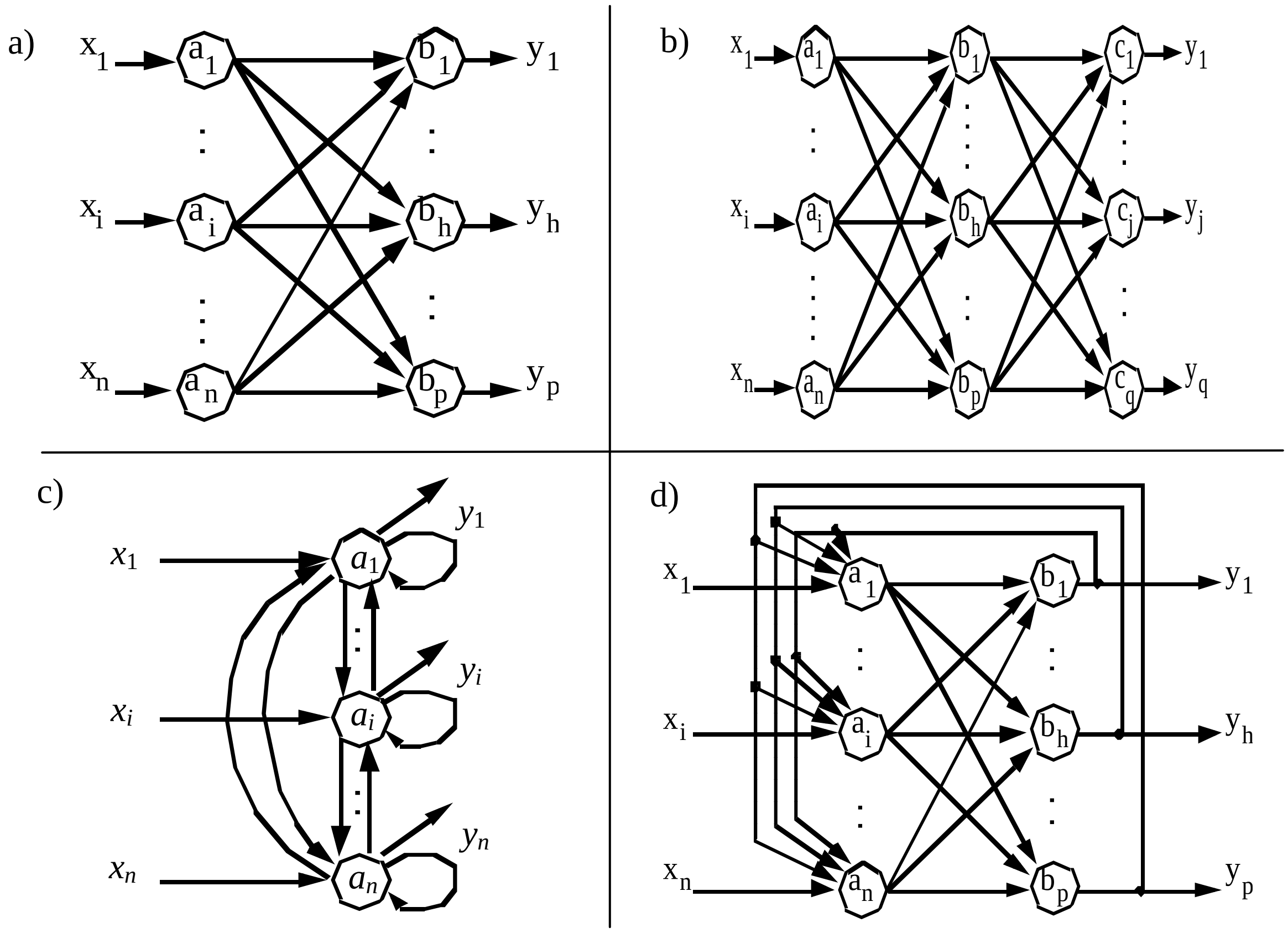

Figure 6-3: Four common ANN architectures: a) two-layer feedforward ANN; b) three-layer feedforward ANN; c) one-layer lateral feedback ANN; d) two-layer feedback ANN.

During feedforward recall, the input cue is passed through the memory, represented by the weights W, and produces an output response in one pass. During feedback recall, the input cue is passed through the memory and produces an output response that is, in turn, fed back into the memory until the cue and response cease to change.

Layer configurations combine layers of neurons, information flow and connection schemes into a coherent architecture. Layer configurations include lateral feedback, layer feedforward, and layer feedback. A layer that receives input signals from the environment is called an input layer and a layer that emits signals to the environment is called an output layer. Any layers that lie between input and output layers are called hidden layers and have no direct contact with the environment. Figure 6-3 illustrates four common ANN topologies.

### 6.2.5. Training/Learning algorithms

In addition to the architectures, the method of setting the values of the weights (learning or training) is an important distinguishing factor of different neural nets. As it has been pointed out in (Hassoun, 1995), in the context of artificial neural networks, the process of learning is best viewed as an optimization process. More precisely, the learning process can be viewed as “search” in a multidimensional (weight) space for a solution, which gradually optimizes a prespecified objective (criterion) function. This view allowed Hassoun to unify a wide range of existing learning rules which otherwise could have looked more like a diverse variety of learning procedures.

All learning methods can be classified into two categories, supervised learning and unsupervised learning, although aspects of each may co-exist in a given architecture. In addition, there are nets whose weights are fixed without an iterative training process.

In *supervised learning* (also called *learning with a teacher*) each input vector, pattern or signal is presented with an associated target output vector. Usually the weights are gradually updated with each step of the learning process so that the error between the desired (given) target and the network’s output is reduced.

On the other hand, *unsupervised learning*, also referred to as self-organization, is a process that incorporates no external teacher. Unsupervised learning involves the clustering or detection of similarities among unlabelled patterns of a given data set. Here, the weights and the outputs of the network are usually expected to converge to representations of the input data.

There is some ambiguity in the labelling of training methods as supervised or unsupervised and some authors find a third category, reinforcement learning or self-supervised learning, useful. *Reinforcement learning* involves updating the network’s weights in response to an “evaluative” teacher signal; this differs from supervised learning, where the teacher signal is the “correct answer”.

In general, however, there is a useful correspondence between the type of training that is appropriate and the type of problem we wish to solve. Some examples of ANN applications are given in the next section.

### 6.2.6. ANN applications

The purpose of this section is to give a sample of various areas of ANN applications and to illustrate a strong preference for using certain types of neural nets to solve certain types of problems.

***Multilayer, feedforward, supervised ANN applications***

Among the supervised learning methods for multilayer neural nets the backpropagation algorithm is by far the most popular. *Backpropagation* and its variations have been applied to a wide variety of problems, including pattern recognition, signal processing, image compression, speech recognition, medical diagnosis, prediction, nonlinear system modelling, and control.

One of the earliest applications of backpropagation was the system known as NETtalk that converts English text into speech (Sejnowski & Rosenberg, 1987). Another example of a multilayer feedforward ANN application is a neural based adaptive interface system, known as Glove-Talk, that maps hand gestures to speech (Fels & Hinton, 1993).

The recognition of handwritten digits is a classic problem in pattern recognition. Specifically, the Postal Service is interested in the recognition of handwritten ZIP codes on pieces of mail. A backpropagation network has been designed to recognize segmented numerals digitized from handwritten ZIP codes that appeared on U.S. mail (Le Cun et al., 1989).

ALVINN (autonomous land vehicle in a neural network) - a backpropagation-trained feedforward network designed to drive a modified Chevy van (Pomerleau, 1991) - is an example of a successful application using sensor data in real time to perform a real-world perception-control task.

Clinical diagnosis is often fraught with great difficulty because multiple, often unrelated disease states can surface with very similar historical, symptomalogic, and clinical data. As a result, physicians' accuracy in diagnosing such diseases is often poor. Feedforward multilayer neural networks trained with backpropagation have been reported to exhibit improved clinical diagnosis over physicians and traditional expert-system approaches (Bounds et al., 1988; Yoon et al., 1989; Baxt, 1990).

One of the major objectives for the management of a water supply and distribution system is the forecasting of the daily demand. The multilayer feedforward ANNs, reported in (Canu et al., 1990; Cubero, 1991), have been used to accomplish this task.

***Feedforward, unsupervised ANN applications***

The best known ANN in this group is the self-organizing map, developed by Kohonen, which has the special property of effectively creating spatially organized "internal representations" of various features of input signals and their abstractions. The self-organizing map has been particularly successful in various pattern recognition tasks involving very noisy signals.

One of the applications demonstrating the power of the map method when dealing with difficult stochastic signals is the area of speaker-independent recognition of speech. The example of the self-organizing map application to speech recognition is the "phonetic typewriter" net (Kohonen, 1988).

Other areas were self-organizing maps have been successfully used include control of robot arm (Graf & LaLonde, 1988; Veelenturf, 1995), EEG signal analysis (Veelenturf, 1995), control of industrial processes, especially diffusion processes in the production of semiconductor substrates (Marks & Goser, 1988).

***Feedback, unsupervised ANN applications***

*Dynamic associative memories* (DAMs), the most representative in this group, are a class of recurrent ANNs that utilize a learning/recording algorithm to store vector patterns as stable memory states. A part of the DAMs are Hopfield networks that have been successfully applied to many combinatorial optimization problems- situations that require the minimization of multiple-constraint cost function to determine the set of optimal system parameters.

An example of the optimization problem, that was addressed in (Hopfield & Tank, 1985) using recurrent neural network, is the classical travelling salesperson problem. A salesperson wants to visit n cities, once each, along a path that ends at the initial city. The problem is to perform this loop in such a way as to minimize the total mileage. An interesting feature of the solution proposed by Hopfield and Tank is the fact that weights are defined by the problem (they are the distances between the cities) and not set using some learning method.

Other than combinatorial optimization applications, the Hopfield ANN's ability to reconstruct entire patterns from partial cues stands out as one of its primary application strengths. In addition, the Hopfield ANN's nearest-neighbour response and fault tolerance qualities are also appealing. Because of these qualities, pattern classification and noise removal from patterns are key Hopfield network's applications.

One more type of ANN and its applications is worth mentioning here: ART (Adaptive Resonance Theory) clustering neural network. This unsupervised ANN has a special interest for us due to the fact that the neural net used in the next chapter is capable to learn cluster structure in a self-organizing, stable manner.

## 6.3. Fuzzy State Clustering and Classification for Operational Control

The operational control of water systems requires the human operator to classify the current operating state (e.g. normal status, leakage) before any control action can be taken. In (Gabrys & Bargiela, 1999) this classification task has been carried out by developing a flexible neural algorithm capable of clustering as well as classifying the state vector with its confidence limits. Their research has shown that the fuzzy state clustering and classification performed by neural networks can copy to a large extent the high level information processing by human operators. They tested their pattern recognition algorithm with examples of state estimates and confidence intervals obtained with the nodal heads equations.

A review of the existent classification and clustering techniques is made before the description of the pattern recognition algorithm proposed by Gabrys and Bargiela (1999) is presented.

### 6.3.1. Pattern clustering review

In many pattern recognition and decision making tasks, there is little prior information available about the data that need to be utilised. Pattern clustering uses the minimum amount of information to organize data into categories such that patterns within a cluster are more similar to each other than patterns belonging to other clusters. There are many different techniques that have been offered for solving the clustering problem. In the following sections is shown a brief review of some traditional, fuzzy and neural network clustering techniques.

1) ***Traditional Clustering:*** There are many clustering algorithms that have been developed to date, including ISODATA, FORGY, WISH, and CLUSTER (Dubes &

Jain, 1976), many of which are commercially sold. Jain (Jain, 1986) has reduced these clustering techniques to two popular methods:

- *Hierarchical Clustering:* A hierarchical clustering technique imposes a hierarchical structure on the data which consists of a sequence of clusters.
- *Partitional Clustering:* A partitional clustering technique organizes patterns into a small number of clusters by labelling each pattern in some way. Unlike hierarchical clustering, which offers several partition of the data, partitional clustering finds a single cluster partition.

In addition to the two techniques cited above, there are also combinations of the two clustering approaches that are employed. There are many books that describe classical approaches to pattern clustering, including (Anderberg, 1973; Everitt, 1974; Hartigan, 1975; Duda & Hart, 1973).

2) ***Fuzzy Clustering:*** Fuzzy sets bring a new dimension to traditional clustering systems by allowing a pattern to belong to multiple clusters to different degrees. Bezdek has organized fuzzy clustering algorithms into five categories:

- *Relation Criterion Functions:* Clustering driven by optimization of criterion function which assesses partitions according to some global property of the grouped data. Ruspini (Ruspini, 1969) was the first to utilize this technique in the fuzzy community and he and Bezdek (Bezdek, 1981) have since considerably extended this pioneering work.
- *Object Criterion Functions:* Clustering directly on the data set A in the n-dimensional feature space according to some objective function is the most popular form of the fuzzy pattern clustering. The fuzzy c-means and fuzzy ISODATA algorithms introduced by Dunn (Dunn, 1974) and generalised by Bezdek (Bezdek, 1981), are the most popular technique for this class of fuzzy clustering algorithms.
- *Convex Decomposition:* The decomposition of a fuzzy partition (a set of fuzzy clusters) into a combination of convex sets. The use of the convex decompositions may provide added insight into data structure that otherwise might be lost. Bezdek & Harris (Bezdek & Harris, 1979) describe three algorithms that can perform this decomposition.
- *Numerical Transitive Closures:* The extraction of crisp equivalence relations from fuzzy transitive similarity relations. This technique is closely related to hierarchical methods based on graph-theoretic models.

- *Generalised Nearest Neighbour Rules:* Although the nearest neighbour algorithm is used mostly for classification, there is a clustering version as well. This technique is primarily used once the data set has already been partitioned using another clustering algorithm such as fuzzy c-means.

3) ***Neural Network Clustering:*** Neural network clustering offers the ability to determine the size, shape, number, and placement of pattern clusters adaptively while intrinsically operating in parallel. In addition, the use of clustering to form sensory maps has strongly biological support. Although there is a large number of neural networks available today there are only two primary neural clustering techniques currently in widespread use:

- *Competitive Learning:* Similar to the c-means clustering algorithm, competitive learning finds the centroids of decisions regions in the n-dimensional pattern space. Although this form of neural network learning seems to have been introduced by Grossberg (Grossberg, 1972; Grossberg, 1976a) and von der Malsburg (von der Malsburg, 1973), it has been most successfully championed by Kohonen (Kohonen, 1984), who has extended the neural dynamics to include topographic constraints.
- *Adaptive Resonance Theory:* Similar to the leader cluster algorithm, adaptive resonance theory nondestructively creates pattern “codes” (clusters). The concept of adaptive resonance was introduced by Grossberg (Grossberg, 1976b) and was first cast into a neural network formalism by Carpenter and Grossberg (Carpenter & Grossberg, 1987). There have been numerous extensions and refinements since (Carpenter & Grossberg, 1987). The most recent results of ART evolution are the algorithms combining ideas of ART and fuzzy logic. These methods, considered as the most flexible (sharp or fuzzy outputs, binary or analogue inputs, supervised or unsupervised learning), seem to be the most appropriate for the purposes of processing fuzzy outputs of confidence limit analysis. The min-max clustering and classification neural networks (Simpson, 1992; Simpson, 1993) seem to be especially interesting because of their representation of classes (clusters) which is a hyperbox in $n$-dimensional pattern space. A hyperbox is completely defined by pairs of min-max points. Gabrys (Gabrys, 1997) found an analogy with the state of the water network after the confidence limit analysis which can be viewed as a hyperbox in $n$-dimensional space defined by upper and lower bounds for each state variable. Other properties like on-line learning, the number of clusters (classes) that grows to meet the demands of the problem were exploited in the neuro-fuzzy classification and clustering algorithm developed by Gabrys for fault detection and identification in operational decision support of water systems.

We end this section by listing here several properties that a good pattern classifier should possess (Simpson, 1992):

**On-Line Adaptation**

A pattern classifier should be able to learn new classes and refine existing classes quickly and without destroying old class information. This property is sometimes referred to as on-line adaptation or on-line learning.

**Nonlinear Separability**

A pattern classifier should be able to build decision regions that separate classes of any shape and size.

**Overlapping Classes**

In addition to pattern classes being nonlinearly separable, they also tend to overlap. A pattern classifier should have the ability to form a decision boundary that minimizes the amount of misclassification for all of the overlapping classes. The most popular method of minimizing misclassification is the construction of a Bayes classifier. Unfortunately, to build a Bayes classifier requires knowledge of the underlying probability density function for each class. This is an information that is quite often unavailable.

**Training Time**

A very desirable property of a pattern classification approach able to learn nonlinear decision boundaries is a short training time.

**Soft and Hard Decisions**

A pattern classifier should be able to provide both soft and hard classification decisions. A hard, or crisp, decision 0 or 1. A pattern is either in a class or it is not. A soft decision provides a value that describes the degree to which a pattern fits within a class.

**Verification and Validation**

It is important that a classifier, neural or traditional, have a mechanism for verifying and validating its performance in some way.

**Tuning Parameters**

A classifier should have as few parameters to tune in the system as possible. Ideally, a classifier system will have no parameters that need to be tuned during training. If there are parameters, the effect these parameters have on the system should be well understood.

**Nonparametric Classification**

Parametric classifiers assume a priori knowledge about the underlying probability density functions of each class. If this information is available, it is possible to construct very reliable pattern classifiers, but often this information is not available. If the classifier is nonparametric, it should be able to describe the underlying distribution of the data in a way that provides reliable class boundaries.

### 6.3.2. The Fuzzy Min-Max Clustering and Classification Neural Network

In the previous chapters the loop based state estimation and CLA algorithms for water distribution networks were presented. These two algorithms are the first two steps on the way from measurement readings to the operational control decisions. Before any control decision can be made the state of the network has to be interpreted - classified. This interpretation task is usually carried out by an experienced, human operator. However, the growing size and complexity of the modern water distribution systems makes this task more and more difficult. The need for the "diagnosis" of a water network state (i.e. normal operating state, leakage between node *i* and node *j* etc.) had prompted Gabrys and Bargiela (Gabrys & Bargiela, 1999) to investigate into classification and clustering neural networks. They developed a new fuzzy neural algorithm based on the concept of the Fuzzy Min-Max Classification and Clustering Neural Network. Their Neural Network combines the functionality of both the Fuzzy Min-Max Classification and Clustering Neural Networks and at the same time a few major changes were made to accommodate the input in a form of the state vector with confidence limits and to improve the effectiveness of the algorithm. The resulted Neural Network was applied to patterns of state estimates and confidence intervals obtained with the nodal heads equations. In the following chapter this Neural Network will be applied for state estimates and confidence intervals obtained with the loop equations. This was actually one of the main conclusions of their work, to study the recognition performance of this Neural Network with different state estimation procedures.

Therefore this section is intended to provide a brief description of their pattern recognition algorithm.

The *fuzzy min-max clustering and classification neural networks* are built using hyperbox fuzzy sets. A *hyperbox* defines a region of the n-dimensional pattern space, and all patterns contained within the hyperbox have full cluster/class membership. A hyperbox is completely defined by its min point and its max point. The combination of the min-max points and the hyperbox membership function defines a *fuzzy set* (cluster). In the case of classification hyperbox fuzzy sets are aggregated to form a single fuzzy set class.

Learning in the fuzzy min-max clustering and classification neural networks consists of creating and adjusting hyperboxes in pattern space as they are received. It is an expansion/contraction process. The learning process begins by selecting an input pattern and finding the closest hyperbox to that pattern that can expand (if necessary) to include the pattern. If a hyperbox cannot be found that meets the expansion criteria, a new hyperbox is formed and added to the system. This growth process allows existing clusters/classes to be refined over time, and it allows new clusters/classes to be added without retraining. One of the residuals of hyperbox expansion is *overlapping hyperboxes*. Hyperbox overlap causes ambiguity. It is reasonable to assume that a pattern can have the same partial membership in more than one cluster/class. It is not reasonable to assume that a pattern can completely belong to more than one cluster/class. In the case of classifying NN the overlap is eliminated for hyperboxes that represent different classes. A contraction process is utilized to eliminate any undesired hyperbox overlaps.

The fuzzy min-max clustering and classification learning algorithm that will be used in the following chapter and it has been originally developed by Gabrys and Bargiela (2000) can be described as follows:

***Initialization***

The pattern recognition algorithm developed by Garbrys and Bargiela has been intended to be used for the water distribution network state classification task. Therefore the information obtained from confidence limit analysis, namely confidence limits for each state variable, had been accommodated by this classification procedure. This requirement had been met by specifying the input to classification/clustering algorithm as a pair of two vectors: $X_h = [X_h^l \; X_h^u]$ - the lower and upper limits for the state vector. In other words instead of a point in n-dimensional space that had to be classified, they obtained a hyperbox with the min point determined by the vector $X_h^l$ and the max point determined by the vector $X_h^u$. When the min and max points are equal the hyperbox shrinks to the point. In conclusion the algorithm is capable of classification/clustering inputs in a form of the n-dimensional vector without any changes to the algorithm because a point in n-dimensional space is simply the special case of a hyperbox with the min and max points equal.

They observed that because of the size of the modern water distribution networks it is impossible to predict and cover all possible combinations of consumption-inflows patterns and anomalies that can occur in the network during day to day operations. Therefore, in order to allow labelled (i.e. normal operating state etc.) and unlabelled inputs to be processed an additional index, $d_h = 0$ meaning that the input pattern is not labelled, had been introduced. A hybrid, supervised (labelled inputs - classification) and unsupervised (unlabelled inputs - clustering), Neural Network had emerged.

***Hyperbox membership function***

The fuzzy hyperbox membership function plays a crucial role in the Fuzzy Min-Max Classification and Clustering algorithms. The decisions whether the presented input pattern belongs to the particular class or cluster, whether the particular hyperbox is to be expanded, depend mainly on the membership value describing the degree to which an input pattern fits within the hyperbox. The *j*-th hyperbox fuzzy set, $B_j$, can be defined by the ordered set:

$$B_j = \{X_h, V_j, W_j, b_j(X_h, V_j, W_j)\}$$

for all *h*=1,2,...,*m*, where $X_h = [X_h^l \; X_h^u]$ is the *h*-th input pattern, $V_j = (v_{j1}, v_{j2}, ..., v_{jn})$ is the min point for the *j*-th hyperbox, $W_j = (w_{j1}, w_{j2}, ..., w_{jn})$ is the max point for the *j*-th hyperbox, and the membership function for the *j*-th hyperbox is $0 \le b_j(X_h, V_j, W_j) \le 1$. The min points are initialized with 1 and the max points with 0.

An investigation has been carried out by Gabrys and Bargiela (Gabrys & Bargiela, 1999) in order to decide the most appropriate form for the membership function. They chose the function so that the membership values of the patterns to decrease steadily with the increasing distance from the hyperbox. The reason for doing so is to eliminate the cases when hyperboxes that represent different classes are overlapping. The chosen function is shown below and it can be described as the minimum value of maximum min-max hyperbox points violations for all dimensions:

$$b_j(X_h) = \min_{i=1..N}(\min([1 - f(x_{hi}^u - w_{ji}, \gamma_i)], [1 - f(v_{ji} - x_{hi}^l, \gamma_i)])) \qquad \text{(Eq. 6.2)}$$

where $x_{hi}^u$ are $x_{hi}^l$ the lower and the upper limits of the *h*-th input pattern specified for each dimension *i*. $w_{ji}$ and $v_{ji}$ are the max and min points of the *j*-th hyperbox. $f(x,\gamma)$ is a two parameter ramp threshold functions which can be written as:

$$f(x,\gamma) = \begin{cases} 1 & if \ x\gamma > 1 \\ x\gamma & if \ 0 \le x\gamma \le 1 \\ 0 & if \ x\gamma < 0 \end{cases}$$

The membership function contains also the parameter $\gamma = [\gamma_1, \gamma_2, ..., \gamma_n]$ that regulates how fast the membership values decrease and it has to be specified for each dimension (i.e. nodal pressures, inflows).

***Hyperbox expansion***

This process can be described briefly as to identify the hyperbox closest to the input pattern that can be expanded and expand it. If an expandable hyperbox cannot be found, add a new hyperbox. A user specified value Θ was introduced to control the size of the hyperbox which can be described as the difference between the max and min value for each dimension. It has been observed (Gabrys, 1997) that keeping the parameter Θ constant during the learning process can have undesired effects on performance or the number of created hyperboxes. Setting Θ big can cause too many misclassifications, especially when there are complex, overlapping classes. On the other hand, it has been observed that when Θ is small too many unnecessary hyperboxes can be created, especially for concentrated, standing alone groups of data forming one class, while small Θ might be needed to resolve other overlapping classes. These problems were addressed by introducing an adaptive maximum size of the hyperbox.

We shall take into account these observations when testing the recognition system with the loop-based state estimates and confidence limits. The other steps performed during the training/learning process of the Neural Network are as follows.

***Hyperbox overlap test***

Determine whether the recent expansion caused any undesired overlap between hyperboxes.

***Hyperbox contraction***

If the overlap test identified overlapping hyperboxes, then contract the hyperboxes to eliminate overlap.

More details about each of these steps can be found in (Gabrys, 1997; Gabrys & Bargiela, 1999). However, we will mention here that the training process is completed when after presentation of all training patterns there have been no misclassification for the training data or the minimum, user specified value of the parameter Θ has been reached.

The neural network that implements the generalized fuzzy clustering-classification algorithm as it has been described above and developed by Gabrys and Bargiela (2000) is shown at Figure 6-4. The topology of this neural network grows to meet the demands of the problem. The input layer has 2*n processing elements, two for each of the n dimensions of the input pattern $X_h = [X_h^l \; X_h^u]$. Each second layer node of this three-layer neural network represents a hyperbox fuzzy set where the connections of first and second layer are the *min-max points* and the transfer function is the hyperbox membership function. The min points are stored in the matrix *V* and the max points are stored in the matrix *W*. The way these connections are adjusted is described in (Gabrys, 1997). A detailed view of the *j*-th second layer node is shown at Figure 6-5. The connections between the second and third layer nodes are binary values. They are stored in the matrix *U*. The equation for assigning the values of *U* is:

$$u_{jk} = \begin{cases} 1 & \text{if } b_j \text{ is a hyperbox for class } c_k \\ 0 & \text{otherwise} \end{cases} \qquad \text{(Eq. 6.3)}$$

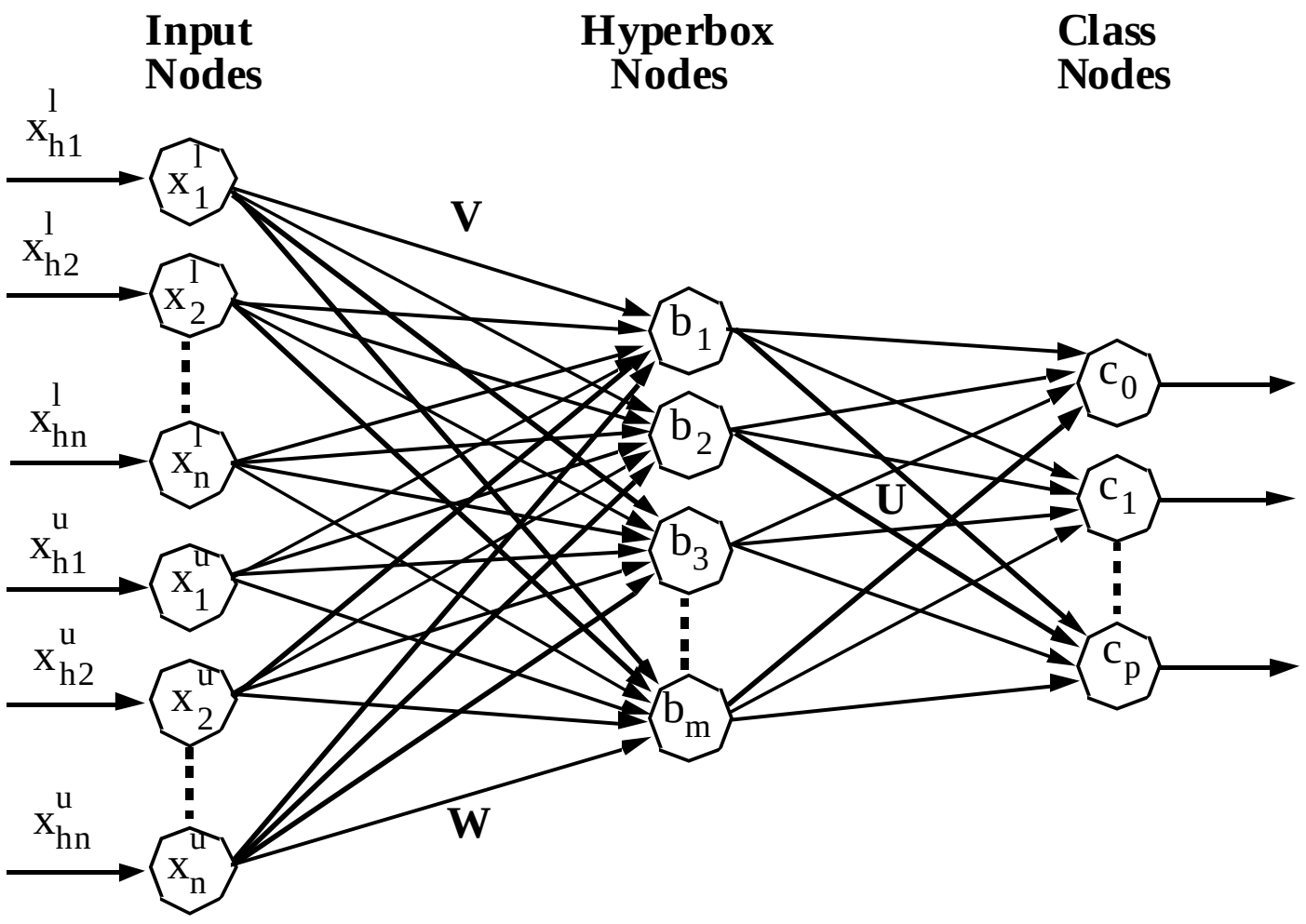

Figure 6-4: The three layer neural network that will implement the clustering-classification algorithm applied to the patterns of ‘loop-based’ state estimates.

Each of the third layer nodes represents a pattern class. The node $c_0$ represents all *unlabelled hyperboxes* from the second layer.

In Figure 6-5 the node with its associated membership function and connections in from of vectors $V_j$ and $W_j$ represents a hyperbox fuzzy set.

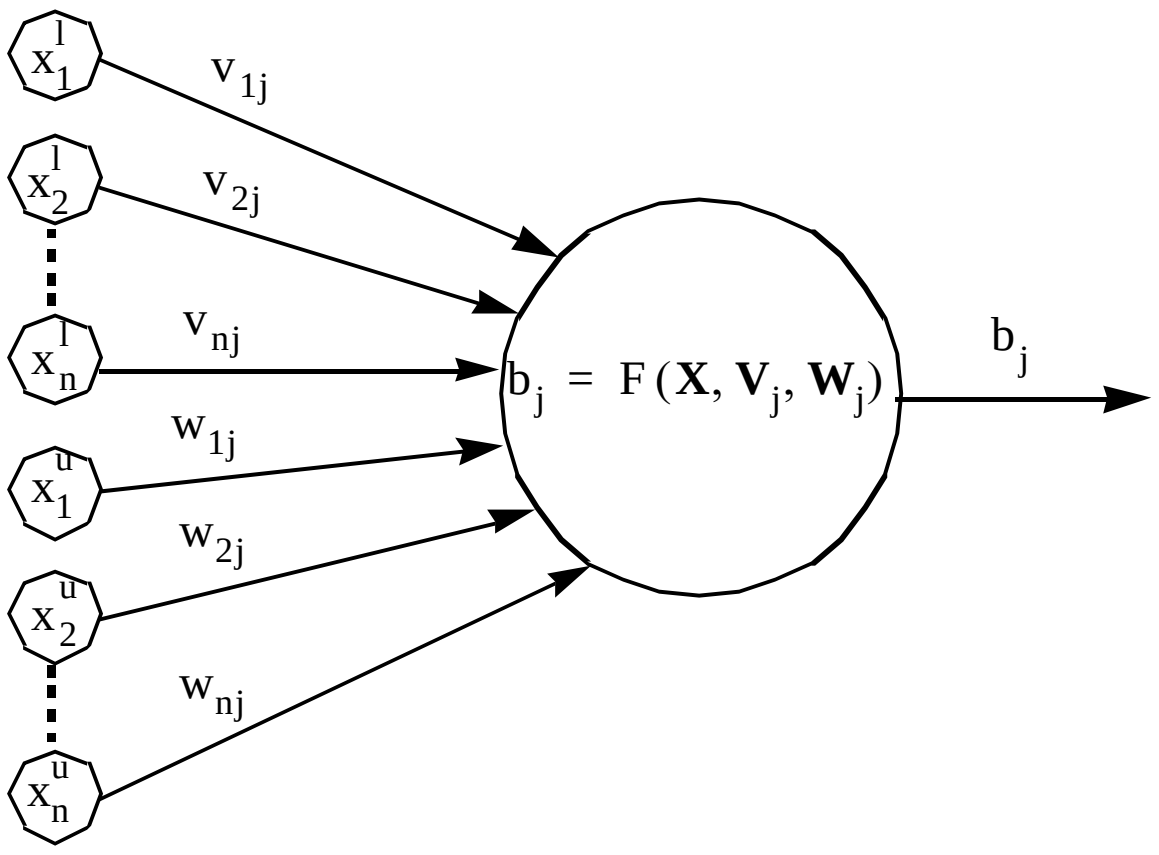

Figure 6-5: A detailed view of the *j*-th second layer node.

The classification and clustering algorithm briefly described here had been tested extensively on different sets of data (both data points and fuzzy labelled and labelled input patterns) and compared to other existent classification algorithms (Gabrys, 1997). We will just say here that Gabrys algorithm dealt successfully with both labelled and unlabelled patterns, in most of the cases resolved all the overlappings between hyperboxes from different classes, which finally resulted in fewer misclassifications compared with several other neural, fuzzy and traditional classifiers (Gabrys & Bargiela, 2000; Gabrys, 1997).

## 6.4. Conclusions

*Pattern recognition* has long been studied in relation to many different applications (and mainly unrelated), such as classifying galaxies by shape, identifying fingerprints or handwriting recognition. Human expertise in these and many similar problems is being supplemented by computer-based procedures, especially neural networks. Pattern recognition has been extremely widely used, often under the names of ‘classification’, ‘diagnosis’, or ‘learning from examples’.

Gabrys and Bargiela (1999) are perhaps the first to use pattern recognition within an operational decision support in water distribution networks and, in particular, to detection and identification of faults based on the fuzzy classification and clustering Neural Networks. Their training data included patterns of water network state estimates and confidence limits obtained by simulating a 34-nodes water network for a complete 24 hours period of operation. An analogy between the information processing by the classification and clustering algorithm, and the human operators had been identified and highlighted in this context.

Studying the performance of the recognition system with patterns of state estimates and confidence limits obtained with the novel loop flows state estimator, is to be addressed in the following chapter.

# Chapter 7.

# The combination of the loop algorithms and the neural classification for fault detection and identification in water systems

## 7.1. Introduction

Two broad categories of faults occurring in water distribution systems are considered in this work. The faults dues of malfunctioning of transducers and telecommunication equipment are referred to as the *measurement errors*. And the faults due to leakages and wrong status of valves, invalidating the system model used in the estimation, are referred to as the *topological errors*.

The crucial difference between these two types of errors is the fact that although both are responsible for poor state estimates, the meter malfunctions do not have any bearings on the actual state of the system while the leakages or the valve status errors directly affect the physical system and can result in service disruptions.

In the case that the measurement errors are uncorrelated and if there is a high enough local measurement redundancy it is often possible to reject erroneous data by using a suitable estimation procedure as described in Chapter 4.

On the other hand, model based errors give rise to correlated changes in groups of incoming signals. In such a case the state estimation procedure trying to compensate for invalid network model may result in a set of errors scattered across the network as it is the case with the nodal heads state estimator. However, it has been shown in the previous chapters a novel state estimator based on the loop corrective flows. It improves the best approximation of the operational status of the water system providing that accurate pressure and flow measurements are available.

It is obvious that in the absence of accurate real measurements, the topological errors not only pose a much greater danger to the safety of water network operation but also are more difficult to locate and eradicate even when reliable and efficient state estimators are available. Depending on the topology of the distribution network and the state estimator used, the topological class of errors form characteristic patterns that can be utilized to classify the state of the network.

The classification of the state of the water network it has been largely investigated in Gabrys (1997) and Gabrys and Bargiela (1999) in the context of the state estimators based on the nodal heads equations. Their approach for diagnosis of leakages and other operational faults occurring in water networks was based on the examination of patterns of state estimates or residuals by a General Fuzzy Min-Max neural network (GFMM). They have shown that both the state estimates with their confidence limits and the residuals with their confidence limits can be successfully used to train the GFMM neural recognition system.

This chapter presents the application of the GFMM neural network to the classification of the state of the water system based on patterns of loop flows state estimates and confidence limits (Bargiela et al., 2002). The investigation will have a twofold intention: first, to build an effective decision system for fault detection and preventive maintenance of water system by using the loop flows state estimator, the confidence limits analysis algorithms and the GFMM neural network. The second attempt will be to search for the advantages that this combination might have over the initial system described in Gabrys (1997).

The chapter is organized in three sections. The review of the previous work on the subject of bad data detection and identification is presented in Section 7.2. This is followed by Section 7.3 which is the main section of the chapter and is concerning the fault detection in water systems based on the combination of the loop algorithms and the GFMM neural network. The aspects of training and testing of the neural network with the 'loop-equations based' state estimates and the variation of nodal demands with confidence limits are discussed. And finally, the closing section of the chapter presents the discussion and conclusions.

## 7.2. Review of the previous work

Very often the algorithms found in the literature and referred to as bad data analysis are concerned with the identification and rejection of erroneous measurements and do not attempt to identify the underlying cause of the bad data. Rather than only asking the questions: Are the state estimates accurate? How to construct the state estimation procedure in order to reject anomalous data?; one would like to know the answers to the questions: What do those state estimates mean? Is the current state a normal operating state of the network? Is there a leakage present that requires a remedial action? etc.

Bargiela introduced the idea of *bad data analysis* in water distribution systems state estimation (Bargiela, 1984). In order to distinguish between the measurement and topological errors his method checks the magnitude and sign of the weighted measurement residuals at each end of a pipe. It was shown that the presence of either a leakage or incorrect status of control valves is equivalent to neglecting a part of the actual network structure thus producing an imbalance at the network nodes adjacent to the pipe in question. The idea therefore was that the topological error can be thought of as a pair of erroneous load measurements for which the error terms (residuals representing the mass balances at those nodes) are carrying information about a type of topology error. Figure 7-1 gives a graphical representation of Barigela's method.

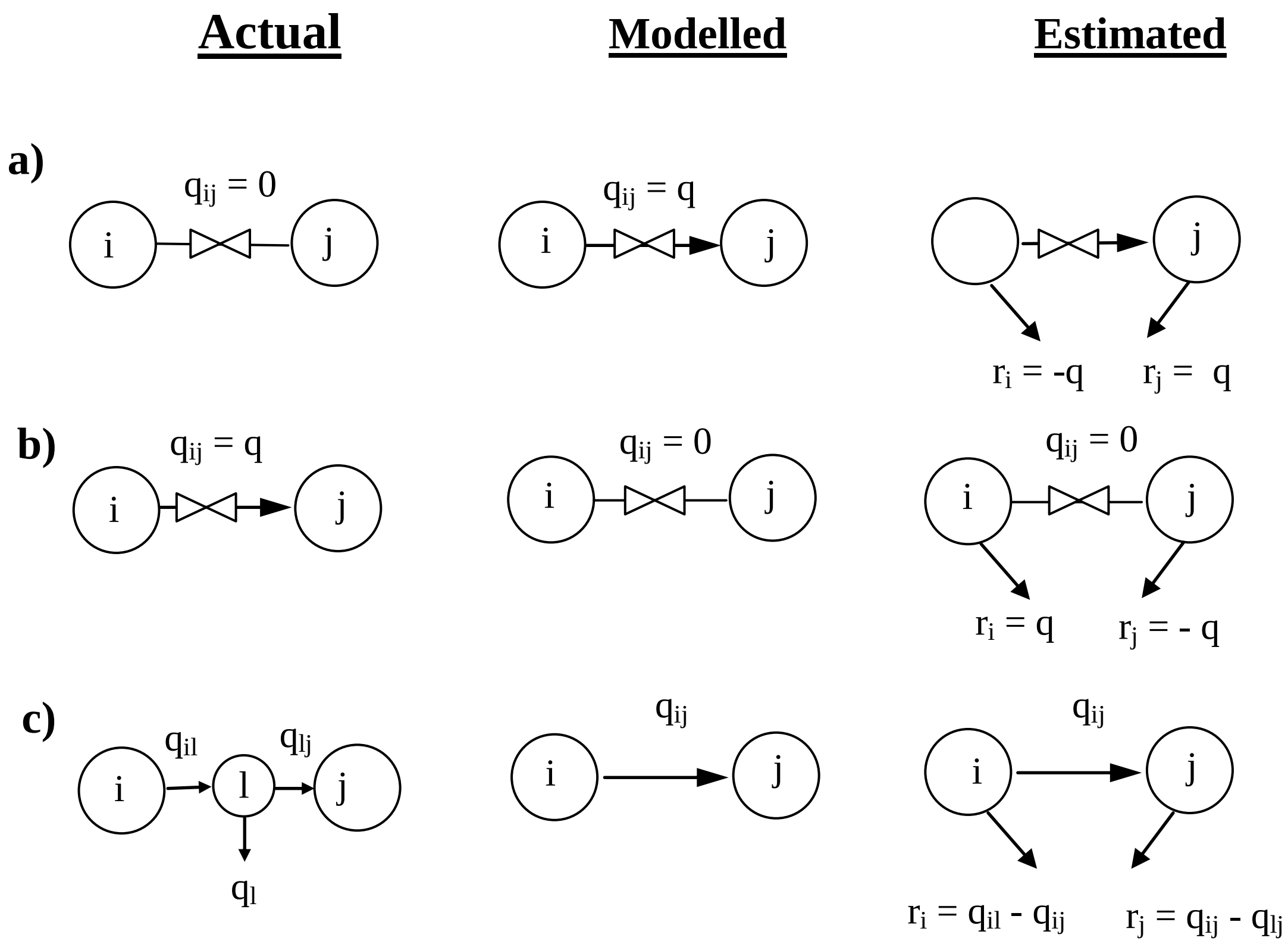

Figure 7-1: Identification of topological errors as presented by Bargiela (1984): a) closed valve monitored as open; b) opened valve monitored as closed; c) leakage.

Although these ideas are very useful they rely on the high local measurement redundancy ratio so that the erroneous data can be rejected. It is not always the case and the effect of topological error occurrence cannot be restricted to the end nodes of effected pipe but is spread in the larger area around the leaking pipe. Perhaps we should mention here the novel state estimator described in the previous chapters that by carefully considering a region in the water network where the state estimation is applied, can reduce the spreading of the topological errors to larger areas in the network (Arsene & Bargiela, 2002a).

These problems have been also recognized by Powell (Powell, 1992) whose method is based on finding paths linking groups of high measurement residuals. Once the connecting paths between high residuals have been identified, heuristics are applied to determine the location and cause of errors. In these heuristics the residuals are sorted by type, direction, magnitude and location. For instance if the leakage is present in the network the pressures in an area near to the leakage will decrease. On the other hand, if there is a blocked pipe in the network the pressure upstream of this pipe will be high and downstream it will be low. Since the changes in pressure are characteristic for different faults one should be able to observe those changes in the residuals representing the pressure measurements in mathematical model of the network.

Other publications on the subject for water distribution system concern only leakage detection studies. Pudar and Liggett (Pudar & Liggett, 1992) attempted the leak detection task by solving an inverse problem. This inverse problem is essentially the state estimation procedure with additional state variables being the unknown leaks. The method assumed that the leaks occur in the nodes and do not change the topology of the network. Furthermore, the locations of suspected leakages are assumes to be known. Unfortunately, both assumptions are a gross oversimplification.

Carpentier and Cohen in their paper (Carpentier & Cohen, 1993) tell the story about 10 years of involvement of their research group in the application of mathematical techniques for the management of complex water supply networks. One of the two main topics discussed is state estimation and leak detection. The leakage detection in this work is based on a comparison of the consumption values estimated on-line, using current, real flow measurements, with the pseudomeasurements of the same consumptions considered as standard values in the normal situation (without any leakage present). These pseudomeasurements are obtained from 24-hour mathematical model of the normal network operations. Throughout this work a heavy emphasis is put on the necessity of having the well calibrated model of the network. The performance of the method was tested on a real subnetwork of the water network of the city of Paris. The leakages were introduced to the network by opening fire-hydrants in some places. The fact that experiments were carried out on the real network give additional weight to the results. The weights in the weighted least squares criterion were chosen in such a way that the errors occur in nodal mass balance equations and represent increase in nodal consumptions. Occurrence of a set of significant errors in some area of the network is treated as a sign of leakage presence in this area. No attempts were made to further process these errors in order to find a reduced number of the most likely pipe(s).

Recently, the fault detection and diagnosis problem in water distribution systems has been attempted based on the examination of patterns of state estimates implemented by a newly developed neurofuzzy recognition system (Gabrys, 1997; Gabrys & Bargiela, 1999). Their approach combines the ability of fuzzy systems to cope with uncertain and ambiguous data with the computational efficiency, learning, and pattern recognition ability of neural networks.

The rationale for this approach is that although the analysis of precise numerical results of state estimation is useful, it also tends to ignore the grater picture of the overall system state, which is something that experienced human operators primarily focus their attention on before analyzing the detail. The pattern recognition based approach to fault diagnosis was thought to mimic the information processing and abstraction forming by human operators.

In particular, it has been shown that both the ‘nodal heads equations’ state estimates with their confidence limits and the residuals with their confidence limits can be successfully used to train the neural recognition system. However, it has been also found that due to the high susceptibility of the residuals to the typical measurement noise, the ability to detect and identify faults by the recognition system based on the state estimates performed better for the data for which it has been trained, due to much lower ration of the noise to the useful signal and larger spatial separation of patterns representing different classes.

The successful application of the fuzzy neural recognition systems to the water network state identification has pointed to a couple of possible further research areas. One of these areas of interest is regarding the study of the recognition system performance in association with different state estimation procedures, and this is addressed here in the context of the loop flows state estimator and the confidence limits algorithms developed in the previous chapters.

## 7.3. The combination of the loop algorithms and the pattern recognition system

### 7.3.1. The significance of confidence limits

In any pattern classifier design problem it is necessary to have a representative set of accurate training examples. Since we would like to utilize the information about confidence limits in the process of constructing our classification system it is absolutely necessary to understand what is the meaning of those confidence limits when calculated for different values of state estimates and if and when they can be used in the training stage without compromising the performance of the pattern classifier.

The significance of confidence limits referred to in the title of this section can be explained by imaging two different experiments.

In the first experiment the estimates are calculated for accurate measurements. If one also assumes that the mathematical model of the process used in the estimation procedure accurately represents the behaviour of the physical system, a true state estimate of the system can be obtained. This is obtained by $x_{i_{acc}}$ in Figure 7-2 for the *i-th* state variable.

However, since the measurements have a finite accuracy it is interesting to know how sensitive is the same estimate to the measurement inaccuracies. In this case, the confidence limits calculated for the true state represent the boundaries within which all estimates of this true state will fall as long as the measurements used are at least as accurate as the ones taken to compute the confidence limits themselves. In Figure 7-2 the lower and upper bound for the *i-th* state variable are denoted $x^{l}_{i_{acc}}$ and $x^{u}_{i_{acc}}$ respectively. In other words, if for purpose of pattern classification the true state was labeled as “the normal operating state” all the estimates falling within its confidence limits could be classified as “the normal operating state”.

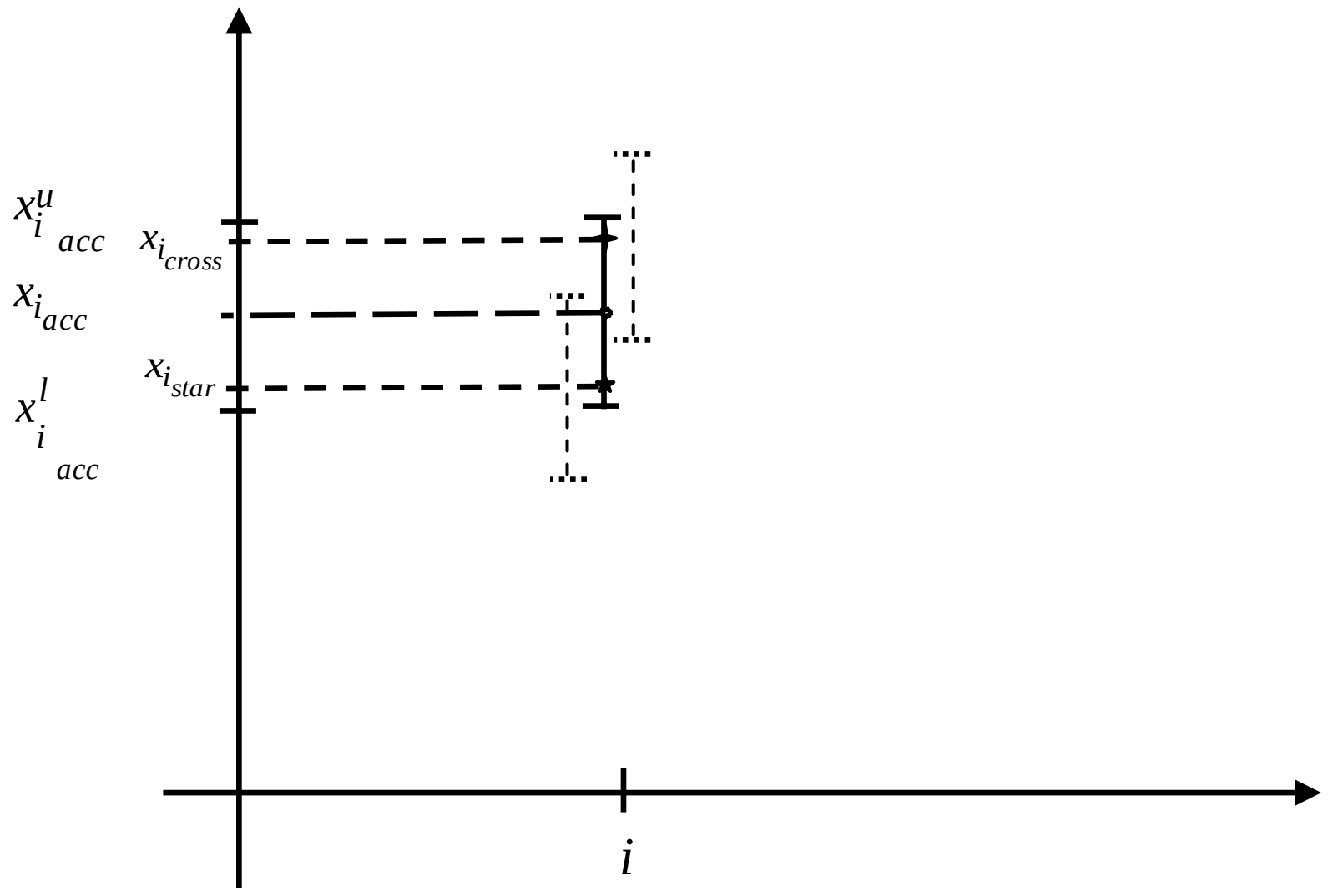

Figure 7-2: Graphical representation of state estimates and confidence limits for accurate and inaccurate measurements. $x_{i_{acc}}$ - estimate for the *i*-th state variable calculated for accurate measurements; $x^l_{i_{acc}}$, $x^u_{i_{acc}}$ - the lower and upper bound for $x_{i_{acc}}$; $x_{i_{star}}$ and $x_{i_{cross}}$ are the two examples of the *i*-th state variable estimate calculated for inaccurate measurements (Gabrys, 1998).

In the second experiment the estimates are calculated for a set of measurements that are measured with some finite error. This is to say that the true state is unknown and for a given set of inaccurate measurements one can only compute the best estimates of this true state. Two examples of the instantaneous estimates of the true state value $x_{i_{acc}}$ are denoted by $x_{i_{star}}$ and $x_{i_{cross}}$ in Figure 7-2. Unlike the confidence limits computed for the true state, the confidence limits found for any of the instantaneous estimates only indicate that the true state value is contained within their range. The confidence limits for $x_{i_{star}}$ and $x_{i_{cross}}$ are depicted in form of dashed vertical lines in Figure 7-2. When in the extreme case the estimated values were equal to $x^l_{i_{acc}}$ and $x^u_{i_{acc}}$ using the confidence limits for such estimates during the training of the classification network would mean the introduction of additional 50% error to the cumulative error resulting from inaccurate measurements.

It follows from the above that when the true state (or a very good estimate of it) can be computed, then the confidence limits found for such an estimate directly give a hyperbox (cluster) without the need to use a large number of instantaneous estimates during the training (which would arrive at the same cluster).

However, when the sufficiently accurate estimate of the true state cannot be found one has to resort to a large number of correctly labeled instantaneous estimates.

### 7.3.2. Generating the training data

While for the well maintained water distribution systems the normal operating state data can be found in abundance the instances of abnormal events are not that readily available. In order to observe the effects of abnormal events in the physical system one sometimes is forced to resort to deliberate closing of valves or opening of hydrants (to simulate leakages) (Carpentier & Cohen, 1993). Although such experiments can be very useful to confirm the agreement between the behaviour of the physical system and the mathematical model, it is not feasible to carry out such experiments for all pipes and valves in the system during the whole day or days as might be required in order to obtain the representative set of labeled data.

It is an accepted practice that, for processes where the physical interference is not recommended or even dangerous, mathematical models and computer simulations are used to predict the consequences of some emergencies so that one might be prepared for quick response. In our case the computer simulations are used to predict the consequences of some emergencies so that one might be prepared for quick response. In our case the computer simulations were used to generate data covering 24 hour period for the water distribution network depicted at Figure 7.3. Such simulations that stretch over longer periods of time are called extended time simulations. The reason for choosing this network is that we will be able to compare the results of training the recognition system with patterns of ‘loop-equations based’ state estimates and confidence limits with the results reported in Gabrys (1997) for the same water network and operational testing conditions in the context of the nodal heads equations.

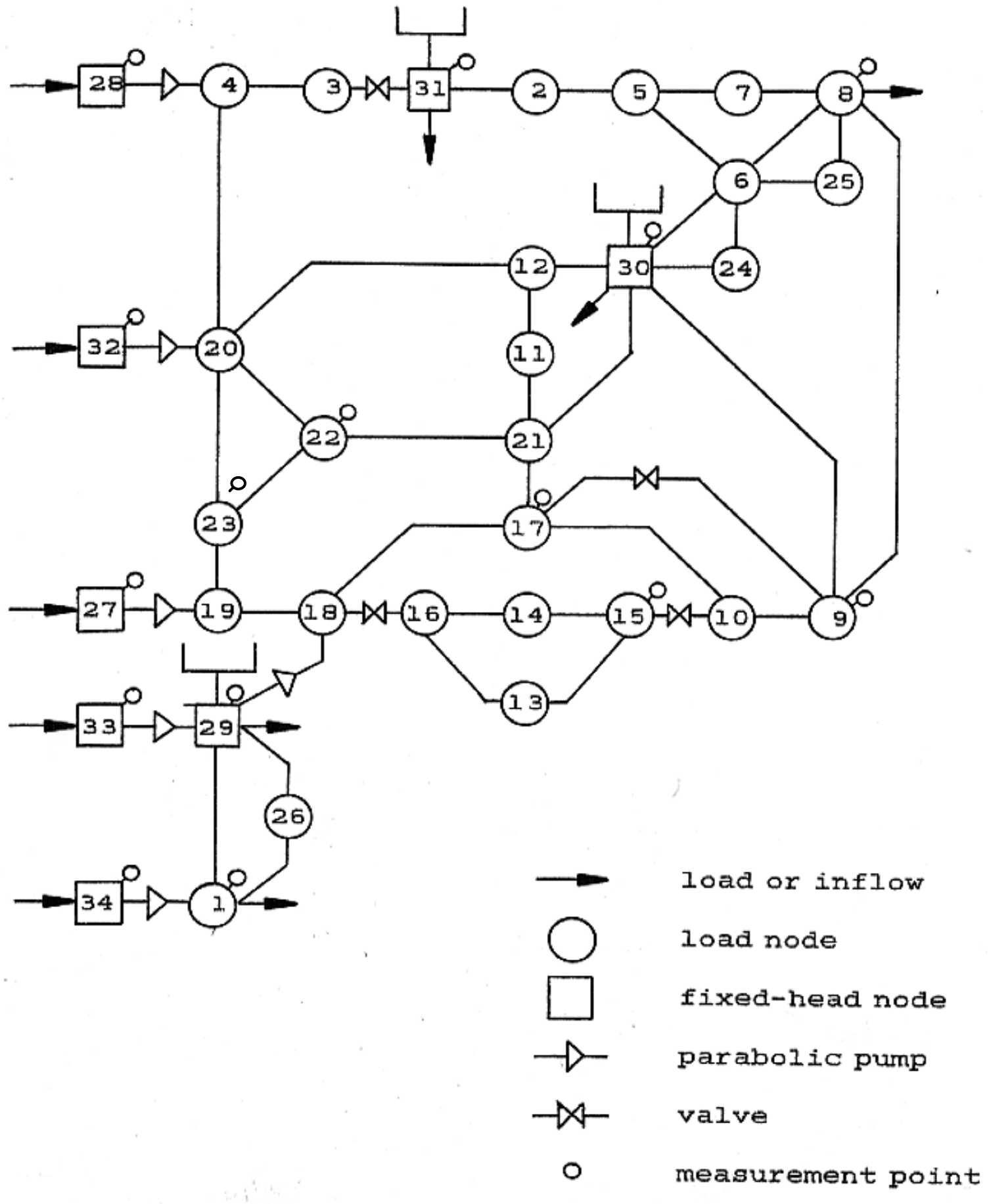

Figure 7-3: 34-node water network used to generate the training data for the pattern recognition system.

The process of generating the training data is shown in the form of block diagram at Figure 7-4. It consists of three major blocks.

The first module is the co-tree flows simulator that is used as a substitute for the physical water distribution network. It is this module where the leakages are simulated by updating the topology information rather than opening hydrants.

In the second module, the loop flows state estimation process is carried out for accurate measurements taken from the simulation module but without knowledge of any anomalous event that might have happened, as would be the case in the real distribution network. In the third module the confidence limits are found for state estimates and the variation of nodal demands calculated at the estimation stage. Additionally to the state estimates with their confidence limits the system's status or label of the current pattern is stored.

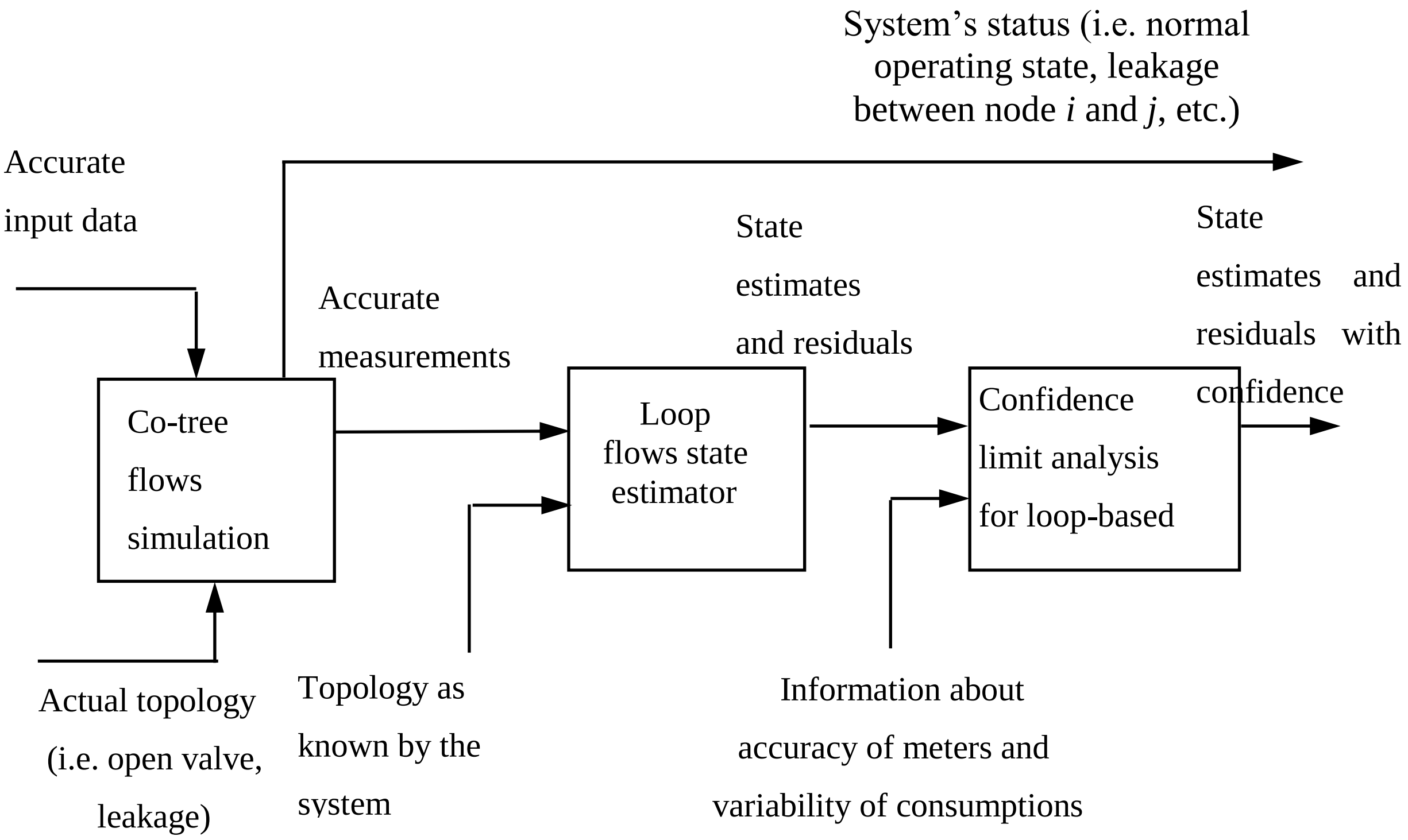

Figure 7-4: Graphical representation of the training patterns generation scheme.

The loop-based numerical algorithms developed in the previous chapters are used here to generate the training data necessary for the classification module. The co-tree flows simulator, the loop state estimator and the confidence limits analysis are the algorithms employed in the block diagram shown at Figure 7-4. However, these algorithms were previously developed as standalone applications that were not communicating each other, as it is the case in an extended time simulation of the block diagram.

It is well known that an extended time simulation of a water network implies different sets of nodal consumptions, in/out flows and head values at the boundary nodes of the network for each simulation of the distribution system.

On the other hand, the co-tree flows simulator and the loop flows state estimator requires as input data the loop and the tree incidence matrixes and the initial pipe flows that have to satisfy the continuity equation. The input data is obtained from a spanning tree which has to be rebuilt at each step of the extended time simulation in order to determine the incidence matrixes and the initial pipe flows.

Rebuilding of the spanning tree may represent a computational drawback for the block diagram shown at Figure 7-4 which may be a disadvantage when compared to the nodal heads variant developed in Gabrys (1997). This is because the block diagram implemented with nodal heads equations does not require any input data that may be computational expensive to obtain.

However, let us assume that $M_{lp}$, $Q_i$ and $T$ are the loop incidence matrix, the initial pipe flows and the tree incidence matrix obtained from the spanning tree as described at Chapter 3. The nodal demands $d$ are given.

The training patterns generation scheme is pursued once. The state estimates with the confidence limits and the status of the water network are stored for subsequent utilization in the classification module.

For the following step in the series of extended time simulations, a new set of nodal demands $d^{'}$ and head values at the boundary nodes of the network are provided. Furthermore, instead of carrying out the time consuming process of rebuilding the spanning tree, the new set of initial conditions (i.e. initial pipe flows, incidence matrixes) are determined with the following equations:

$$Q_i^{'} = T^{-1} d^{'} \qquad \text{(Eq. 7.1)}$$

where $Q_i^{'}$ are the initial pipe flows used in the next extended time simulation.

The loop and tree incidence matrixes are obtained function of the initial pipe flows. Therefore, where the direction of initial flows $Q_i^{'}$ changes due to the new set of nodal demands $d^{'}$, then the loop and the tree incidence matrixes are updated as follows:

$$M_{lp}^{'}(:,k) = (-1)\, M_{lp}(:,k) \qquad \text{(Eq. 7.2)}$$

$$T^{'}(:,k) = (-1)\, T(:,k) \qquad \text{(Eq. 7.3)}$$

where $k$ is the pipe with the reversed flow, $M_{lp}^{'}$ and $T^{'}$ are the new loop and tree incidence matrixes used in the next extended time simulation.

By means of the equations (Eq. 7.1 –Eq. 7.3), the block diagram shown at Figure 7-4 has been successfully run for a 24 hours extended time simulation. The Central Processing Unit (CPU) times were similar with the times obtained for the implementation based on nodal heads equations. The 24 hour profiles of consumptions and inflows that characterize the normal operating states throughout the day are similar with the ones reported in Gabrys (1997) and are shown below for completeness.

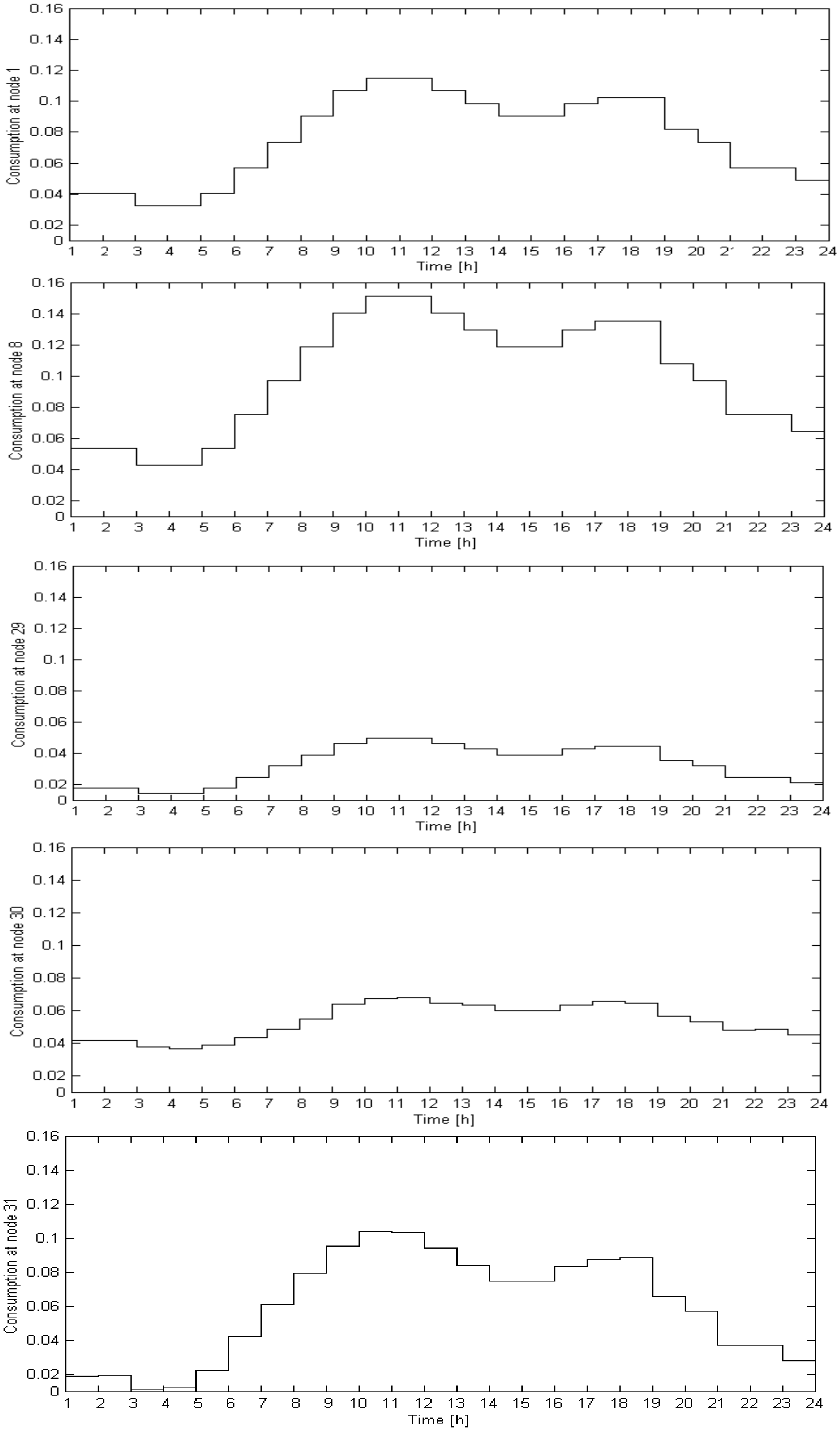

Figure 7-5: 24 hour profiles of consumptions at nodes 1,8, 29, 30 and 31.

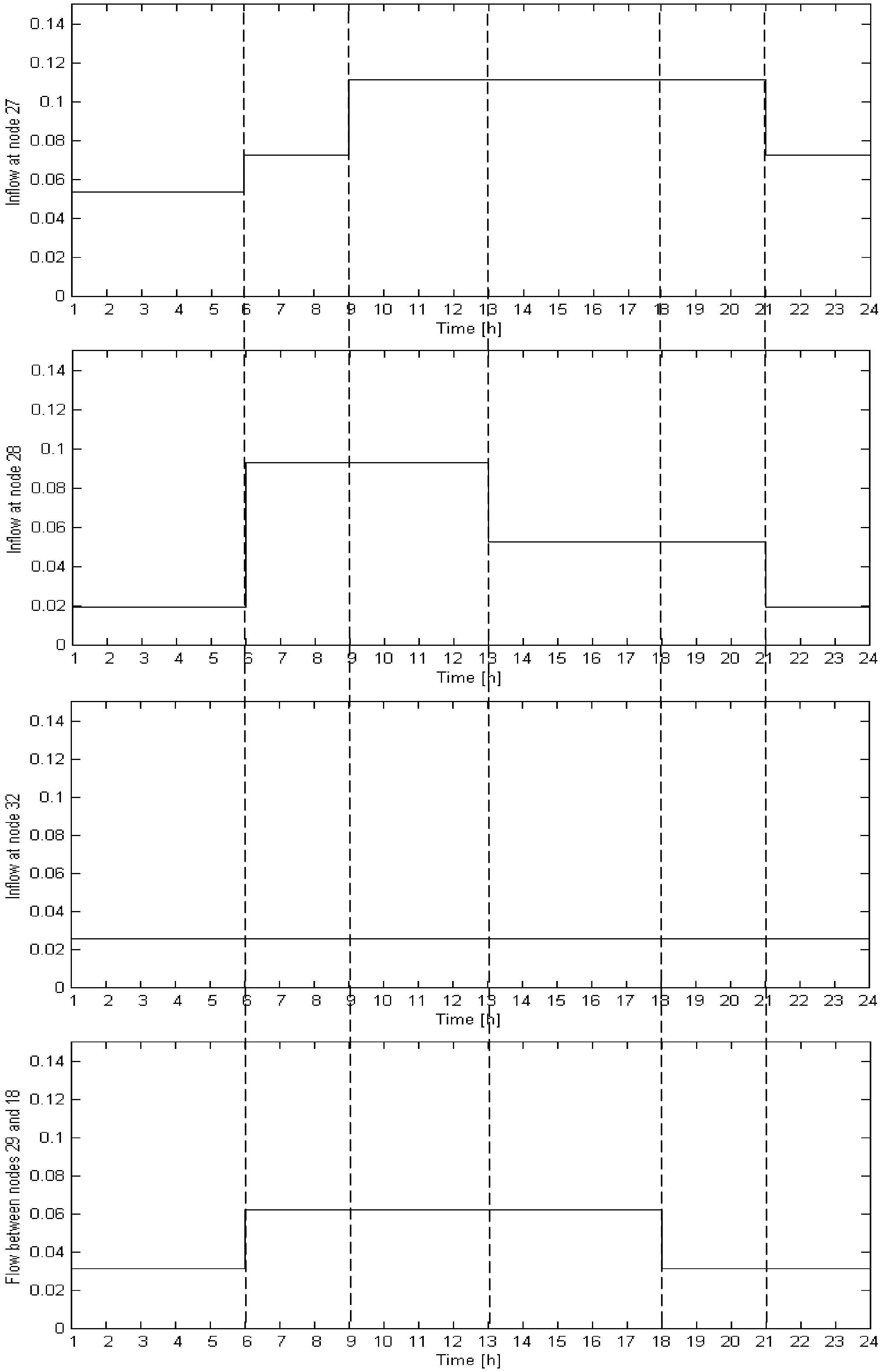

Figure 7-6: 24 hour profiles of inflows at fixed head nodes 27, 28, 32 and booster pump between nodes 29 and 18.

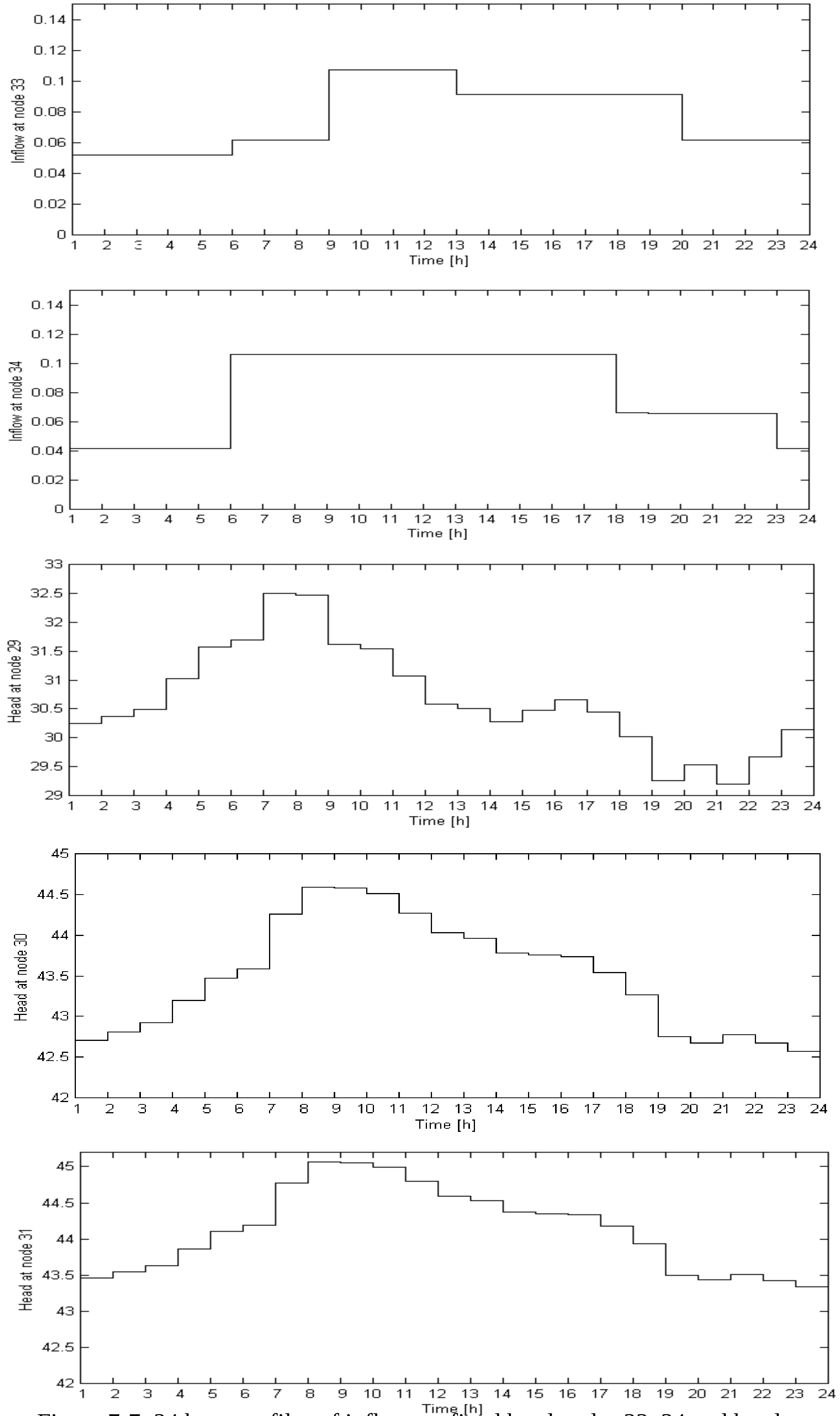

Figure 7-7: 24 hour profiles of inflows at fixed head nodes 33, 34 and heads at reservoir nodes 29, 30, 31.

***Simulation of leakages***

In the physical system simulator, the leakage is modeled as an additional demand lying midway between the two end nodes of a pipe. The additional demand is not modeled as a pressure dependent variable and thus can be set to any desired value.

The spanning tree for the 34-node water network is shown at Figure 7-8. The main root node is node 30 and a pseudo-loop is added between the fixed head-node 31 and the main source node. The inflows to the other fixed-head nodes 27, 28, 29, 32, 33 and 34 are maintained constant. This means that the pumping stations represented by links 32-20, 27-29, 28-4, 33-29, 29-19 and 34-1 are assumed to produce a constant inflow and are not affected by leakage. Therefore the inflows at the reservoirs 30 and 31 will be adjusted during the Newton-Raphson method so that to cover the additional demand resulting from the leakage. Please notice also that new labels are assigned to nodes so that the tree incidence matrix to become upper triangular.

However, before simulating the leakages perhaps we should observe that the link 29-18 is a pump with a constant flow that is not included in any loop. Unfortunately, as other authors have observed (Bounds, 2002), such situations can not be solved by using a simulator based on the loop corrective flows. Moreover, Gabrys (1997) observed in the context of the simulator based on the nodal heads equations, that is possible even for very small leakages, to restrict the possible leakage area to the three pipes connecting nodes 1, 26 and 29. This is because of the constant flow in the pump between nodes 29 and 18 that separates nodes 1, 26, 29, 33, and 34 from the rest of the network. It has been concluded that since changes in the lower part of the network had no bearings on the other part, only the upper part will be used for fault detection and identification (Gabrys, 1997).

We should take in consideration the last observation and by systematically working through the network, ten levels of leaks are introduced, one at a time, in every single pipe for every hour of the 24 hour period. Since there are 38 pipes multiplied by 10 levels of leakages and plus the normal operating status gives 381 patterns of state estimates for each hour. For a full day this will become a training set of data consisting of 9144 labeled patterns of state estimates computed for accurate measurements and leakages ranging from 0.002 to 0.029 [$m^3$/s].

However, since an additional consumption is used in order to simulate the leaks, this would require modifying the incidence matrixes and the initial pipe flows for the loop algorithms. Therefore rebuilding the spanning tree for each of the 9144 patterns of data would represent a computational drawback for the training patterns generation scheme shown at Figure 7-4. This would be also a disadvantage when compared to the implementation based on the nodal heads equations, which does not require the recalculation of the incidence matrixes.

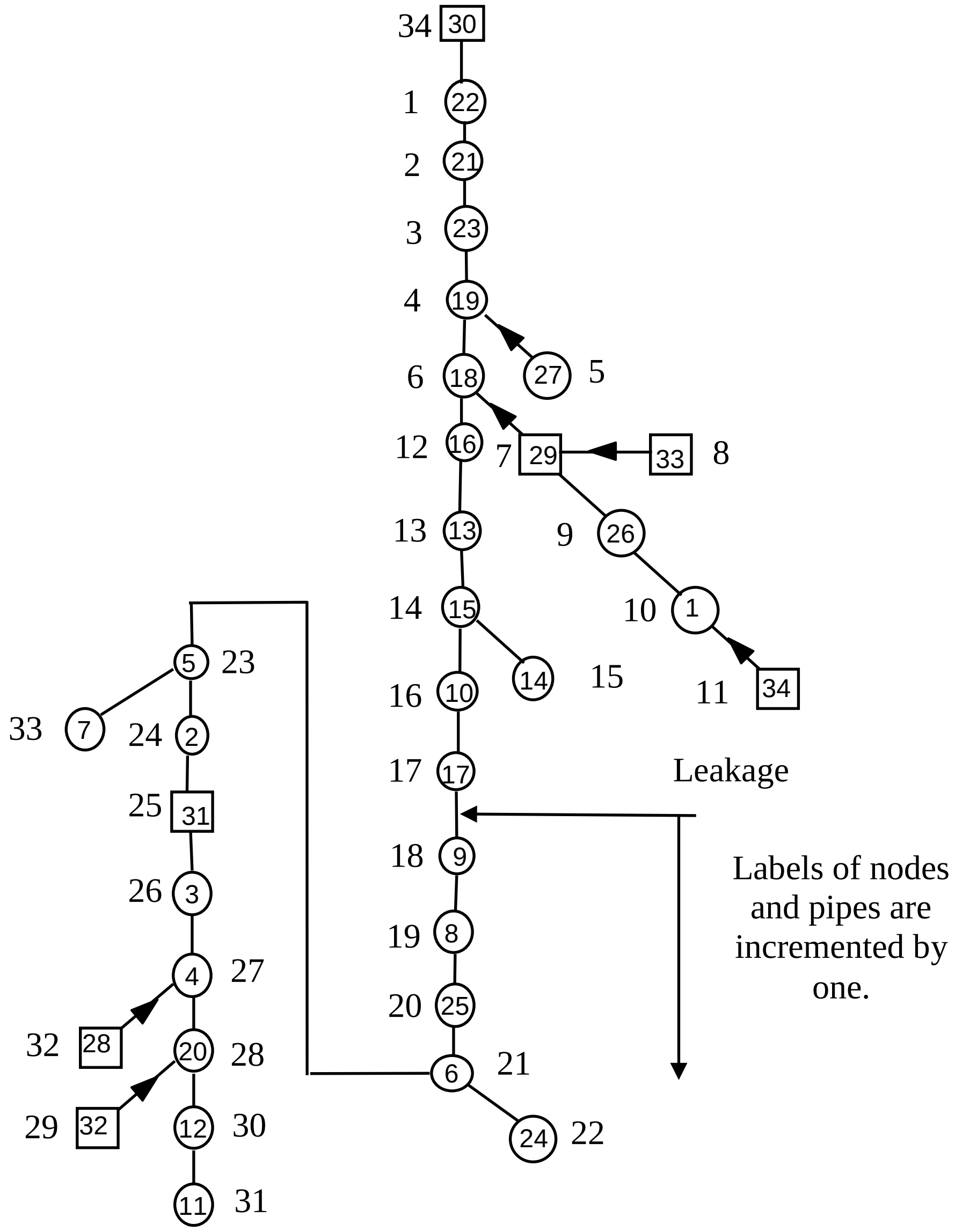

Figure 7-8: Graphical representation of the spanning tree for the 34-node water network.

The solution adopted here is to modify the initial spanning tree built for the normal operating state of the water network so that to account for the additional water consumption that models the leak. Furthermore, a new set of incidence matrixes and initial flows are determined for the simulator algorithm that avoids the time consuming process of rebuilding the spanning tree.

Let us now simulate a leakage in the pipe between nodes 17 and 18. In order to simulate the 35-node water network (i.e. the original 34-node water network plus the leakage modeled as an additional consumption) the incidence matrixes and the initial flows are recalculated.

The labels for the nodes and pipes that are situated in the spanning tree below the leakage location, are incremented by one so that to preserve the upper form of the tree incidence matrix. It should be observed that although the lower part of the network (i.e. nodes 26, 29, 33, and 24) has been included in this process, no leaks will be considered for this part of the network.

The new vector of nodal demands $d^{''}$ comprises the initial nodal demands $d$ plus the leakage that is introduced as a distinct element in the vector of water consumptions.

One column and one row are introduced in the incidence matrixes (loop and topological) so that to take into account the incidence of the two half-pipes resulted from the additional demand. Following this, the new initial pipe flows and the loop and the tree incidence matrixes are obtained through simple matrix operations (Eq.7.1 – Eq.7.3) more efficient to use in terms of computational time than to reconstruct a spanning tree for the 36–node water network.

We can say that the computational time required to rebuild the spanning tree and assign new labels for each of the 9144 labeled patterns of data is roughly 15 minutes. This is unfavorable when compared to less of 40 seconds obtained by using the graph and matrix operations described above. It is worth mentioning also that the computational time obtained during the rebuilding of the spanning tree increases steadily with the size of the network (i.e. the larger is the size of the network in terms of pipe and nodes, more time is required to build the spanning tree and assigned new labels). By contrast, the solution presented here is based on a couple of basic matrix operations that are almost insensitive to the size of the network.

Finally, the whole set of parameters used during the generation of the training set are shown at Table 7-1.

| | |
|---|---|
| Head measurements | 1, 2, 4, 8, 11, 15, 17, 19, 22, 29, 30, 31 |
| Fixed-head inflow measurements | 27, 28, 29, 30, 31, 32, 33, 34 |
| Water consumptions | All nodes |
| Fixed-head measurements | 27, 28, 29, 30, 31, 32, 33, 34 |
| Leak levels | 0.002, 0.005, 0.008, 0.011, 0.014, 0.017, 0.020, 0.023, 0.026, 0.029 [$m^3/s$] |
| Parameters used in confidence limit analysis | |
| Accuracy of head measurements at load nodes | +-0.1[m] |
| Accuracy of inflow measurements | +-1% |
| Variability of consumptions | +-10% |

Table 7-1: Parameters used during generation of the training data set.

### 7.3.3. State estimates and classification system design

In order to design the recognition system based on state estimates the set of 9144 *training patterns* representing 37 categories were used. The training data spanned across 24 hour period of water network operation. The 37 categories stand for normal operating state and leakages in 36 pipes of the upper part of the network shown at Figure 7-3. The indexes $d_h$ of classes (see the description of the pattern recognition system in Chapter 6) were chosen the same as in the original algorithm (Gabrys, 1997): $d_h$ =1 – normal operating state; $d_h$ =2 – leakage in pipe between nodes 3 and 4; $d_h$ =3 – leakage in pipe between nodes 4 and 20, etc.

The training data has to be first scaled in order to be contained in the range (0,1) as required by the pattern recognition system. The range of values for the nodes' head state variables was chosen to be between 2 and 50 [mH2O], and for inflows between –0.2 and 0.2 [$m^3/s$]. There are 6 state variables (heads in fixed-head nodes 27, 28, 32, 33, 34 and inflow at node 32) that do not change during the 24 hour simulation period and since they do not introduce any additional information that could be used to distinguish between patterns from different classes they are excluded from the training set.

Furthermore since we will carry out the analysis only for the upper part of the water network, the fixed-head node 29, nodes 26 and 1, and the inflows at nodes 29, 33, 34 will also be excluded from the training set.

In the original work based on patterns of state estimates calculated with the nodal heads equations (Gabrys & Bargiela, 1999), it has been observed the existence of multiple classes with full membership for a large number of testing patterns (i.e. patterns belonged to classes representing leakages in different pipes). Therefore a two level *recognition system* has been proposed as a means of solving the problem (Figure 7-9) (Gabrys, 1997).

The purpose of the first level of the recognition system was to distinguish between different typical behaviour of the water system (i.e. night load, peak load etc.) while the second level components were responsible for detection of anomalies for some characteristic load patterns. The second level was viewed as “experts”. By doing so, the distinctive variations in the typical network behaviour for different days of the week or seasons of the year, could be accommodated without the need to retrain the existing networks. In exchange, a new expert network was added to the second level and the size of the first level network was increased accordingly. Thus the general fuzzy min-max Neural Network was able to grow so that to meet the demands of the problem.

Moreover, the dimensions if the input patterns processed by neural networks in the first and second level are reduced in comparison to the full pattern. Furthermore, the fact that only one of the “experts” is selected for further processing also means that the other n-1 “experts” are not active. This way another dimensionally reduction is achieved since each of the second level networks covers only small part of the day rather than 24 hour period.

Input to the first level network will consist in our case (the same as in the original system, (Gabrys, 1997)) of inflow to nodes 27 and 28, flow between nodes 29 and 18 and heads at reservoir nodes 30 and 31. Hence the dimension of the input vectors to the second level of neural networks becomes 26.

Six characteristic inflow patterns can be found for six periods during 24 hour water network operation and they are marked by dashed vertical lines at Figure 7-6.

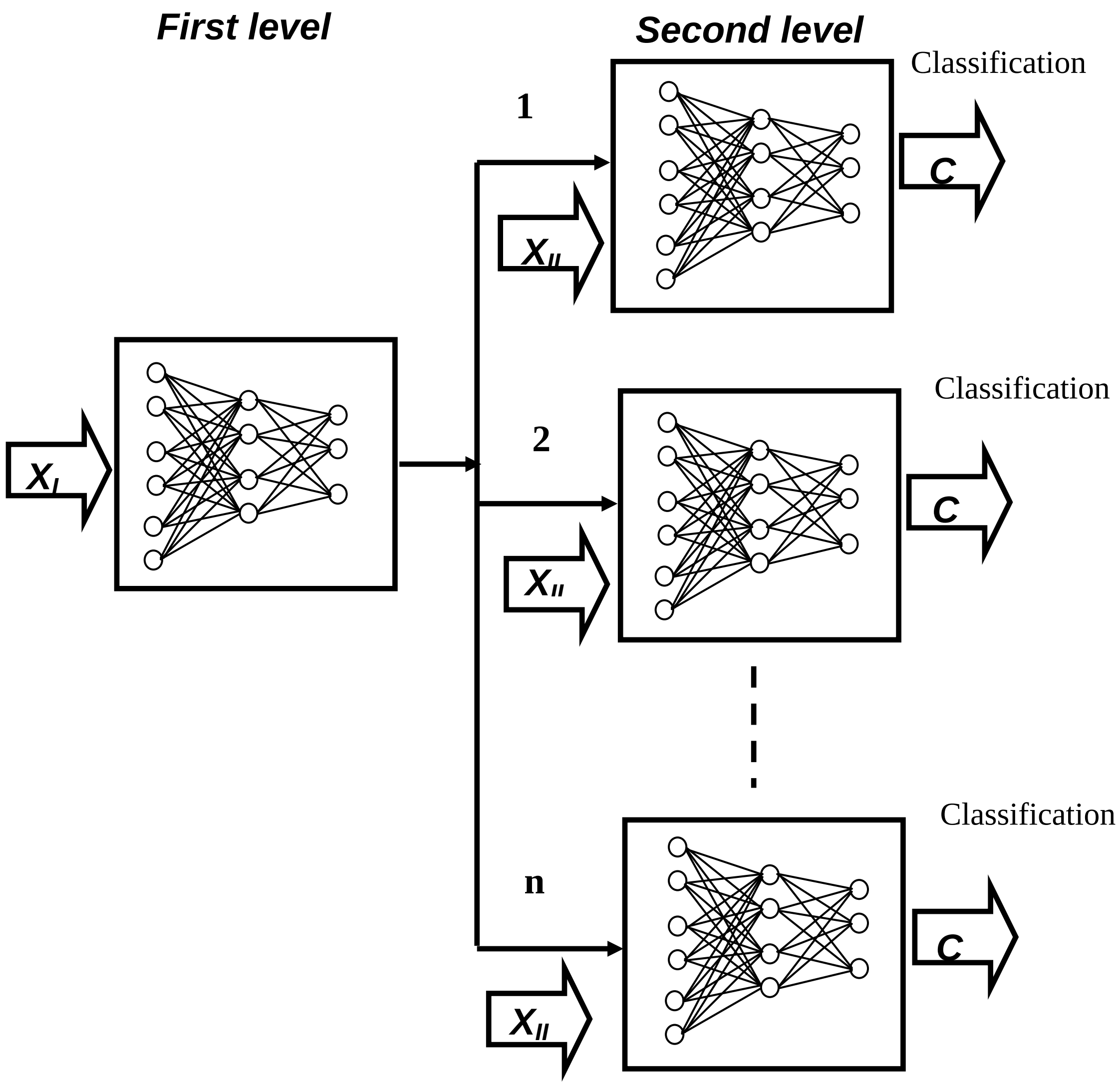

Figure 7-9: Two level recognition system proposed in (Gabrys, 1997). First level consists of one neural network of the type shown at Figure 6-4 and its purpose is to select one of the n second level "experts". Input to the first level NN, $\mathbf{X_I}$, comprises all the variables not affected by occurrence of anomaly. Second level consists of n NNs. They are called "experts" since each of them is trained using only a part of training set and covers a distinctive part of 24 hour operational period. Input to the second level NNs, $\mathbf{X_{II}}$, comprises all the variables sensitive to occurrence of anomaly. The output of the second level NNs is the classification of the water network state.

We are going to use the two level recognition system shown above for 9144 labeled patterns of data obtained with the loop flows state estimator and the confidence limits.

The six distinct periods during 24 hours are 1-5, 6-8, 9-12, 13-17, 18-20 and 21-24 and they are shown at Figure 7-6.

The misclassification rates for the testing set consisting of 9144 examples of loop flows state estimates with and without confidence limits computed for accurate measurements is shown at Table 7-2.

| Training set | Parameter Θ | Misclassification rates | | | |
|---|---|---|---|---|---|
| | | Highest member-ship | Top 2 alterna-tives | Top 3 alterna-tives | Top 5 alternatives |
| Loop flows state estimates computed for accurate measurements without confidence limits | 0.2 | 33.41 | 21.63 | 15.63 | 8.69 |
| | 0.1 | 10.12 | 4.62 | 2.6 | 2.39 |
| Loop flows state estimates computed for accurate measurements including confidence limits | 0.2 | 7.83 | 6.51 | 5.39 | 3.71 |
| | 0.1 | 1.01 | 0.73 | 0.70 | 0.32 |
| | Variable* | 0.002 | 0.001 | 0 | 0 |

Table 7-2: Misclassification rates for a test set consisting of 9144 examples of loop flows state estimates computed for accurate measurements.

* Parameter Θ was determined separately for each dimension of each of the six subsets of the training set and was set to the value of the largest input hyperbox for each of these six subsets.

The first interesting result is the comparison of the performance of the recognition system trained for patterns of loop flows state estimates with the performance of the recognition system trained for patterns of nodal heads state estimates. The *misclassification rates* in our case are slightly higher with 2-4% on average (see the similar table reported in (Garbys, 1997)). This is due to the high sensitivity of the state estimates calculated with the loop flows state estimator to the available pressure and flow measurements. For this case we used a large number of pressure measurements compared to the reduced size of the water network shown at Figure 7-3. Consequently, the loop flows state estimates used for training in Table 7-2 define a space of patterns of data, which are overlapping, making difficult to resolve them in a robust way.

The second observation is regarding the training of the recognition system with patterns of loop flows state estimates and confidence limits. The obtained misclassification rates compares well with what has been reported in the context of the training set consisting of nodal heads state estimates with confidence limits. However, after an examination of the number of hyperboxes obtained during the training process, it has been observed that in order to solve all the *overlapings*, there were necessary a number of hyperboxes equal to the number of patterns of loop flows state estimates and confidence limits.

| Training set | Parameter Θ | Top 5 alternatives | | |
|---|---|---|---|---|
| | | Misclassification rates | Number of hyperboxes | Patterns of data |
| Loop flows state estimates computed for accurate measurements including confidence limits | 0.2 | 3.71 | 6411 | 9144 |
| | 0.1 | 3.2 | 7062 | 9144 |
| | 0.009 | 2.9 | 8597 | 9144 |
| | 0.008 | 0.1 | 8700 | 9144 |
| | 0.005 | 0 | 8777 | 9144 |

Table 7-3: Number of hyperboxes and misclassification rates for different parameters Θ.

Since the attempt to solve all the overlapping for the training set has resulted in an unacceptable number of hyperboxes representing identical classes of operation (i.e. leakages), further efforts to train the recognition system with patterns of loop flows state estimates and confidence limits has been abandoned. Instead, the training has been performed for patterns of variation of load measurements and confidence limits. This has given excellent recognition rates.

### 7.3.4. Classification of the water network state based on patterns of variation of load measurements and confidence limits

Bargiela (1984) in the context of the nodal heads LS state estimator introduced the idea that the topological errors can be thought of as a pair of erroneous load measurements for which the error terms (residuals representing the mass balances at those nodes) are carrying information about a type of topology error.

Since for an accurate model of the network and accurate measurements all the residuals should be zero irrespectively of the operating state (e.g. night load, peak load etc.), this property would make the use of the residuals in bad data analysis very desirable since the presence of any anomalies could be detected by monitoring the deviation of the residuals from the zero reference point.

Gabrys (1997) used the previous ideas and trained the recognition system with patterns of residuals and confidence limits. Unfortunately, the simulations that were carried out showed a very poor recognition rate with a high number of input patterns representing large leakages being misclassified. The poor performance was due to the inability of the training algorithm to resolve overlappings in a robust way. The overlappings were caused by the fact that in the nodal heads LS state estimator, for typical variabilities of consumptions and inaccuracies of meters encountered in water distribution systems, the ratio of noise (quantified as confidence limits) to the useful signal (value of the residual that would result from the occurrence of an anomaly) was very high. For the recognition system it meant a large number of input hyperboxes concentrated around the zero reference point with big overlapping regions.

In the case of the loop flows LS state estimator, it has been shown that for a pressure measurement the mismatch between the actual measurement and the value of the measured quantity as computed by our estimation algorithm tends to be zero (i.e. the residual is zero) (i.e. referred to the Chapter 4 the section with the Hessian matrix).

It means that the variation of the load measurements (nodal consumptions) due to the presence of a pressure measurement will carry out information about the possible existence of topological errors. In terms of the recognition system, this ensures that the hyperboxes representing topological errors of different magnitudes will be moved away from the *zero reference point* representing the normal operating point for the 24 hour operational period. This, in turn, will assure the conditions to deal with the overlappings in a robust manner.

To conclude, the presence of topological error in the vicinity of a pressure or flow measurement would result in the alteration of the nodal demands $\Delta d$ located in the respective region of the water network. By using patterns of variations of nodal demands, we can classify the operational state of the water network. Similarly we dealt with the residuals from the nodal heads LS state estimator.

For an accurate model and accurate measurements in the loop flow state estimator all the variations of nodal demands should be zero irrespectively of the operating state.

The variation of nodal demands with the corresponding confidence limits have been used. The training data have been scaled and mapped onto the [0,1] range.

A single neural network of the type shown at Figure 6-4 has been used for the entire operational period of 24 hours.

The testing showed excellent recognition rates for both patterns of variations of nodal demands as well as patterns of variations of nodal demands with confidence limits. The initial maximum size of hyperbox was set to the value $\Theta = 0.1$. The training was completed after one run through the entire training data of 9144 examples of variations of nodal demands and confidence limits. There were no misclassifications.

In Table 7-4 there are shown the number of hyperboxes created during the training process. Since no information about the level of leakage has been included in the training set, then by increasing the size of the hyperbox we could eventually obtain a single hyperbox that is representing all the levels of leakage from a pipe.

| Training set | Parameter $\Theta$ | Number of hyperboxes | Patterns of data | Missclasification rates |
|---|---|---|---|---|
| Variations of nodal demands computed for accurate measurements without confidence limits | 0.2 | 53 | 9144 | 0 |
| | 0.25 | 47 | 9144 | 0 |
| | 0.5 | 39 | 9144 | 0 |
| Variations of nodal demands computed for accurate measurements including confidence limits | 0.2 | 62 | 9144 | 0 |
| | 0.25 | 52 | 9144 | 0 |
| | 0.5 | 39 | 9144 | 0 |

Table 7-4: The significance of the parameter $\Theta$ on the number of hyperboxes and misclassification rates for a test set consisting of 9144 examples of variations of nodal demands and confidence limits.

To understand the reasons for the excellent recognition rates of the classification system let us show a couple of examples of the behaviour of the variations of nodal demands for different levels of leakages between nodes 3 and 4 (Figure 7-10) and between nodes 15 and 10 (Figure 7-11).

Figure 7-10a presents an example of variation of nodal demand at node 3 found in the course of state estimation carried out for accurate measurements. As one can see, even for a small leakage of 2 [/s] the variation of nodal demand at node 3 is distinctive from the zero reference point ("normal operating point"). At Figure 7-10b we can see

the influence of the random measurement errors on the variation of nodal demand at node 3. The monotonic trend caused by the leakage is not too much distorted by the measurement noise. Finally, Figure 7-10c shows effective ranges within which the variation of nodal demand at node 3 can vary. The same experiment is carried out for 10 levels of leakage between nodes 15 and 10 and the variation of nodal demand at node 15 is shown at Figure 7-11a.

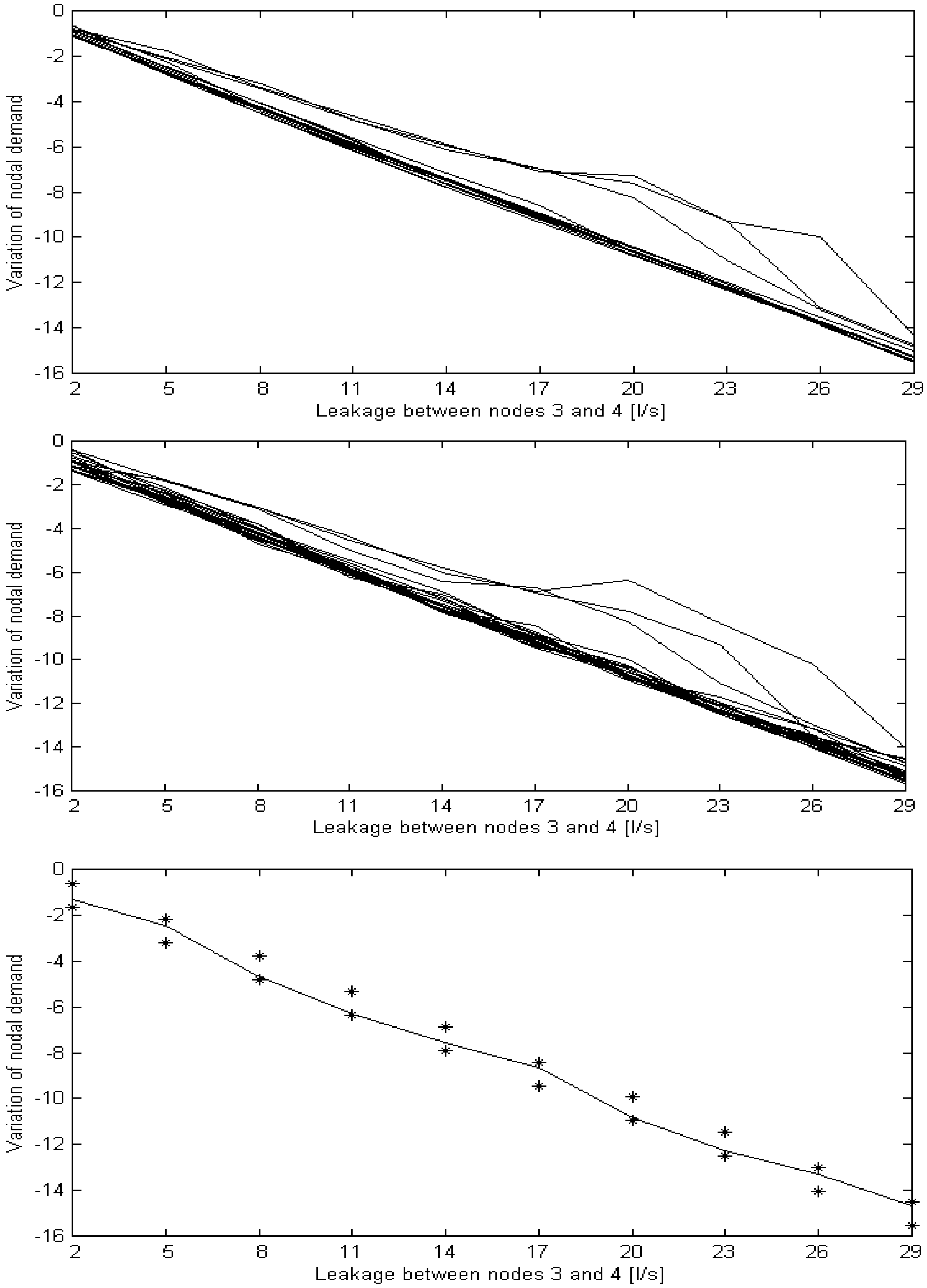

Figure 7-10: Examples of variation of nodal demands for different levels of leakage between nodes 3 and 4; a) variation of nodal demand at node 3 for accurate measurements; b) variation of nodal demand affected by typical measurements inaccuracies; c) examples of tight confidence limits marked with "*" for variation of nodal demand at node 3 represented by solid line.

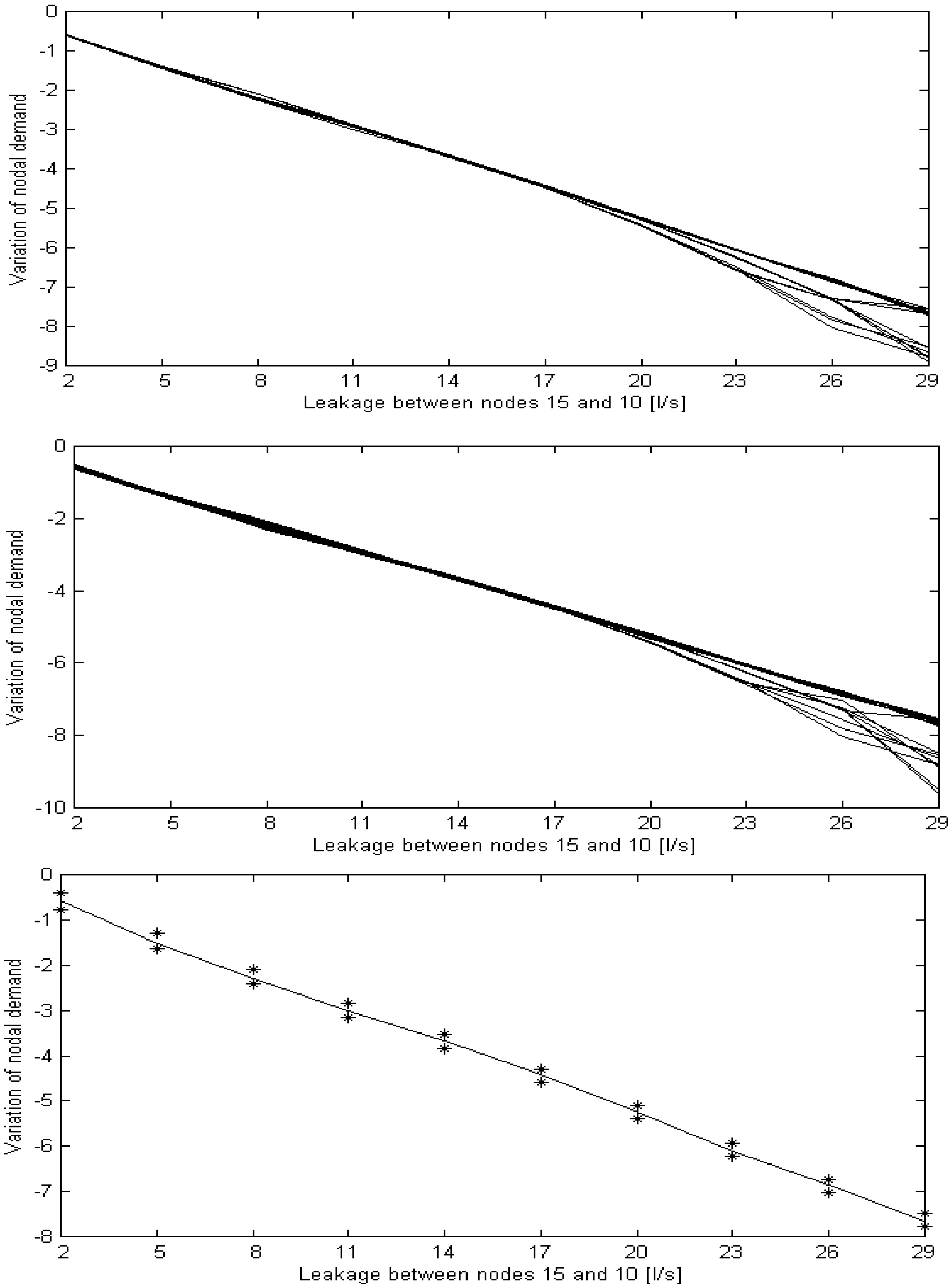

Figure 7-11: Examples of variation of nodal demands for different levels of leakage between nodes 15 and 10; a) variation of nodal demand at node 15 for accurate measurements; b) variation of nodal demand affected by typical measurements inaccuracies; c) examples of tight confidence limits marked with “*” for variation of nodal demand at node 15 represented by solid line.

Two main conclusions can be drawn up from the last experiments:

- The ability of the loop flows state estimator to make full use of the pressure measurements has produced patterns of variation of nodal demands that are distinctive from the zero values corresponding to the normal operating point. Consequently this helped the identification of the topological errors that have been deliberately introduced in the system.
- Tight confidence limits on the variation of nodal demands can be also attributed to the previous bullet point and it further explains the similar recognition rates for the testing set consisting of 9144 examples of variations of nodal demands and confidence limits shown at Table 7-4.

In particular our water network contains a high number of pressure measurements, which makes it easier to spot the leakages or other types of malfunctions (i.e. wrong status of valves). For the purposes of comparison, we have kept the same number of measurements and accuracy of the measurements as the measurement data used in the original study performed in (Gabrys, 1997) in the context of patterns of state estimates obtained with the nodal heads state estimator. In Figure 7-12 it is shown the location of the pressure measurements together with the new labels for nodes, and with dashed lines are shown the chord pipes.

The conclusions from above are reinforced in Figure 7-13 where the confidence limits on the variations of nodal demands representing normal operating states are pictured in form of dashed lines, while examples of the input patterns representing different levels of leakage in the pipe between nodes 27 and 26 (4 and 3 on the old notation), are marked by '*'. Once again, the tight confidence limits for the normal operating state together with the distinctive variation of the nodal demands located in the vicinity of the leakage are the reason for the excellent performance of the detection system based on patterns of variation of load measurements.

It can be observed that for ease in interpretation, for the variation of nodal demands that are located near the leakage, we have used the labels for nodes and pipes from Figure 7-9, which are shown again below. This assures some consistency between the proximity of the nodal demands to the position of the leakage and the information presented in Figure 7-13.

A similar simulation is carried out for three levels of leakage (29 [l/s], 17 [l/s] and 2 [l/s]) in pipe between nodes 14 and 16 (nodes 15 and 10 with the original labels), for which the results are shown at Figure 7-14. It is easy to identify the presence of the leakage by looking to the variation of the nodal demands that are situated outside of the upper and the lower bounds displayed with dashed lines. The examples shown at Figures 7-13 and 7-14 were obtained for the same operational time period, which explains the similarity of the confidence intervals for the normal operating state.

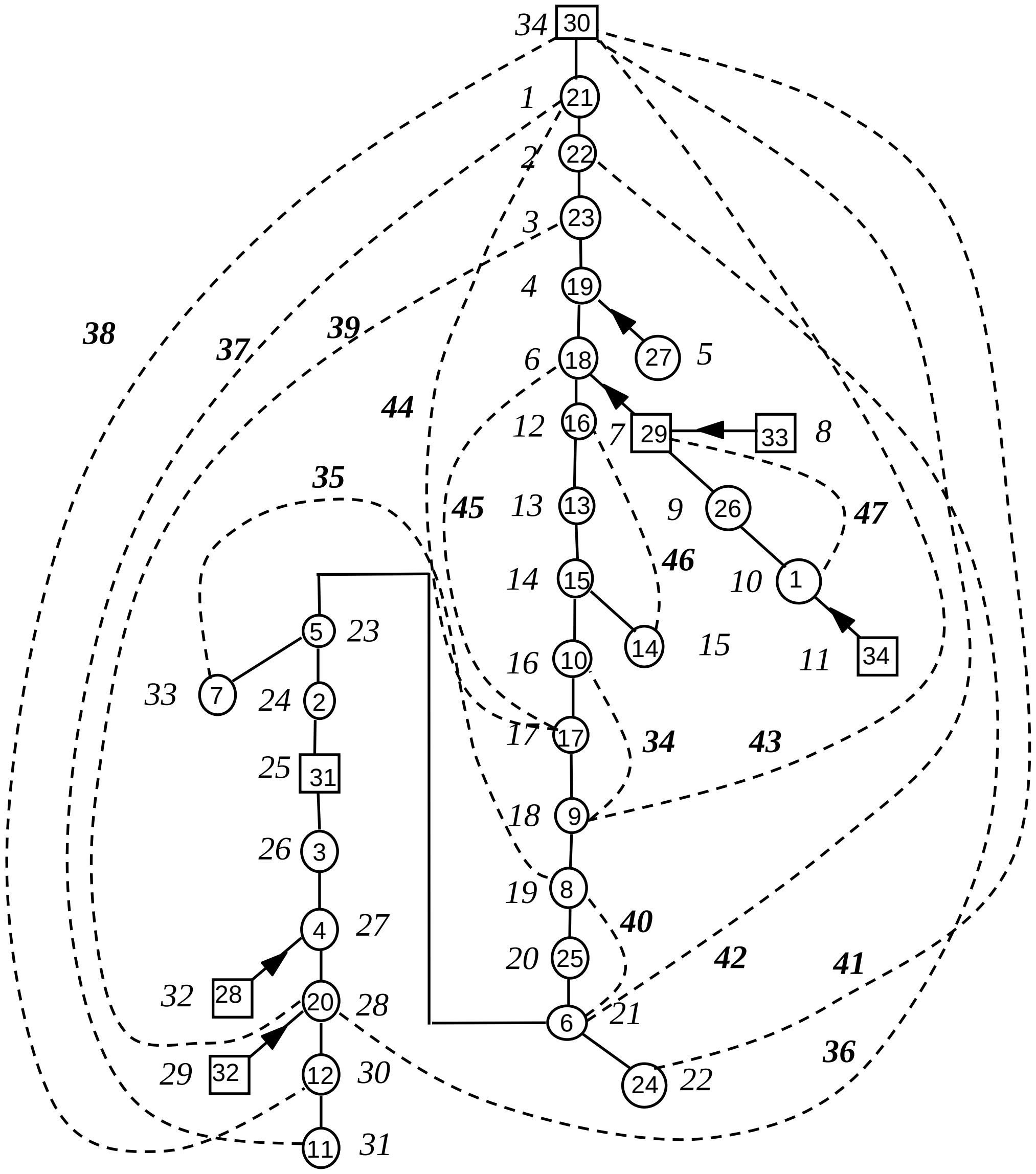

Figure 7-12: New labels for nodes which forms an upper triangular incidence matrix; with dashed lines are shown the co-tree pipes which close the loops.

Since the testing set consisting of patterns of variation of load measurements and confidence limits has shown excellent recognition rates, a similar study carried out on a larger amount of pattern examples would come to confirm again the findings presented in this section.

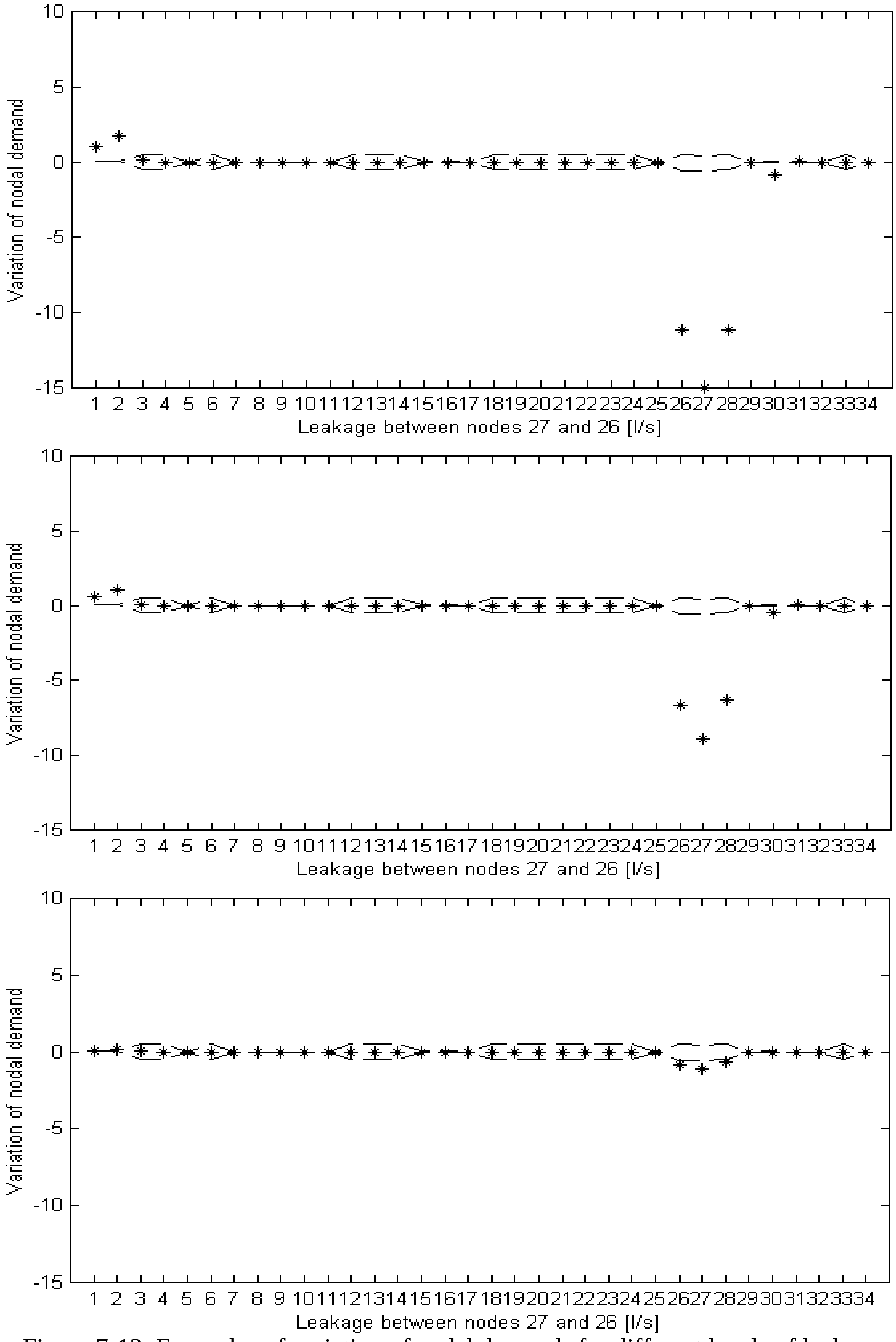

Figure 7-13: Examples of variation of nodal demands for different levels of leakage between nodes 27 and 26 (nodes 3 and 4 if you refer to the original notations); a) leakage of 29 [l/s]; b) leakage of 17 [l/s]; c) leakage of 2 [l/s].

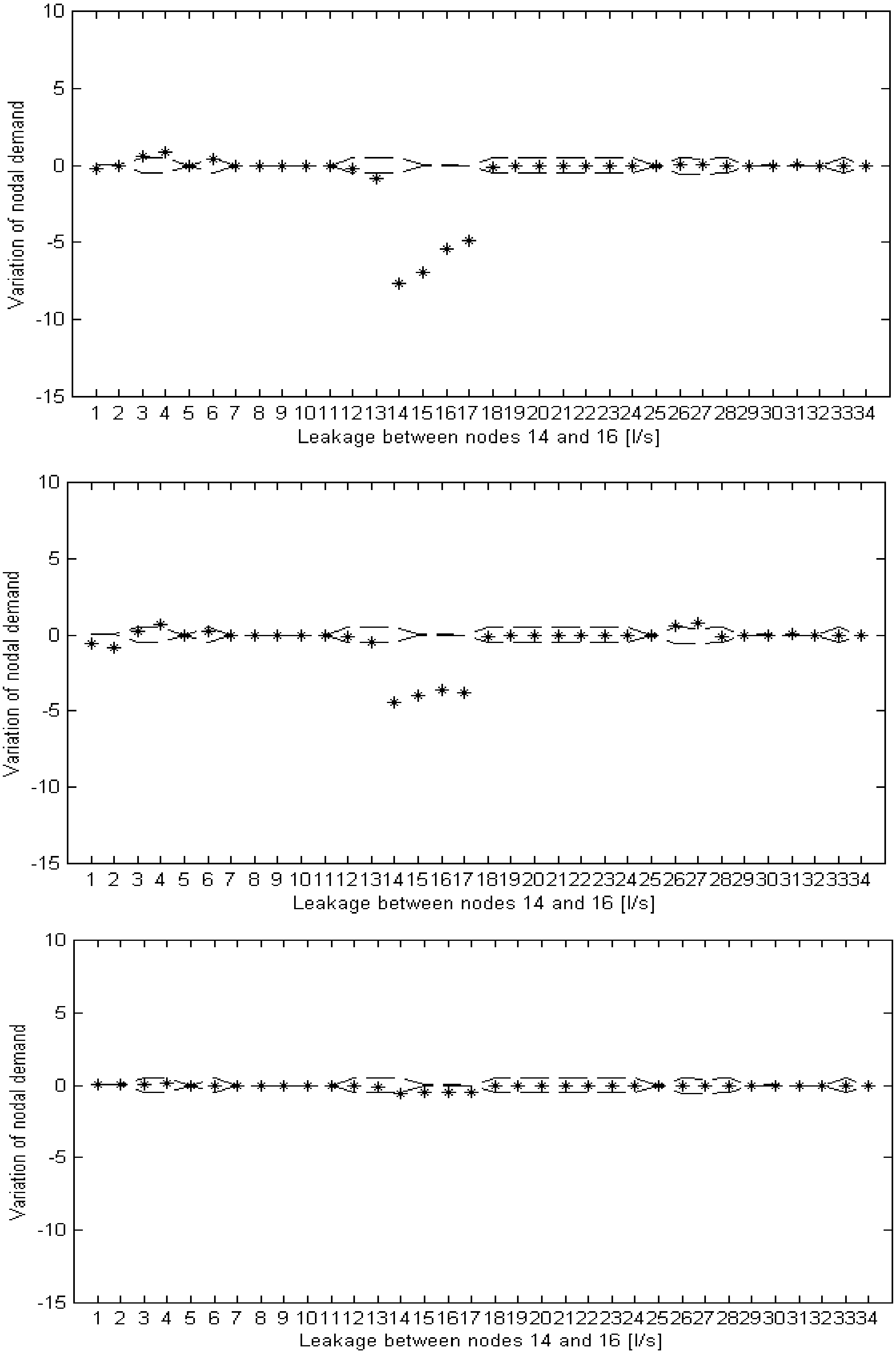

Figure 7-14: Examples of variation of nodal demands for different levels of leakage between nodes 14 and 16 (nodes 15 and 10 if you refer to the original notations); a) leakage of 29 [l/s]; b) leakage of 17 [l/s]; c) leakage of 2 [l/s].

## 7.4. Conclusions

The suitability of interpretation/classification of the water network state by using a fuzzy neural network for pattern recognition has been once again confirmed.

The assumption that fault diagnosis can be based on pattern analysis without a need to employ any heuristic or specialist knowledge has been again highlighted within the context of patterns of ‘loop-equations based’ state estimates and confidence limits and patterns of variation of nodal demands and confidence limits.

It has been shown that the ‘loop-equations based’ state estimates can be successfully used to train the neural recognition system. Slightly higher misclassification rates have been obtained when compared to the training of the recognition system with nodal heads based state estimates. This is because of smaller separation of patterns representing different classes (topological errors, operational time periods) and due to a much higher ratio of the sensitivity of the nodal heads to the existing set of pressure and flow measurements.

The recognition system based on confidence limits for ‘loop-equations based’ state estimates has performed somehow better for the data for which it has been trained for. However, this came at the expense of a high number of hyperboxes necessary to cover the space of input patterns.

On the other hand, the classification of the water network state based on patterns of variation of load measurements and confidence limits has given excellent results.

The neural network used for fault detection in this case has been essentially a simpler version of the recognition system used above. It consisted of a single neural network of the type shown at Figure 6-4, which has been used for the entire 24 hour period of operations of the realistic water distribution network.

Remarkable recognition rates have been obtained for the detection of anomaly based on patterns of variation of the nodal demands and confidence limits. The overlapping regions of different classes could be resolved in a robust way. This was due to the fact that the presence of topological error of different magnitudes have been reflected in the variation of the nodal demands which sprung up the hyperboxes from the zero reference point representing the normal operating point for the entire 24 hour operational period.

While the overlappings could be resolved in a robust manner, it also kept the number of hyperboxes representing different classes to a small number. This is because identical topological errors happening at different operational times have resulted in similar variations of the nodal consumptions, which in turn were represented by the same hyperbox.

The use of the ‘loop-equations based’ state estimates and confidence limits, and the variation of nodal consumption with confidence limits to train the neuro-fuzzy recognition system for detection of faults in a water distribution system has been proved successful.

# Chapter 8.

# Conclusions and further research

## 8.1. Conclusions

The purpose of this book was to investigate the implications of the loop equations formulation of the state estimation procedure for the implementation of decision support systems in the operational control of water networks. The nonlinear models and large scale of the water distribution systems made them both a challenging problem to be tackled and a very good validation example for a prototype decision support system useful in other utility systems.

We divided the book in two distinctive parts. In the first part, we used the loop equations for the implementation of a co-tree flows simulator algorithm and developed a novel loop flows state estimator. A particular emphasis has been placed on the fast calculation of the initial input data (the incidence matrixes and the initial pipe flows), enhancement of the results and good convergence properties for the numerical algorithms.

The second part of the book was concerned with uncertainty based reasoning in modeling and simulation of water networks (confidence limit analysis) as a prelude to an existing module for interpretation and classification of the water system state based on a fuzzy neural network pattern recognition system.

All the developed modules have been integrated into an efficient operational decision system used for fault detection and identification for a realistic 34-node water network.

Short summaries of the problems uncovered and solutions found in the course of investigations as well as the main conclusions of the book are presented below.

### 8.1.1. Loop equations and simulator algorithm

Steady state analysis of flows and pressures in a distribution system has been a major issue for hydraulic engineers involved in the design, management or planning of water distribution system. This interest has led to the development of many methods of analysis (simulator algorithms) using various types of decomposition (i.e. the independent variables used to build the network equations).

In this work the loop corrective flows were used to express the network equations. In particular, a co-tree flows formulation, which is derived from the basic loop corrective flows algorithm, has been developed. The emphasis was put on the fast determination of the input data required by the simulator (the loop and the topological incidence matrixes and the initial flows) as well as the quick calculation of the nodal heads at the end of the simulation. It has been shown that the spanning trees can provide the means for obtaining this information.

Due to the properties highlighted above, the co-tree flows formulation of the simulator algorithm has some advantages over the original formulation based on loop corrective flows. In relation to other simulators (e.g. nodal heads simulator), the co-tree flows simulator is numerically stable and has a superior rate of convergence.

The developed simulator algorithm served as a preamble to a novel state estimation techniques based on loop corrective flows.

### 8.1.2. Loop equations and water network state estimation

Over the last two decades state estimators gradually became the key utility for the implementation of monitoring and control of large scale public systems such as water, gas or electric power distribution systems.

In the context of the state estimation of water distribution systems, a couple of problems have been addressed. One of these problems is the ill-conditioned problem which can appear for example in particular water networks like ones displaying low pipe flows. In those cases the numerical stability of the state estimator may suffer. A solution to this problem is to employ the more stable loop flows state estimation techniques.

Therefore a new formulation of the standard least square (LS) criterion for water networks has been developed in which the loop corrective flows and the variation of nodal demands are the state variables.

In spite of the increased size of the state vector (loop corrective flows plus variation of nodal demands > nodal heads), a satisfactory convergence is obtained through an enhancement in the Jacobian matrix for the loop corrective flows. Hence, the convergence of the new state estimator is comparable with the convergence obtained for the nodal heads variant of the LS state estimator for similar water networks and testing conditions.

However, the novel LS state estimator has exhibited the same lack of numerical stability as the nodal heads LS state estimator when it has been tested on difficult examples (e.g. networks with low pipe flows). We have focused our attention on the tree incidence matrix $T$ which expresses the incidence of nodes and pipe in the spanning tree. In our state estimator this matrix relates the variation of nodal demands to the tree pipe flows. It has been shown that by zeroing some columns of the tree incidence matrix, it is possible to run the state estimation problem for some parts of the water network while for the rest of the network the simulation problem is carried out. By constraining the state estimation procedure to the regions of the water network where the real measurements are located, a limitation of the spreading of the variation of nodal demands has been obtained. Thus, it was possible to avoid the lack of numerical stability characteristic to the nodal heads LS state estimators.

Ultimately some of the intrinsic properties of the novel loop flows state estimator, were revealed. It has been shown that because of the way the network equations are constructed, the introduction of accurate pressure measurements can significantly improve the accuracy of the state estimates. Otherwise the corrupted pressure measurements may affect the final state estimates and eventually mislead the human operator.

The inflows at the fixed head nodes could be also used as measurement data. Alternatively, the head values of the fixed nodes could be used to form the pseudo-loops in which case the inflows are used only to calculate the initial pipe flows. It has been also observed that flow measurements may improve the accuracy of the state estimates but not at the extent of a pressure measurement.

An efficient and effective loop flows LS state estimator has been developed that has been tested successfully on realistic water networks.

### 8.1.3. Loop equations and confidence limit analysis

Bargiela and Hainsworth (1989) were perhaps the first to investigate the precise nature and level of the measurements uncertainty impact on the accuracy to which state estimates can be calculated. It is believed that the safety of the operational control can be enhanced when operators are given not only the information about estimates of the current operating state but also an indication of how reliable these estimates are for a given set of measurements at a particular operating state. In confidence limit analysis this information is provided in the form of upper and lower bounds for each state estimate variables.

Bargiela and Hainsworth (1989) found that in the context of the nodal heads LS state estimator, the Jacobian matrix can be used as a sensitivity matrix between the measurement vector and changes in the state vector (nodal heads). This formed the basis of a sensitivity matrix approach to confidence limit analysis. It has been also observed that the method produces results that compare well with the Monte Carlo results while the computation time is much shorter. The Monte Carlo method is the best known and mathematically the most reliable method of quantifying the state uncertainty which is based on repeated simulation for a large number of parameters with random variations.

In this work new confidence limit analysis algorithms based on the novel loop flows state estimator were developed.

First, it is shown that the Jacobian matrix from the co-tree flows simulator algorithm can not act as a sensitivity matrix between the loop corrective flows and the pseudo-measurements because of the non-realistic way the initial loop head losses are calculated. This has a negative impact on the calculation of the confidence limits that are much bigger than expected it.

Instead, a sensible number of simulations have been used, one for each measurement modified with its defined maximum variability, in order to determine an experimental sensitivity matrix. The experimental sensitivity matrix has been used to determine the nodal heads and the inflows for a random error in the measurement data. The confidence limits obtained with the experimental sensitivity matrix are comparable with the ones produced with the pseudo-inverse of the Jacobian matrix from the nodal heads state estimator when no real measurements are introduced. The Experimental Sensitivity Matrix (ESM) method provides a trusted reference point against which other 'loop-equations based' confidence algorithms can be tested. However, the ESM method required a large number of simulations equal to the number of real measurements and pseudo-measurements, which can represent an unrealistic proposition for the real-time applications especially for large water networks. An alternative method has been developed, the Error Maximization (EM) method, which requires only an extra simulation in order to derive the confidence limits. An additional simulation is carried out for the estimated measurement vector instead of the observed measurement vector, which is modified with the highest level of inaccuracies. Following this the confidence limits are calculated by subtracting the resulted state vector from the optimal state vector obtained for the observed measurement data. Finally the set of confidence limits are compared with the ESM method and it shows a very good similarity. The computational efficiency of the EM method renders it suitable for online decision support applications.

### 8.1.4. Classification of the water network state based on patterns of 'loop-equations based' state estimates and confidence limits

The appropriateness of interpretation/classification of the water network state by using a fuzzy neural network for pattern recognition has been once again confirmed.
The assumption that fault diagnosis can be based on pattern analysis without a need to employ any heuristic or specialist knowledge has been again highlighted within the context of patterns of 'loop-equations based' state estimates and confidence limits.

It has been shown that the 'loop-equations based' state estimates can be successfully used to train the neural recognition system. The emphasis has been put on the task of detection and accurate location of topological errors. Thus the performance of the recognition systems has been tested for a large number of topological errors (i.e. 10 different levels of leakages have been simulated for all pipes) over the 24 hour period of operations of a realistic water distribution network.

Slightly higher misclassification rates have been obtained when compared to the training of the recognition system with nodal heads based state estimates. This is because of smaller separation of patterns representing different classes (topological errors, operational time periods) and due to a much higher ratio of the sensitivity of the nodal heads to the existing set of pressure and flow measurements.

On the other hand, the recognition system based on confidence limits for 'loop-equations based' state estimates has performed somehow better for the data for which it has been trained. However this came at the expense of an increased number of hyperboxes necessary to cover the space of input patterns. Eventually it has been observed that for a given operational period of time in order to solve all the misclassifications it might be necessary to have a number of hyperboxes equal to the number of input patterns of data (in this case a hyperbox is equal to a fuzzy input pattern described by the lower and upper limits on each input vector variable). This is due to the high non-linearity of the input space, which consists of regions of different classes that are overlapping, making difficult to separate them without increasing the number of hyperboxes.

The non-linearity of the input space has been attributed to the properties of the loop flows state estimator of making full use of the existing pressure and flow measurements. This resulted in patterns of state estimates with confidence limits, corresponding to different operational time periods and topological errors, which were overlapping, and so it has made it difficult to identify and classify them correctly except by increasing the number of hyperboxes (obtained by decreasing the parameter $\Theta$ that controls the maximum size of the hyperbox).

In conclusion, the training of the fuzzy neural recognition with patterns of 'loop-equations based' state estimates and confidence limits although successful has revealed the following fact: the high sensitivity of the nodal heads and inflows to the available set of real measurements in the loop flows state estimator results in patterns of state estimates and confidence limits that are overlapping which makes it difficult to identify and classify the data patterns without decreasing the size of the hyperbox which further results in increasing the number of hyperboxes that are covering the input pattern space.

However, the successfully application of the fuzzy neural network for pattern recognition to the water system identification task based on patterns of ‘loop-equations based’ state estimates and confidence limits, has confirmed the previous findings that the high level of information processing by human operators can be mimicked, to a large extent, by a suitable neural based recognition system.

### 8.1.5. Classification of the water network state based on patterns of variation of load measurements and confidence limits

By far the classification of the water network state based on patterns of variation of load measurements and confidence limits performed the best in our study.

The great advantage of processing the variation of load measurements (nodal consumptions) is the fact that irrespectively of the operating state, when there is no anomaly present in the system, the variation of the nodal consumptions is zero, or in practical terms, it is contained within the confidence limits. This represents a universal reference point (hyperbox) labeled as normal operating state. All the other patterns representing leakages and other possible malfunctions can be mapped into the space around this zero reference point.

The neural network used for fault detection in this case has been essentially a simpler version of the recognition system used in the previous paragraph. It consisted of a single neural network of the type shown at Figure 6-4, which has been used for the entire 24 hour period of operations of the realistic water distribution network.

Excellent recognition rates have been obtained for the detection of anomaly based on patterns of variation of the nodal demands and confidence limits. The overlapping regions of different classes could be resolved in a robust way. This was due to the fact that the presence of topological error of different magnitudes have been reflected in the variation of the nodal demands which sprung up the hyperboxes from the zero reference point representing the normal operating point for the entire 24 hour operational period.

While the overlappings could be resolved in a robust manner, it also kept the number of hyperboxes representing different classes to a low limit. This is because identical topological errors happening at different operational times have resulted in similar variations of the nodal consumptions, which in turn were represented by the same hyperbox.

Finally, the use of the 'loop-equations based' state estimates and confidence limits, and the variation of nodal consumption with confidence limits to train the neuro-fuzzy recognition system for detection of faults in a water distribution system has been proved successful.

It is postulated the combination of the training of the recognition algorithm with patterns of nodal heads state estimates and confidence limits, and patterns of variation of nodal demands and confidence limits obtained with the loop flows state estimator would result in excellent recognition rates.

## 8.2. Further research

As the simulation studies presented in this book have shown it is possible to use the loop equations in developing robust state estimation techniques for solving water networks. However, the full potential of these numerical algorithms can only be realized when implemented on a real-time monitoring and control operational decision system for water distribution networks.

The GFMM classification and clustering neural network has performed well for different training and testing patterns of data consisting of 'loop-equations based' state estimates and confidence limits as well as the variation of nodal demands and confidence limits, respectively. However, similar to the training of the recognition algorithm with nodal heads state estimates, the hyperboxes created in the training stage with patterns of 'loop-equations based' state estimates depend on the order of presentation of the training patterns. This reinforces the previous observation referred to the initial fuzzy neural recognition system, with regard to the optimization of the GFMM algorithm so that the final partitioning of the pattern space is independent of the order of presentation of the training patterns.

Since application of the loop equations to the operational control of water systems has shown to be successful the following topics deserve further research effort:

- Including of various non-linear hydraulics elements (pressure reducing valves, non-return valves, pressure sustaining vales) into the numerical algorithms developed.
- Developing an operational decision system for real-time control and monitoring of water system by integrating seamlessly the modules developed in this book with other information systems such as Geographical Information Systems (GIS), Supervisory Control, Automation and Data Acquisition Systems (SCADA), databases and customer billing systems.

Since the study of the recognition system performance in association with different state estimation procedures has been successfully, the following bullet points may worth further investigation:

- Modification of the neural pattern recognition system to the detection of multiple malfunctions.
- Comparison of the performances of the fuzzy neural recognition system to other architectures of neural networks (e.g. radial basis neural networks, self-organizing map).

# Index